\documentclass[traditabstract]{aa}
\usepackage{graphicx}
\usepackage{txfonts}
\usepackage{hyperref}
\usepackage{multirow}
\usepackage{xcolor}
\usepackage{float}
\usepackage{soul} % to remover afterwards
\hypersetup{
     colorlinks   = true,
     citecolor    = blue
}

\begin{document}
\title{High resolution spectral imaging of CO(7-6), [CI](2-1) and continuum of three high-z lensed dusty star-forming galaxies using ALMA}

   %\subtitle{I. Overviewing the $\kappa$-mechanism}
	\authorrunning{Gururajan et al.}
	\titlerunning{High-resolution [CI](2-1), CO(7-6) and dust continuum of 3 lensed DSFGs with ALMA.}
	
   \author{G. Gururajan
          \inst{1},
          M. Béthermin
          \inst{1},
          P. Theulé 
          \inst{1},
          J. S. Spilker
          \inst{2}
          \and
          M. Aravena
          \inst{3}
          \and
          M. A. Archipley
          \inst{4,5}
          \and
          S. C. Chapman
          \inst{6,7,8}
          \and
          C. De Breuck
          \inst{9}
          \and
          A. Gonzalez
          \inst{10}
          \and
          C. C. Hayward
          \inst{11}
          \and
          Y. Hezaveh
          \inst{12,11}
          \and
          R. Hill
          \inst{6}
          \and
          S. Jarugula
          \inst{4}
          \and
          K. C. Litke
          \inst{13}
          \and
          M. Malkan
          \inst{14}
          \and
          D. P. Marrone
          \inst{13}
          \and
          D. Narayanan
          \inst{10,15,16}
          \and
          K. A. Phadke
          \inst{4}
          \and
          C. Reuter
          \inst{4}
          \and
          J. D. Vieira
          \inst{4,5,17}
          \and
          D. Vizgan
          \inst{4,16}
          \and
          A. Wei\ss
          \inst{18}
          }

   \institute{Aix Marseille Univ, CNRS, CNES, LAM, Marseille, France,
              \email{gayathri.gururajan@lam.fr}
         \and
             NHFP Hubble Fellow, University of Texas at Austin, 2515 Speedway Stop C1400, Austin, TX 78712, USA
        \and
            Núcleo de Astronomía, Facultad de Ingeniería y Ciencias, Universidad Diego Portales, Av. Ejército 441, Santiago, Chile
        \and
            Department of Astronomy, University of Illinois at Urbana-Champaign, 1002 West Green St., Urbana, IL 61801, USA
        \and
            Center for AstroPhysical Surveys, National Center for Supercomputing Applications, Urbana, IL, 61801, USA
        \and
            Department of Physics and Astronomy, University of British Columbia, 6225 Agricultural Road, Vancouver, V6T 1Z1, Canada
        \and
            National Research Council, Herzberg Astronomy and Astrophysics, 5071 West Saanich Road, Victoria, V9E 2E7, Canada
        \and
            Department of Physics and Atmospheric Science, Dalhousie University, 6310 Coburg Road, B3H 4R2, Halifax, Canada
        \and
            European Southern Observatory, Karl Schwarzschild Stra\ss e 2, 85748 Garching, Germany
        \and
            Department of Astronomy, University of Florida, 211 Bryant Space Sciences Center, Gainesville, FL 32611 USA
        \and
            Center for Computational Astrophysics, Flatiron Institute, 162 Fifth Avenue, New York, NY, 10010, USA
        \and
            Département de Physique, Université de Montréal, Montreal, Quebec, H3T 1J4, Canada
        \and
            Steward Observatory, University of Arizona, 933 North Cherry Avenue, Tucson, AZ 85721, USA
        \and
            Department of Physics and Astronomy, University of California, Los Angeles, CA 90095-1547, USA
        \and
            University of Florida Informatics Institute, 432 Newell Drive, CISE Bldg E251, Gainesville, FL 32611, USA
        \and
            Cosmic Dawn Center (DAWN), DTU-Space, Technical University of Denmark, Elektrovej 327, DK-2800 Kgs. Lyngby, Denmark
        \and
            Department of Physics, University of Illinois Urbana-Champaign, 1110 West Green Street, Urbana, IL, 61801, USA
        \and
            Max-Planck-Institut f¨ur Radioastronomie, Auf dem H¨ugel 69 D-53121 Bonn, Germany}

  % \date{}

\abstract {High-redshift dusty star-forming galaxies with very high star formation rates (500 -- 3000 M$_{\odot}$\,yr$^{-1}$) are key to understanding the formation of the most extreme galaxies in the early Universe. Characterising the gas reservoir of these systems can reveal the driving factor behind the high star formation. Using molecular gas tracers like high-J CO lines, neutral carbon lines and the dust continuum, we can estimate the gas density and radiation field intensity in their interstellar media. In this paper, we present high resolution ($\sim$0.4$^{\prime\prime}$) observations of CO(7-6), [CI](2-1) and dust continuum of 3 lensed galaxies from the SPT-SMG sample at $z\sim$ 3 with the Atacama Large Millimeter/submillimeter Array. Our sources have high intrinsic star-formation rates ($>$850\,M$_{\odot}$\,yr$^{-1}$) and rather short depletion timescales ($<$100\,Myr). Based on the L$_{[\rm CI](2-1)}$/L$_{\rm CO(7-6)}$ and L$_{[\rm CI](2-1)}$/L$_{\rm IR}$ ratios, our galaxy sample has similar radiation field intensities and gas densities compared to other submillimetre galaxies. We perform visibility-based lens modelling on these objects to reconstruct the kinematics in the source plane. We find that the cold gas masses of the sources are compatible with simple dynamical mass estimates using ULIRG-like values of the CO-H$_2$ conversion factor $\alpha_{\rm CO}$ but not Milky Way-like values. We find diverse source kinematics in our sample: SPT0103-45 and SPT2147-50 are likely rotating disks while SPT2357-51 is possibly a major merger. The analysis presented in the paper could be extended to a larger sample to determine better statistics of morphologies and interstellar medium properties of high-$z$ dusty star-forming galaxies.}

%  % context heading (optional)
%   % {} leave it empty if necessary  
%   {}
%   % aims heading (mandatory)
%   {}
%   % methods heading (mandatory)
%   {}
%   % results heading (mandatory)
%   {}
%   % conclusions heading (optional), leave it empty if necessary 
%   {}


\maketitle
%--------------------------------------------------------------------
\section{Introduction}

% [CI]($^3P_2 - ^3P_1$) (henceforth : [CI](2-1))

% GIVE YOUR COSMO AND IMF (CHABRIER)!!!!

% . 

% OUTLINE : 

% Main aim is to clearly explain why we did the following analysis and what are its implications. Try to make it clear as to why this is interesting and justify the methodology used. 

% parah 1 
% Massive DSFGs, what are they, why are they interesting? 

%parah 2 
% Gas reservoirs of massive DSFGs. Composition, size, depletion time? ISM 

% parah 3 
% Tracers: cant use CO 1-0, alternative - CI, radiation field and density tracer, ratio analysis

%parah 4 
% Dynamics and morphologies : mergers, disks

% parah 5 
% unlensed sources, disadvantages, lensed sample, examples, SPT-SMG sample

% parah 6
% outline of the paper and the IMF and cosmology used. 

\begin{table*}
\centering
\caption{\label{tab:obs_source}Our sample is introduced below. The redshifts were calculated from the observed CO(7-6) frequencies and are in agreement with the spectroscopic redshift measurements in \citet{Reuter20}. The CO(7-6) and [CI](2-1) frequencies given are the observed frequencies.}
\begin{tabular}{ccccccc}
\hline
\hline
&&&&&&\\
Source&Project &R.A.  &Dec. &Redshift&CO(7-6) & [CI](2-1) \\
&Number&(J2000)&(J2000)&&(GHz)&(GHz)\\
&&&&&&\\
\hline
&&&&&&\\
SPT0103-45&2017.1.01018.S&01:03:11.50&-45:38:53.90&3.089&
197.27&197.93\\
SPT2147-50&2018.1.01060.S&21:47:19.05&-50:35:54.00&3.760&169.46&170.03\\
SPT2357-51&2017.1.01018.S&23:57:16.84&-51:53:52.90&3.070&
198.15&198.81\\
\hline
\end{tabular}
\end{table*}

\begin{table*}
\centering
\caption{\label{tab:obs_details} Observational details of our sample. PWV is the precipitable water vapour of the atmosphere measured during the observations.}
\begin{tabular}{ccccccc}
\hline
\hline
&&&&&\\
Source&Observation&Date&Starting&Duration of &Number of&PWV\\
&number&&time&observation (min)&antennas&(mm)\\
&&&&&&\\
\hline
&&&&&&\\
SPT0103-45&1&18/09/2018&05:04:14&75&42&0.6\\
&2&18/09/2018&07:55:43&75&43&0.5\\
\hline
&&&&&&\\
&1&19/11/2018&23:52:38&73&45&1.3\\
SPT2147-50&2&25/11/2018&02:12:40&74&46&0.7\\
&3&27/11/2018&23:51:21&73&45&2.0\\
\hline
&&&&&&\\
SPT2357-51&1&16/09/2018&05:07:09&71&45&1.2\\
&2&16/09/2018&06:18:07&71&45&1.2\\
\hline
\end{tabular}
\end{table*}
%--------------------------------------
\begin{figure}
\centering

\includegraphics[width=9cm]{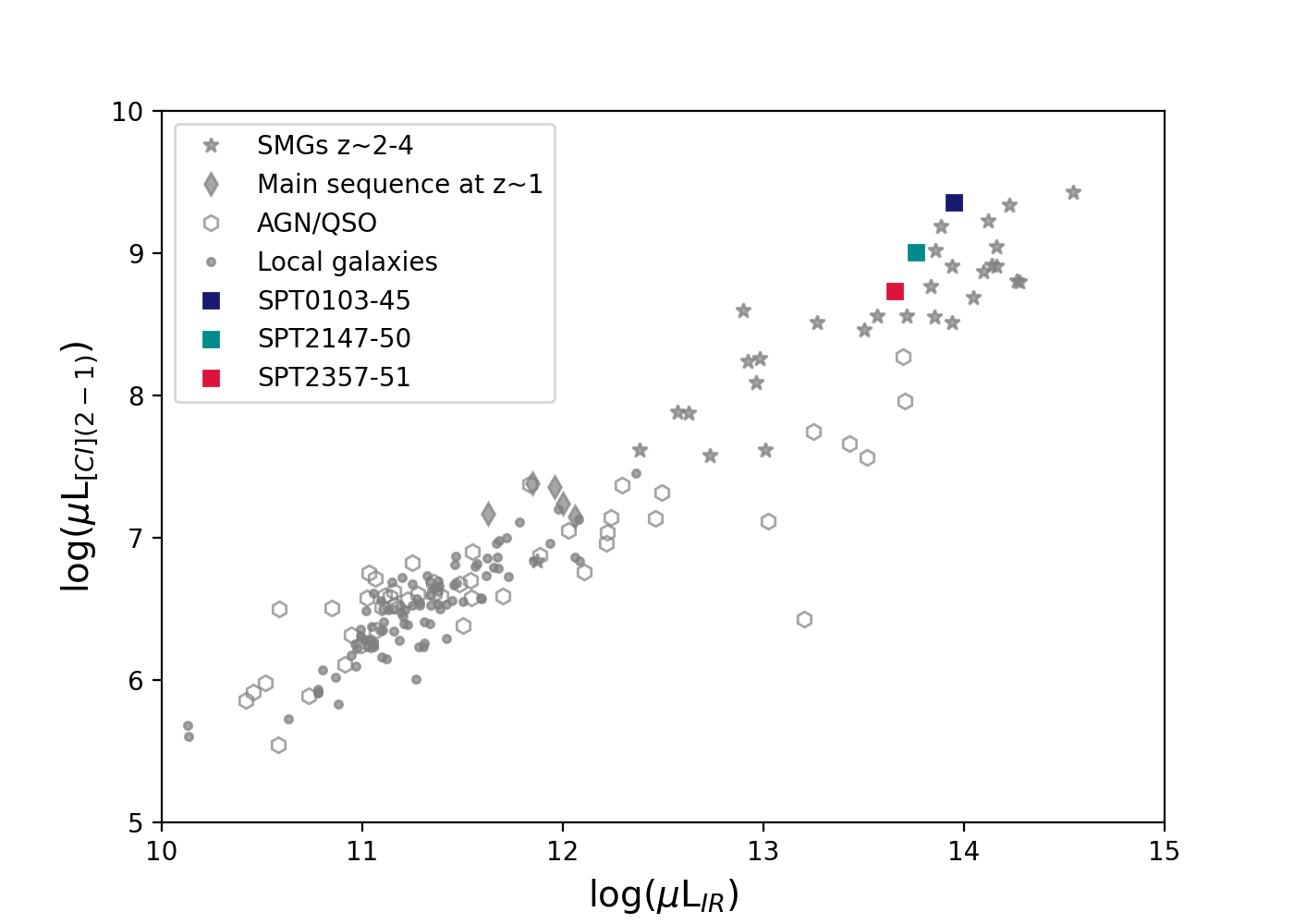} 
\includegraphics[width=9cm]{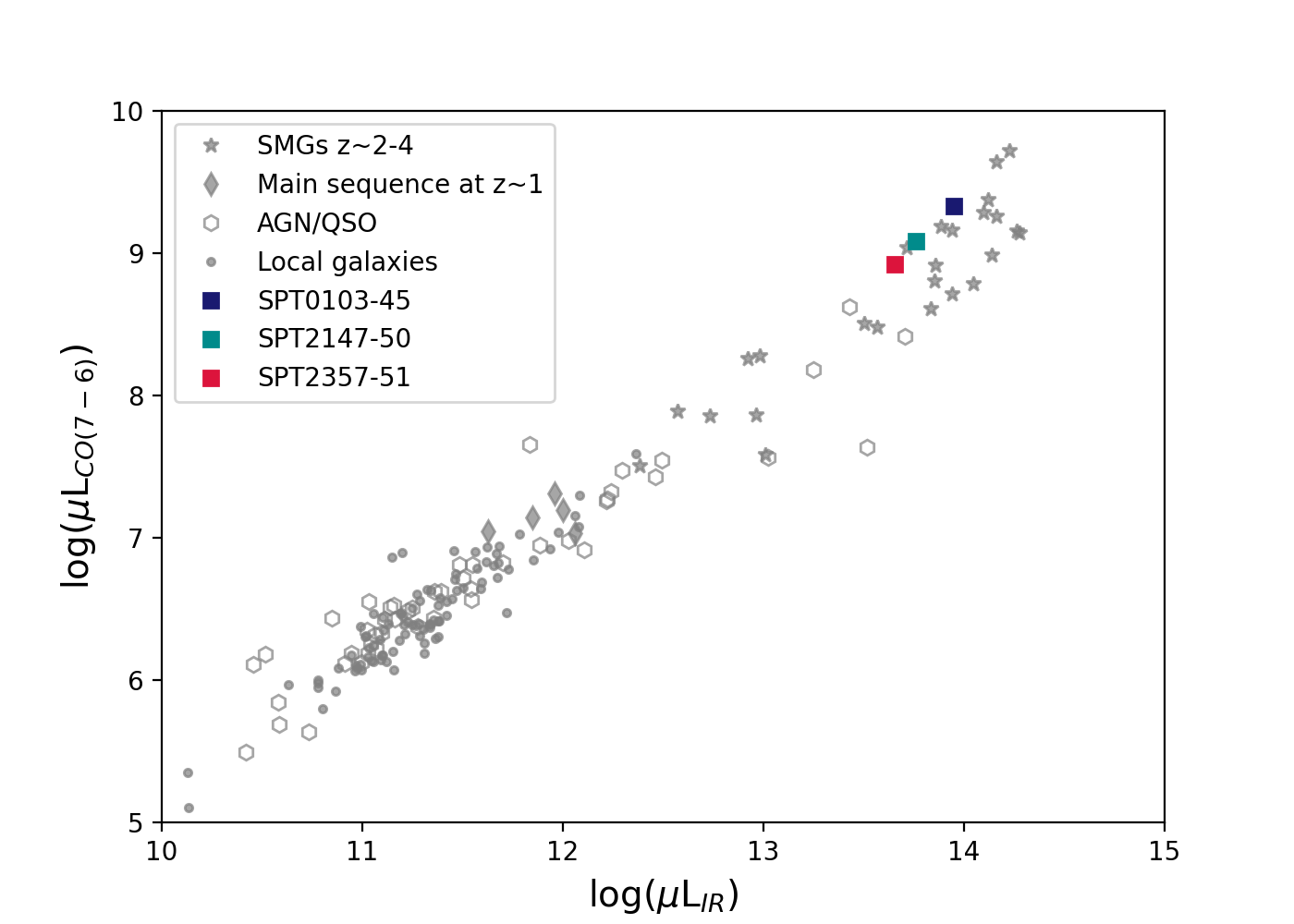} 

\caption{\label{fig:sample_selection} Line luminosity of [CI](2-1) (top-panel) and CO(7-6) (bottom-panel) versus IR luminosity. All the luminosities in this plot are not corrected for magnification. Our sources (filled squares) are compared with the SMGs, (stars) and main-sequence at z$\sim$1 (diamonds), AGN/QSO (hexagons), local galaxies (dots) compilation of \citet{Valentino20}.}
\end{figure}
%--------------------------------------
Large populations of dusty star-forming galaxies (DSFGs) have been observed in the high-redshift Universe \citep[e.g.,][]{ Greve12,Casey14,daCunha15}. They can host star-formation rates (SFR) as high as $\sim$1000 M$_{\odot}$\,yr$^{-1}$. DSFGs account for roughly $\sim$20$\%$ of the star formation density and $\sim$10$\%$ of the molecular gas content at  $z\sim$ 1-5 \citep{Swinbank14}. Typically selected at infrared and millimetre wavelengths, DSFGs are bright in submillimetre wavelengths as their ultraviolet and optical emission is mainly absorbed by the dust and re-emitted in rest-frame far-infrared (redshifted to the submillimetre). The rapid star formation in these systems seem to be driven by not only massive gas reservoirs, but also an increase in the star formation efficiency, SFE\,$\equiv$\,SFR/$\rm M_{\rm H_2}$ or equivalently, short gas depletion timescales $\rm t_{\rm dep} \equiv \rm M_{\rm H_2}/$\,SFR \citep[e.g.,][]{Tacconi13,Dessauges-Zavadsky15,Bethermin15,Aravena16}. Understanding the formation of such systems could provide key insights into the evolution of the most extreme galaxies at early times.

% Probing into the gas content of such systems could tell us whether they have a large gas reservoir driving the SFR (CITE!!!!) or they have a greater star formation efficiency (SFE, defined here as the ratio between SFR and gas mass) induced by mergers \citep[][]{Genzel10,Daddi10a}. 

At $1<z<4$, most massive galaxies have SFEs similar to local spirals, but with larger gas reservoirs \citep[e.g.,][]{Tacconi10,Daddi10a,Saintonge13,Dessauges-Zavadsky15,Bethermin15} possibly driven by rapid accretion of cold gas from the cosmic web \citep[e.g.,][]{ Dekel09,Kleiner17,Kretschmer20,Chun20}. However, the most rapidly star-forming systems (SFR$>$500\,M$_\odot$\,yr$^{-1}$) usually have both a large gas reservoir \textit{and} a high SFE \citep{Harris12, Ivison12,Tan14,Aravena16,Ciesla20,Jarugula21}. These high SFEs also seem to be associated with high dense gas fractions \citep{Oteo17,Bethermin18}. Measuring the composition and physical properties of the gas reservoir is thus key to understanding the physical mechanisms leading to the extreme star formation observed in DSFGs.

The observed dust continuum of DSFGs can be used to derive the total infrared (IR) luminosity (L$_{\rm IR}$), which can then be used to estimate the obscured SFR \citep{Kennicutt98,Kennicutt12}. The contents of the Interstellar medium (ISM) gas reservoirs can also be observed and quantified using various tracers. The carbon monoxide (CO) emission lines and particularly the low-J transitions are common tracers of the molecular H$_2$ gas  \citep[e.g.,][]{Greve05, Solomon05, Omont07,CarilliWalter13,Bothwell13, Bolatto13}. Measuring the low-J CO emission thereby provides a good estimate of the total molecular gas mass of the system \citep[e.g.,][]{Daddi10a, Genzel10}. Together, the low-J CO lines and the IR luminosity can then be used to estimate the SFE. High-J (J$\ge$7) CO lines trace warm and dense molecular gas in active star-forming regions \citep[e.g.,][]{Weiss05,Weiss07,CarilliWalter13,Yang17}. 

Measurements of the molecular gas mass through low-J CO lines are limited by the need to assume $\alpha_{\rm CO}$, the molecular gas mass-to-CO luminosity conversion factor, which is known to vary with density and metallicity \citep{Bolatto13}, but it can also be affected by mergers \citep[e.g.,][]{Daddi10b,Genzel10}. At $z>3$, the CO(1-0) line is difficult to detect, so most studies focus on the detection of J$\ge$3 transitions. Because $\alpha_{\rm CO}$ is based on the CO(1-0) transition, these studies must then assume or estimate the CO line excitation in order to convert the observed line luminosity to the luminosity of the ground state transition \citep{Narayanan14,Tunnard16}. The CO line excitation depends on the gas temperature and density, which in turn also influence many other aspects of galaxies such as the instantaneous SFR.

Alternatively, the molecular gas mass can also be traced by the two neutral carbon fine structure lines, C$_{\rm I}$($^3\rm P_2 - ^3\rm P_1$) and C$_{\rm I}$($^3\rm P_1 - ^3\rm P_0$), henceforth the [CI](2-1) and [CI](1-0) lines, respectively. Although initially, the [CI] emission was thought to be arising only from a thin layer between the molecular carbon and the ionised carbon regions of the photodissociation region (PDR) \citep{Langer76,Tielens85}, subsequent detection of [CI] and CO in the galactic molecular clouds, revealed that the [CI] was more widely distributed across the cloud \citep{Keene85}. This began the new prospects of using [CI] as a tracer of extended molecular gas in the galaxies \citep{Papadopoulos04, Papadopoulos04b}. 

Unlike the CO transitions, [CI] is optically thin and thus a good tracer of the gas mass if the carbon abundance, $X_{CI}$ is known or can be assumed \citep{Bisbas15, Dunne21}. It is also less affected by physical conditions like cosmic rays, unlike CO(1-0) \citep[e.g.,][]{Weiss03,Papadopoulos04,Papadopoulos04b,Walter14,Tomassetti14,Bothwell17} and the line excitation has only a small impact on the derived gas mass for typical line excitation temperatures \citep[e.g.,][]{Walter11, Harrington21}.

%Initial observations of [CI](2-1) at high redshift ($z\sim$2.5) was done by \citet{Weiss03}.} 

%However, CO(1-0) has many shortcomings. The relative CO abundance to H$_2$ varies with metallicities and the destruction of CO due to cosmic rays produced by star formation or active galactic nuclei (AGN)  activity \citep[eg,][]{Bisbas15,Bisbas17}. 

%[CI] lines have a higher emergent flux densities per H$_2$ column density than low-J CO lines \citep{Alaghband-Zadeh13}. They are also not affected by cosmic rays and have lower excitation temperatures thereby tracing the molecular gas of galaxies in a large look-back time (\textbf{REFERENCE, plus I'm not sure if this is okay}). 

Various analyses have been conducted on the high-J CO / [CI] line ratios \citep[e.g.,][]{Yang17, Andreani18, Valentino20}. This ratio has shown to be a good tracer of the density of the ISM. \citet{Andreani18} advocate that [CI](2-1)-dominated objects (CO(7-6)/[CI](2-1) $\le$ 1 ) are extended gas-rich disk structures with modest star-formation efficiencies, while CO(7-6)-bright objects are predominately merging systems. A significant fraction of DSFGs are shown to have multiple components \citep[e.g.,][]{Hodge13,Karim13} often interpreted as evidence of merging systems. The CO / [CI] ratio could be an alternative diagnostic tool for the merging status of a DSFG.

High-resolution imaging of DSFGs typically requires long integration times. Surveys like ALMA-ALPINE \citep{Le_Fevre20,Bethermin20,Faisst20} observed the very bright [CII] line emission, but even this survey could only marginally spatially resolve the sources. CO samples have mainly been observed at $z<3$ and only a very small fraction of them are spatially resolved \citep[e.g.,][]{Tacconi13,Dessauges-Zavadsky15,Magdis17}.
In contrast, gravitationally-lensed samples can provide better resolution due to the lensing magnification in shorter observation time \citep[e.g.,][]{Bussmann15,Spilker16}. Many samples of lensed DSFGs have been built in the recent years using the \textit{Herschel/}SPIRE \citep{Negrello10,Bussmann15}, Atacama Cosmology Telescope (ACT, \citet{Marsden14}), South Pole Telescope (SPT, \citet{Vieira10,Carlstorm11})  and \textit{Planck} \citep{Canameras15} surveys. Large follow-up campaigns were then performed using interferometers such as the Atacama Large Millimeter/sub-millimeter Array (ALMA), NOrthern Extended Millimeter Array (NOEMA) and Sub-Millimeter Array (SMA).
%--------------------------------------------------------------------
\begin{table*}
\centering
\caption{\label{img perf}Imaging performance. The resolution is the synthesised beam size of the continuum and line maps. The estimation of the $\sigma_{\rm channel}$ and $\sigma_{\rm continuum}$ are described in Sect.\,\ref{sec: imaging_perf}. }
\begin{tabular}{cccccccc}
\hline
\hline
&&&&&&&\\
Source&Resolution&Resolution&$\sigma_{\rm continuum}$&$\sigma_{\rm channel}$&$\sigma_{\rm channel}$&Channel\\
&continuum&line&& mean&range&width\\
&&&(mJy beam$^{-1}$)&(mJy beam$^{-1}$)&(mJy beam$^{-1}$)&(km s$^{-1}$)\\
&&&&&&&\\
\hline
&&&&&&&\\
SPT0103-45&
0.43$^{\prime\prime} \times$0.36$^{\prime \prime}$&
0.48$^{\prime\prime} \times$0.39$^{\prime\prime}$&
0.018&
0.225&
0.195 - 0.374&
12\\
SPT2147-50&
0.43$^{\prime\prime}\times$0.35$^{\prime\prime}$&
0.47$^{\prime\prime}\times$0.39$^{\prime\prime}$&
0.013&
0.096&
0.086 - 0.107&
14\\
SPT2357-51&
0.48$^{\prime\prime}\times$0.41$^{\prime\prime}$&
0.47$^{\prime\prime}\times$0.42$^{\prime\prime}$&
0.025&
0.283&
0.254 - 0.341&
12\\
\hline
\end{tabular}

\end{table*}

\begin{table*}[]
\centering  
\caption{\label{tab:mom0_fluxes} Continuum fluxes and integrated line intensities are tabulated below. Continuum fluxes are estimated from the continuum maps produced for the sources (Sect.\,\ref{Sec:cont_imaging}). Line fluxes are estimated from the integrated intensity maps of the lines (Sect.\,\ref{sec: moment maps}). The uncertainties on these fluxes are the statistical uncertainties combined with a $10\%$ absolute calibration uncertainty. APEX intensities are estimated by summing the fluxes for all the channels in a manually-defined velocity range. The fourth and fifth columns are the fluxes estimated in a velocity window estimated from the APEX data (narrow window, Sect.\,\ref{sec:apex}), while the eighth and night columns correspond to the flux estimated using the same velocity range as in the ALMA spectra (broad window, Sect.\,\ref{int_spec}). All the errors given are 1\,$\sigma$ errorbars, except for the upper limits which are 3\,$\sigma$.}
\begin{tabular}{ccccccccc}
\hline
\hline
&&&&&&&\\
Source    & Continuum &Continuum &$\rm I_{\rm [CI](2-1)}$&$\rm I_{\rm CO(7-6)}$& $\rm I_{\rm [CI](2-1)}$&$\rm I_{\rm CO(7-6)}$ & $\rm I_{\rm [CI](2-1)}$&$\rm I_{\rm CO(7-6)}$ \\
&frequency&flux&APEX&APEX&ALMA&ALMA&APEX&APEX\\
&&&(Narrow)&(Narrow)&&&(Broad)&(Broad)\\
&(GHz)& (mJy) & (Jy km\,s$^{-1}$) & (Jy km\,s$^{-1}$)& (Jy km\,s$^{-1}$) & (Jy km\,s$^{-1}$) & (Jy km\,s$^{-1}$) & (Jy km\,s$^{-1}$) \\
&&&&&&&\\
\hline
&&&&&&&\\
SPT0103-45& 208.93 & 25.52$\pm$2.73 &  7.59$\pm$2.34 & $<$13.72 (3$\,\sigma$) &15.29$\pm$1.75&14.38$\pm$1.43
&$<$21.6 (3$\,\sigma$)& $<$18.0 (3$\,\sigma$) \\

SPT2147-50& 173.91 & 6.94$\pm$0.84 & < 8.17 (3$\,\sigma$) & < 9.21 (3$\,\sigma$) &4.94$\pm$0.83&5.92$\pm$0.90
& < 7.29 (3$\,\sigma$) & 8.95$\pm$1.88 \\

SPT2357-51& 190.99 & 8.18$\pm$1.11 &< 9.34 (3$\,\sigma$)& 8.71$\pm$2.14 &3.63$\pm$0.60&5.71$\pm$0.80 & 5.88$\pm$2.14 & 10.86$\pm$2.42
\\

\hline

\end{tabular}

\end{table*}

%--------------------------------------------------------------------

In this paper, we take advantage of the gravitational lensing magnification of three lensed DSFGs from the SPT sample \citep{Vieira13,Spilker16,Strandet16,Reuter20} in order to study the resolved properties of the ISM using high resolution spectral and spatial imaging of the CO(7-6), [CI] (2-1) line and dust continuum emission with ALMA. With this data, we analyse the kinematics and morphology of the sources. We also explore the resolved ISM properties of our sources through line and continuum ratio analysis. We perform lens modelling of the line and continuum emission of our sources to understand their source-plane morphology, explore potential differential magnification effects \citep[e.g.,][]{Serjeant12,Hezaveh12} and reconstruct the intrinsic kinematics of the sources.

The paper is structured as follows. Section\,\ref{sec: sample and obs} presents our sample and the ALMA observations of the sources. The data reduction and imaging are described in Sect.\,\ref{data red}. The observed source morphologies and the spectra of the lines are described in Sect.\,\ref{spec and maps}. Section\,\ref{sec: kinematics} presents the analysis of the kinematics of our sample. The lens modelling is presented in Sect.\,\ref{sec: lens_modelling}. The resolved line and continuum ratio analyses are presented in Sect.\,\ref{Sec:ratios}. Finally, a brief discussion and conclusion of our analyses are presented in Sections\,\ref{sec: discussions} and \ref{sec: conc}, respectively.

Throughout this paper, we adopt a $\Lambda$CDM cosmology with $\Omega_m = 0.3$, $\Omega_{\Lambda} = 0.7$ and $H_0 = 70$ km\,s$^{-1}$\,Mpc$^{-1}$ and a \citet{2003PASP..115..763C} initial mass function (IMF).

\section{Sample and observations \label{sec: sample and obs}}

\subsection{SPT-SMG APEX/SEPIA [CI] sample \label{sec:apex}}

Our sample of three galaxies was selected from an initial program (PIs: Béthermin and Strandet, project number: 097.A-0973) targeting CO(7-6) and [CI](2-1) in 8 DSFGs from the SPT-SMG sample with the APEX/SEPIA instrument (on-source observation time are 492 mins for SPT0103-45, 559 mins for SPT2147-50 and 412 mins for SPT2357-51), a precursor of the ALMA band-5 receivers \citep{Belitsky18}.

We reduced the data using the \texttt{CLASS} package of the \texttt{GILDAS} software\footnote{\url{https://www.iram.fr/IRAMFR/GILDAS/}}. The data quality was estimated using an automatic procedure (Zhang et al. in prep.) and the bad data were removed from the analysis. We produced spectra combining all the other data after masking the region of the line and subtracting the baselines. 

%The entire dataset will be presented in a future paper together with another complementary sample observed with the Atacama compact array (ACA).

We use an antenna gain of 38 Jy K$^{-1}$ to obtain the APEX fluxes. We first estimate the line fluxes from the APEX spectra by summing the fluxes for all the channels in a manually-defined narrow range of velocities around the line peak. The noise is estimated by computing the standard deviation of the signal ($S_\nu \Delta v$ in Jy\,km/s) in line-free channels multiplied by $\sqrt{\rm N}$, where N is the number of channels used to estimate the line flux. [CI](2-1) is tentatively detected (3.24\,$\sigma$) in SPT0103-45, but not CO(7-6). No line is detected in SPT2147-50. For SPT2357, we tentatively detect CO(7-6) at 4.1\,$\sigma$, but not [CI](2-1). The results are summarised in Table\,\ref{tab:mom0_fluxes} and are compared to ALMA fluxes in Sect.\,\ref{int_spec}.

% We estimate the intensities of the lines from the spectra by summing the fluxes for all the channels encompassing the line (Fig.\,\ref{fig:int_spectra}).  

%--------------------------------------------------------------------

\subsection{ALMA sample selection}

%These sources were selected from the five ones with at least one line clearly detected.

%---------------------------

%------------------------------

In the APEX sample, 5 out of 8 DSFGs have at least one line tentatively detected. From these, we selected the sources with a reliable lens model to aid the interpretation \citep{Spilker16}. Three objects match our criteria: SPT0103-45, SPT0125-50 and SPT2357-51. These sources exhibit very diverse properties. SPT0103-45 has a bright [CI](2-1) line, SPT0125-50 is brighter in CO(7-6), whereas for SPT2357-51, the CO(7-6) line is unusually broad in the APEX/SEPIA spectra.\footnote{Note that the [CI](2-1) line was also tentatively detected and broad in the initial data reduction of the APEX spectra used to write the ALMA proposal.} The APEX spectra are shown in Fig.\,\ref{fig:int_spectra} (grey dotted line). We also required the sources to be observable with ALMA receivers. Unfortunately, the lines of SPT0125-50 were between the ALMA Band 4 and Band 5 frequencies and thus could not be observed. We thus observed two sources from this sample, SPT0103-45 and SPT2357-51 at high-resolution with ALMA, with the Band-5 receivers (PI: Béthermin, 2017.1.01018.S).  The target sensitivity on these observations is derived from the flux estimated on the APEX spectra (see, Sect.\,\ref{sec:apex}).

Our third source, SPT2147-50, was not detected with APEX despite the second data reduction, showing a weak signal for CO(7-6) (Fig.\,\ref{fig:int_spectra}). However, SPT2147-50 is a part of the Targeting Extremely Magnified Panchromatic Lensed Arcs and Their Extended Star formation (TEMPLATES) James Webb Space Telescope (JWST) Early Release Science program\footnote{\url{https://www.stsci.edu/jwst/observing-programs/approved-ers-programs/program-1355}}. The source was thus observed with ALMA in CO(7-6), [CI](2-1) and dust continuum to aid in the interpretation of the upcoming JWST data (PI: Vieira, 2018.1.01060.S).

Figure\,\ref{fig:sample_selection} shows  the line luminosity of [CI](2-1) and CO(7-6) against the IR luminosity. We compare our sources with the sample compilation presented in \citet{Valentino20}. Our sources are at the bright end of the SMG cloud.
Our final sample thus consists of three sources: SPT0103-45, SPT2147-50 and SPT2357-51. Our sources are presented in Table\,\ref{tab:obs_source}. 

%--------------------------------------------------------------------

\begin{figure*}[h]
\centering

\begin{tabular}{ccc}
\includegraphics[width=6cm]{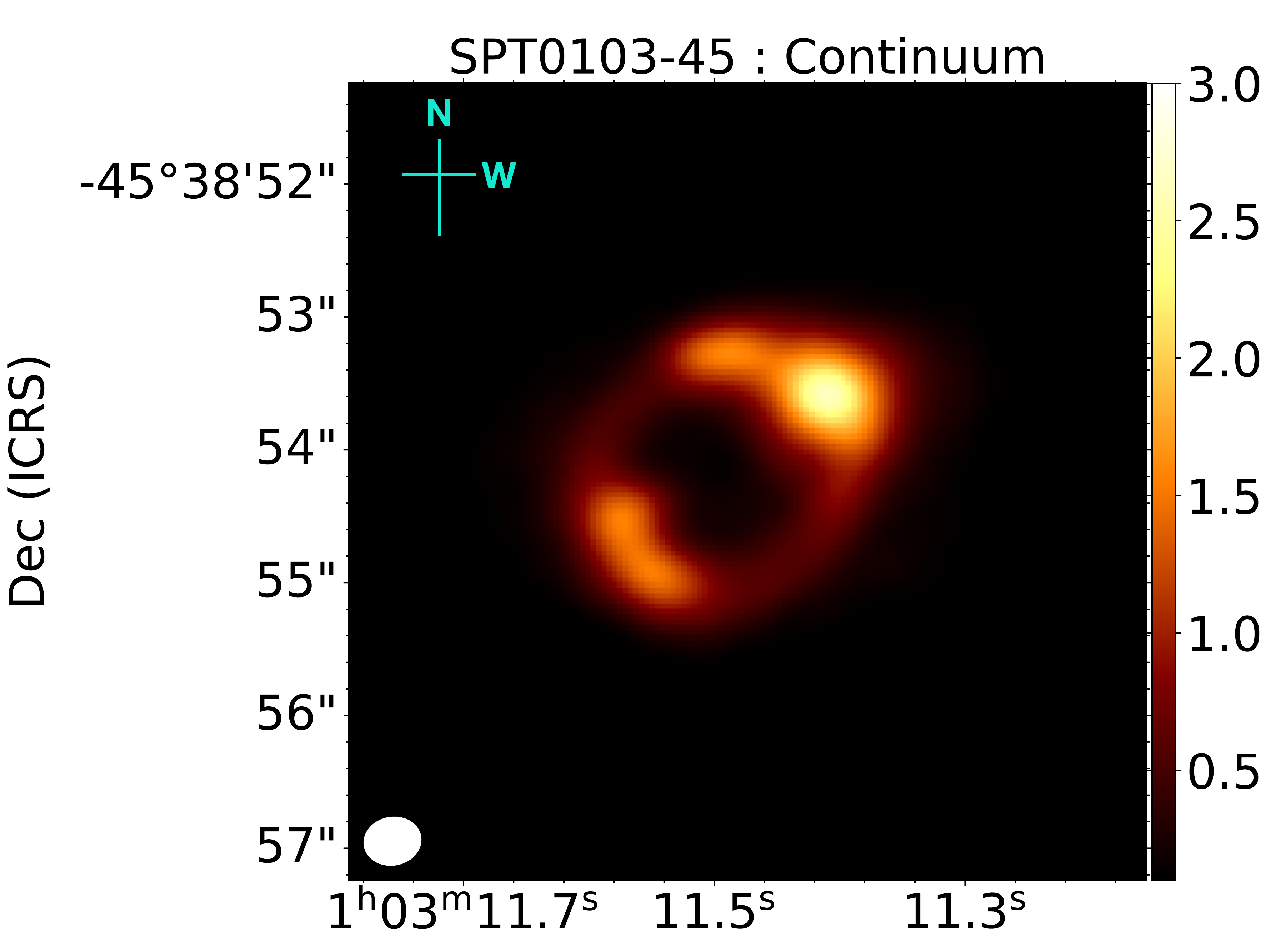} &  \includegraphics[width=5.8cm]{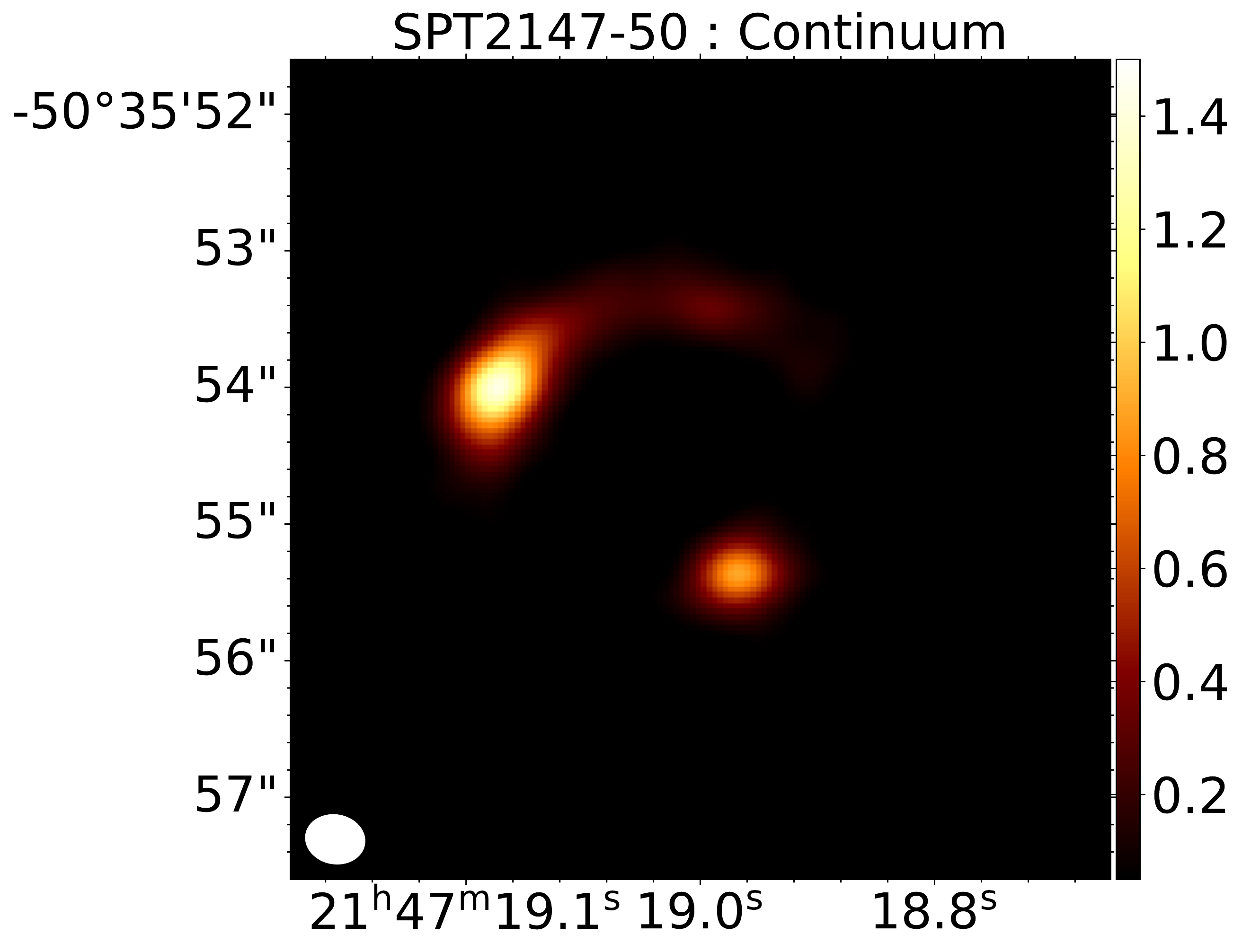} & \includegraphics[width=6cm]{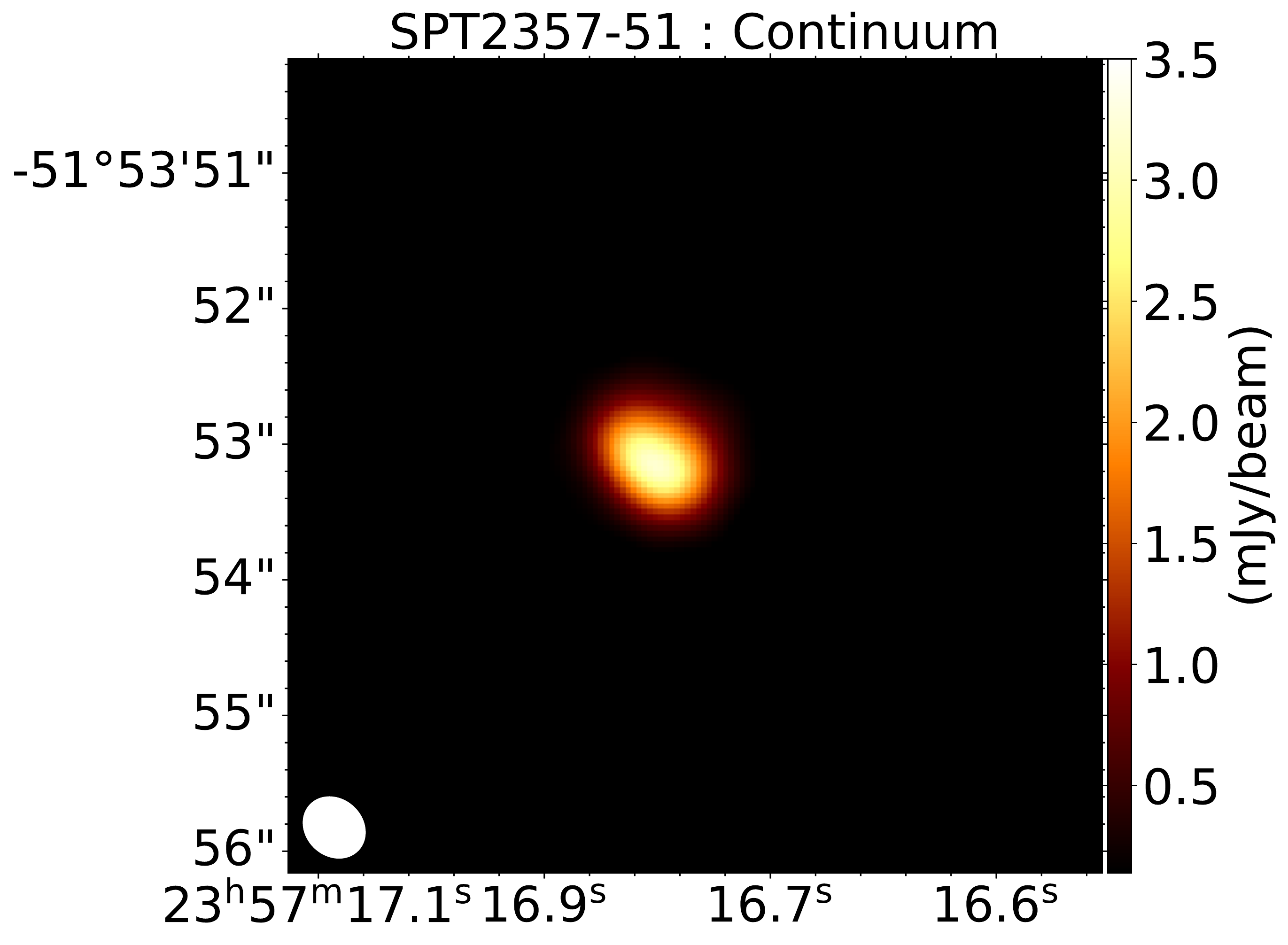} \\
\includegraphics[width=6cm]{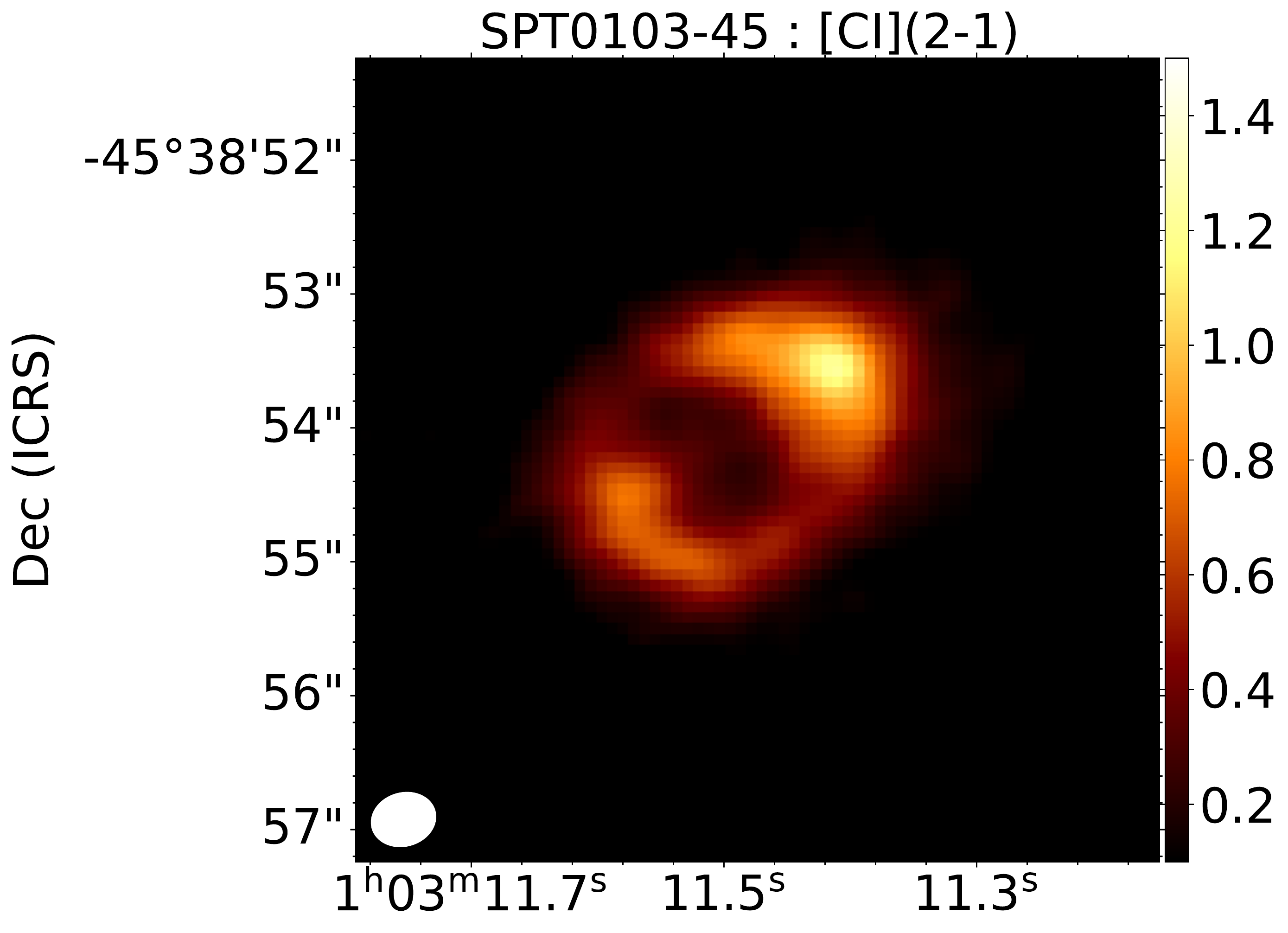} &  \includegraphics[width=5.8cm]{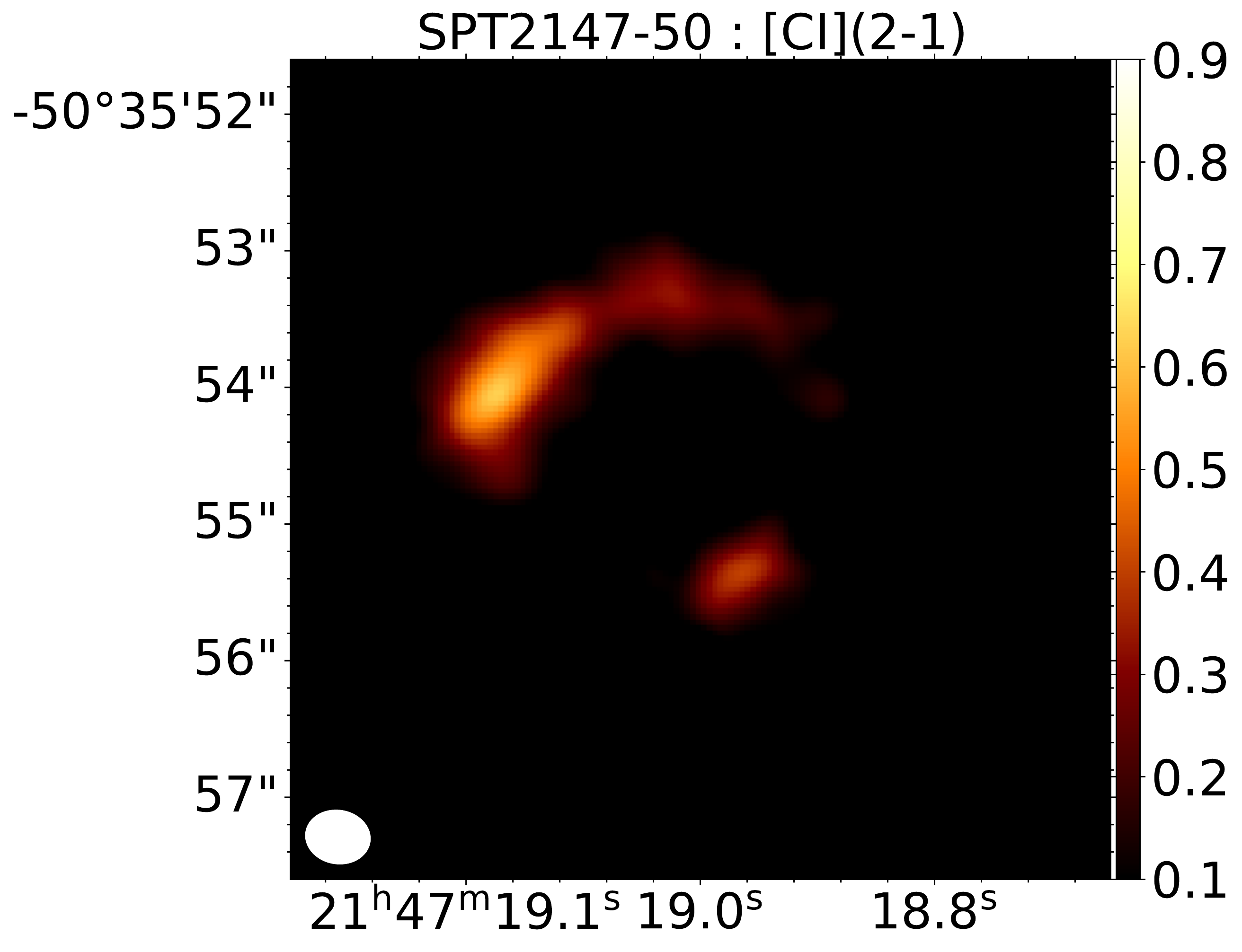} & \includegraphics[width=6cm]{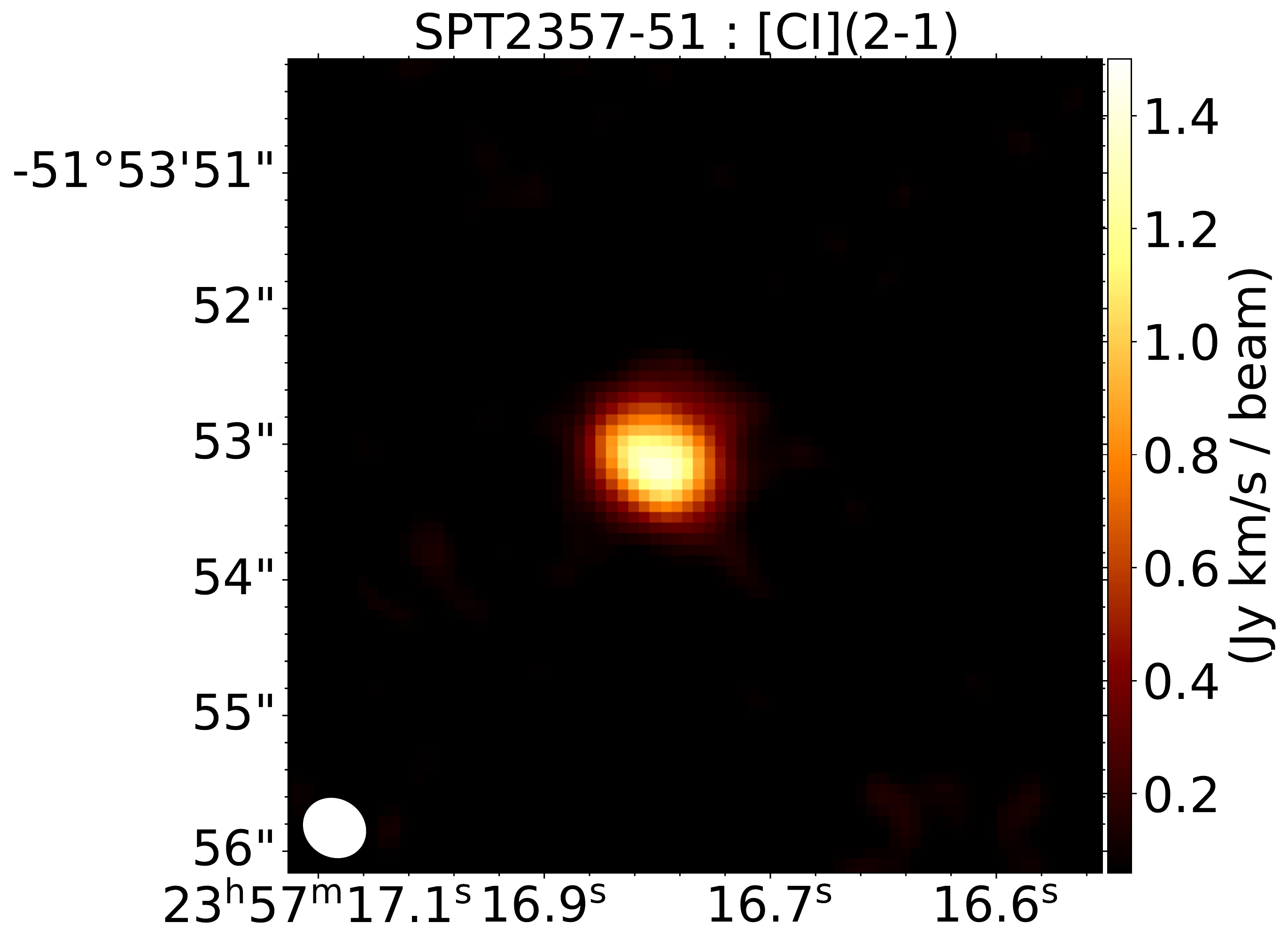} \\
\includegraphics[width=6cm]{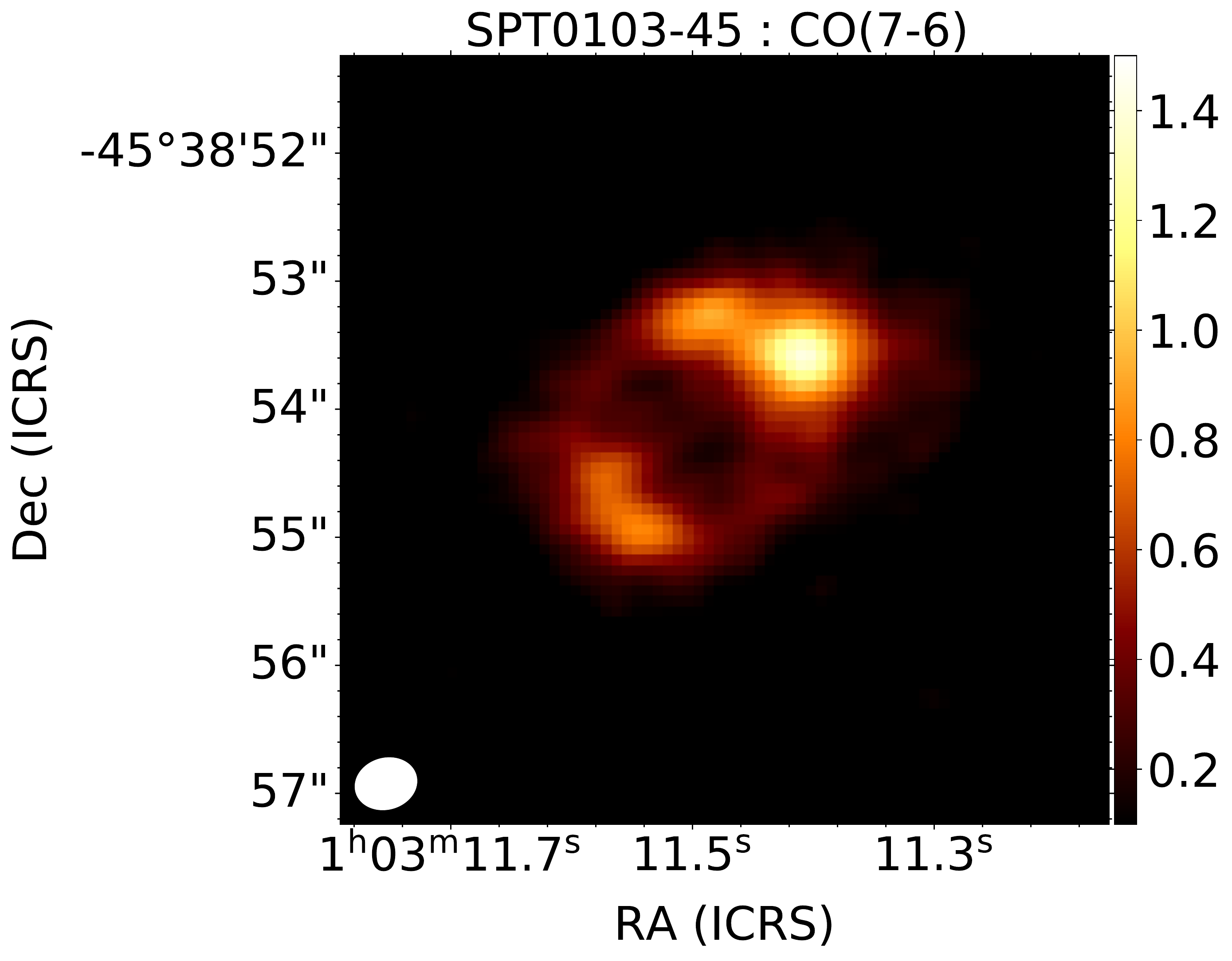} &  \includegraphics[width=5.8cm]{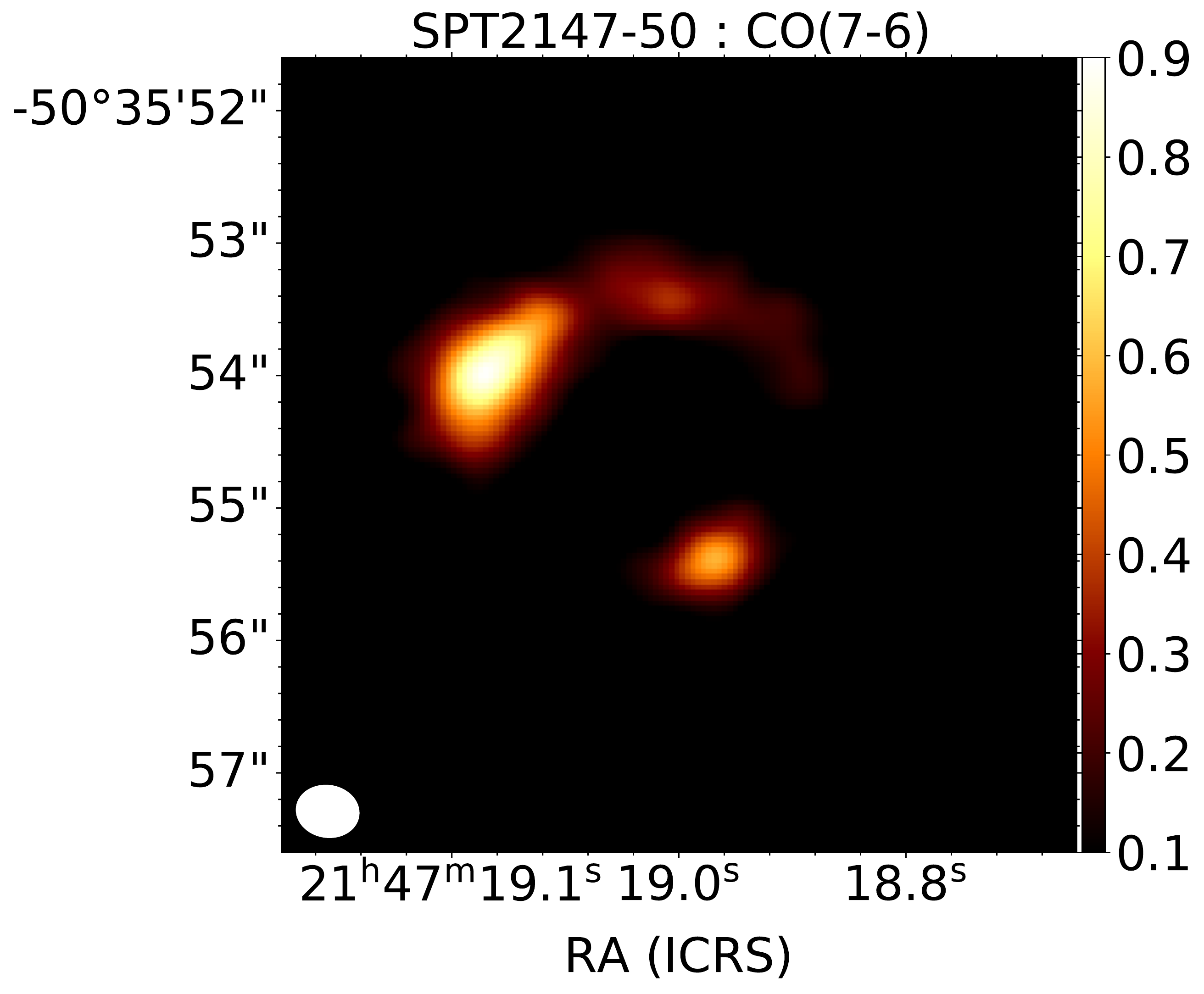} & \includegraphics[width=6cm]{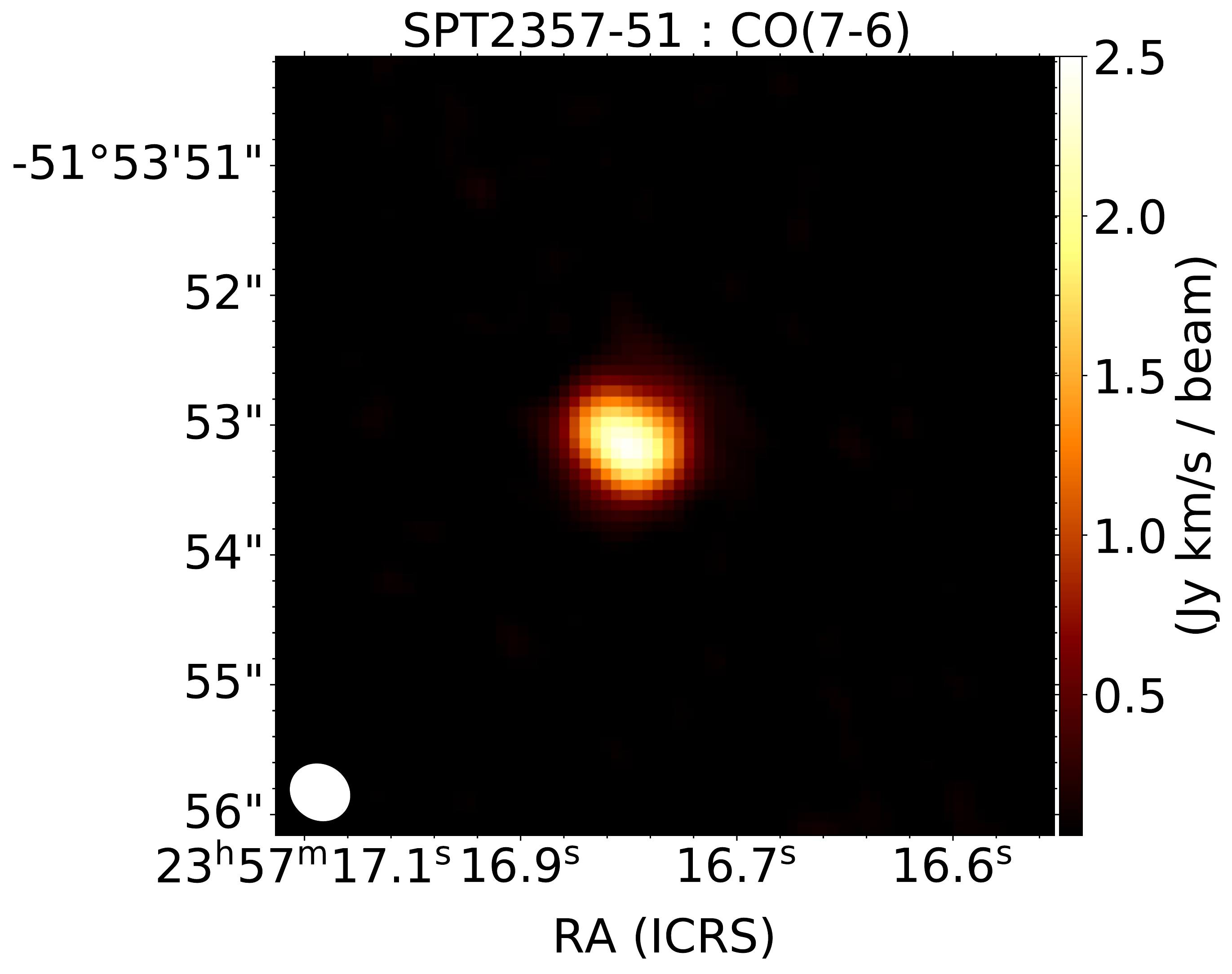} \\
\end{tabular}
\caption{\label{fig:maps} Continuum and moment-0 maps of our sample. The first row shows the continuum maps of (from left to right) SPT0103-45, SPT2147-50 and SPT2357-51 (see Sect.\,\ref{Sec:cont_imaging}). The synthesised beam is represented as the white ellipse on the bottom left of every map. The second row is the integrated [CI](2-1) intensity maps and the third row is the integrated CO(7-6) intensity maps of our sources (see Sect.\,\ref{sec: moment maps})}
\end{figure*}
%--------------------------------------------------------------------

\subsection{ALMA observations \label{sec:ALMA obs}} 

All our targets were observed with the ALMA 12-metre array using the Band-5 receivers and in the C43-5 configuration (shortest baseline: 15\,m; longest baseline: 1400\,m). The requested angular resolution was 0.3" for SPT0103-45 and SPT2357-51 and 0.4" for SPT2147-50. This was the best possible angular resolution allowed in Band 5 during Cycle 5 and allows us to resolve the sources over several synthesised beams.

%These resolutions are a compromise, which allows us to resolve the sources into several synthesised beams, without spreading the flux of our sources into too many of them, making the signal too difficult to detect. 

To observe CO(7-6) and [CI](2-1), two spectral windows with 1.875\,GHz bandwidth and 7.813\,MHz resolution (corresponding to 11--13\, km s$^{-1}$ velocity resolution) were used\footnote{ SPT2147-50 has only one spectral window centered on the lines with 7.813\,MHz resolution. The second spectral window in the sideband has a coarser resolution (31.25\,MHz).}. For each source, we also placed two other spectral windows in the other sideband to measure the continuum. The same spectral resolution was used for SPT0103-45 and SPT2357-51, while 31.25\,MHz was used for SPT2147-50. 

To characterise the dynamics of these systems, we need to securely constrain the position of the peak of at least one line in each synthesised beam. For SPT0103-45 and SPT2357-51, we chose the target sensitivity to obtain a 10$\,\sigma$ detection when integrated over the full line width. We estimated the signal and the line width from the APEX spectra (Sect.\,\ref{sec:apex}) and divided the flux into the number of synthesised beams covered by the source from the previous 870\,$\mu$m continuum imaging and lens model \citep{Spilker16}. We derived a target sensitivity of 80\,$\mu$Jy beam$^{-1}$ in a 320\,km s$^{-1}$ channel for SPT0103-45 and 67$\mu$Jy beam$^{-1}$ in a 510\,km s$^{-1}$ channel for SPT2357-51. For SPT2147-50, the sensitivity was estimated by rescaling the observed continuum profile to the estimated CO and [CI] flux, aiming for a 6$\,\sigma$ detection over 1/3 of the line width at the peak of the detection, which is equivalent to 10.4$\,\sigma$ over the full line width. This corresponds to 100\,$\mu$Jy beam$^{-1}$ in a bandwidth of 100\,km s$^{-1}$ (1/3 of the line width) or 58\,$\mu$Jy beam$^{-1}$ in a bandwidth of 300\,km s$^{-1}$. The target sensitivities for the three sources are thus rather similar in the end.
The details of the observations are presented in Table\,\ref{tab:obs_details}.

%--------------------------------------------------------------------
%--------------------------------------------------------------------

\section{Data processing \label{data red}}

\subsection{Calibration}
%Talk about decovolution, parameters that you used in clean, and all that
Initial calibrations were done by the observatory using the standard ALMA pipeline based on the Common Astronomy Software Applications (\texttt{CASA}) \citep{CASA}. Since our sources are bright, we attempted to self-calibrate the phase solutions. However, many long baselines were poorly fit and were thus flagged. The beam size was consequently degraded without a significant gain in sensitivity, so we discarded the self-calibration.
%--------------------------------------------------------------------
\subsection{Continuum imaging \label{Sec:cont_imaging}}
To image the data, we used the \texttt{CLEAN} routine in \texttt{CASA}. The data were imaged using a pixel size of 0.04$^{\prime \prime}$ in order to finely sample the synthesised beam ($\sim$10 pixels per beam). A Briggs weighting \citep{Briggs95} of the visibilities, with a robust parameter of 0.5 was chosen to give the best compromise between S/N and angular resolution. 

To measure the continuum, we combine the two spectral windows in the line-free sideband and the channels without line contamination in the sideband where CO(7-6) and [CI](2-1) are located. These channels are identified using a first preliminary line imaging (see Sect.\,\ref{sect:line_imaging}).
The continuum images are then generated using a multi-frequency synthesis (MFS) method \citep{Conway90}. A first imaging is made to estimate the noise level (hereafter $\sigma_{\rm noise}$). We then perform a second imaging using 3\,$\sigma_{\rm noise}$ from the previous imaging as the \texttt{Clean} threshold. The first row of Fig.\,\ref{fig:maps} shows the resulting continuum maps of our sources.

\begin{figure}
\centering

\includegraphics[width=9cm]{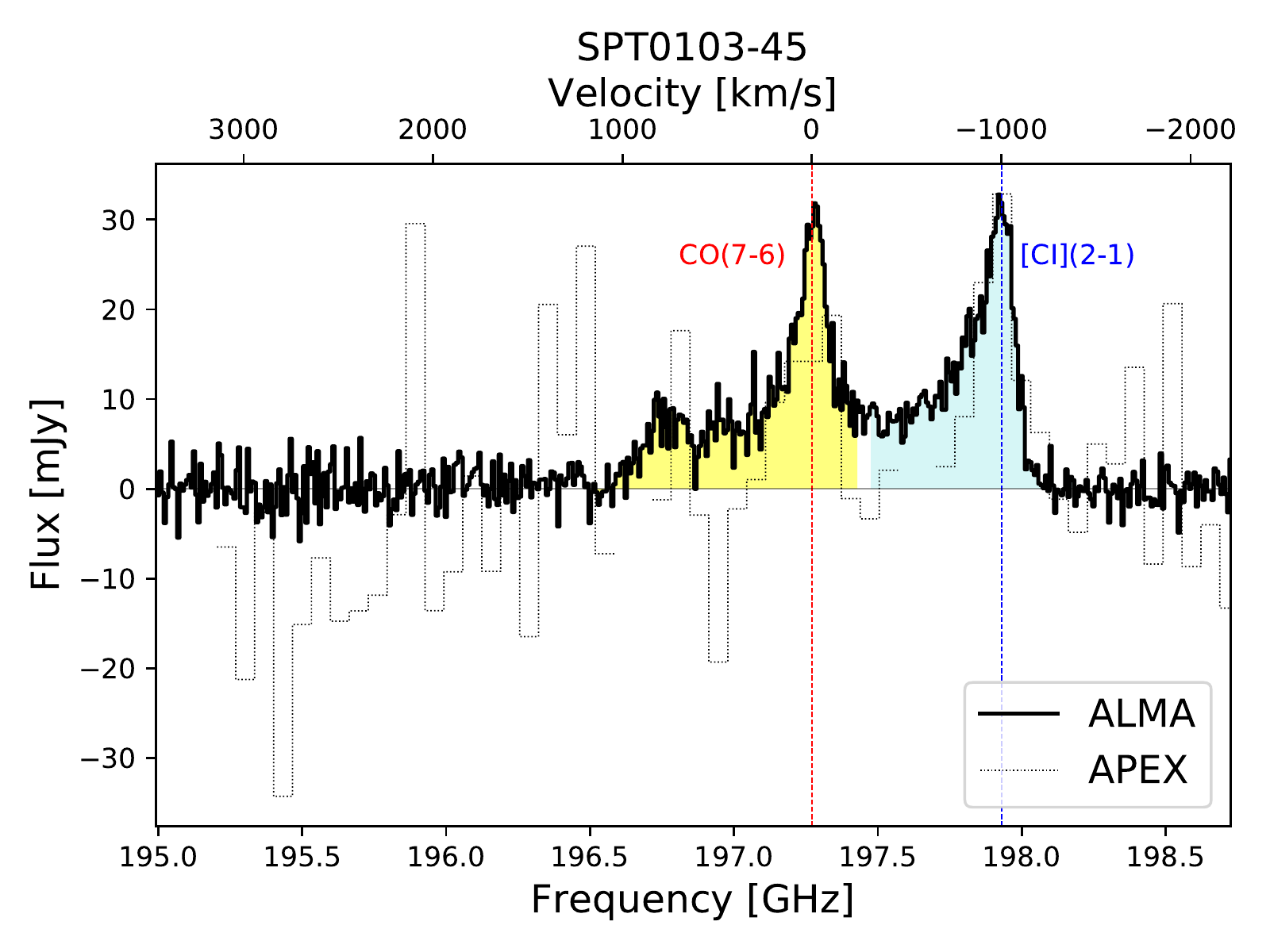} 
\includegraphics[width=9cm]{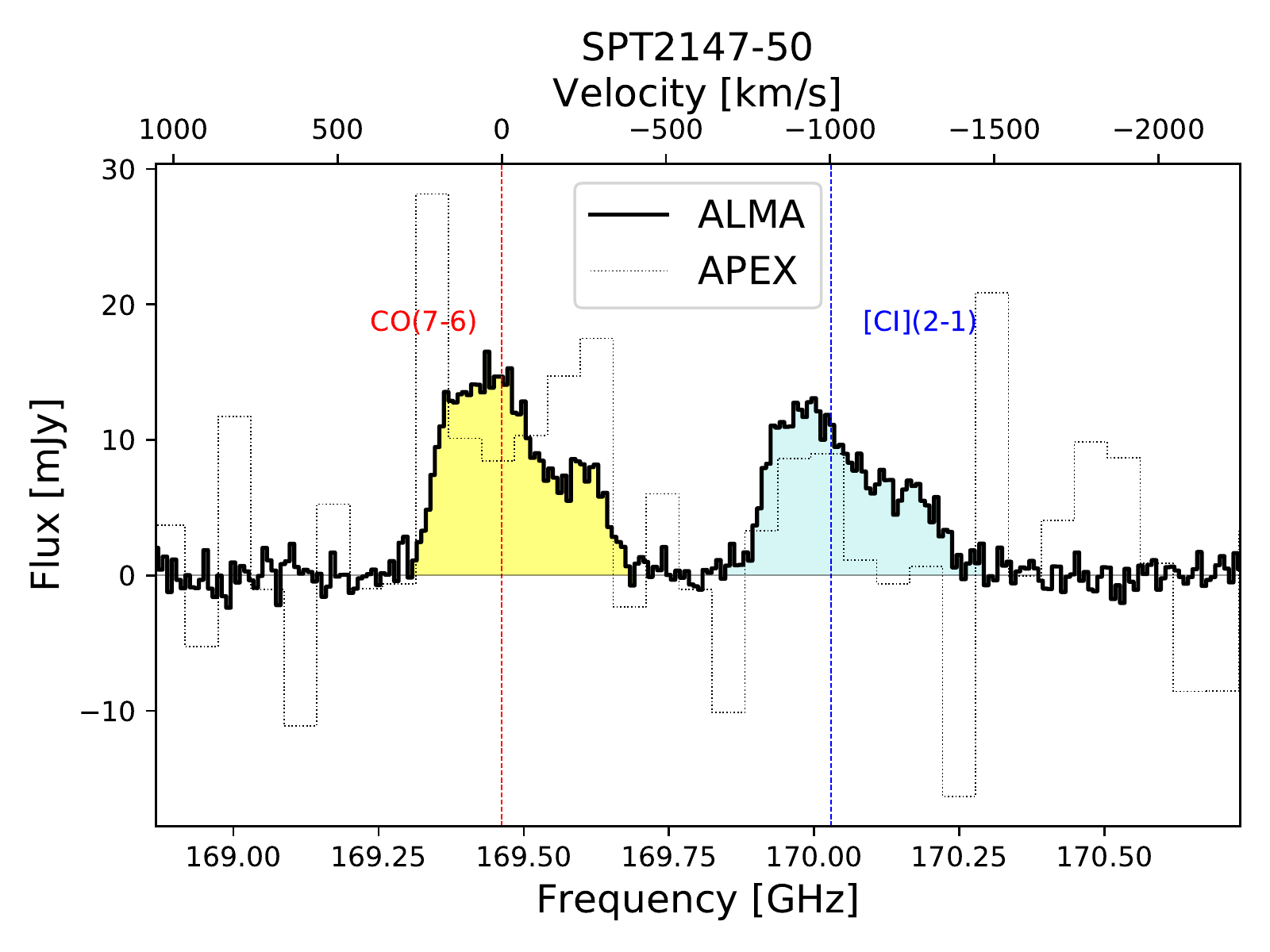} 
\includegraphics[width=9cm]{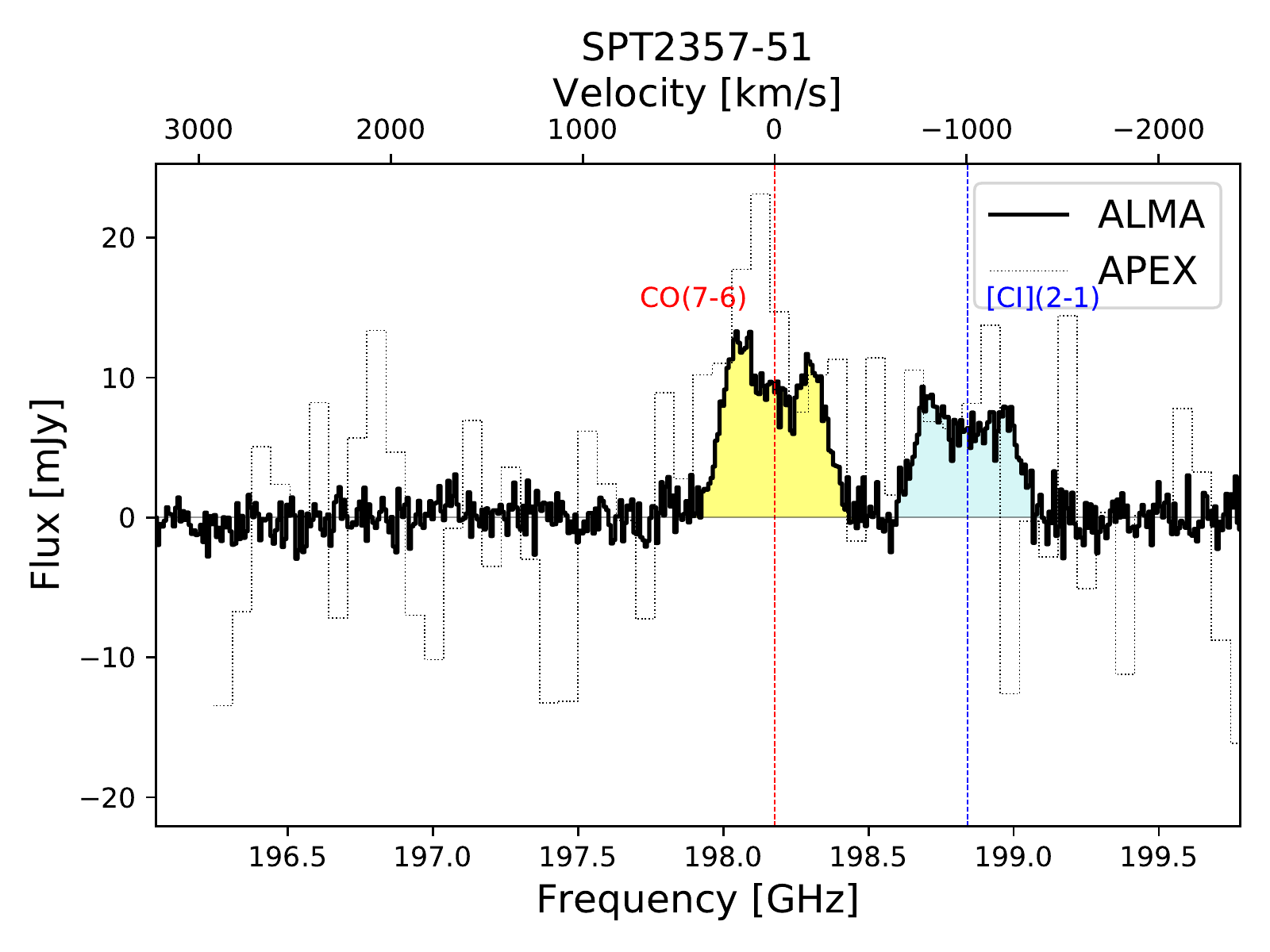} 

\caption{\label{fig:int_spectra} Integrated spectra of SPT0103-45 (top panel), SPT2147-50 (middle panel) and SPT2357-51 (lower panel). The black solid line is the spectrum extracted from the ALMA observations (see Sect.\,\ref{sec:ALMA obs}). The light gray line is the APEX/SEPIA spectrum (see Sect.\,\ref{sec:apex}). Red and blue vertical dashed lines indicate the position of CO(7-6) and [CI](2-1) lines, respectively, at the reference redshift of the source. The yellow and cyan shaded areas show the integration windows used to compute integrated fluxes of the CO(7-6) and [CI](2-1) lines, respectively (see Sect.\,\ref{sec: moment maps}). The zero-velocity corresponds to the observed frequency of CO(7-6) line for each of the sources.}
\end{figure}

%--------------------------------------------------------------------
\subsection{Line imaging}

\label{sect:line_imaging}

Using similar parameters as for the continuum, we first image the  spectral window(s) covering CO(7-6) and [CI](2-1). We then subtract the continuum directly in the \textit{uv} plane using the \textit{uvcontsub} \texttt{CASA} task. The channels that are line-free in our initial image cube are used to fit and subtract the continuum. We assume a constant continuum level (as a function of the frequency) for SPT2357-51 and a first-order polynomial for the two other sources as the first-order polynomial gave a better fit of the continuum. We then re-imaged the data to obtain a continuum-subtracted line cube.  No rebinning from the native $\sim12$\,km/s spectral resolution was used in the process, since the S/N per channel is sufficiently high.

%--------------------------------------------------------------------
\subsection{Imaging performance \label{sec: imaging_perf}}

 For the continuum, the synthesised beam major axis size is between 0.43$^{\prime \prime}$ and 0.48$^{\prime \prime}$, while the minor axis size is between 0.35$^{\prime \prime}$ and 0.41$^{\prime \prime}$. The axis ratio varies from 1.17 to 1.21.

Concerning the line cubes, for SPT0103-45 and SPT2147-50 the synthesised beam is slightly larger than for the continuum, with the major axis ranging from 0.47$^{\prime \prime}$ to 0.48$^{\prime \prime}$ and the minor axis ranging from 0.39$^{\prime \prime}$ to 0.42$^{\prime \prime}$. This is because the spectral lines are located in the lower sideband of the frequency coverage. Conversely for SPT2357-51, the continuum has a coarser resolution because the lines are in the upper sideband, so the continuum-only data is at a lower average frequency. The beam axis ratio varies from 1.12--1.22. Overall, all three sources have similar, nearly circular beams.
We estimate the noise level at the phase center by computing the standard deviation in the non-primary beam corrected map after masking the source. The average continuum sensitivity varies from 13--25\,$\mu$Jy\,beam$^{-1}$. The same procedure is used channel by channel for the line cubes. The mean sensitivity per channel and the range of sensitivities per channel for every source are summarised in Table\,\ref{img perf}.

\begin{figure*}[h]
\centering

\includegraphics[width=18cm]{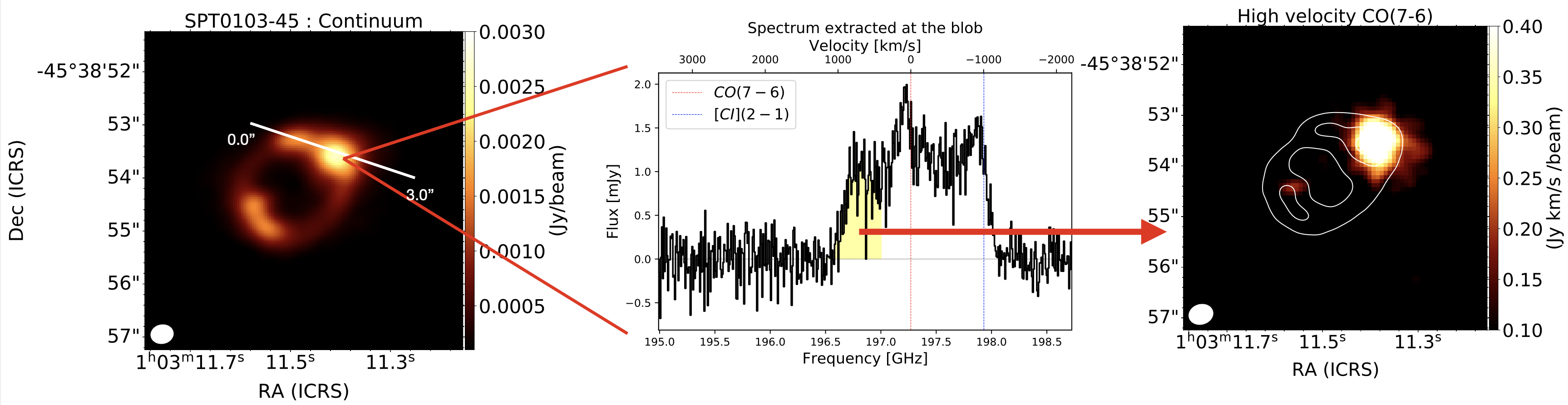}
\caption{\label{fig:bright blob}Properties of the `gem' in SPT0103-45. The left panel is the continuum map as described in Sect.\,\ref{Sec:cont_imaging}. The white line shows the direction along with the slice for the PV diagram is extracted (Fig.\,\ref{fig:pv 0103}). The spectrum shown in the center panel is extracted at the coordinates of the gem (Sect.\,\ref{sect:line_imaging}). The frequency range represented by the yellow region is used to compute the integrated intensity of the high-velocity emission, represented in the right panel. The white contours are the 5, 10 and 15$\,\sigma$ contours of the continuum emission.}

\end{figure*}

%-----------------------------------------------------------
\begin{figure}
\centering

\includegraphics[width=9cm]{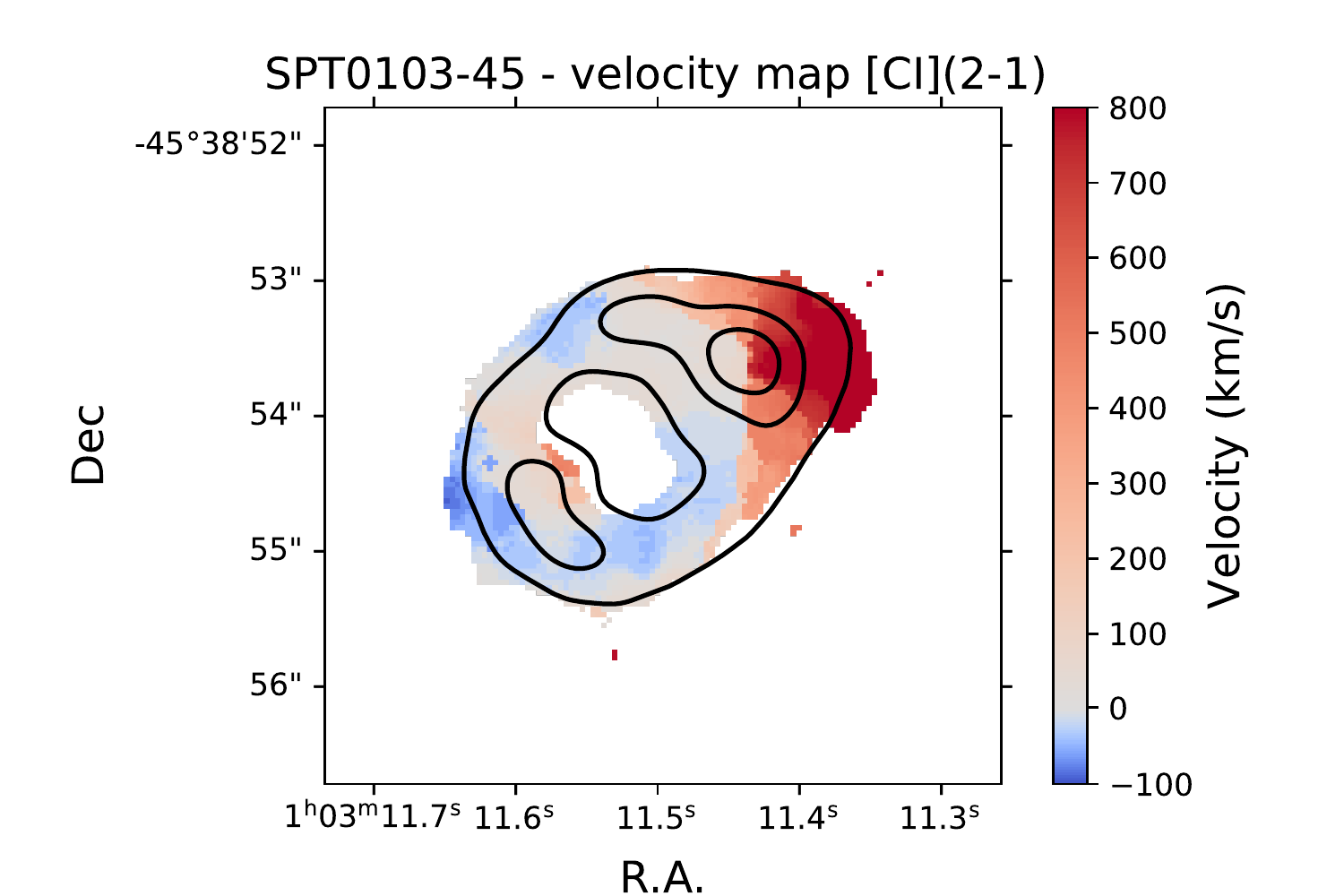}  \includegraphics[width=9cm]{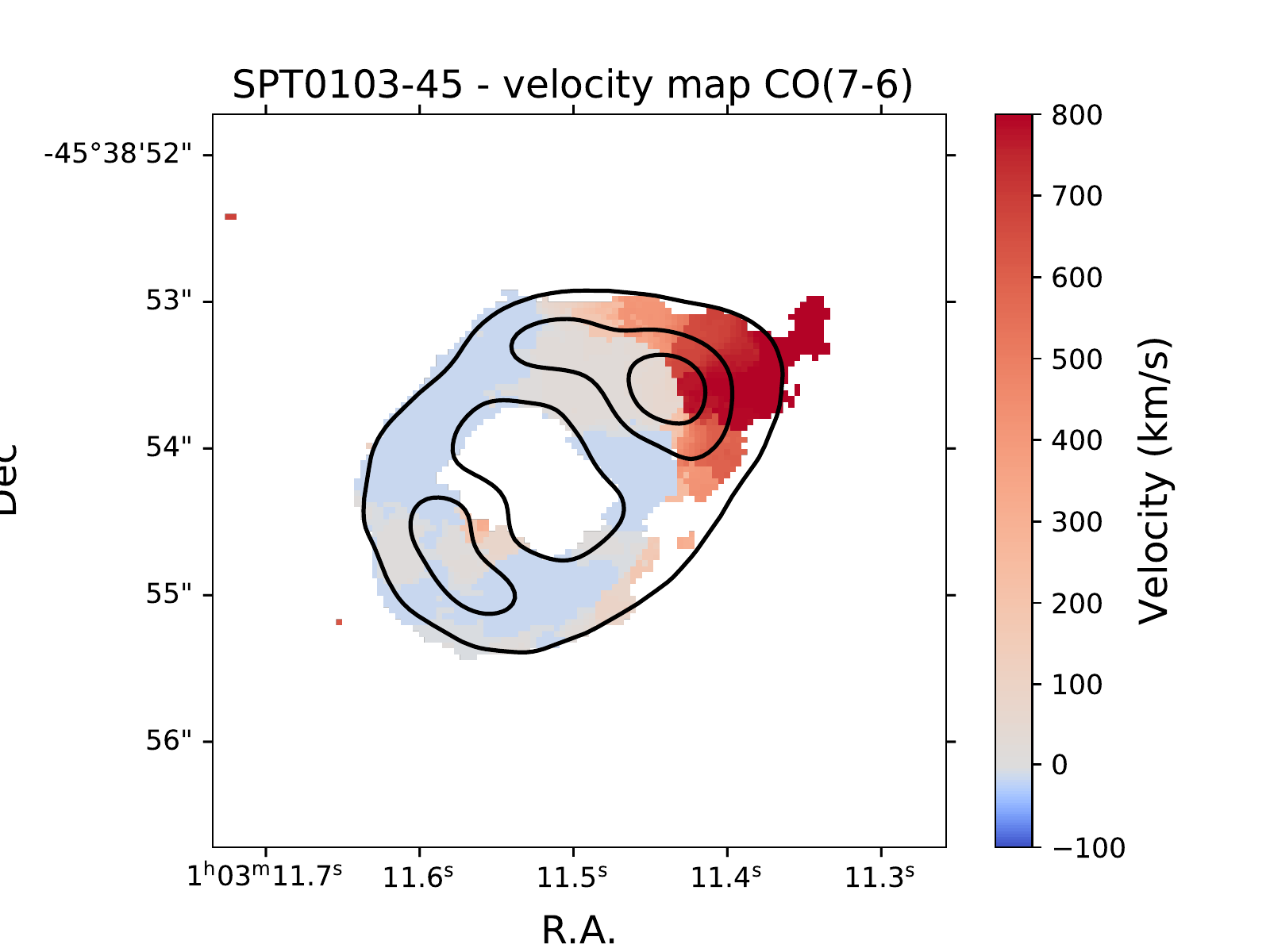} 

\caption{\label{fig:decomp 0103}Pixel-wise peak velocity decomposition of SPT0103-45. The top row shows the velocity map of the [CI](2-1) line and bottom row shows the CO(7-6) line. The 5, 10 and 15$\,\sigma$ contours of the continuum emission are plotted on all velocity maps. The construction of these maps is described in Sect.\,\ref{kinematics_0103}.}
\end{figure}

%-----------------------------------------------------------
\section{Integrated spectra and source morphologies \label{spec and maps}}

%----------------------------------------------------------------
\subsection{Continuum morphology}

The dust continuum images are shown in the upper panels of Fig.\,\ref{fig:maps}. 

SPT0103-45 is ring-shaped, which is characteristic of strongly lensed sources \citep{Spilker16}. The brightest region is located in the northwest (henceforth, the \textit{gem}) and another bright region can be seen in the southeast. They correspond to two counter-images of the same region of the source (see Sect.\,\ref{sec: lens_modelling}).

SPT2147-50 exhibits an extended arc in the northeast and a compact counter-image in the southwest (see Sect.\,\ref{sec: lens_modelling}). The northeast arc is brighter in its most eastern part and there is a small gap with no detected continuum at the most northern part of the arc. 

SPT2357-51 is also gravitationally lensed and has a very compact continuum emission compared to the other sources. The compact emission is due to the small Einstein radius of the foreground lens galaxy ($\theta_{\rm E} \sim 0.2^{\prime\prime}$, \citealt{Spilker16}). The resolved structure of SPT2357-51 is difficult to see visually in our default continuum image using all the line-free channels and a Briggs weighting with a robustness parameter of 0.5. In order to demonstrate the lensing geometry of this source more clearly, we re-imaged the continuum selecting only visibilities on baselines $>$200\,k$\lambda$, shown in Fig. \,\ref{fig:highres_2357}. We note that this image is for demonstration purposes only, as our lens modelling procedure (Sect.\, \ref{sec:continuum_lens}) fits directly to the interferometric visibilities and is not influenced by choices in imaging parameters. 

The total continuum flux density is estimated from these maps by integrating the flux in an elliptical region encompassing the entire source. The continuum fluxes are tabulated in Table\,\ref{tab:mom0_fluxes}. To estimate the uncertainties of these fluxes, we select a polygonal region near the source and measure the noise RMS in this region. We then re-scale this noise by the factor of $\sqrt{\rm N}$, where N is the number of synthesised beams in the elliptical region where the source flux is estimated. We combine this statistical uncertainty with the estimated 10$\%$ absolute flux calibration typical of ALMA data.

\begin{figure}
\centering

\includegraphics[width=9cm]{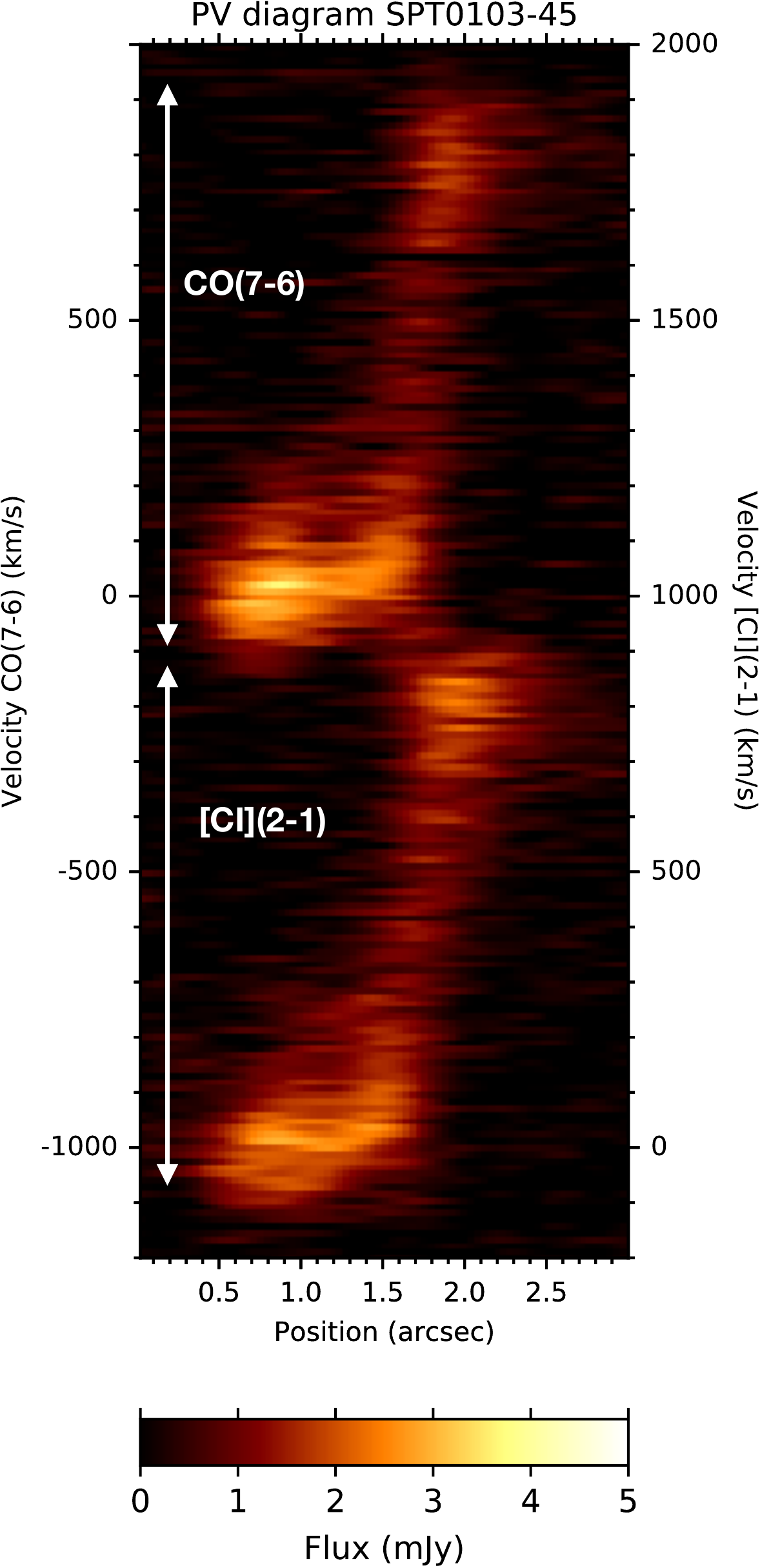} 
\caption{\label{fig:pv 0103} Position-velocity diagram of SPT0103-45. The axis used to make this diagram is shown in the first panel of Fig.\,\ref{fig:bright blob}. The velocity axis on the left y-axis has the zero-velocity of the CO(7-6) observed frequency and the right y-axis has the zero-velocity centered at the [CI](2-1) observed frequency.} % CO(7-6) emission is located between -100 and 900\,km/s on the CO(7-6) reference velocity scale, and [CI](2-1) is located between -100 and 900\, km/s on the [CI](2-1) reference velocity scale.
\end{figure}

%-----------------------------------------------------------

\subsection{Integrated spectra \label{int_spec}}

We extract the integrated spectra from the line cubes after manually selecting the emission region. This process was repeated multiple times to ensure the robustness of the method. Figure\,\ref{fig:int_spectra} shows the integrated spectra of all our sources. There is a clear detection, S/N > 14, for both the CO(7-6) and [CI](2-1) lines in all sources.

The line profiles of SPT0103-45 show a well-defined peak along with a redshifted tail in both CO and [CI] (Fig.\,\ref{fig:int_spectra}, top panel). These tails extend to about $+$1000\, km\,s$^{-1}$. The uni-directionality of the high-velocity tail is interesting and will be further explored in Sect.\,\ref{kinematics_0103}. The [CI](2-1) tail is slightly blended with the CO(7-6) peak emission and cannot be well isolated. We assessed the impact of this line blending by fitting the spectrum of SPT0103-45 with the sum of three Gaussian profiles each, for CO and [CI], fixing the center velocity relative to each line's systemic velocity and the width of each pair of components (i.e. with a total of 12 free parameters to fit both lines simultaneously: three center velocities, three Gaussian line widths and six amplitudes). We find that accounting for the line blending increases the [CI]/CO line ratio by $\sim$5\% compared to the simple division illustrated in Fig\,.\ref{fig:int_spectra}. We increase the uncertainty on the measured line ratio by this amount, added in quadrature to the statistical uncertainty from the noise in the data. The deblending procedure and results are further demonstrated in Appendix.\,\ref{deblending}. 

In SPT2147-50, both the lines are broad (FWHM $\sim$ 550\,km\,s$^{-1}$) and asymmetric, with the redshifted side brighter than the blueshifted side of the spectra (see Fig.\,\ref{fig:int_spectra}, middle panel).

% Both the lines exhibit a secondary intensity peak at $\sim$ -200 km\,s$^{-1}$, somewhat more prominently in CO than [CI]. 

For SPT2357-51, the line profiles show double peaks in both lines (see Fig.\,\ref{fig:int_spectra}, bottom panel). %\textbf{(FWHM $\sim$ 580\,km\,s$^{-1}$), for each component}. 
The blueshifted peak is slightly brighter ($\sim$2\,mJy) than the redshifted peak. The nature of this double peak feature will be further explored in the Sect.\,\ref{sec:kinematics 2357}. 

In Fig.\,\ref{fig:int_spectra}, we compare the ALMA spectral profile with the APEX/SEPIA observations. We also re-estimate the APEX line fluxes using a broader integration window corresponding to the entire width of the line seen by ALMA (blue and yellow shaded regions of Fig.\,\ref{fig:int_spectra}). The results are tabulated in Table\,\ref{tab:mom0_fluxes}. While the shape of the APEX and ALMA spectra generally agree, there are differences up to 2$\,\sigma$ in the integrated line fluxes. It is unlikely to be an interferometric effect (e.g., filtered-out large-scale emission) since the largest difference is found for SPT2357-51, the most compact source of our sample. It is more likely that the APEX spectrum is simply of lower quality than the ALMA spectra.

%-----------------------------------------------------------
\subsection{Line morphologies \label{sec: moment maps}}

We construct integrated intensity (i.e. moment-0) maps of the [CI](2-1) and CO(7-6) lines by summing the flux of every channel covering the line using the \texttt{CASA} \texttt{immoments} task. We define a frequency range for each line as the integration window to compute the integrated intensity from the continuum-subtracted datacubes. The integration windows are shown in Fig.\,\ref{fig:int_spectra}, with CO(7-6) in yellow and [CI](2-1) in blue. For SPT0103-45, the most redshifted CO(7-6) emission is blended with the most blueshifted [CI](2-1) emission, as noted above. We thus use a narrower range to integrate the [CI](2-1) emission in order to avoid the contamination of CO(7-6) emission, but note that the line blending introduces some unavoidable uncertainty into the moment maps.

The moment-0 maps for the [CI](2-1) and CO(7-6) lines are shown in the second and third rows of Fig.\,\ref{fig:maps} respectively. The line emission is more extended than the continuum emission for all sources. The compactness of the dust continuum with respect to spectral lines that trace cold gas is very common in DSFGs \citep[e.g.,][]{Spilker15,CalistroRivera18,Dong19,Apostolovski19}. 
% The overall morphology of the [CI](2-1) and CO(7-6) emission of our sources is fairly similar to their continuum emission, a result of the well-constrained geometry of gravitational lensing and the fact that dust and gas are broadly co-spatial in galaxies.

We estimate the line intensities of the sources by selecting the region encompassing the source manually and obtaining the intensities in this region. The uncertainties are estimated similarly to the continuum emission (Sect.\,\ref{Sec:cont_imaging}). The intensities are tabulated in Table\,\ref{tab:mom0_fluxes}.

%----------------------------------------------------------------------
%----------------------------------------------------------------------
\begin{figure*}[h]
\centering

\begin{tabular}{cc}
\includegraphics[width=8.5cm]{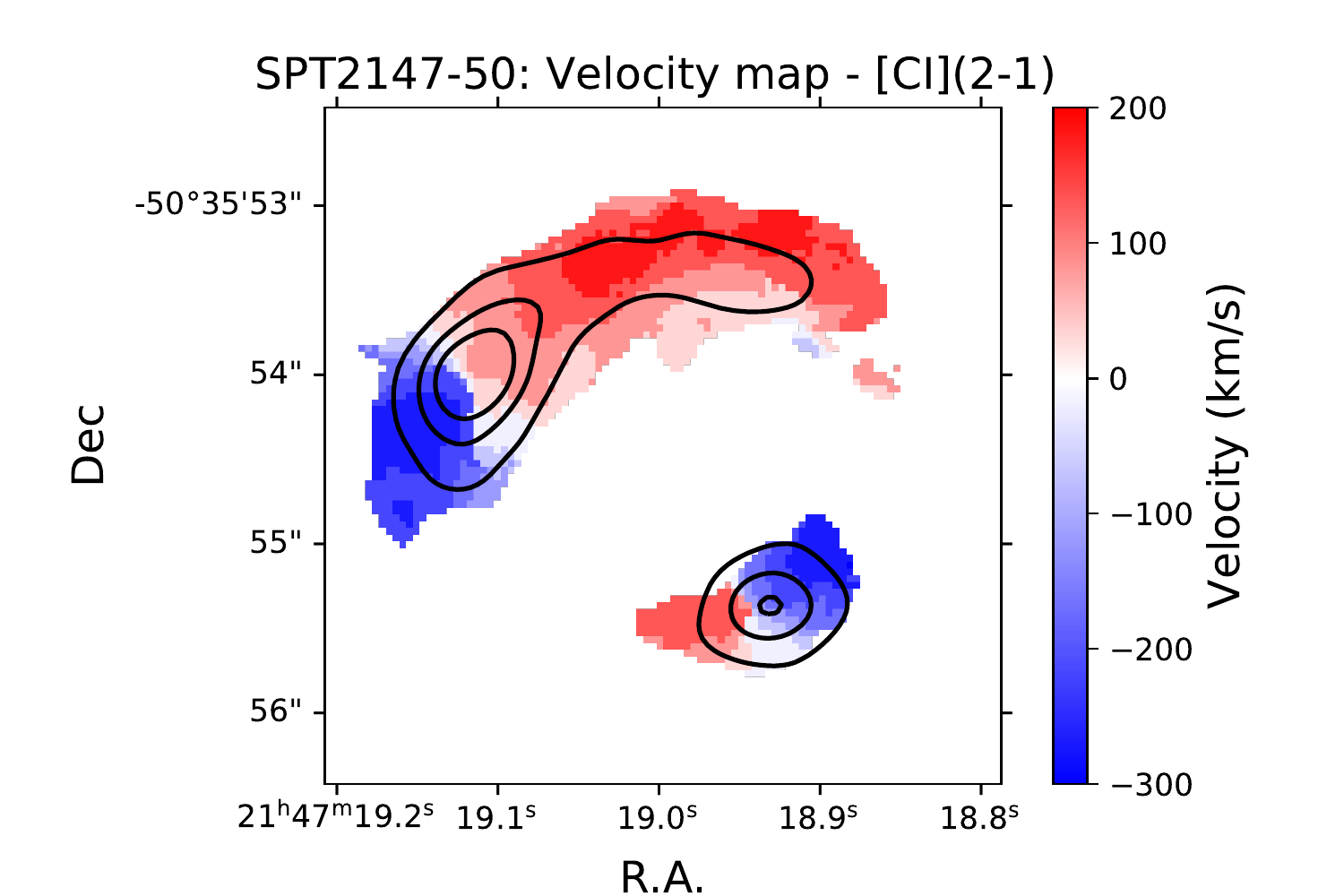} &  \includegraphics[width=8.5cm]{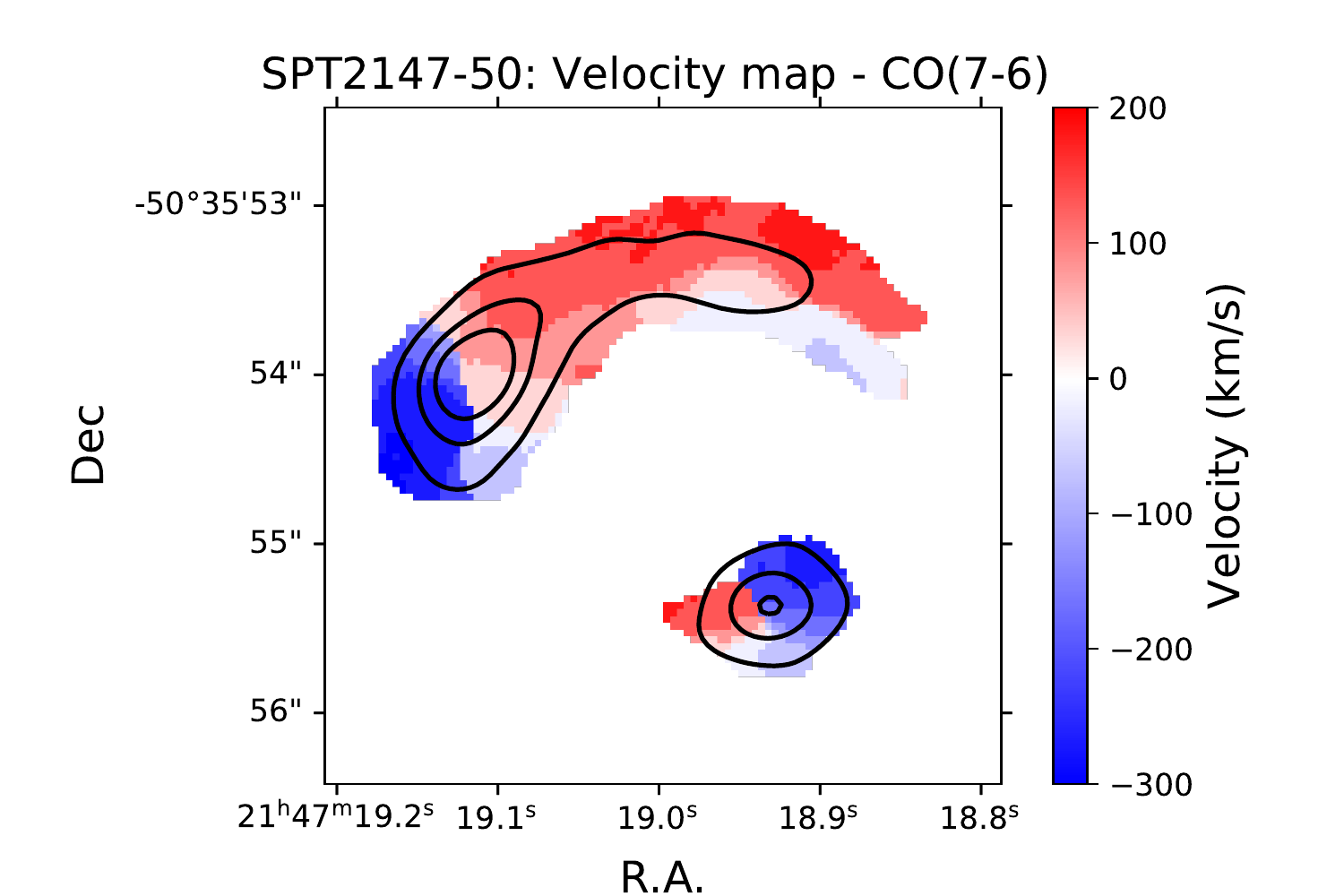} \\ 
\end{tabular}
\caption{\label{fig:vel map 2147} Pixel-wise peak velocity decomposition of SPT2147-50. The velocity maps of [CI](2-1) (left) and CO(7-6) (right) are shown. The 3, 5 and 10$\,\sigma$ contours of the continuum are overplotted on the velocity maps. The construction of these maps are described in Sect.\,\ref{Sec: kinematics_2147}}
\end{figure*}

\begin{figure*}[h]
\centering

\begin{tabular}{cc}
\includegraphics[width=9cm]{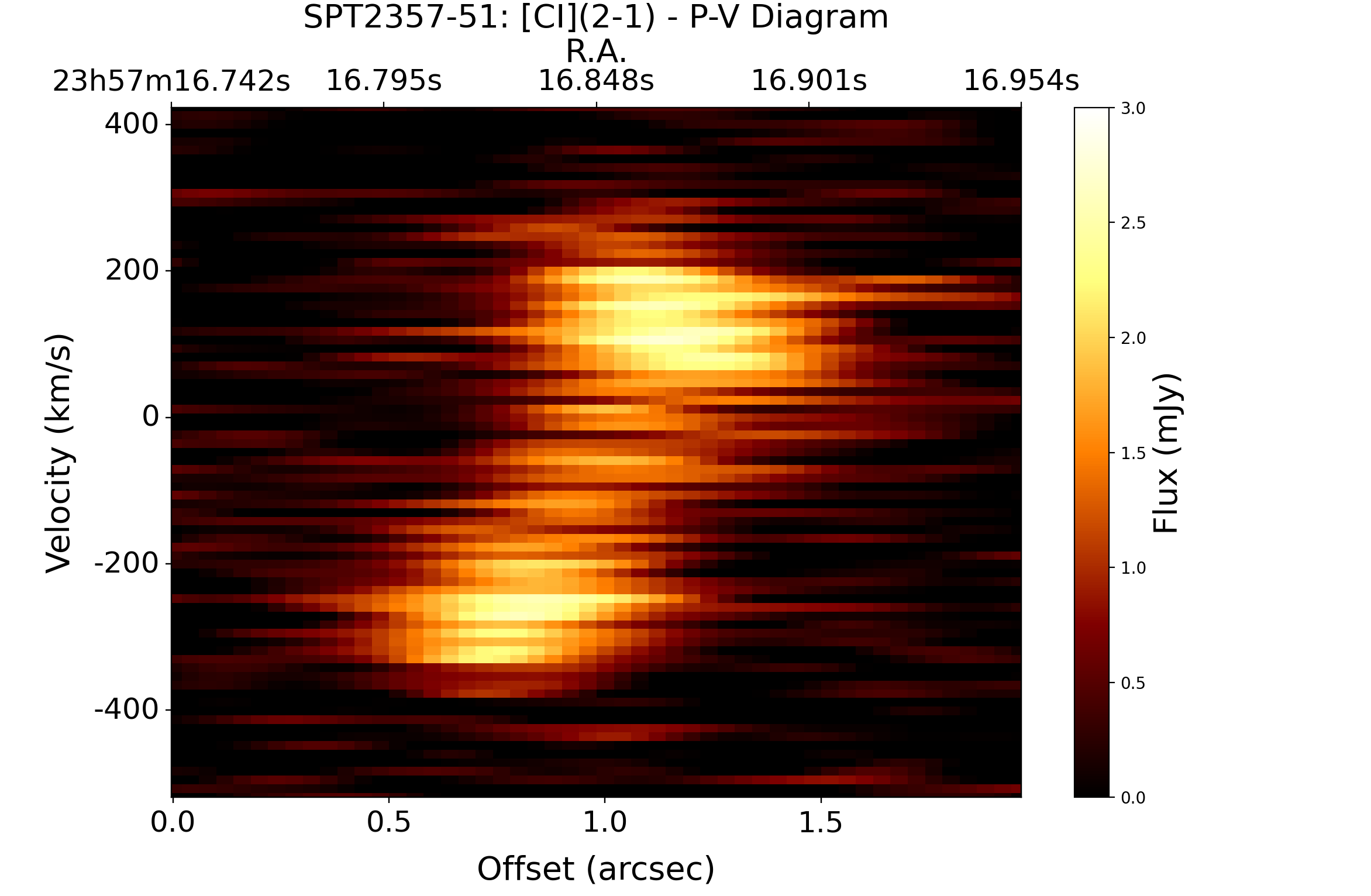} &  \includegraphics[width=9cm]{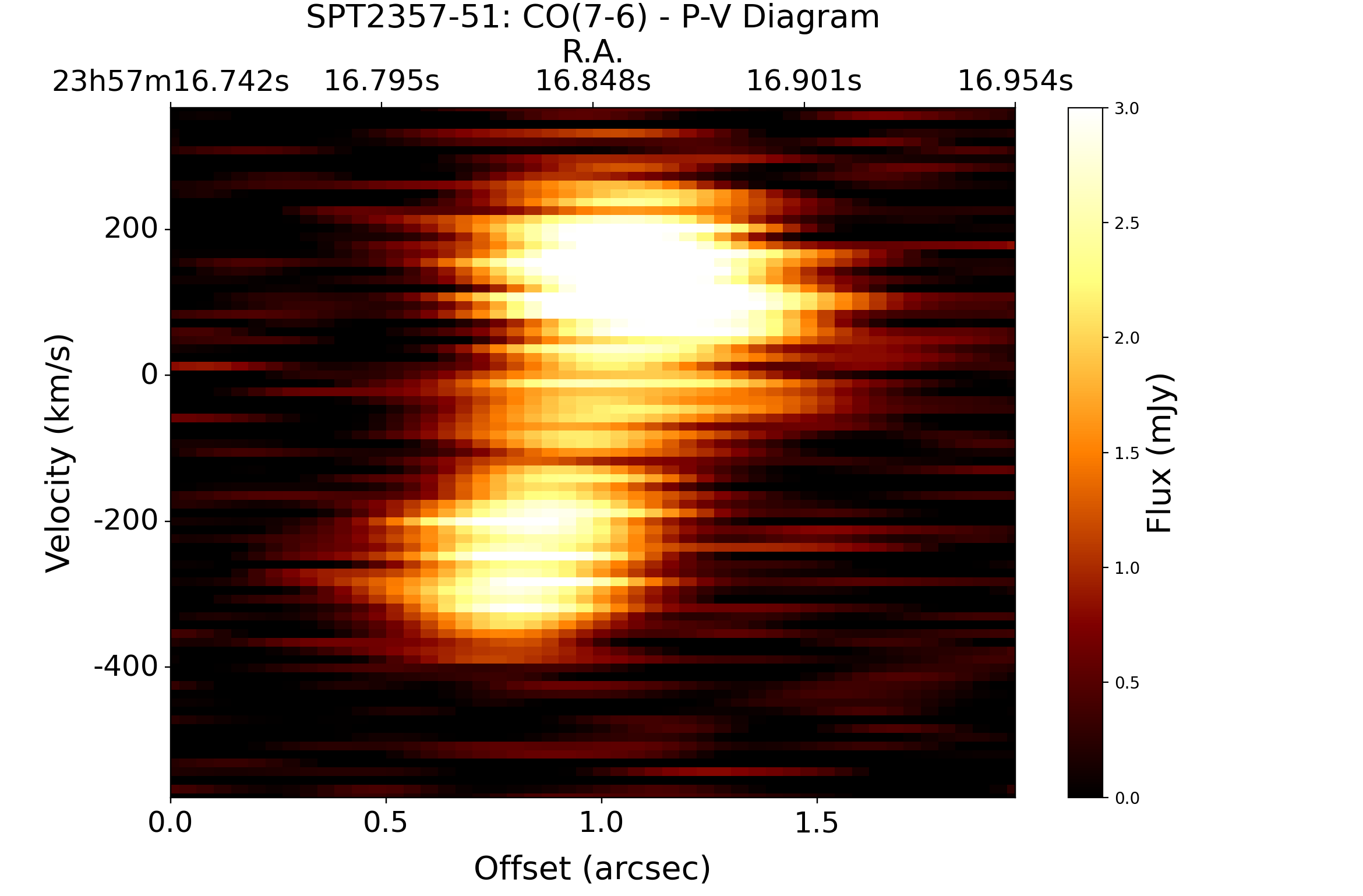} \\ 
\end{tabular}
\caption{\label{fig:pv_diag} The position-velocity diagram of the [CI](2-1) and CO(7-6) lines of SPT2357-51. The slice was taken along the R.A. axis.}
\end{figure*}

%-----------------------------------------------------------------------
%-------------------------------------------------------------------
\section{Image-plane kinematics \label{sec: kinematics}}

\subsection{SPT0103-45 \label{kinematics_0103}}

The two most intriguing features in this source are the high-velocity redshifted tail visible in both lines (see Fig.\,\ref{fig:int_spectra}, top panel) and the presence of a particularly bright continuum feature, which we call the `gem'. 

Such high-velocity tails are often seen as signatures of molecular outflows driven by AGNs \citep[e.g.,][]{Maiolino12,Cicone14,Florez21,Pantoni21} and/or supernovae \citep{Ginolfi20a,Spilker20a,Spilker20b}. However, we see high-velocity tails only in one direction which makes it unlikely to be a signature of an outflow. We further explore other explanations such as magnification effects that could result in distorting the line profiles (see Sect.\,\ref{sec: diff mag}).

To try to understand the nature of SPT0103-45, we extract the spectra at the pixel where the continuum flux density is maximal, i.e. at the position of the gem. Figure\,\ref{fig:bright blob} shows the spectrum at the position of the gem, clearly showing very broad and blended lines. The high-velocity tail is almost as bright as the peak of the emission near the systemic velocity. We also reconstruct the integrated intensity map of the high-velocity emission of CO(7-6) using the integration range shown in yellow in the center panel of Fig.\,\ref{fig:bright blob}. The 3, 5 and 10\,$\sigma$ contours of the continuum are also plotted on the high-velocity intensity map shown in the bottom-right panel of Fig.\,\ref{fig:bright blob}. This reveals a small spatial offset between the high-velocity regions and the brightest continuum region.

We then build the velocity map of the source. To avoid artifacts caused by magnification or high-velocity components, we mapped the peak velocity of each line for each line of sight instead of a moment-1 map. To do that, we extract the spectrum at each pixel and fit a linear spline to reduce the impact of the noise. We then identify the peak of emission and the velocity corresponding to it. We search for the lines in the 197.5--198.1\,GHz range  (727.1 to -257.5\,km\,s$^{-1}$) for [CI](2-1) and 196.5--197.38\,GHz (866.2 to -45.6\,km\,s$^{-1}$) for CO(7-6). These intervals were chosen to avoid cross-contamination between the two lines in the region, where low-velocity CO(7-6) and high-velocity [CI](2-1) could overlap. To ensure that this peak is real, we require that the maximal flux determined by the spline fit should be above 3\,$\sigma$. 

In Fig.\,\ref{fig:decomp 0103}, we show maps of the mean velocity of each line (i.e. moment-1 maps). To allow a better visualisation of the geometry of the source, the 3, 5 and 10\,$\sigma$ contours of the continuum are plotted on the velocity maps. For both lines, we observe a very similar velocity gradient from east to west. However, the high-velocity regions corresponding to the redshifted tail are all located west of the gem.

To further analyse the kinematics of the source, we construct position-velocity (PV) diagrams along the bright arc passing through the gem. The slice is indicated as the white line in the left panel of Fig.\,\ref{fig:bright blob}. The PV diagram is shown in Fig.\,\ref{fig:pv 0103}. The zero velocity corresponds to the observed frequency of the CO(7-6) line.  The PV diagrams of both lines are similar in profile. From a position 0 to 1.5\,arcsec, the velocity is almost flat followed by a quick increase between 1.5 and 2\,arcsec and finally a second plateau around $\sim$800\,km\,s$^{-1}$. The second plateau is much fainter, which could be due to differential magnification (see Sect.\,\ref{sec: lens modelling 0103}).

%----------------------------------------
\begin{figure*}[h]
\centering

\begin{tabular}{cc}
\includegraphics[width=9cm]{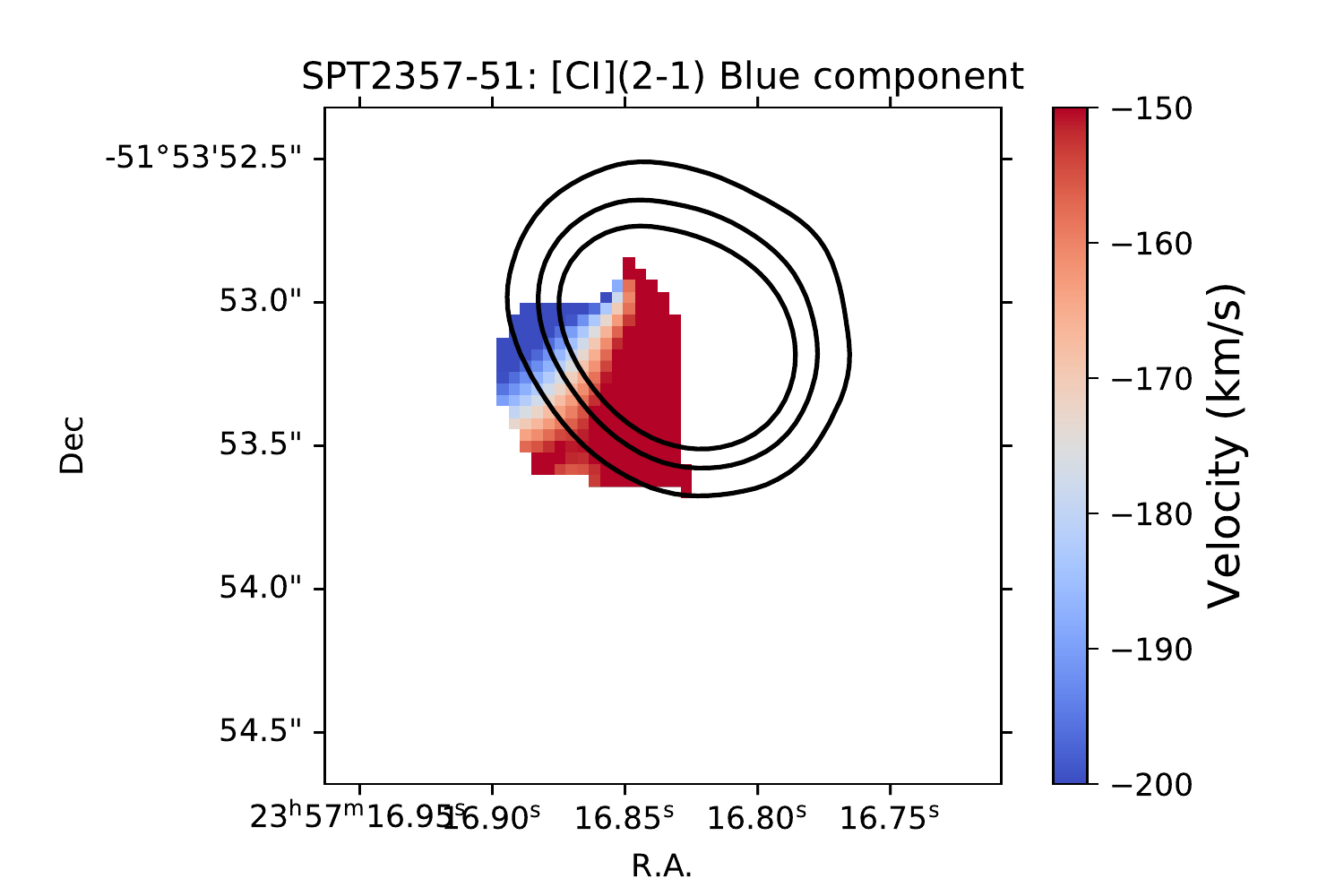} &  \includegraphics[width=9cm]{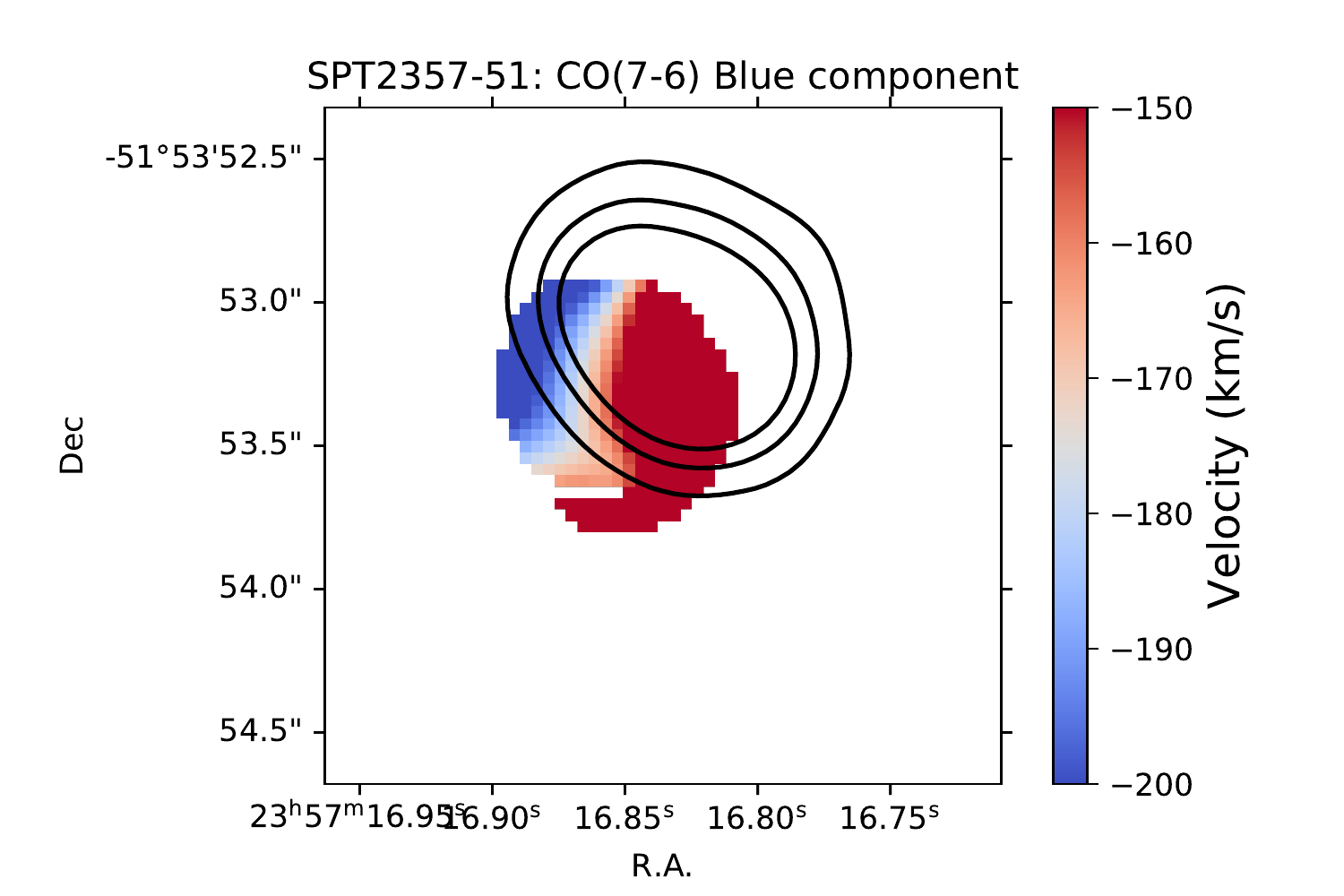} \\ \includegraphics[width=9cm]{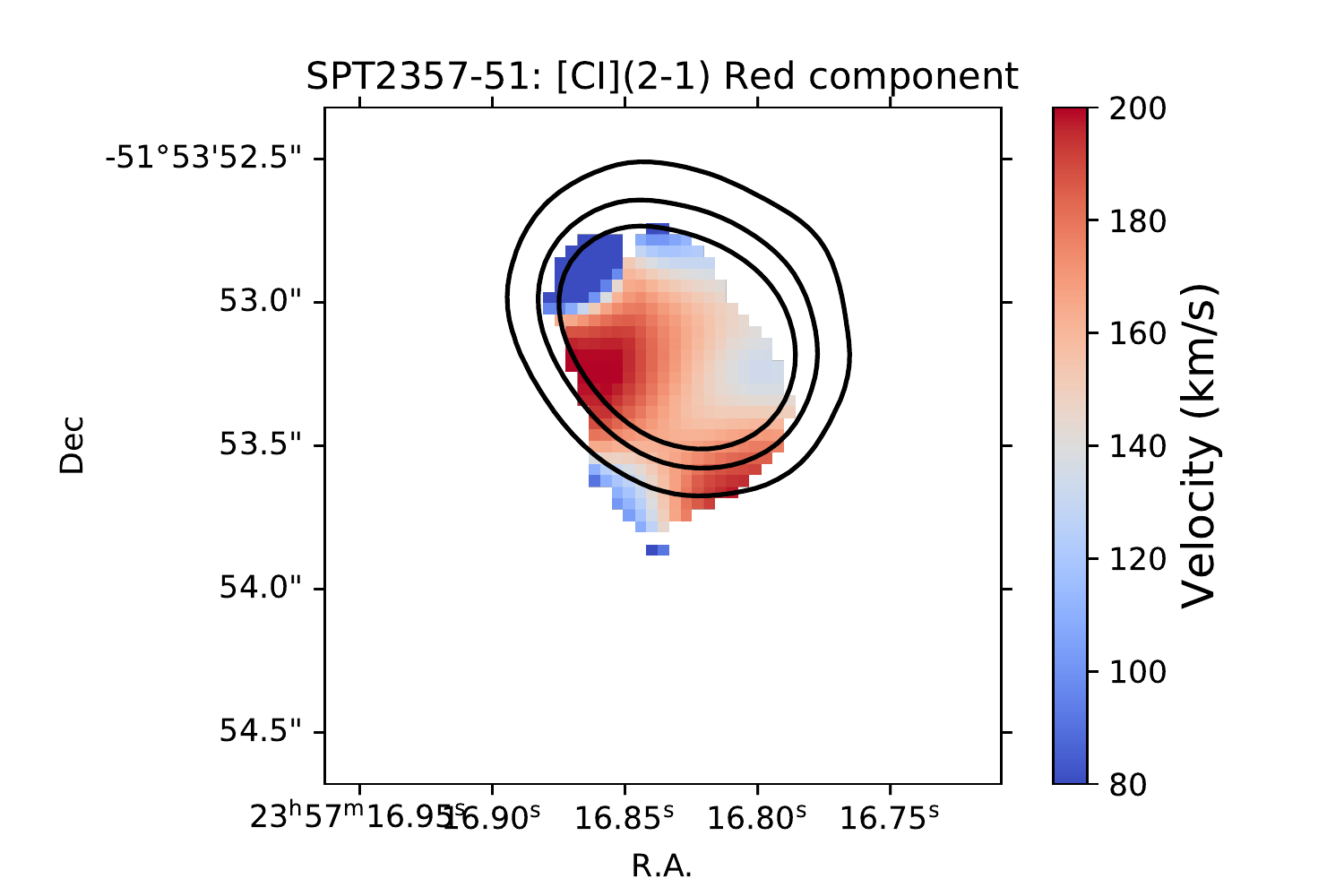} &
\includegraphics[width=9cm]{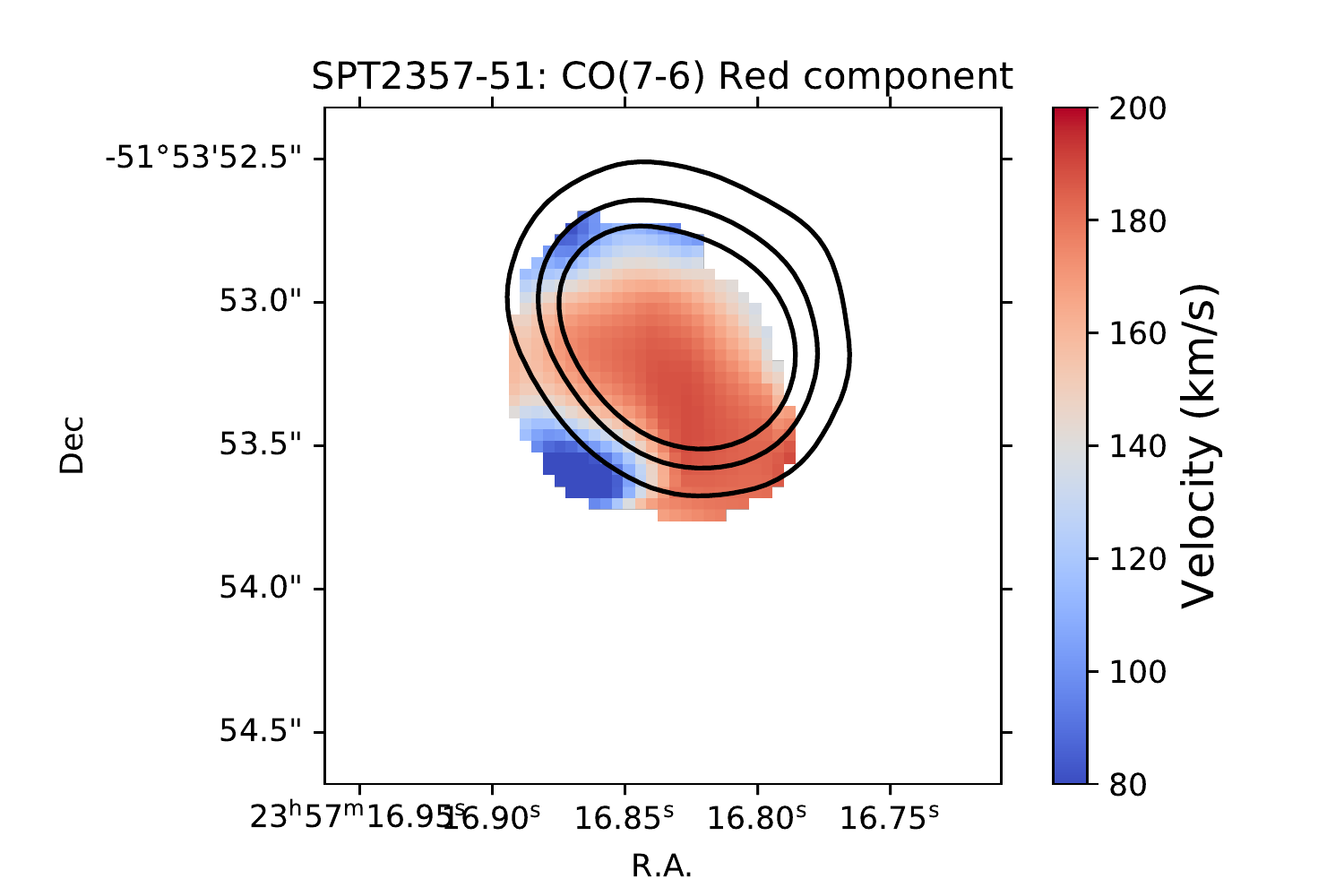} \\ 
\end{tabular}
\caption{\label{fig:decomp 2357}Decomposition of the CO(7-6) and [CI](2-1) lines in SPT2357-51. The central velocity of each component from the Gaussian fitting is shown above. The fitting procedure is described in Sect.\,\ref{sec:kinematics 2357}. The 3, 5 and 10\,$\sigma$ contours from the continuum map are also plotted on the maps. We see a smooth velocity gradient for the blue component, but no discernible pattern in the red component. }
\end{figure*}
%--------------------------------------------------------------------
\subsection{SPT2147-50 \label{Sec: kinematics_2147}}

The [CI](2-1) and CO(7-6) lines of SPT2147-50 are broad and asymmetric with more flux on the redshifted side (see Fig.\,\ref{fig:int_spectra}). To understand the origin of the asymmetric and broad lines, we construct a pixel-wise velocity map. 

Since the lines are not always symmetric even in a given pixel, we locate the line peak velocity using a similar approach as for SPT0103-45 (see Sect.\,\ref{kinematics_0103}). As only one single peak per line can be found along any given line of sight, we define a single velocity range from -400 to +400km\,s$^{-1}$ to search for the peak.

These peak velocities are then used to construct the velocity maps seen in Fig.\,\ref{fig:vel map 2147}. We also plotted the 3, 5 and 10\,$\sigma$ contours of the continuum emission to better visualize the geometry of the source. The velocity maps measured independently for each line are strikingly similar. The peak velocity varies from $-$300 to 150 km\,s$^{-1}$. We observe a smooth velocity gradient from east to west in the northern arc and the opposite in the south arc, as expected for a gravitationally-lensed arc and counterimage (see further discussions in Sect.\,\ref{sec: lens modellling 2147}).

%--------------------------------------------------------------------

%--------------

\begin{figure*}
    \centering
    \begin{tabular}{c}
    \includegraphics[width=18cm]{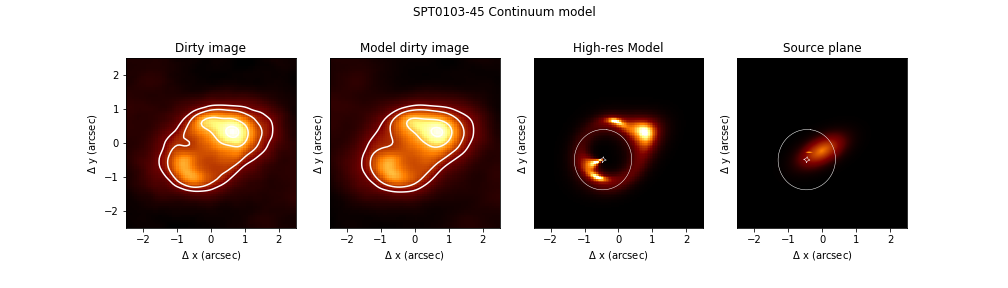}\\
    \includegraphics[width=18cm]{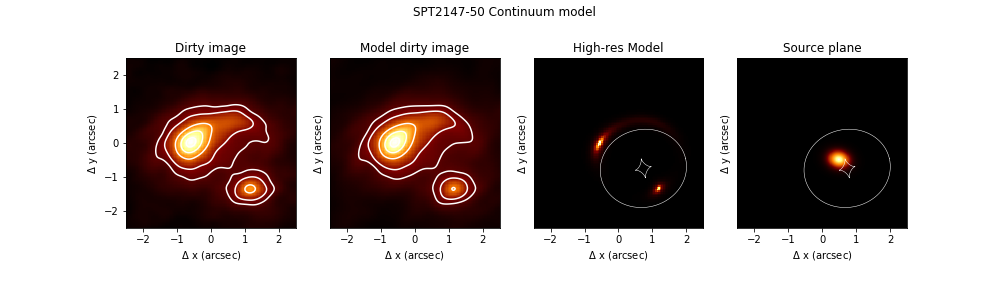} \\
    \includegraphics[width=18cm]{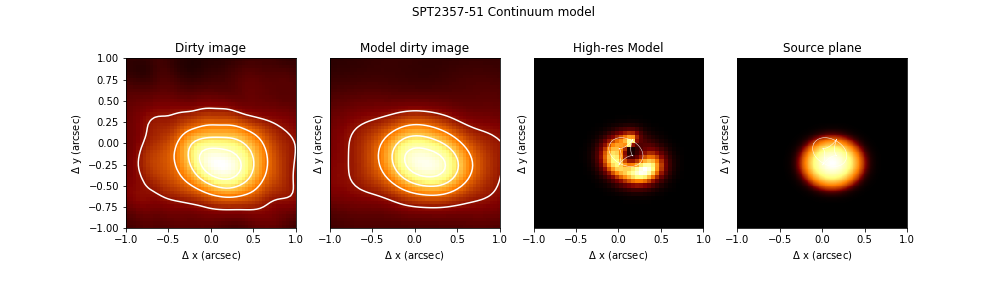}
    \end{tabular}
    
    \caption{\label{fig:cont lens } Modelling the continuum emission of our sources. SPT0103-45 is in the first row, SPT2147-50 in the second row and SPT2357-51 in the third row. For every source, the first panel (from the left) is the dirty image of the continuum, the second is the dirty image produced by best-fit model, third is a `high-resolution' image-plane model, i.e. not sampled by the $uv$ coverage of the data and the fourth panel is the best-fit model in the source plane. In the third and fourth panels, the lensing caustics are shown in white.}
    
\end{figure*}
%-----------------------------
\subsection{SPT2357-51 \label{sec:kinematics 2357}}
%pv diag

The line profile of both lines has a doubly-peaked profile (see Fig.\,\ref{fig:int_spectra}). A similar profile was observed for the [CII] line of SPT0346-52, which has been identified as a major merger \citep{Litke18}. To assess whether SPT2357-51 may also be a merging system, we first examine the position-velocity diagram extracted along the right ascension axis (Fig.\,\ref{fig:pv_diag}), chosen to maximize the spatial separation between the two components. It shows two clearly-separated components: a western component at blueshifted velocities (henceforth, the blue component) and an eastern component at redshifted velocities (henceforth, the red component). Their peaks of emission are separated spatially by about $\sim$1$^{\prime \prime}$ and spectrally by $\sim$400km\,s$^{-1}$. 

%Gaussian decomposition
To characterise both components seen in the position-velocity diagram, we performed a pixel-wise decomposition of the spectra into two Gaussian components for each line.

We extract the continuum-subtracted spectrum for each pixel and fit it with four Gaussian profiles (two for each line). To ensure a physical and meaningful result, we applied several constraints. The amplitude is forced to be positive, since we do not expect to see CO(7-6) or [CI](2-1) in absorption in such a system. The line width is allowed to vary from 0 to 250km\,s$^{-1}$ to avoid unphysically broad components. Finally, we define a velocity range associated with each component in order to avoid overfitting one component with two Gaussians if one component is very faint. The bounds for the central velocity of the peaks were defined from the velocity separation seen in the position-velocity diagram. The red component is fitted in a velocity range between 0 and +250km\,s$^{-1}$ and the blue component between $-$150 and $-$300km\,s$^{-1}$ for both [CI](2-1) and CO(7-6) lines. 

To derive uncertainties on the amplitude, width and position of our Gaussians, we measure the noise in each channel and take it into account with the fitting tool (\texttt{curvefit} from the \texttt{scipy} package). In the rest of the analysis, we only consider results for which the significance of the amplitude is higher than 3$\,\sigma$.  The residuals are below 2.5$\,\sigma$ for every fit.

Figure\,\ref{fig:decomp 2357} shows the velocity map for each of the components of the [CI](2-1) and CO(7-6) lines along with the 3, 5 and 10$\,\sigma$ contours from the continuum map. The velocity gradients obtained for CO(7-6) and [CI](2-1) are remarkably similar. The blue component shows a smooth velocity gradient suggestive of rotation. The velocity structure of the red component is less clear, showing essentially no coherent structure. 

The continuum is centered on the red component while the blue component is an offset from the peak of the dust continuum. Since these two components have similar apparent line luminosities, this could indicate that the red component has a larger dust emission and/or stronger obscured star formation. This could originate from different gas metallicities or temperatures between the two components. This difference of ISM properties between the two velocity components is a clue that this object could be a merger between two galaxies. It is unlikely that differential magnification could cause the low dust emission of the blue component. Lensing is a purely geometric effect, so any given individual region must have the same magnification for the lines and the continuum. Further discussion on the geometry of this source based on lens modelling will be done in Sect.\,\ref{fig:line lens 2357}. 

%--------------------------------------------------------------------
%-------------------------------------------------------------------
\section{Lens modelling \label{sec: lens_modelling}}

To understand the source-plane morphology and derive the intrinsic properties of the galaxies in our sample, we perform lens modelling of both the continuum and CO/[CI] line emission. Modelling spectral velocity channels separately allows us to characterise the dynamics of these systems and quantify potential differential magnification effects both across the line profiles and in comparison to the continuum \citep{Serjeant12,Hezaveh12,Paraficz18,Dong19}. 

\subsection{Method}

We use the lens modelling code \texttt{visilens} from \citet{Spilker16}. It directly models Fourier-plane interferometric visibilities instead of the inverted image plane to limit the biases caused by correlated noise generated by the imaging algorithm. This also helps us better account for the antenna delay and the mismatched absolute flux calibration that could arise due to combining multiple observations taken on different days, which is the case for SPT2147-50. The angular resolution of our data does not allow us to make a reliable pixelated reconstruction of the source-plane.

Our  model describes the foreground lens with one singular isothermal ellipsoid (SIE) profile. The profile is parametrised by the following free parameters: the position of the lens relative to the phase center ($x_L$,$y_L$), the angular Einstein radius $\theta_{E,L}$, the ellipticity $e_L$ and the position angle of the major axis $\phi_L$. The background sources are modelled with \citet{Sersic68} and/or Gaussian profiles, for both the continuum emission and each spectral channel. The position of the source ($x_S$,$y_S$) is defined relative to the lens position. The other parameters are total flux density ($S$), effective half-light major axis ($a_S$), axis ratio ($b_S$/$a_S$), Sérsic index ($n_S$) and position angle ($\phi_S$). 
We note that models with substantially more freedom in the source morphology, such as a pixellated source plane \citep[e.g.,][]{Dye15,Hezaveh16} are not warranted for our moderate-resolution data. \texttt{Visilens} uses a Monte Carlo Markov chain (MCMC) fitting procedure to sample the parameter space \citep{Spilker16}. The best-fit parameters are then determined after the chains converge. We finally produce model dirty images, high-resolution models and source plane models from the best-fit model (see Fig.\,\ref{fig:cont lens }).

\begin{figure}[h]
\centering

\includegraphics[width=9cm]{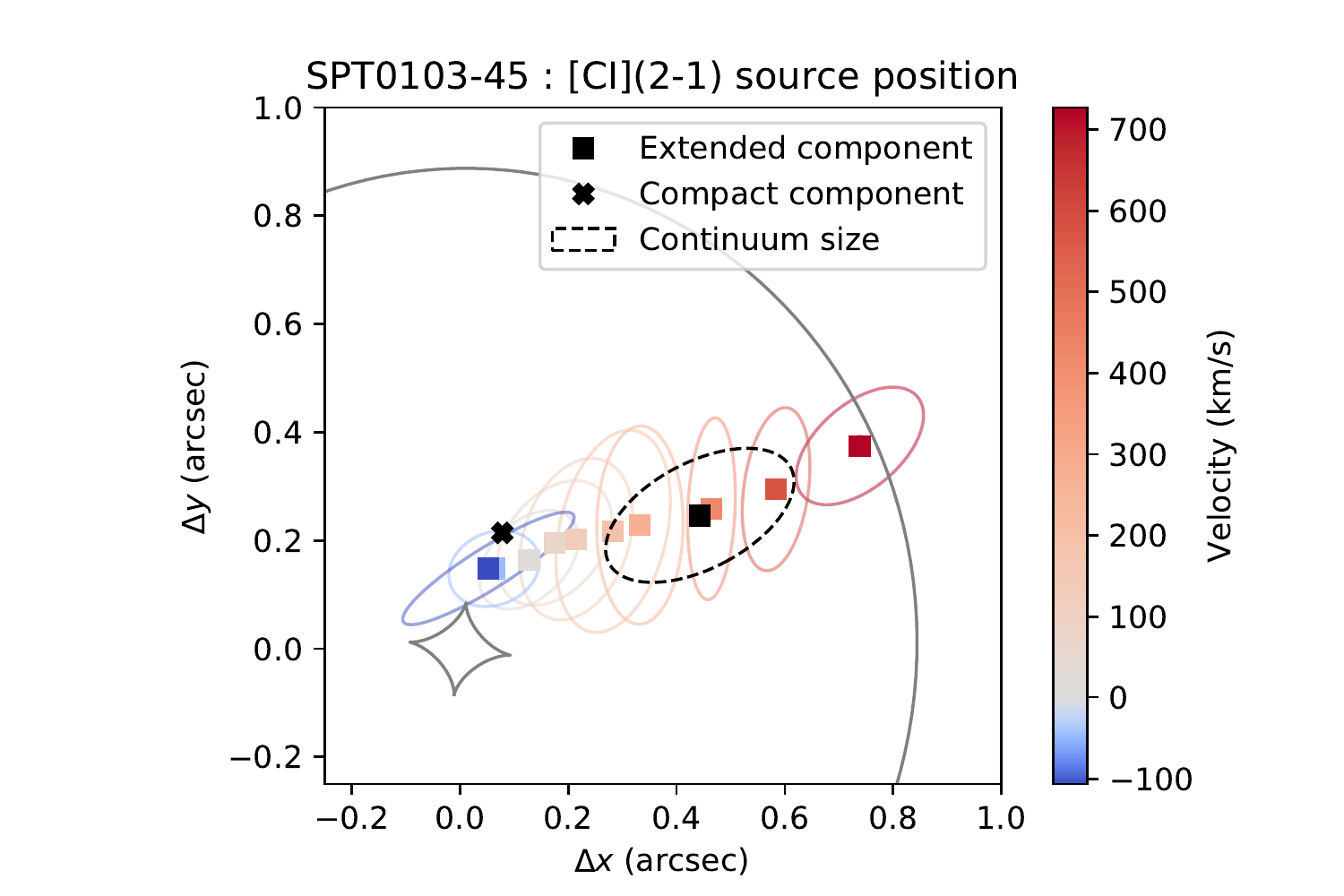}   \includegraphics[width=9cm]{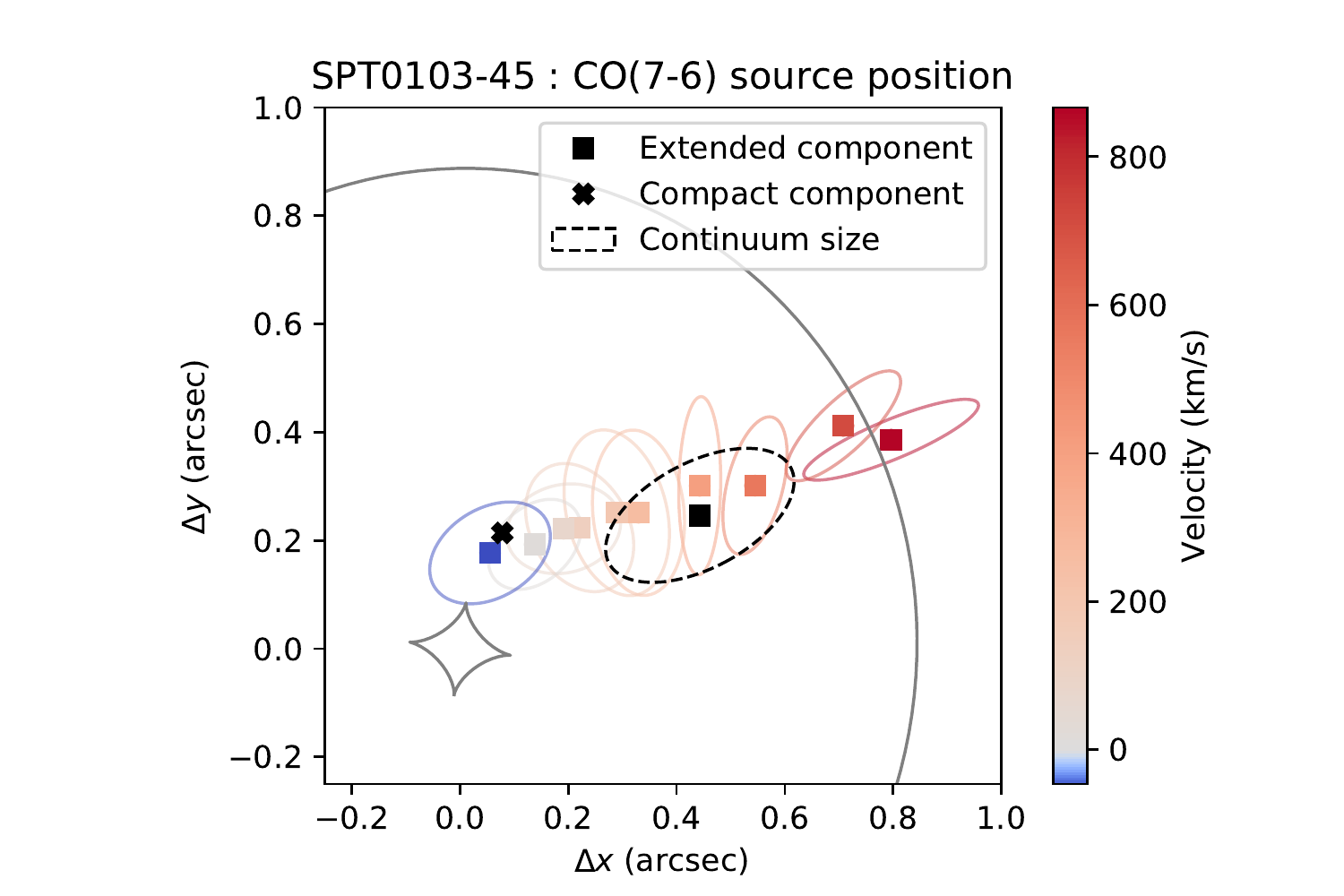} 

\caption{\label{fig:line lens 0103}Plot of the best-fit source positions of the channelized models of [CI](2-1) (top) and CO(7-6) (bottom) in SPT0103-45. The velocity scaled is the central velocity of each of the bins. The black square and cross represent the positions of the extended and compact component from the continuum model, respectively. The ellipses represent the half-light ellipses of each of the model. The lensing caustics are shown in grey.  }
\end{figure}

\subsection{Continuum modelling \label{sec:continuum_lens}}
%------------------------------------------------------------------
To perform the lens modelling of the continuum, we select only the data from the line-free spectral windows. We then average the two orthogonal polarisations. We choose a first estimate of the lens and source parameters using the previous modelling performed on low-resolution data by \citet{Spilker16} and define loose reasonable priors (e.g., positive fluxes, $0<n_S<4$). We use a single component for the source profile of SPT2147-50 and SPT2357-51 but two components for SPT0103-45, as the model gave us lower residuals. We use 50 chains of 10\,000 usable steps and 1000 burn-in steps to run the MCMC chains and verify the chains converged. We also check that the marginalized parameter distributions do not return a high probability for values close from the prior limits. The best-fit lens parameters for all our sources are given in Table\,\ref{tab:lens_params}. These parameters are in agreement with the best-fit lens parameters from \citet{Spilker16}.

%-----------------------------------------------------------------
\subsection{Line modelling}

In order to model the line emission of [CI](2-1) and CO(7-6), we use the continuum-free visibilities described in Sect.\,\ref{sect:line_imaging}.  We divide the lines into frequency bins with widths that ensure that the line emission is detected at a peak S/N$\gtrsim$8 in each channel. For SPT2147-50 and SPT2357-51, we use 50\,MHz bins ($\sim$80\,km\,s$^{-1}$). For SPT0103-45, we use 100\,MHz bins ($\sim$150\,km\,s$^{-1}$) for the redshifted tail where the emission is faint and narrower bins of 40\,MHz width ($\sim$50\,km\,s$^{-1}$) at the line peak where the S/N is high. 

Since the S/N of the continuum is much higher than any of the lines, we simply fix the parameters of the lensing potential for all velocity channels of each source to those found in the continuum modelling. The source parameter starting values are also derived from the continuum analysis, but these parameters are allowed to vary. We use 40 chains of 10\,000 usable steps and 1000 burn-in steps for each of the runs. We verify that the chains converged for every run. The best-fit parameters for the line modelling of SPT0103-45, SPT2147-50 and SPT2357-51 are listed in Table\,\ref{source_params_0103}, Table\,\ref{tab:source params 2147} and Table\,\ref{tab:source params 2357} respectively.

%----------------------------------
\subsection{SPT0103-45\label{sec: lens modelling 0103}}

%-------------------------------------------------------------

%continuum modelling

%\textcolor{red}{MODELLING NEED TO BE RE-RUN. PLEASE IGNORE THIS SUBSECTION...}

For the continuum of SPT0103-45, a single source component was insufficient to properly fit the visibilities, thus we use two S\'ersic components to model the source. The best-fit model converged to a solution with an extended component ($n_S = 0.994 \pm 0.011 $, see Table\,\ref{source_params_0103}) and a second very compact component ($r_{eff} = 0.055 \pm 0.002$ arcsec). The maximum continuum residuals were $\sim$8.3\,$\sigma$, approximately 4\% of the peak S/N in the data. This illustrates how difficult it is to model our very high S/N continuum data even with two components and suggests that a future approach with more freedom in the source plane (e.g. pixellated modelling) could be more successful. The dirty image of the data and best-fit model are shown in Fig.\,\ref{fig:cont lens } (top-right panel) together with a high-resolution model (neglecting the Fourier-plane sampling) and the source-plane emission. The best-fit parameters of both the components are given in Table\,\ref{source_params_0103}. The intrinsic flux ratio of these two components is about 17, with the extended component being brighter. The nature of this second component is unclear and we cannot conclude based on our data if it is a clump in the disk, a compact neighbour, or even an artifact of the model to mimic an asymmetry of the disk in the source plane. 

To model the CO(7-6) and [CI](2-1) line data in each velocity bin, a single S\'ersic source was sufficient. We start our MCMC chains from the best-fit parameters of the extended continuum component. The highest residuals are < 5$\,\sigma$, except for one velocity bin (centered at $\sim 136$ km\,s$^{-1}$) for which the maximum residual was $\sim 7\,\sigma$. In Fig.\,\ref{fig: counter images CI 0103} and Fig.\,\ref{fig: counter images CO 0103}, we compare the best-fit model and the observed data for the [CI](2-1) line and the CO(7-6) line, respectively.

In Fig.\,\ref{fig:line lens 0103}, we plot the best-fit source position and size for every velocity bin for each of the lines. We can see a smooth velocity gradient across the source, consistent for both the lines. Emission at blueshifted velocities is close to the inner diamond caustic, while the most redshifted emission is near the outer caustic. The results thus favor a scenario in which SPT0103-45 is a rotating disk.

\begin{figure}[h]
\centering

\includegraphics[width=9cm]{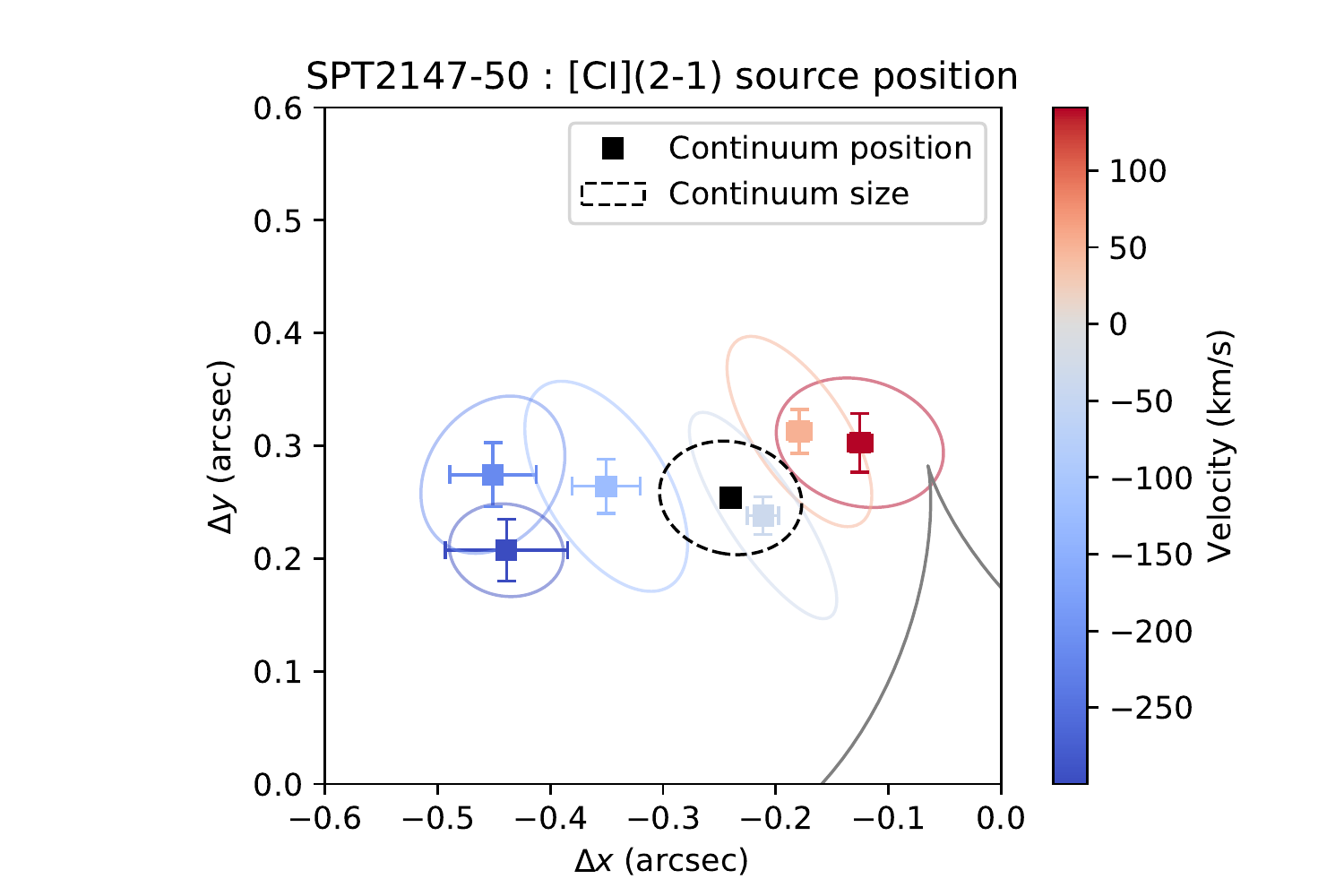}   \includegraphics[width=9cm]{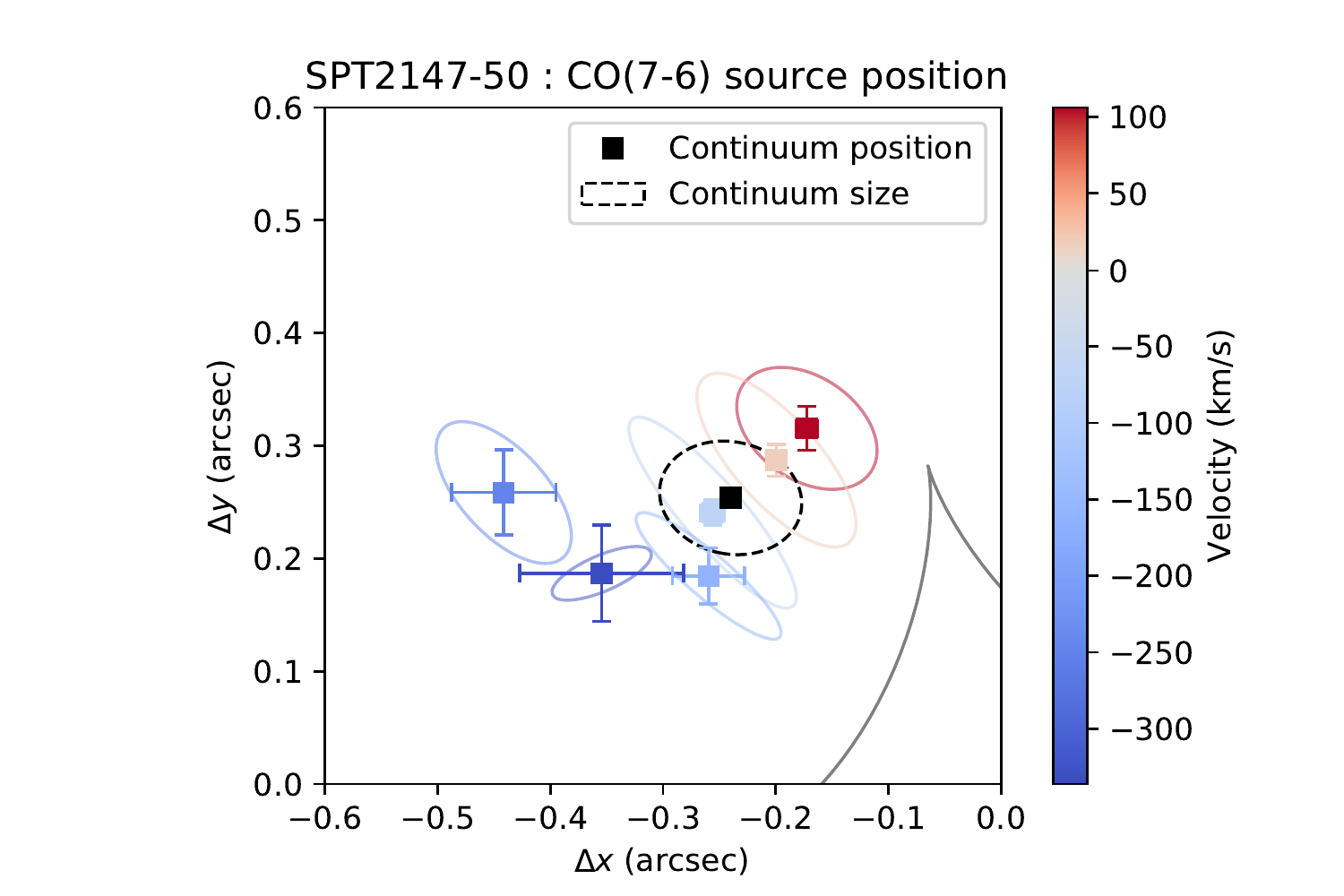} 

\caption{\label{fig:line lens 2147}Plot of the best-fit source position of SPT2147-50 for both [CI](2-1) (top) and CO(7-6) (bottom). The velocity scale is the central velocity of each of the bins. The black square represents the position of the continuum. The ellipses represent the half-light ellipses of each model. The lensing caustics are shown in grey.}
\end{figure}

%----------------------------------
\subsection{SPT2147-50\label{sec: lens modellling 2147}}
%-------------------------------------

In the case of SPT2147-50, the source was observed on three different days (see, Table\,\ref{tab:obs_details}). At each step of the MCMC code, we separately compute the likelihood of the data corresponding to each day and then combined them. To take into consideration the different atmospheric conditions and the small calibration offsets that could occur in these observations, we introduce two multiplicative amplitude re-scaling as free parameters for the latter two observations. In simple terms, these parameters account for the fact that SPT2147-50 is detected with S/N$\gg$20 while the absolute flux scale of ALMA data is quoted by the observatory to be accurate at the $\sim5-10$\% level.

For both the continuum and the line models, we use a single component with a Sérsic profile. This relatively simple model is sufficient to fit our source. The continuum residuals are lower than 1.5$\,\sigma$ and the residuals for the line modelling are always below 4$\,\sigma$ in any velocity channel. In Fig.\,\ref{fig:cont lens } (middle row), we show the continuum dirty image and the modelled dirty images together with the high-resolution model and the source plane model. The comparison between the model and the dirty image in each velocity channel is presented in Appendix\,\ref{fig: counter images CI 2147} and Appendix\,\ref{fig: counter images CO 2147}.

The best-fit value of the Sérsic index for the continuum is 1.36$\pm$0.05 (see Table\,\ref{tab:source params 2147}). This index, close to 1, is compatible with the hypothesis that the object is a rotating disk (see also Sect.\,\ref{Sec: kinematics_2147}). The best-fit parameters for the line and continuum modelling are presented in Table\,\ref{tab:source params 2147}. The source plane model (see Fig.\,\ref{fig:cont lens }) shows that part of this disk-like profile is near the inner caustic. Significant differential magnification effects (see Sect.\,\ref{sec: diff mag}) can thus be expected.

In Fig.\,\ref{fig:line lens 2147}, we present the source-plane reconstruction of the CO and [CI] line emission. We see a clear shift of the source position  with velocity. The continuum center of emission (in black) coincides with velocities nearest the systemic velocity. The eastern region is blue shifted, while the western region near the caustic is redshifted. The gradient is similar for both lines. Overall, the center of emission regions at the various velocities are well aligned. This provides an additional evidence that the source could a rotating disk. 

%----------------------------------

\begin{figure}[h]
\centering

\includegraphics[width=9cm]{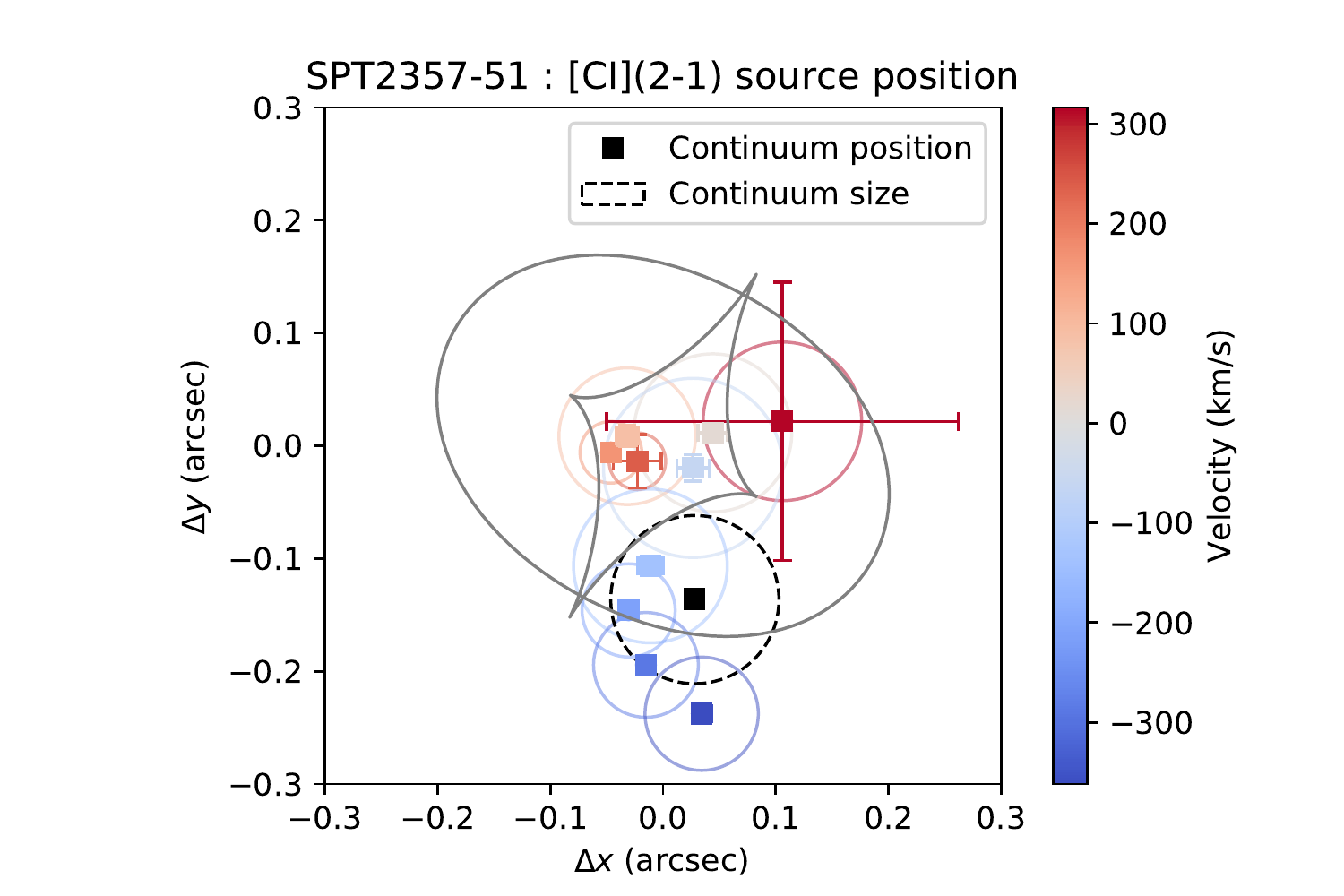}   \includegraphics[width=9cm]{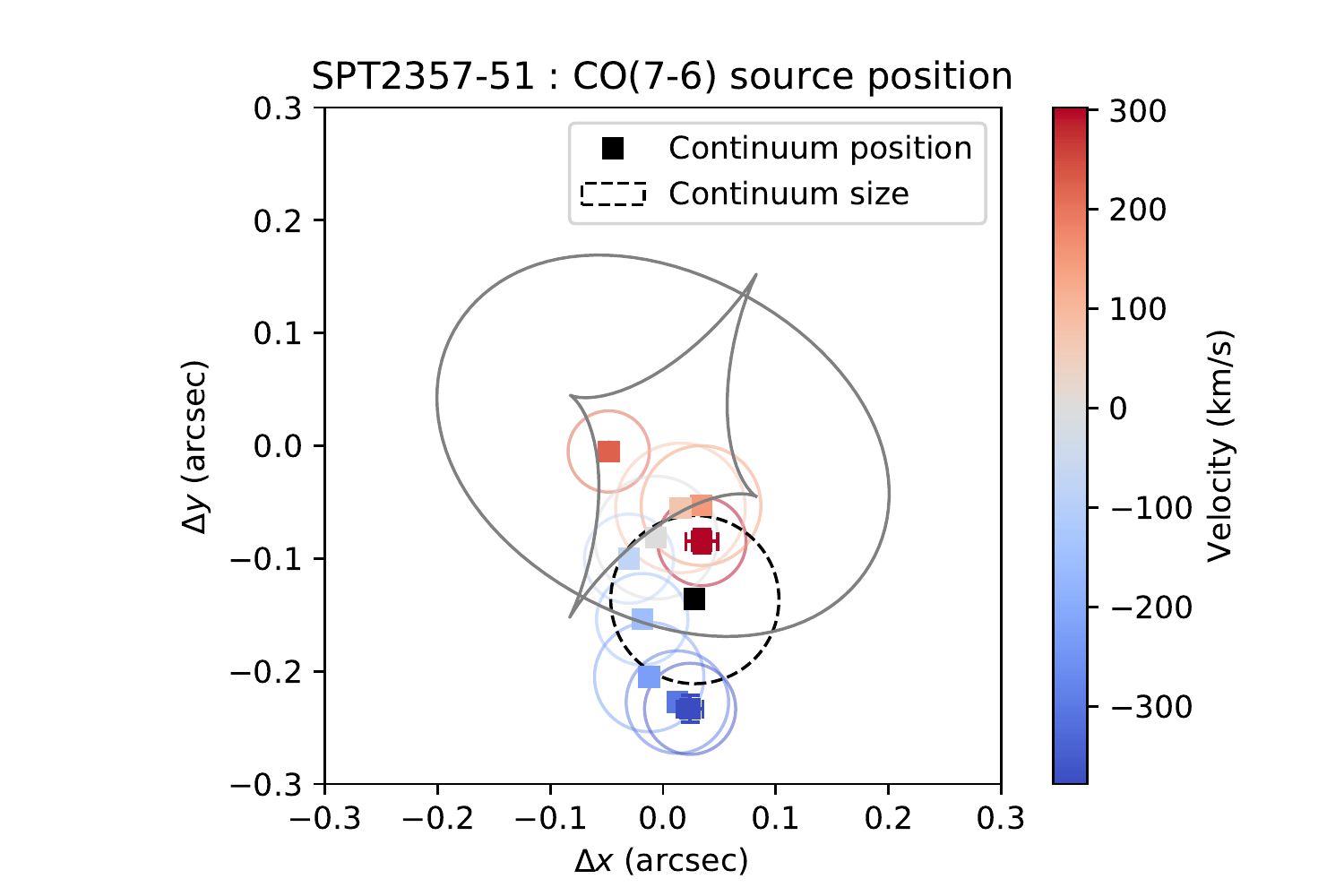} 

\caption{\label{fig:line lens 2357} Plot of the best-fit source position of SPT2357-51 for both [CI](2-1) (top) and CO(7-6) (bottom). The velocity scale is the central velocity of each bin. The black square represents the position of the continuum. The ellipses represent the half-light ellipses of each model. The lensing caustics are shown in grey. }
\end{figure}

%---------------------------------------------------------------------------------
\subsection{SPT2357-51 \label{sec: lens modelling 2357}}
%---------------------------------------------------------------

%---------------------------------------------------------------
%cont modelling

For SPT2357-51, the decomposition of the spectrum along the various line of sight was suggestive of a two-component system (see Sect.\,\ref{sec:kinematics 2357}). However, a single source-plane Gaussian component was sufficient to model the data and gave low residuals (1.72$\,\sigma$), likely due to the low spatial resolution of our data and the compact lensing geometry. The best-fit parameters of the continuum lens model are presented in Table\,\ref{tab:source params 2357}. In Fig.\,\ref{fig:cont lens } (lower panels), we present the results of our best-fit continuum model.

The same source model is used in each velocity bin. The largest residuals for each of these bins are  $<$ 5$\,\sigma$.  In Fig.\,\ref{fig:line lens 2357}, we plot the position of the source for different velocities. The geometry of this source is more disturbed than the two other sources. While, there is a velocity gradient for the velocities corresponding to the blue component, the geometry associated to the red component ($>$0\,km s$^{-1}$) is much more disturbed. The kinematics from our source-plane reconstruction are only marginally consistent with the expectation for rotation, though it is hard to constrain the emission region for the red component because the signal is emitted very close to caustics. Similar to the image-plane kinematic analysis (Sect.\,\ref{sec:kinematics 2357}), we also find that the continuum is offset from the main region of gas emission. Our source-plane reconstruction thus favors the scenario of a disturbed kinematics possibly linked to a merger.

\begin{table}[]
\caption{\label{tab: diff mag}Differential magnification. The effective magnification of the [CI](2-1) and CO(7-6) lines compared with the continuum magnification for all our sources. For SPT0103-45 continuum, the flux-weighted average magnification of the extended and compact component is given. The effective magnification of the lines is taken to be the flux-weighted average magnification summed over all velocity channels.}

\begin{tabular}{cccc}
\hline
\hline
&&&\\
Source&$\mu_{\rm eff}$&$\mu_{\rm eff}$&$\mu$\\
&[CI](2-1)&CO(7-6)& Continuum\\
&&&\\
\hline
&&&\\
SPT0103-45 & 6.2$\pm$0.2& 5.6$\pm$0.1& 4.7$\pm$0.6\\
SPT2147-50& 6.7$\pm$0.3& 6.9$\pm$0.4 & 6.9$\pm$0.2 \\
SPT2357-51& 3.1$\pm$0.2& 3.3$\pm$0.1& 2.8$\pm$0.4 \\
\hline
\end{tabular}

\end{table}

\begin{figure}
\centering

\includegraphics[width=9cm]{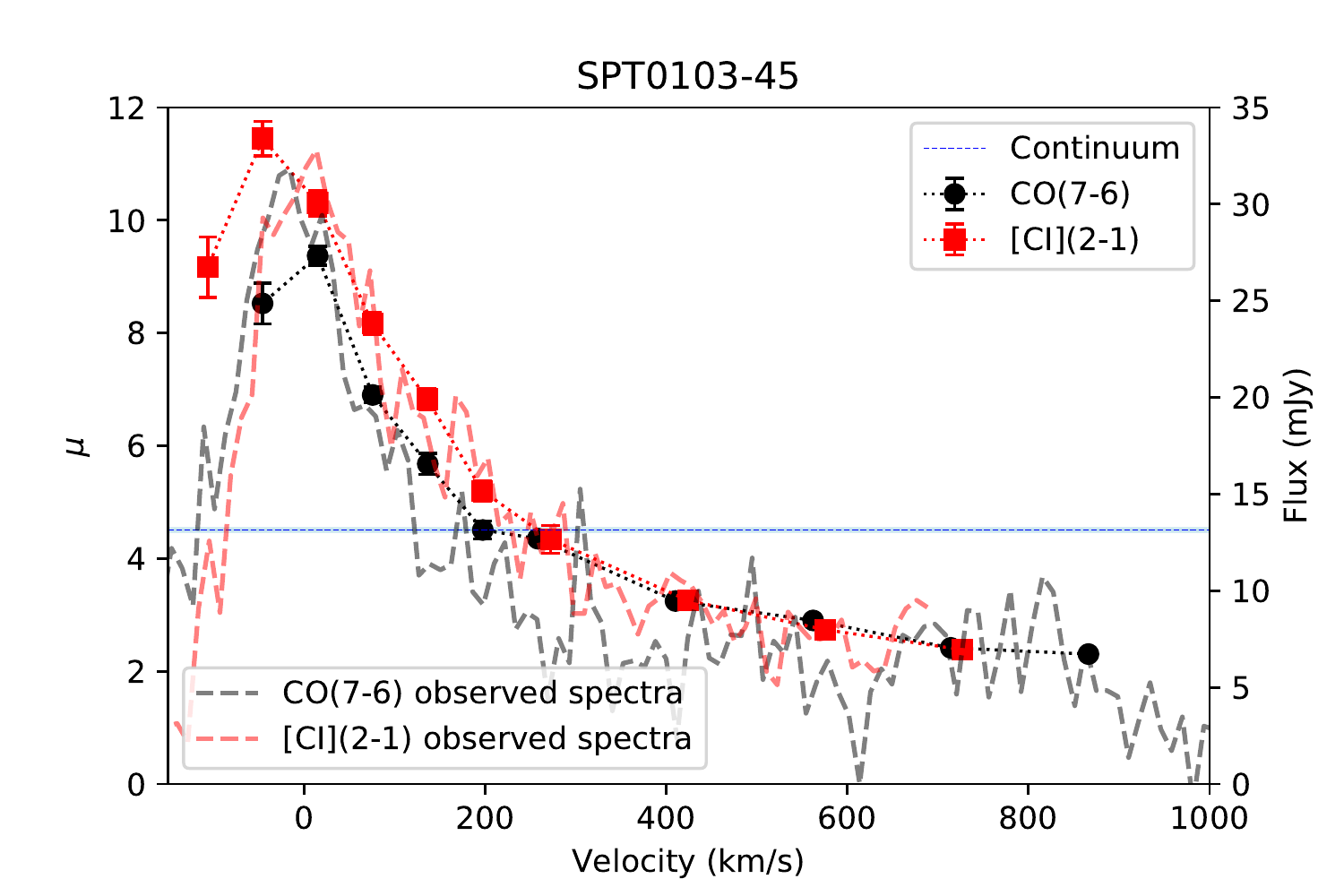}  \includegraphics[width=9cm]{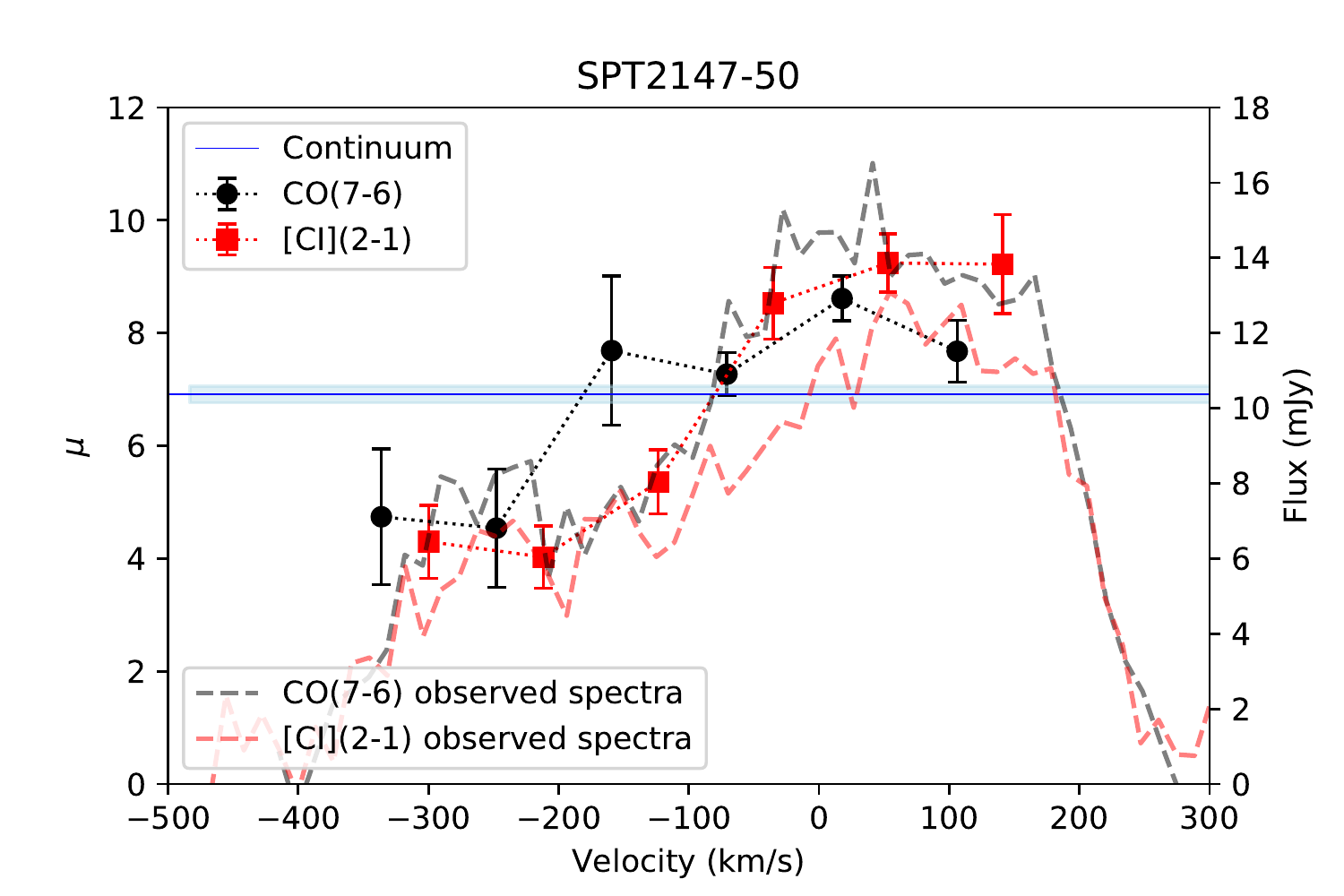} 
\includegraphics[width=9cm]{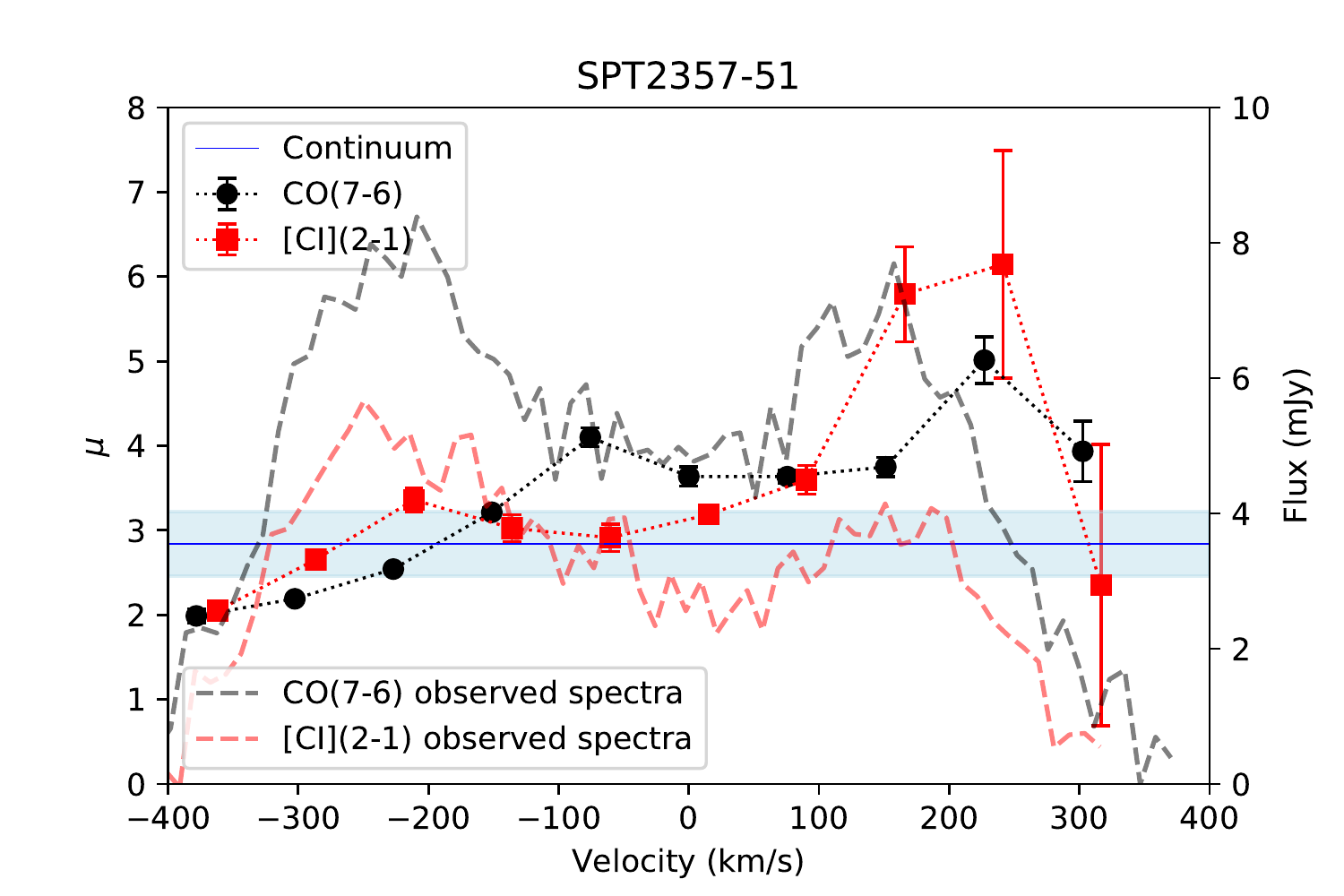} 

\caption{\label{fig:mag vs vel} The variation of the magnification as a function of velocity for the [CI](2-1) line (red filled squares) and the CO(7-6) line (black filled circles). The blue line is the continuum magnification and the blue shaded region indicates the 68\% confidence interval. The top panel shows the results for SPT0104-45. The middle panel shows the plot for SPT2147-50 and the bottom panel for SPT2357-51. We also overplot the observed spectra using a dashed-line, red for [CI](2-1) and black for CO(7-6).}
\end{figure}

\subsection{Differential magnification \label{sec: diff mag}}

From the results of the lens modelling for our sources, we find differences in the magnification between the continuum and the line emission as well as across the CO and [CI] line profiles. Figure\,\ref{fig:mag vs vel} shows the magnification as a function of velocity for both CO(7-6) and the [CI](2-1) along with the continuum magnification (horizontal blue dotted line). We note that the systemic redshifts of the sources were measured from the apparent spectra (not the intrinsic spectra).
We also plot the CO(7-6) and [CI](2-1) observed spectra along with the magnifications to differentiate the intrinsic features from those arising due to magnification effects. 

% In SPT0103-45, the lines are more magnified on the peak of emission than in the redshifted tail. To evaluate how this difference of magnification could generate the asymmetric line profile discussed in Sect.\,\ref{sect:line_imaging} (see also Fig.\,\ref{fig:int_spectra}), we overplot the magnification and the spectra of the two lines in Fig.\,\ref{fig:mag vs vel}. The observed line profiles (dashed-lines) match the shape of the magnification gradient (points and dotted lines).  The asymmetry of these lines and the presence of a redshifted tail of emission likely arises due to differential magnification.

For SPT0103-45, the agreement between the observed line profile (dashed lines) and the magnification profile (points and dotted lines) in Fig.\,\ref{fig:mag vs vel}, top panel, demonstrate that the velocity-dependent magnification is largely responsible for the line shape in this object.

\begin{table*}[]
    \centering
    \caption{\label{tab:luminosities}The line and IR luminosities of the sources are given below. The integrated luminosities are corrected for magnification provided from our lens modelling (Table\,\ref{tab: diff mag}) and the ratios are computed using these corrected quantities. The calculation of these values are described in Sect.\ref{Subsec:ratio_maps}.  Note that the calibration uncertainties (up to 10\,\%) were not taken into account. Note that the ratios are computed from the full precision values of the luminosities and not the truncated values provided in the table.}
    \begin{tabular}{cccccc}
    \hline
    \hline
    &&&&&\\
    Source & L$_{\rm IR}\times 10^{13}$& L$_{[\rm CI](2-1)}\times 10^{8}$ & L$_{\rm CO(7-6)}\times 10^{8}$ & \multirow{2}{*}{$\frac{\rm L_{[\rm CI](2-1)}}{\rm L_{\rm IR}}\times 10^{-5}$}& \multirow{2}{*}{$\frac{\rm L_{[\rm CI](2-1)}}{\rm L_{\rm CO(7-6)}}$} \\
         & (L$_{\odot}$ ) & (L$_{\odot}$) &(L$_{\odot}$)& & \\
    &&&&&\\
    \hline
    &&&&&\\
    SPT0103-45&1.90$\pm$0.60&3.67$\pm$0.12&3.84$\pm$0.06&
    1.91$\pm$0.60 &0.96$\pm$0.04\\
    SPT2147-50&0.83$\pm$0.20 & 1.52$\pm$0.13& 1.76$\pm$0.13& 1.80$\pm$0.50 &0.87$\pm$0.10\\
    SPT2357-51&1.62$\pm$0.26 &1.75$\pm$0.15& 2.54$\pm$0.12& 1.08$\pm$0.20 &0.69$\pm$0.07\\
    
    \hline
    \end{tabular}
    
\end{table*}

\begin{figure*}[h]
\centering

\begin{tabular}{ccc}
\includegraphics[width=6cm]{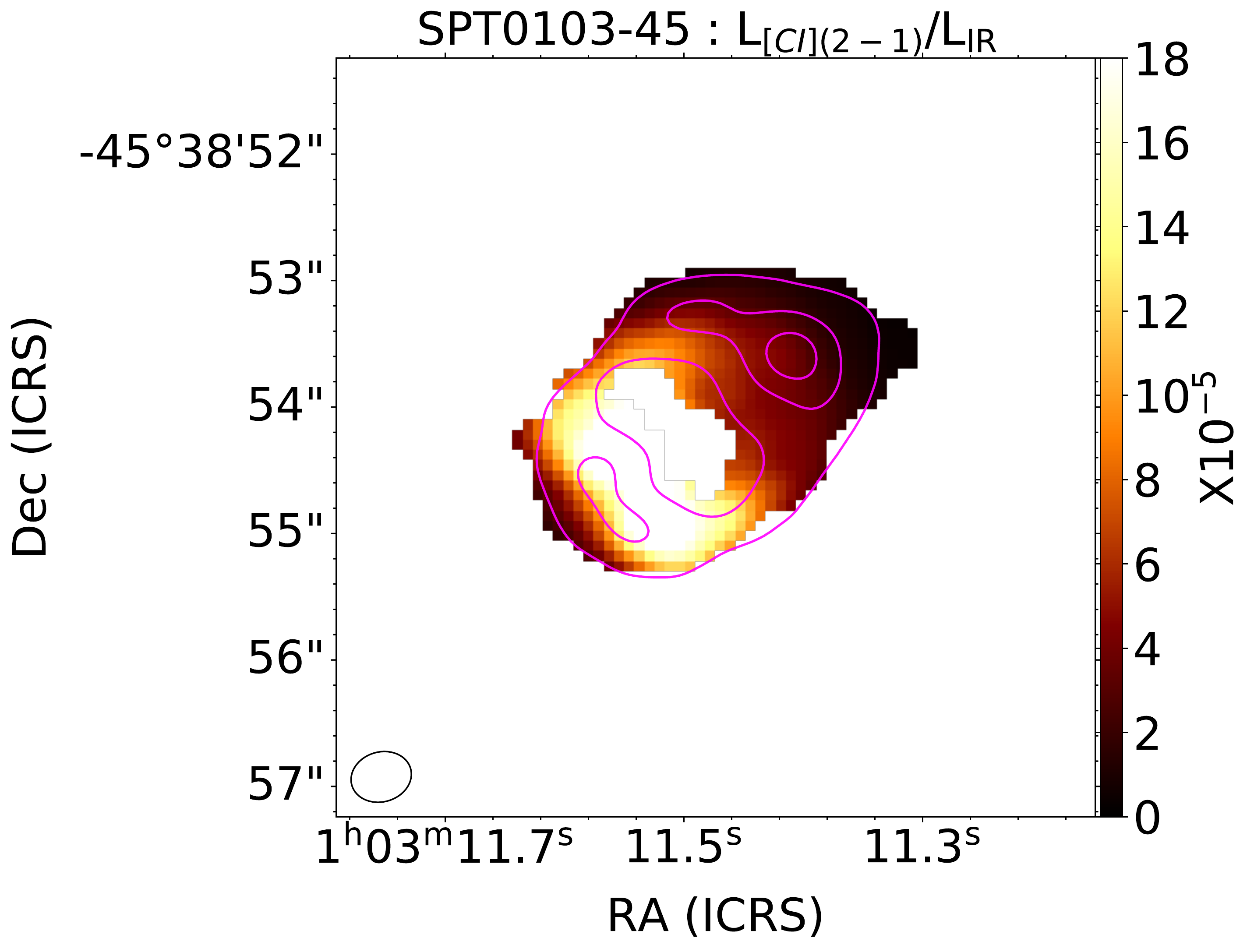} &  \includegraphics[width=6cm]{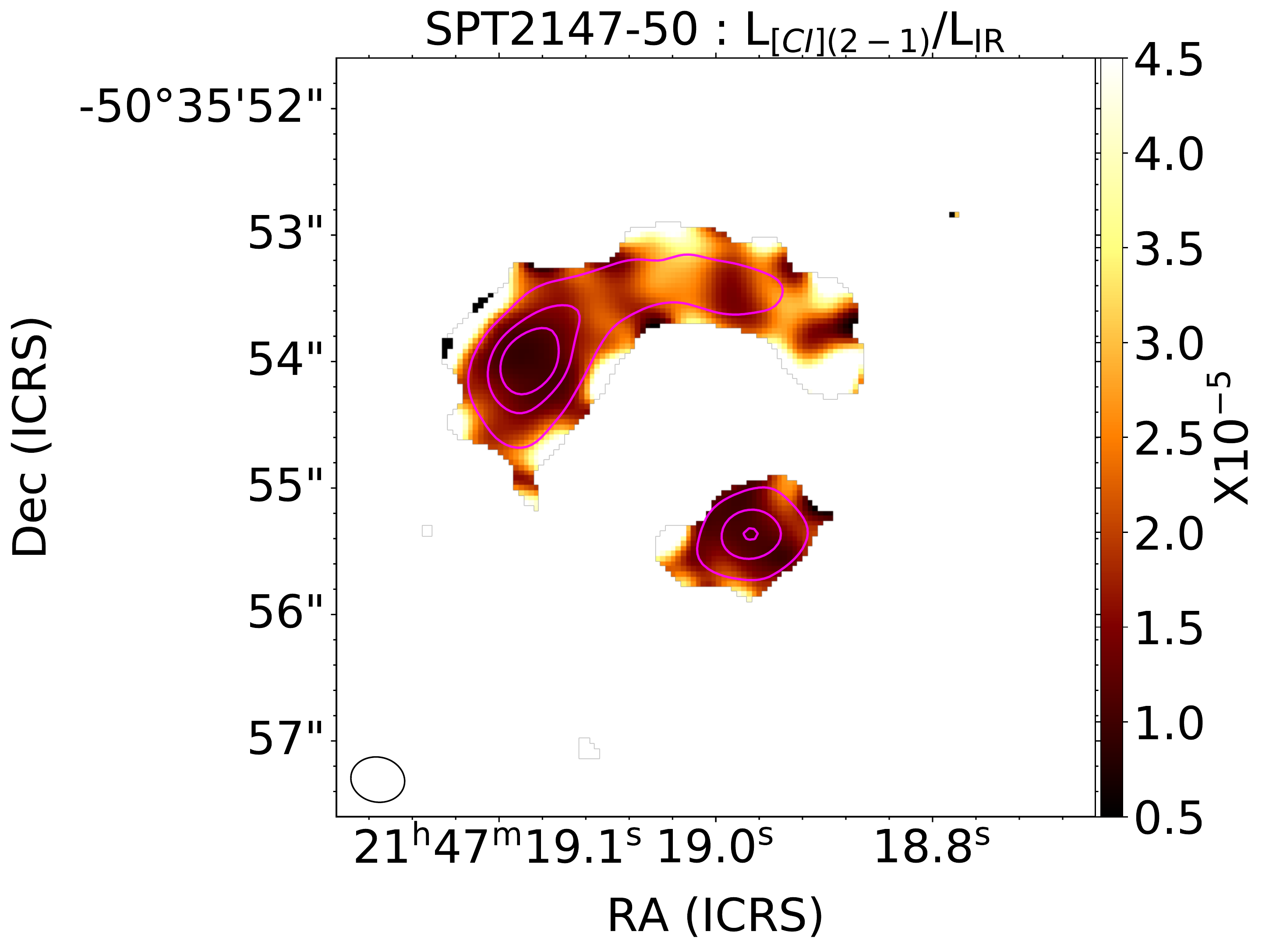} & \includegraphics[width=6cm]{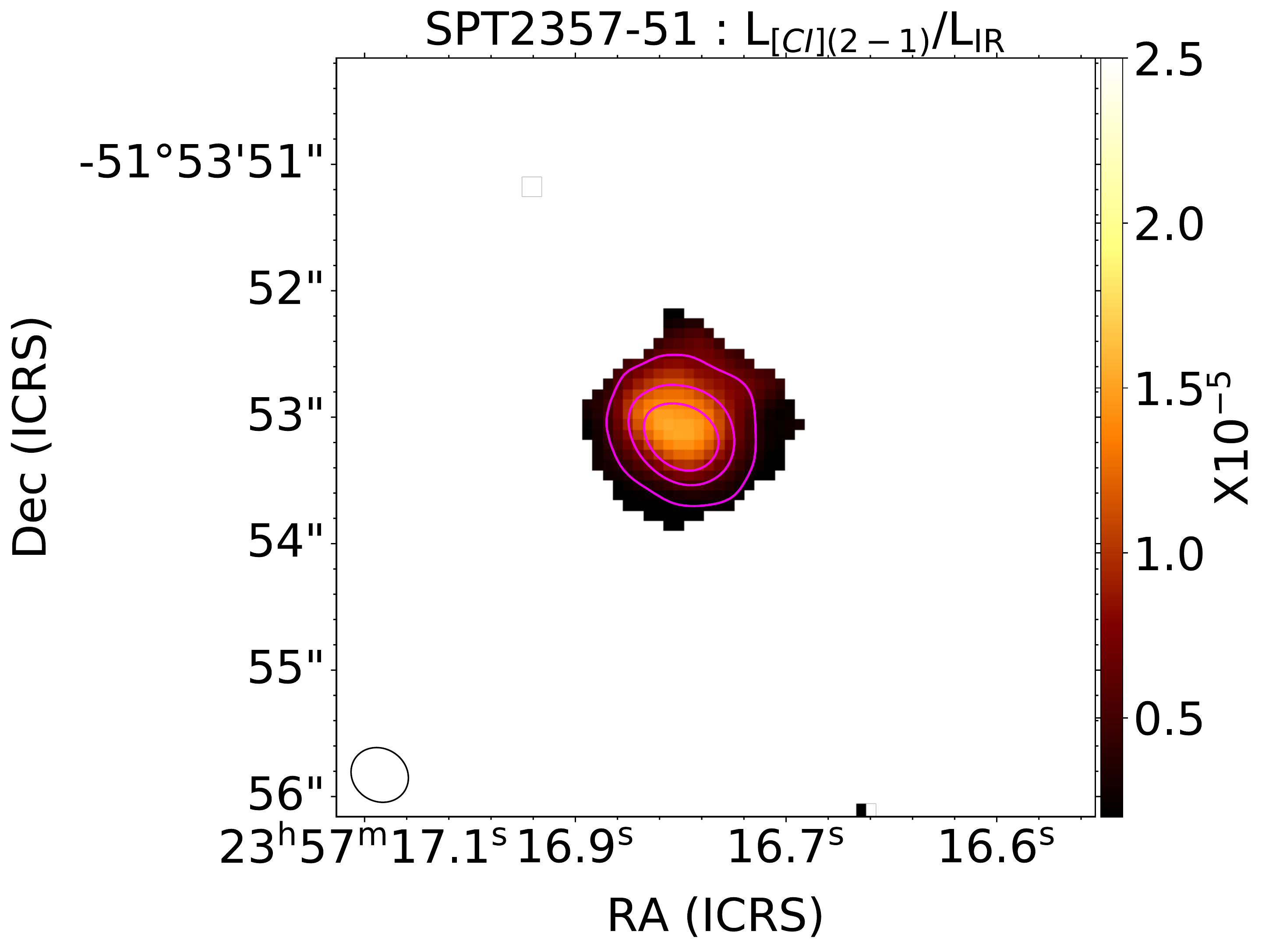} \\
\includegraphics[width=6cm]{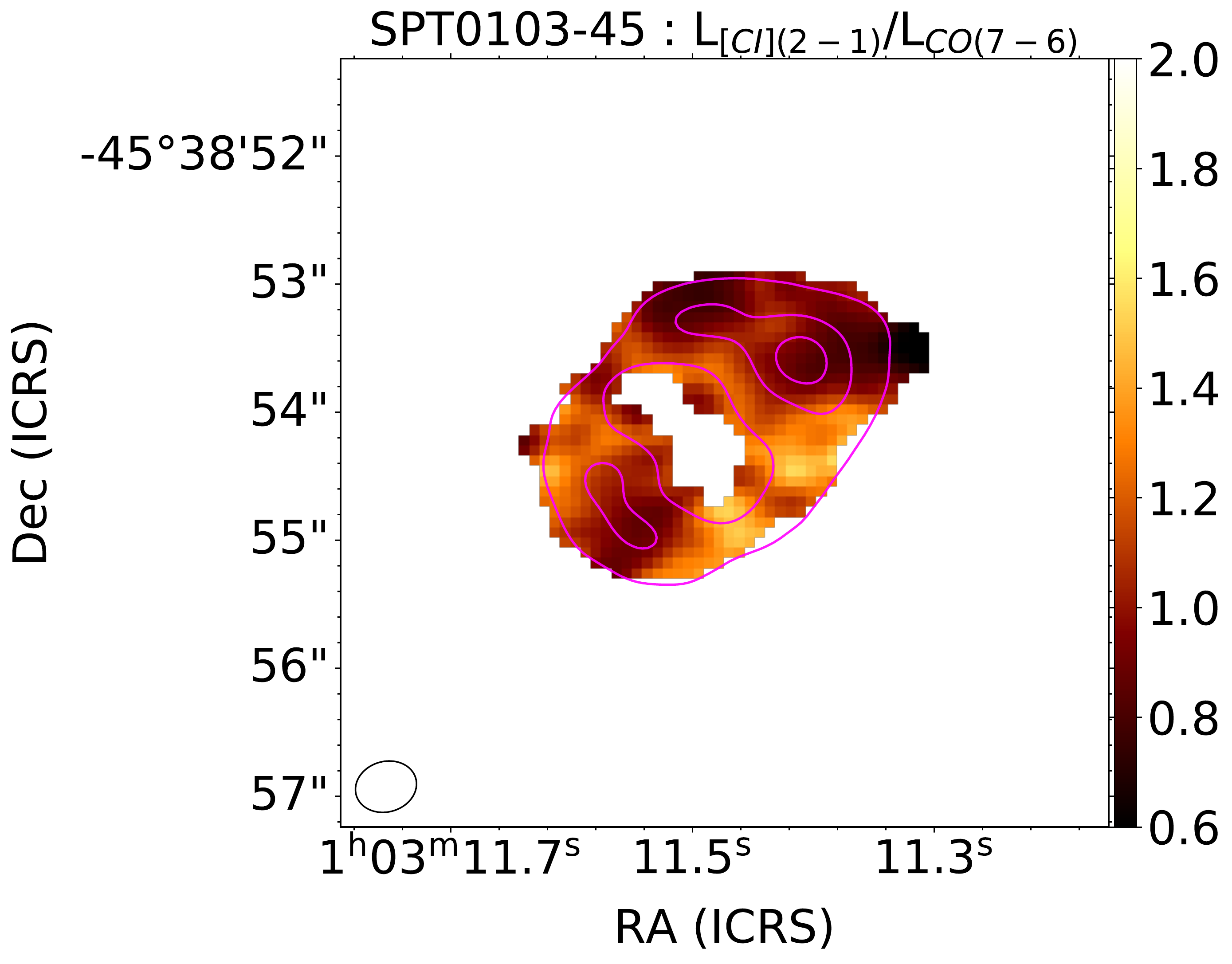} &  \includegraphics[width=6cm]{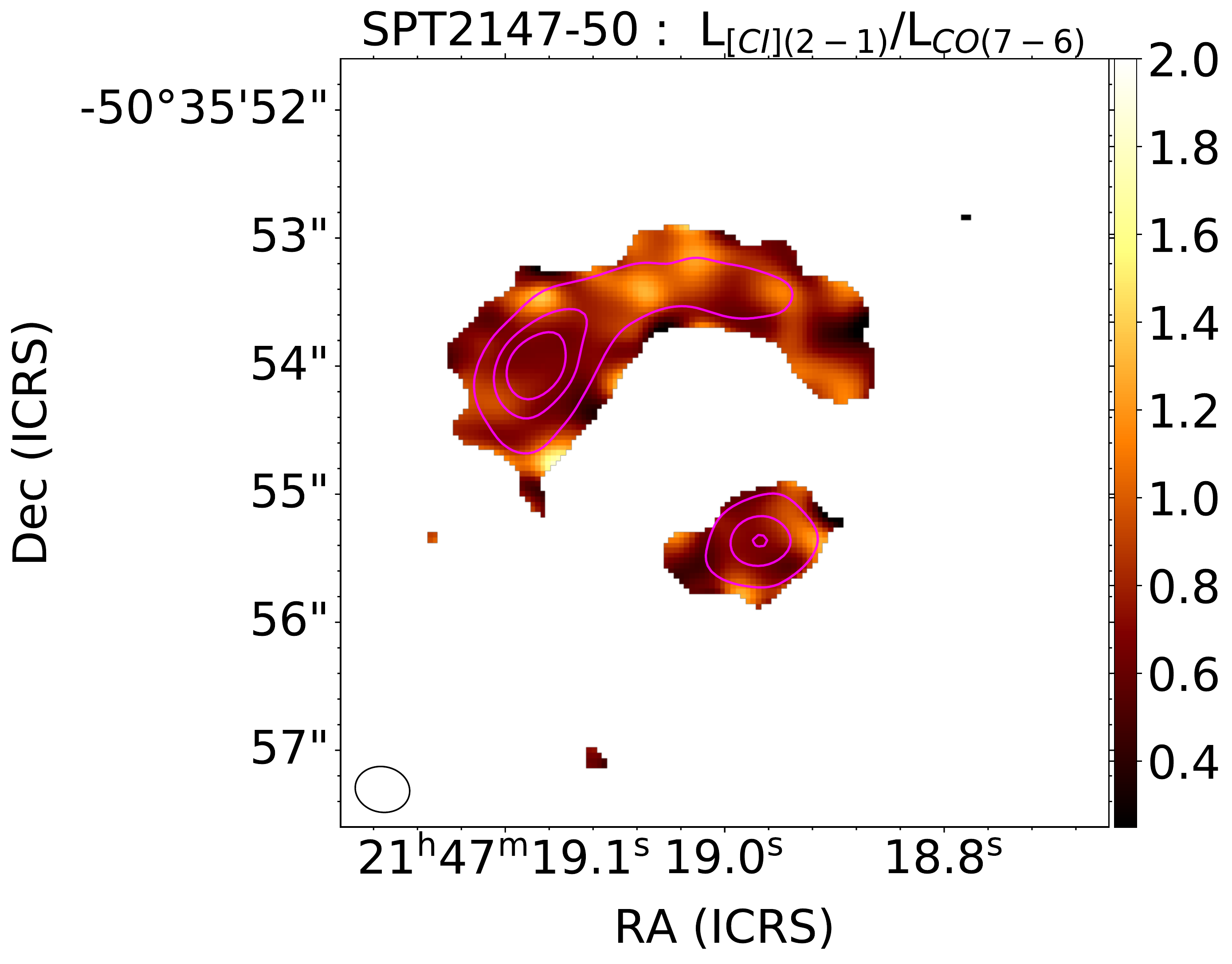} & \includegraphics[width=6cm]{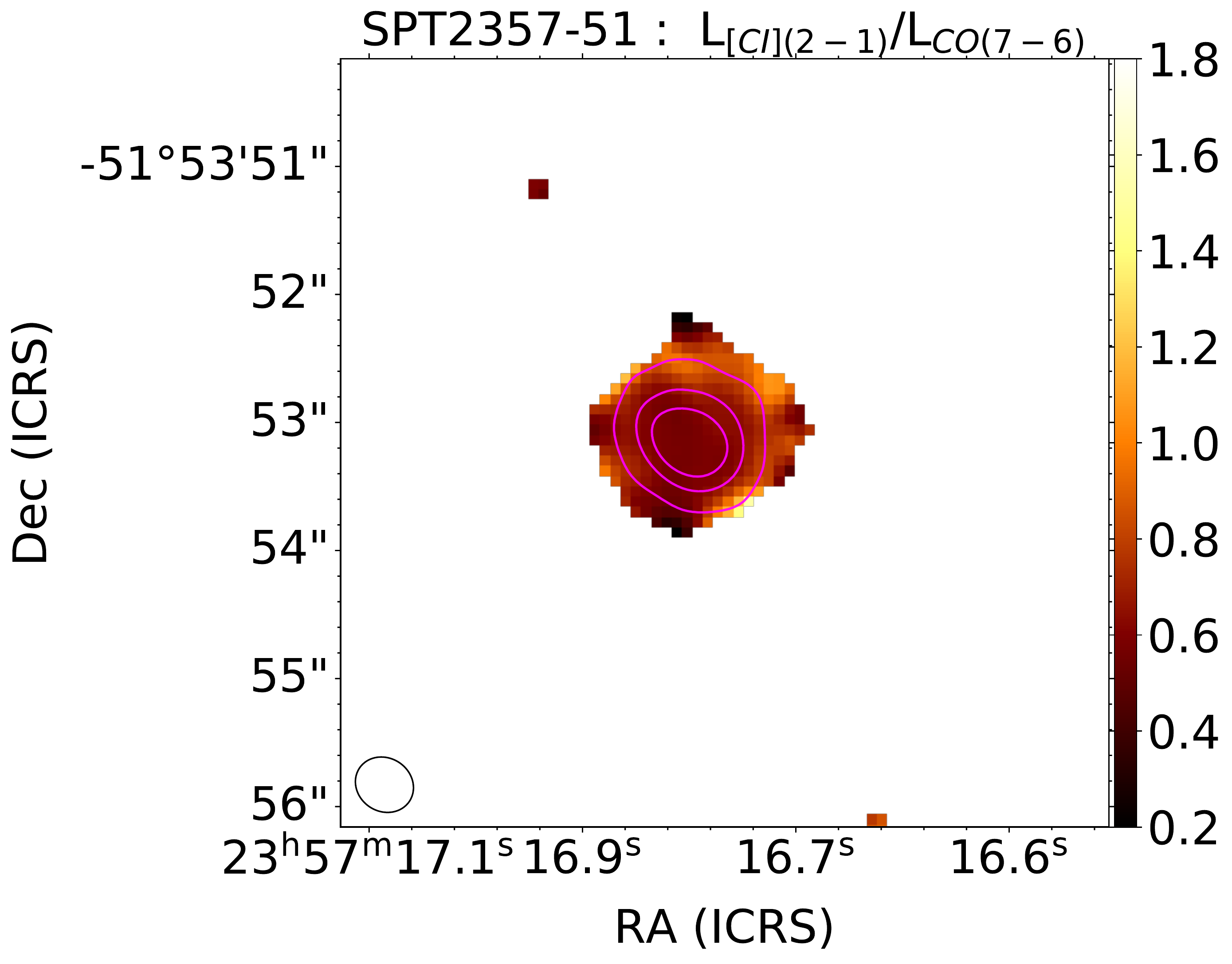} \\

\end{tabular}
\caption{\label{fig:ratio maps} The [CI]/continuum ratio maps are in the upper panels and the [CI]/CO line ratio maps are in the lower panels for SPT0103-45, SPT2147-50 and SPT2357-51, from left to right. The 5, 10 and 15$\,\sigma$ contours of the continuum are plotted on the ratio maps (in magenta). The computation of these ratios are described in Sect.\,\ref{Subsec:ratio_maps}. The synthesised beam is represented as the black ellipse in the bottom left of every map.}
\end{figure*}

For both lines, the magnification of SPT2147-50 increases toward redshifted velocities.  The line profile of SPT2147-50 is similar to SPT0103-45 with respect to the magnification versus velocity, except that in this case the redshifted side is more magnified. The asymmetry of the integrated line profile is again likely due to differential magnification. 

The red component of SPT2357-51, which was brighter in the continuum (see Sect.\,\ref{sec:kinematics 2357} and Fig.\,\ref{fig:decomp 2357}), is more magnified than the blue component. Since they have similar apparent line fluxes, the red component is thus intrinsically fainter.  

For each source, we calculated the effective magnification ($\mu_{\rm eff}$) of each line. It is computed as the ratio between the sum of the apparent flux in all velocity bins and the sum of the intrinsic fluxes in the same bins:
\begin{equation}
\centering
    \mu_{\rm eff} = \frac{\sum_{\rm bins}\mu \Delta \upsilon S_{\nu}}{\sum_{\rm bins} \Delta \upsilon S_{\nu}},
\end{equation}
where $\mu$ is the magnification in each of the bins, $\Delta \upsilon$ is the width of the bin and $\rm S_{\nu}$ is the intrinsic flux density. 
In our lens model, the magnification and the intrinsic flux are correlated because the total apparent flux density is very well constrained by the data. We thus do not estimate uncertainties on $\mu_{\rm eff}$ by combining the marginalised uncertainties on the input quantities. We instead calculate the effective magnification for all MCMC steps and compute the median and the standard deviation of the obtained values. The effective magnifications are tabulated in Table\,\ref{tab: diff mag}.

For SPT2147-50 and SPT2357-51, the CO(7-6), [CI](2-1) and the continuum magnifications are similar with variations $<$15\%. There is thus no evidence for strong differential magnification on integrated flux measurements. For SPT0103-45, we find a larger difference between the [CI](2-1) and continuum magnification ($\sim$24\%) and between the CO(7-6) and the continuum ($\sim$16\%). Traditionally, luminosity ratios are used to mitigate magnification effects in lensed sources. This assumes implicitly that the luminosity of all the lines and the continuum are magnified in the same way. Our analysis suggests that it is a reasonable assumption in general, but can introduce small biases depending on the lensing configuration of the individual sources. In contrast, our results show that spectral line profiles can be significantly distorted by differential magnification. Differential magnification effects are more severe across the line profiles than between the dust continuum and molecular gas tracers.

%----------------------------------------------------------------------

\begin{figure}[H]

\centering

\includegraphics[width=0.44\textwidth]{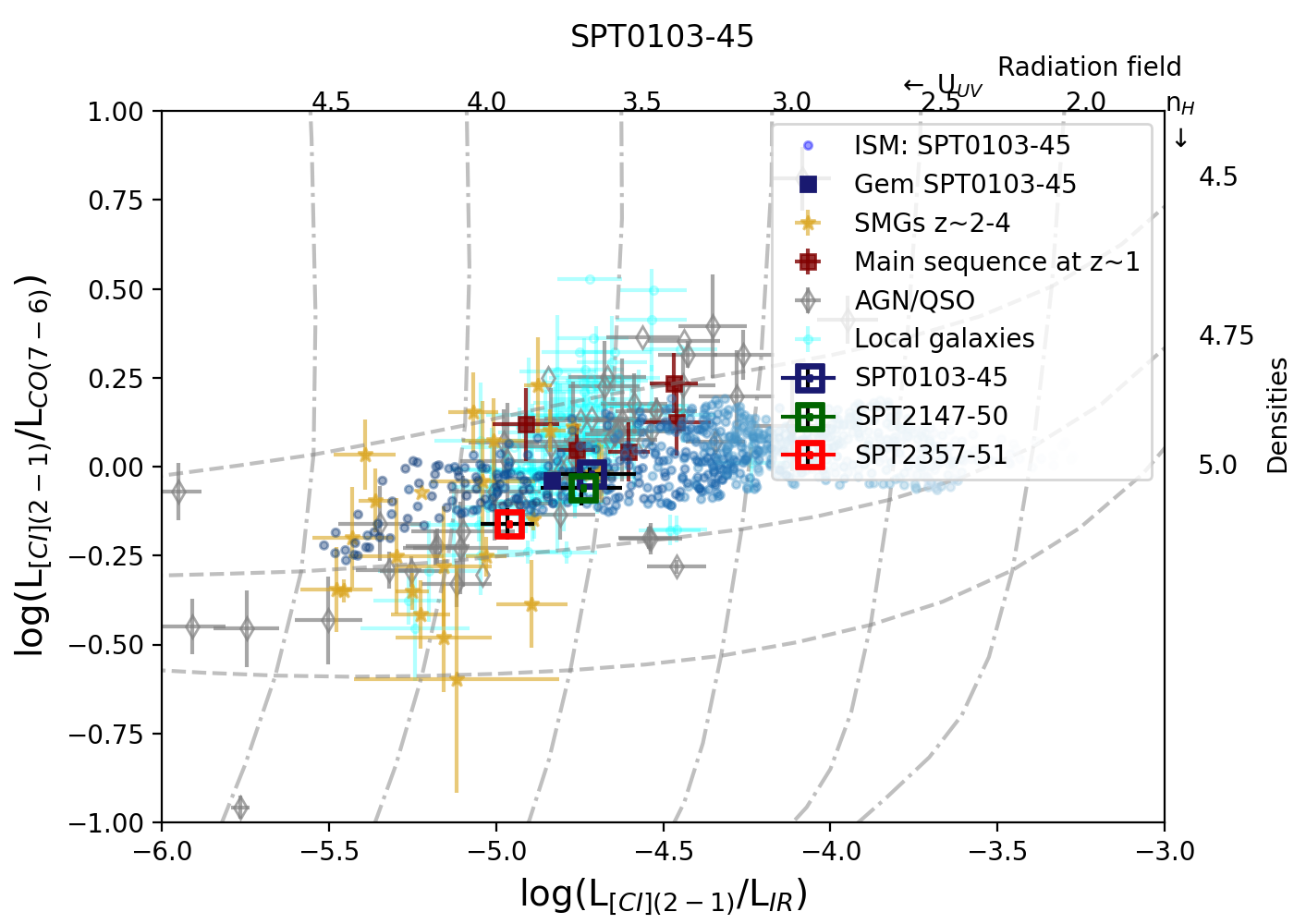}  \includegraphics[width=0.44\textwidth]{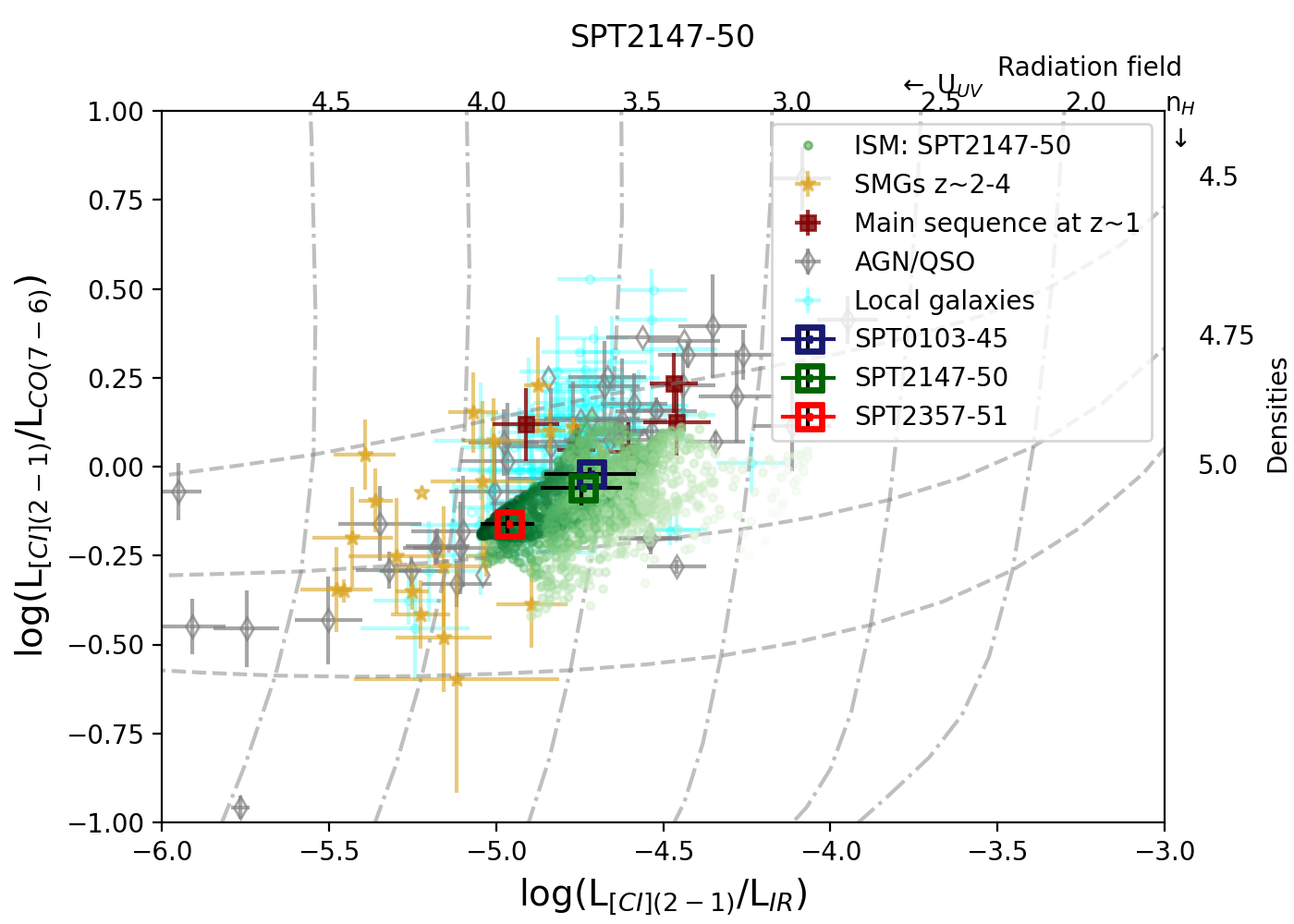} \includegraphics[width=0.44\textwidth]{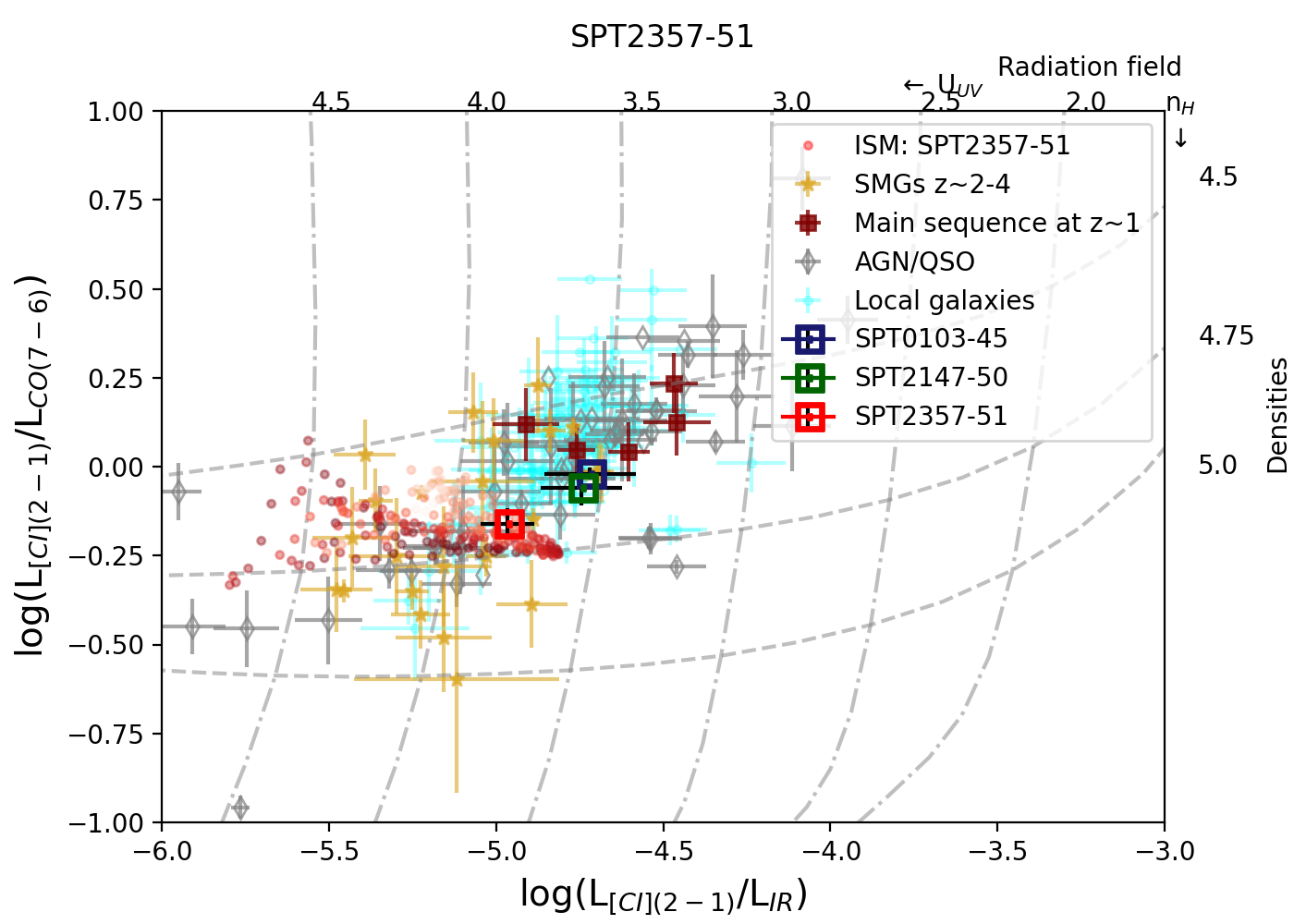}

\caption{\label{fig:fv plots}$L_{\rm [CI](2-1)}/L_{\rm CO(7-6)}$ versus $L_{\rm [CI](2-1)}/L_{\rm IR}$ for our sources, SPT0103-45 (top) SPT2147-50 (middle) and SPT2357-51 (bottom). The global ratios of our sources are represented as squares. The pixel-wise ratios of our sources are represented as small dots. Note that since we use the data in each pixel and not each beam, the data points are not independent. The pixel-wise values are color-coded based on the L$_{\rm IR}$ of each pixel, with the darkest corresponding to the most luminous pixels. We compare our sources with the compilation presented in \citet{Valentino20}. The SMGs at $z\sim$ 2-4 are represented by golden stars \citep{Walter11,Alaghband-Zadeh13,Bothwell17,Yang17,Andreani18,Canameras18,Nesvadba19,Dannerbauer19,Jin19}. The main-sequence galaxies at z$\sim$1 are represented by maroon squares \citep{Valentino18,Bourne19,Valentino20}. The local FTS samples of star-forming galaxies are represented by aqua dots \citep{Liu15,Kamenetzky14} and local FTS samples with AGN signatures are represented by grey diamonds \citep{Veron-cetty10,Liu15,Kamenetzky14}. The grey dot-dashed and dashed lines are the isocontours of the UV radiation field ($U_{UV}$) and the hydrogen density ($n_H$) from the PDR models of \citet{Kaufman99} in logarithmic values.}

\end{figure}

%------------------------------------------------------

%------------------------------------------------------

\section{\label{Sec:ratios}Resolved line and continuum ratios}

%----------------------------------------------------------------------
\subsection{\label{Subsec:ratio_maps}Ratio maps}

%--------------------------------------------------------

Analysing the resolved line and continuum ratios can help to understand the ISM properties of our sources. The bolometric infrared luminosity between 8 and 1000\,$\mu$m (L$_{\rm IR}$) is a well-known tracer of the SFR of galaxies \citep{Kennicutt98}. The [CI](2-1) line can be used as a tracer of the total molecular gas \citep[e.g.,][]{Papadopoulos04} and the high-J CO lines are a tracer of the dense molecular gas \citep[e.g.,][]{Weiss07}. Combining these three tracers, we can map both the resolved star formation and gas properties in our objects. We continue our ratio analysis in the image plane. This avoids the introduction of any artifacts in the line ratio due to our parameterised models for the source-plane emission. As gravitational lensing conserves surface brightness, an image-plane analysis is also equivalent to the same analysis in the source plane, with the caveat that multiply-imaged regions of the source plane appear multiple times in the image plane.

%\textbf{We continue our ratio analysis in the image-plane. Since the differential magnification on the full integrated fluxes are less than 20$\%$ for our sources, we can reasonably assume that the differential magnification is negligible in a given synthesised beam. Furthermore, we tried to map these ratios in the source plane, by dividing the two Sersic models and found strong artifacts due to the simplicity of these models. Analysing the ratio in the source plane would thus be too complicated with our data.}

To compute the continuum to line ratios accurately and avoid resolution effects, our continuum and line maps must have comparable resolutions. In Table\,\ref{img perf}, we notice that the continuum has a slightly smaller synthesised beam size than the lines. This is a result of the frequency coverage of our data, since most data used for the continuum imaging comes from the other sideband, 12\,GHz higher in frequency, that samples the $uv$ plane at slightly higher spatial frequencies. Hence, we produce another set of continuum maps using only the line-free continuum channels from the same sideband as the lines to more closely match the synthesised beam size between lines and continuum.

To obtain physically meaningful ratios, we convert both continuum and line fluxes in each beam into apparent luminosities. For the continuum maps, we use the following conversion to derive the apparent infrared luminosity per beam ($\mu L_{\rm IR(beam)}$): 
\begin{equation}
    \mu \rm L_{\rm IR(beam)} = \frac{\mu \rm L_{\rm IR (total)}}{S_{\rm \nu (total)}}\times S_{\nu,\textrm{(beam}) } \quad \, (\textrm{L}_{\odot}\, \textrm{beam}^{-1}),
\end{equation}
where $S_{\nu,\textrm{(beam})}$ is the continuum flux density per beam measured in any pixel of the continuum map, $S_{\nu \rm (total)}$ is the total continuum flux density integrated over the entire source and $\mu \, L_{IR}$ is the total apparent infrared luminosity determined by \citet{Reuter20} from \textit{Herschel} and ground-based photometry. 

This conversion implicitly assumes that the shape of the spectral energy distribution (SED) is the same everywhere in the galaxy. This is an approximation. Using simulations, \citet{Cochrane19} estimated the dust temperature ($T_{\rm dust}$) gradient in star-forming galaxies ($\gamma$ in $T_{\rm dust} =  T_{\rm center}\, r^{\gamma}$,\footnote{Note that the factor relating to the spatial distribution of dust is referred to as $\beta$ in \citet{Cochrane19}. We chose to refer to it using the symbol $\gamma$ to avoid confusions with the dust emissivity index.} where $r$ is the distance from the center in kpc). They found a range of $\gamma$ between -0.3 and -0.05. For the smallest $\gamma$ and the largest source (SPT0103-45), we obtain a temperature variation by a factor of 1.4 between the center and the half-light radius. Assuming a simple $L \propto T^4$ scaling, this corresponds to a factor of 3.5 in luminosity at constant flux. For $\gamma=-0.05$, the effect is much smaller, with only a 24\,\% luminosity difference.

For the line flux maps, we use the conversion formula from \citet{Solomon92}
\begin{equation}
    \mu \rm L_{\rm line (beam) } = 1.04 \times 10^{-3}\, S_{\rm line (beam)}\Delta\upsilon  \times D_L^2 \, \nu_{\rm obs} \quad \, (\textrm{L}_{\odot}\, \textrm{beam}^{-1}),
\end{equation}
where $S_{\rm line (beam)} \Delta \upsilon$ is the value obtained from the line flux maps derived in Sect.\,\ref{sec: moment maps}, in the units of Jy km\,s$^{-1}$\, beam$^{-1}$, $D_L$ is the luminosity distance of the source in Mpc and $\nu_{\rm obs}$ is the observed line frequency in GHz. The total line luminosities and IR luminosity corrected for magnification for all our sources are tabulated in Table\,\ref{tab:luminosities}. We use the effective magnifications of the continuum or the associated line computed in Sect.\,\ref{sec: diff mag} and listed in Table\,\ref{tab: diff mag}. The uncertainties on the line luminosity ratio of SPT0103-45 are the combined uncertainties derived from the moment map luminosities and the luminosities estimated by deblending the lines (see Appendix.\,\ref{deblending}).

We then create $\rm L_{[\rm CI](2-1)}/\rm L_{\rm IR}$ and $\rm L_{[\rm CI](2-1)}/\rm L_{\rm CO(7-6)}$ ratio maps, shown in Figure\,\ref{fig:ratio maps} using a 3$\,\sigma$ threshold on the IR and line luminosities. SPT0103-45 has a lower [CI]/continuum ratio and marginally lower [CI]/CO ratio at the position of the gem. In SPT2147-50, the northeast part of the main arc and its southern counter-image show a lower [CI]/continuum than the western part of the arc, while no obvious features are seen in the [CI]/CO ratio map.

The [CI]/continuum ratio of SPT2357-51 exhibits a higher value in the center than on the outskirts. This may be counter-intuitive, since galaxy sizes measured from gas tracers are usually larger than from the continuum \citep{Spilker15,Dong19, Apostolovski19, Fujimoto20}. This central region could correspond to the overlap of the two components of this candidate merger system (see Sect.\,\ref{sec:kinematics 2357}). In this interacting region, the continuum luminosity could possibly be underestimated if the dust is warmer in the core of this object. We again see no clear pattern in the [CI]/CO ratio map. For this source, higher-resolution data would be necessary to make stronger conclusions, given the compact lensing configuration of this source.

We next compare the luminosity ratios of our sources with other lensed SMGs. \citet{Andreani18} found L$\rm _{CO(7-6)}$/L$\rm _{[CI](2-1)}$ ratios of 2.5 and 4.0\footnote{after converting from L$^\prime$ (K\,km/s\,pc$^2$) to L (L$_\odot$) units.} for their sample.  \citet{Yang17} found that the line ratio vary from 0.8 to 3.0 in their sample of \textit{Herschel}-selected SMGs. These ratios are typically higher than we measure in our sources, i.e. 0.93$\pm$0.02 for SPT0103-45, 1.2$\pm$0.1 for SPT2147-50 and 1.6$\pm$0.1 for SPT2357-51. Our ratios are consistent with the ratios seen in local galaxies by \citet[][from 0.4 to 2]{Kamenetzky14} and close to the value for the Milky Way center (0.91) measured by \citet{Fixsen99}. Our systems thus have less extreme line ratios than the \textit{Herschel}-selected lensed galaxies of \citet{Andreani18}. Those authors claim that the L$_{\rm CO(7-6)}$/L$_{[\rm CI](2-1)}$ ratio can be indicative of whether a system is disk-dominated (line ratio $\sim$1) or merger-driven (up to $\sim$10). According to the dynamics and the lensing reconstruction of our sources performed in Sect.\,\ref{sec: kinematics}, \ref{sec: lens_modelling} and \ref{sec: discussions}, SPT0103-45 and SPT2147-50 seem to be dominated by a rotating and extended component, consistent with their low [CI]/CO line ratios $\lesssim$1. In contrast, SPT2357-51 has a slightly higher ratio and is a possible merger candidate. These results are thus an additional indication that the L$_{\rm CO(7-6)}$/L$_{[\rm CI](2-1)}$ ratio could be used to diagnose the kinematic nature of high-$z$ sources. However, it may be less accurate than  initially suggested by \citet{Andreani18} and larger samples with high-resolution kinematic information will be necessary to draw stronger conclusions.

\begin{figure}
\centering

\includegraphics[width=9cm]{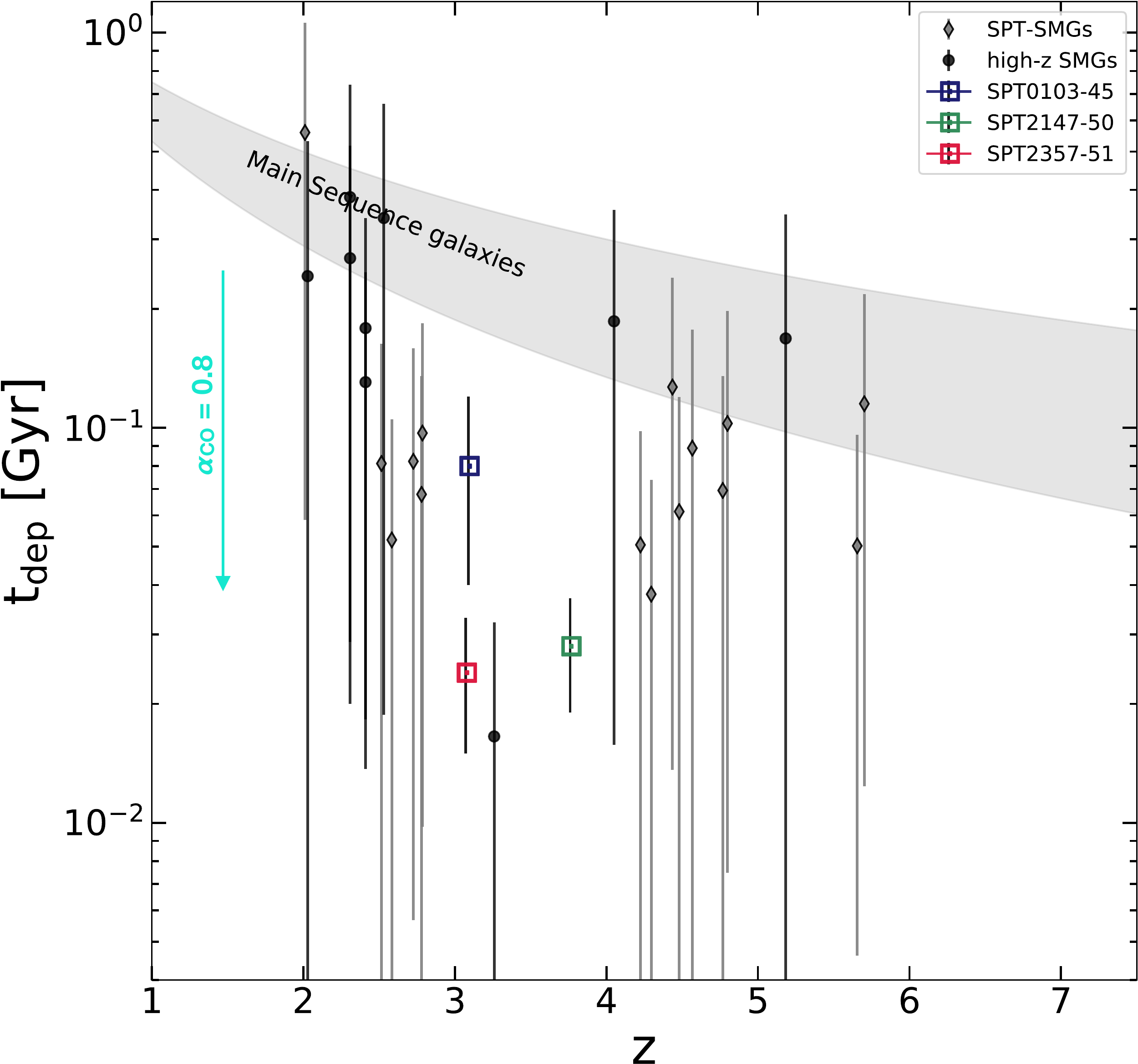}
\caption{\label{fig:tdep_vs_z} The depletion timescale versus redshift of our sources, compared with other SPT-SMGs and other high-z SMGs. The depletion timescales for our sources are computed with the gas mass estimated from low-J CO fluxes with an $\alpha_{\rm CO} = 0.8$. The grey shaded region represents the evolution of the depletion timescale with redshifts for main sequence galaxies. This plot is adapted from \citet{Jarugula21}. The high-redshift gravitationally lensed SPT-SMGs are from \citet{Aravena16} and \citet{Reuter20} and the other high-redshift SMGs are taken from \citet{Carilli10,Walter12,Fu12,Ivison13,Fu13,Alaghband-Zadeh13}. The main sequence galaxies are described in \citet{Saintonge13} as $\rm t_{\rm dep} = 1.5\,(1 + z)\, \alpha$, where $\alpha$ is from $-$1.5 \citep{Dave12} to $-$1.0 \citep{Magnelli13}, which is shown as the
grey shaded region.}

\end{figure}

%----------------------
\subsection{Resolved ISM properties \label{Sec: resolved ism}}

To identify the physical origin of the variations of these ratios across our sources, we plot $\rm L_{[\rm CI](2-1)}/\rm L_{\rm CO(7-6)}$ against $\rm L_{[\rm CI](2-1)}/\rm L_{\rm IR}$ for each pixel of our ratio maps (Fig.\,\ref{fig:fv plots}) to explore the heterogeneity of the ISM in our sources. In Fig.\,\ref{fig:fv plots}, we compare our sources with sample compilation presented in \citet{Valentino20}.

% \textbf{The differential magnification on the full integrated fluxes are less than 20$\%$ for our sources. Although this impact is rather small, we have corrected for this effect. In order to do so, we correct the integrated fluxes of the line and continuum with their corresponding magnifications (Table\,\ref{tab: diff mag}). We then compute the ratios on these de-magnified fluxes (see, Table\,\ref{tab:luminosities}). The results are represented as empty squares in Fig.\,\ref{fig:fv plots}. In the case of the ratios in individual line of sight, this effect should be negligible and thus we do not correct them for differential magnification.}

Along with the ratios, in Fig.\,\ref{fig:fv plots}, we plot the radiation field and hydrogen volume density iso-contours from the photo dissociation region (PDR) modelling of \citet{Kaufman99}. The iso-density contours are mostly horizontal in this parameter space, indicating that the ratio between [CI](2-1) and CO(7-6) can be a proxy of the gas density. Similarly, the radiation-field iso-contours are almost vertical, indicating that the [CI](2-1)-to-$\rm L_{\rm IR}$ ratio is a proxy of the radiation field strength.

However, we should note that comparing the line ratios to PDR models does have caveats. There may be additional heating mechanisms that cannot be represented only with line ratios. The hot and dense PDRs can reproduce the line ratios but may not accurately reproduce the L$_{\rm IR}$ as a part of the L$_{\rm IR}$ may arise from the HII region. The internal properties and heating mechanisms of the sources may also be driven by the IR emission arising from the warm dust \citep{Valentino18}. 

The globally-integrated ratios of our sources suggest that they have similar radiation field and densities to the compilation presented in \citet{Valentino20}. SPT0103-45 has the lowest density and radiation field strength of our three targets, while SPT2357-51 is our most extreme target in both parameters.

Turning to the pixel-wise ratios, we see that SPT0103-45 has a large variation in radiation field strength with almost uniform density, a result of the wide range in [CI]/$\rm L_{\rm IR}$ across the source. The gem region has a slightly higher radiation field strength and density compared to the global average. This region is likely a compact star-forming region close to the center of the object (see Fig.\,\ref{fig:line lens 0103}). Inspection of Fig.\,\ref{fig:ratio maps} suggests that the regions closest to the gem have a stronger radiation field, possibly due to an elevated dust temperature or a high SFE producing more young stars and UV photons.

In SPT2147-50, we observe a fairly large dispersion in the gas density and the radiation field ($\sim 0.5$\,dex). The kinematics and lensing reconstruction of the source (Sect.\,\ref{sec: kinematics} and Sect.\,\ref{sec: lens_modelling}) suggest that the source could be a rotating disk. These variations of density could thus be caused by dynamical instabilities in this gas-rich system \citep{Bournaud09a,Bournaud09b,Zanella15}. 

The core of SPT2357-51 dominates the flux of the source (Fig.\,\ref{fig:ratio maps}) and the associated pixels are very similar to the globally-integrated ratios. In contrast, the outskirts of the system appear to have similar or lower densities, but higher radiation fields. Several explanations are possible. The center of the object could be dustier, thereby preventing UV photons from penetrating the clouds. There could also be a dust temperature gradient from the outskirts to the center, which could bias our estimate of L$_{\rm IR}$ and thus the radiation field strength (see Sect.\,\ref{Subsec:ratio_maps}).

%----------------------------------------------------------------------

\section{Discussion \label{sec: discussions}}

\subsection{SPT0103-45 \label{disc:0103}}

%The most striking feature of this source is the gem observed in its continuum map. Through our analysis, we try to understand the spatial and spectral nature of the region responsible for this bright continuum emission. Fig.\,\ref{fig:bright blob}, shows that the gem is mainly emanating from a higher velocity component seen in the line. The velocity map in Fig.\,\ref{fig:decomp 0103} shows a gradient indicating that the source could be a rotator.

Despite having a complicated morphology in the image plane (Fig.\,\ref{fig:maps} and \ref{fig:bright blob}) and an asymmetric observed integrated line profile (Fig.\,\ref{fig:int_spectra}), SPT0103-45 seems to be compatible with a rotating disk from our lensing reconstruction (Sect.\,\ref{sec: lens modelling 0103}), with the asymmetry of the line profiles arising from differential magnification (Sect.\,\ref{sec: diff mag}).

%The properties of this gem a little different from the source itself. The gem is has a higher CO(7-6) content than the bulk of the source indicating it could be a compact star forming region. In the Fig.\,\ref{fig:fv plots}, we can see the difference in the density and the radiation field of the gem from that of the source, suggesting that it may be chemically different from the rest of the source. 

%The source plane morphology and the intrinsic properties were probed with the help of lens modelling. For SPT0103-45, the continuum was modelled with a two-source system, whereas the line emission was sufficiently modelled with a single source. The intrinsic source positions at every velocity range shows a clear gradient (see Fig.\,\ref{fig:line lens 0103}) , further suggesting that the source could be a rotating disk. We also see strong effects of differential magnification for the emission lines. We obtain the magnification of the components from the lens modelling and we can thus estimate the intrinsic SFR with the L$_{\rm IR}$ presented in \citet{Reuter20}. 

The intrinsic SFR of SPT0103-45, after correcting for the continuum magnification derived in Sect.\,\ref{sec: lens_modelling}, is $1900 \pm 600  \, \rm M_{\odot} yr^{-1}$. Furthermore, using our continuum size estimates of the extended component, we compute the SFR surface density ($\Sigma_{\rm SFR}$) and find $69 \pm 20 \, \rm M_{\odot} yr^{-1} kpc^{-2}$. This is similar to typical SFR surface densities observed in SMGs \citep[e.g.,][]{Daddi10b,Casey14}.

We can also estimate the molecular gas mass from CO(3-2) observations from \citet{Reuter20}. We assume a CO(3-2)/CO(1-0) line ratio of 0.7 $\pm $ 0.1  from \citet{Harrington21} and use an $\alpha_{\rm CO} = 3.4\,\, \rm M_{\odot} (K \, km/s \, pc^{2})^{-1}$ (where $\alpha_{\rm CO}$ is the CO-to-H$_2$ conversion factor) from \citet{Jarugula21}. We note that the quoted value and uncertainty on the line ratio we assume is derived from and applicable to high-redshift SMGs only, the uncertainty would increase if we also considered other galaxy populations when making this assumption. In any case, we expect the uncertainty on $\alpha_{\rm CO}$ to dominate over the uncertainty in CO(3-1)/CO(1-0) line ratio. We find M$_{\rm gas} = (6.8 \pm 2.1) \times 10^{11} \, \rm M_{\odot}$ and a gas depletion timescale $\rm t_{\rm dep} = 360 \pm 140 $ Myr. Instead, assuming a lower $\alpha_{\rm CO} = 0.8$ \citep{DownesSolomon98}, we find M$_{\rm gas} = (1.6 \pm 0.5) \times 10^{11} \, \rm M_{\odot}$ and a gas depletion time scale $\rm t_{\rm dep} = 80 \pm 40 $ Myr. This short depletion timescale is typical of starbursts \citep[e.g.,][]{Daddi10b,Bethermin15}. 

Furthermore, we estimate the dynamical mass of the system using a simple approach. Assuming a simple circular Keplerian motion, the enclosed dynamical mass $\rm M_{\rm dyn}$ can be computed using the following equation: 
\begin{equation}\label{dyn mass}
    \rm M_{\rm dyn} = \frac{\rm R\,\rm V_{\rm obs}^2}{\rm G \rm \sin^2(i)}
%2.33 \times 10^{5} \frac{R\,V_{\rm obs}^2}{\rm sin^2(i)} = 
\end{equation}
where R is the radius, which we take to be half of the distance between the lowest and highest velocity components in the source plane (see Table\,\ref{source_params_0103}). We estimated the inclination $i$ of the system using the b/a ratio of the continuum model ($\cos i = b/a$).  $\rm V_{\rm obs}$ is the projected component of the circular velocity along the line of sight and is computed as half of the velocity difference between the two extreme velocity bins. This approach is an intermediate between the crude estimate based on the line width and the size \citep[e.g.,][]{Aravena16,Bothwell17,Dessauges-Zavadsky20} and a full dynamical approach \citep[e.g.,][]{Rizzo20,Jones21} done on  higher resolution data or unlensed objects. Contrary to the crude approach, we can take into account the inclination of our objects and obtain the circular velocity in the outskirts of the objects instead of an integrated line width. Of course, it relies on the strong assumption that these systems are dynamically relaxed disks.

We estimate the uncertainties on the result by computing $\rm M_{\rm dyn}$ for all realizations of MCMC chains from the lens model, though we expect that systematic uncertainties in e.g. the assumed source geometry are dominant. We find a dynamical mass $\rm M_{\rm dyn}$ of $(1.2 \pm 0.5) \times 10^{11} \, \rm M_{\odot}$ using the CO(7-6) radius and velocities and ($1.55 \pm 0.4) \times 10^{11} \, \rm M_{\odot}$ using the [CI](2-1) radius and velocities.

We find compatible values between this dynamical mass and the molecular gas mass using the lower $\alpha_{\rm CO} = 0.8$ value. Under this assumption there is only a small difference between the dynamical and gas masses, implying that the stellar mass of the source is relatively low. We note that the dynamical mass is calculated with the radius of the mid-J CO transition which requires higher excitation temperatures, typically found in the galactic centers. In Fig.\,\ref{fig:tdep_vs_z}, the depletion timescale of SPT0103-45 is similar to the bulk SMG population. However, the short depletion time and the high SFR density corresponds to a starburst-like star formation mode, which is usually not expected in isolated disks. An $\alpha_{\rm CO} = 3.4$ ($\sim$4) leads to a longer depletion timescale usually found in disk galaxies, but this would also result in a strong tension with the dynamical mass. Only higher resolution data could allow us to identify a potential interaction triggering the starburst. We also cannot exclude processes linked to disk instabilities \citep[e.g.,][]{Bournaud07}.

\subsection{SPT2147-50}

%From the integrated spectra (Fig.\,\ref{fig:int_spectra}), we notice an asymmetry in both the CO(7-6) and [CI](2-1) lines. The origin of this asymmetry and whether or not it is intrinsic to the source was a question. But with our lens reconstruction, we can clearly see that the asymmetry arises due to the magnification effects. In Fig.\, \ref{fig:mag vs vel}, we can see the magnification and the observed spectra together and the tail region is less magnified than the peak of emission. Without this bias, the line would not show asymmetry but would be a broad line instead. 

%The kinematic analysis of the lines on the image plane suggest that the source could be a rotator. Fig.\, \ref{fig:vel map 2147} shows the velocity map of both the lines, indicating a smooth gradient. The robustness of this method can also be verified as the trends of the two lines show good agreement. Furthermore, we confirm the rotation in our source plane analysis. Fig.\, \ref{fig:line lens 2147} shows the intrinsic position of source in every velocity range and we obtain a smooth gradient. 

The dynamics of SPT2147-50 are similar to SPT0103-45, with our source plane modelling being compatible with rotation. SPT2147-50 has a lensing-corrected SFR of $830 \pm 200  \, \rm M_{\odot} \rm yr^{-1}$ and a $\Sigma_{\rm SFR}$ of $30 \pm 7 \, \rm M_{\odot} yr^{-1} kpc^{-2}$. \citet{Bothwell17} estimated the molecular gas mass using [CI](1-0) and found $\rm M_{\rm mol} = (6.8 \pm 0.5) \times 10^{10} \, \rm M_{\odot}$ (after updating their magnification by our [CI](2-1)-based measurement). \citet{Aravena16} found a smaller molecular gas mass of $(2.3 \pm 0.5) \times 10^{10} \, \rm M_{\odot}$ using CO(2-1) ($\alpha_{\rm CO} = 0.8$, $\rm L'_{\rm CO(2-1)}/\rm L'_{\rm CO(1-0)} = 0.8$ and updating the magnification using our CO(7-6)-based value). We derive a depletion timescale $\rm t_{\rm dep} = 82 \pm 7$ Myr using [CI](1-0) and $\rm t_{\rm dep} = 28 \pm 9 \, \rm Myr$ using CO(2-1). This short depletion timescale is usually associated to a starburst. If instead we compute the gas mass using a Milky Way-like $\alpha_{\rm CO} = 3.4$, we find a gas mass of $(9.9 \pm 0.5) \times 10^{10} \, \rm M_{\odot}$ and a depletion time $\rm t_{\rm dep} = 119 \pm 29 \, \rm Myr$.

We can also compute the dynamical mass of the system using the same method as SPT0103-45. We find $\rm M_{\rm dyn} = (2.5 \pm 0.4) \times 10^{10} \, \rm M_{\odot}$ from the CO(7-6) data and $\rm M_{\rm dyn} = (3.4 \pm 0.4) \times 10^{10} \, \rm M_{\odot}$ using [CI](2-1). This is very similar to a comparable dynamical estimate from \citet{Aravena16} using only the line width and the continuum effective radius. We note that the estimate of the gas mass from CO(2-1) and using $\alpha_{\rm CO} = 0.8$ seems to agree better with the dynamical mass than the [CI]-derived ones. In Fig.\,\ref{fig:tdep_vs_z}, we see that the depletion timescale is somewhat shorter compared to the bulk of the SMGs. Similar to SPT0103-45, using $\alpha_{\rm CO} = 3.4$ leads to tension between the gas and dynamical masses.

%------------------------------------------------------------------

\subsection{SPT2357-51}

The most interesting feature of this compact source is the double-peak profile of the spectrum (Fig.\,\ref{fig:int_spectra}, bottom panel) seen for both lines. The PV diagram of the source (see Fig.\,\ref{fig:pv_diag}), the image plane spectral modelling of the source with two components (Sect.\,\ref{sec:kinematics 2357}) and the lens reconstruction (Sect.\,\ref{sec: lens modelling 2357}) suggest a merger between two components. The intrinsic flux ratio between the two components are 0.5:1 and 0.84:1 in CO(7-6) and [CI](2-1), respectively, with the blue component being brighter.

The blue component shows a smooth velocity gradient for both lines suggesting possible rotation or tidal streaming motions, whereas the red component exhibits a much more disturbed velocity structure. Comparing the positions of these components with the continuum emission, we notice that the red component is more centered on the continuum than the blue component. This could indicate a difference in their dust properties, with the red component being more dusty than the blue one. This could indicate a different chemical maturity of the two components which might be merging. However, it is surprising that the dustier component is the less massive one, since usually more massive galaxies are also dustier \citep{Heinis14,Fudamoto20}. %The Fudamoto is the ALPINE paper.

%DISCUSSED IN THE LENS MODELLING The lensing reconstruction was done using a single-Gaussian source profile for both the lines and the continuum emission. The source positions for different velocity bins is shown, where we can see a gradient. There seems to be a trend with the red component, slightly varying with the blue component.  There are no significant evidences of differential magnification. The blue-component is intrinsically brighter than the red component. 

%ALREADY DISCUSSED IN THE RATIO SECTION The ratio analysis of the source reveals a rather strange behaviour. We see a deficit of continuum in the center of the source. It is counter-intuitive, since the size measured with gas tracers are usually smaller than the continuum one \citep{Spilker16,Dong19,Fujimoto20}. This central region corresponds to the overlap of the two component of this candidate merger system (see Sect.\,\ref{sec:kinematics 2357}). In this interacting region, the line could be more excited. The continuum luminosity could be also underestimated if the dust is much warmer in the core of this object. We do not observed any significant variation of the CO/[CI] ratio, suggesting that there is no strong suppression of CO due to cosmic rays in the center of the system \citep{Papadopoulos04,Papadopoulos18}. But we do not overinterpret this as we still do not have a significant resolution. 

Using the magnification factor obtained from our continuum best-fit lens model and the L$_{\rm IR}$ from \citet{Reuter20}, we compute an intrinsic SFR of $1620 \pm 260 \rm \, M_{\odot} \rm yr ^{-1}$. Based on the continuum source size from the lens modelling (Sect.\,\ref{sec: lens modelling 2357}), we estimate the SFR density $\Sigma_{\rm SFR}$ to be $59 \pm 10 \, \rm M_{\odot} yr^{-1} kpc^{-2}$, which is similar to the other sources and typical of SMGs. 

We also estimate the gas mass of the system from the CO(3-2) flux measurement from \citet{Reuter20}. Using a CO(3-2)/CO(1-0) conversion factor of 0.7 $\pm$ 0.1 from \citet{Harrington21} and $\alpha_{\rm CO} = 3.4$, we obtain $\rm M_{\rm gas} = (1.6 \pm 0.5) \times 10^{11} \, \rm M_{\odot}$ and a depletion timescale $\rm t_{\rm dep} = 99 \pm 29$ Myr. Alternatively, using an  $\alpha_{\rm CO} = 0.8$, we obtain $\rm M_{\rm gas} = (3.9 \pm 1.3) \times 10^{10} \, \rm M_{\odot}$ and a very short depletion timescale $\rm t_{\rm dep} = 24 \pm 9$ Myr. This is much lower than the depletion timescales of the bulk of the SMG population shown in Fig.\,\ref{fig:tdep_vs_z}.
Such short depletion times have also been measured in some extreme lensed SMGs but remain unusual \citep{Bethermin16,Ciesla20}. This reinforces the hypothesis that SPT2357-51 is an extreme starburst possibly triggered by a major merger. We do not calculate the dynamical mass of the system as we cannot assume that it is a dynamically relaxed system. 

%-----------------------------------------------------------------

%------------------------------------------------------------------

\section{Conclusions \label{sec: conc}}

In this paper, we presented high-resolution ALMA spectral imaging of a sample of three DSFGs: SPT0103-45, SPT2147-50 and SPT2357-51. We characterised the gas reservoirs, kinematics and morphology of the sample. We performed pixel-wise velocity decomposition, visibility-based lens modelling and ratio analyses of the line and continuum emission of our sources. Our main conclusions are:
% \textcolor{green}{would you like to make a table (in the discussion or in the conclusion) summarizing the different parameters (SFR, Mmol,tdep,...) for the three sources, so the reader can grasp the total results in a glance ?}
\begin{itemize} 
    \item All the sources have high intrinsic SFRs > 800 $\rm M_{\odot}$ yr$^{-1}$. 
    \item Our sources have comparable ISM radiation field intensities and densities as other SMGs. 
    \item The CO(7-6)/[CI](2-1) luminosity ratios are lower than in the lensed SMGs presented in \citet{Andreani18} and \citep{Yang17}. 
    \item SPT0103-45 and SPT2147-50 are consistent with rotating starbursts with short depletion scales ($<$100 Myr). However, since our tracers predominantly trace cold gas, we cannot exclude the possibility of merging with a mostly ionised or low-magnification component.
    \item The dynamical mass is compatible with the gas mass of these two sources estimated assuming $\alpha_{\rm CO} = 0.8$ but not Milky Way like values. A higher $\alpha_{\rm CO}$ results in the gas mass being much larger than the dynamical mass, which is not physical. 
    \item SPT2357-51 could possibly be an ongoing merger. The red and the blue components seen in the Gaussian decomposition are intrinsically different. This source is a strong starburst with a short depletion timescale of 26 $\pm$ 5 Myr. 
\end{itemize}
This small sample demonstrates how powerful spatially and spectrally resolved multi-line ALMA observations can be in unveiling the nature of high-$z$ lensed starbursts. Our analysis reveals a diversity in the morphology and dynamics of SMGs, in agreement with simulations \citep[e.g.,][]{Hayward11}. With larger samples, more emission lines and higher spatial resolution, we will be able to better characterise the ISM of such massive galaxies and probe into the diversity of the population.

%--------------------------------------------------------------------
\begin{acknowledgements}
We thank Francesco Valentino for providing the data associated with his [CI] data compilation. We thank Zhi-Yu Zhang for sharing his GILDAS/CLASS script to automatically flag bad APEX data. This paper makes use of the following ALMA data: ADS/JAO.ALMA$\#$2017.1.01018.S, ADS/JAO.ALMA$\#$2018.1.01060.S. ALMA is a partnership of ESO (representing its member states), NSF (USA) and NINS (Japan), together with NRC (Canada), MOST and ASIAA (Taiwan) and KASI (Republic of Korea), in cooperation with the Republic of Chile. The Joint ALMA Observatory is operated by ESO, AUI/NRAO and NAOJ. The National Radio Astronomy Observatory is a facility of the National Science Foundation operated under cooperative agreement by Associated Universities, Inc. This publication is based on data acquired with the Atacama Pathfinder Experiment (APEX). APEX is a collaboration between the Max-Planck-Institut fur Radioastronomie, the European Southern Observatory and the Onsala Space Observatory. Based on observations made with ESO Telescopes at the La Silla Paranal Observatory under programme ID 097.A-0973. This work was supported by the Programme National “Physique et Chimie du Milieu Interstellaire” (PCMI) of CNRS/INSU with INC/INP co-funded by CEA and CNES. This work was supported by the Programme National Cosmology et Galaxies (PNCG) of CNRS/INSU with INP and IN2P3, co-funded by CEA and CNES. JSS is supported by NASA Hubble Fellowship grant \#HF2-51446  awarded  by  the  Space  Telescope  Science  Institute,  which  is  operated  by  the  Association  of  Universities  for  Research  in  Astronomy,  Inc.,  for  NASA,  under  contract  NAS5-26555. MA acknowledges support from FONDECYT grant 1211951, ANID+PCI+INSTITUTO MAX PLANCK DE ASTRONOMIA MPG 190030, ANID+PCI+REDES 190194 and ANID BASAL project FB210003. MAA and JV acknowledge support from the Center for AstroPhysical Surveys at the National Center for Supercomputing Applications in Urbana, IL. JV acknowledges support from the Sloan Foundation.

\end{acknowledgements}

% style aa.bst
\bibliographystyle{aa} 
\bibliography{bibliography.bib}

\begin{appendix}

%---------------------------------------------------------------
\section{High-resolution continuum image of SPT2357-51}
\begin{figure}
\centering

\includegraphics[width=9cm]{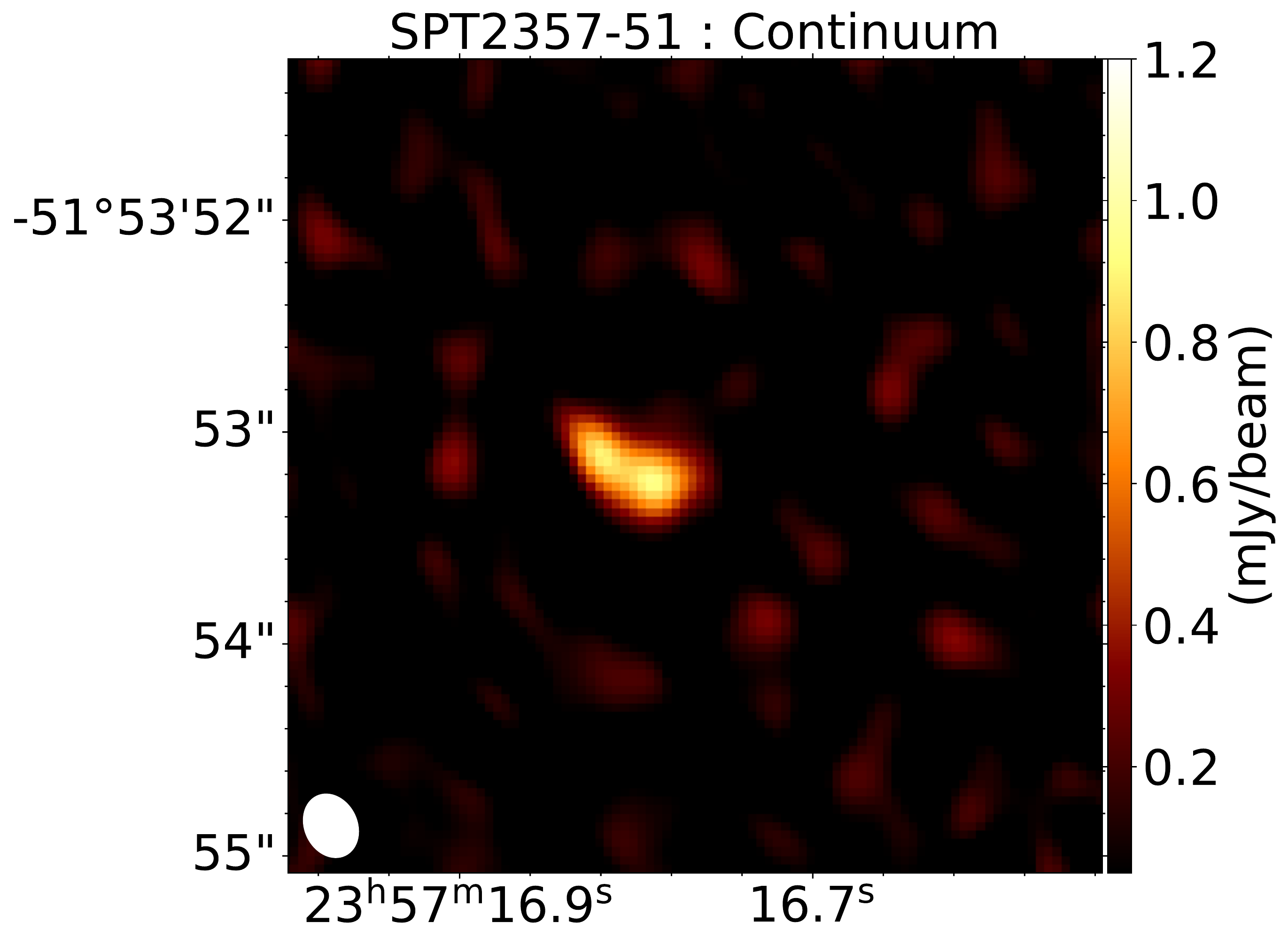}

\caption{\label{fig:highres_2357} The continuum image of SPT2357-51 imaged using the data from baselines $>$200\,k$\lambda$, demonstrating the resolved structure of the galaxy and its agreement with the results of our gravitational lens modelling.}
\end{figure}

To visualise the continuum emission of SPT2357-51 at higher resolution than our default imaging scheme, we imaged the continuum data using only the visibilities on the longest baselines, as this sample the structure of the galaxy on the smallest scales. Figure~\ref{fig:highres_2357} shows the continuum image of SPT2357-51 created using only the data from baselines $>$200\,k$\lambda$. The spatial resolution of this image is 0.31$\times$0.24$^{\prime\prime}$. We note that this image is merely for illustrative purposes because our lens modelling procedure fits directly to the visibilities, which are not affected by choices in imaging parameters.

%---------------------------------------------------------------
\section{De-blending the spectra of SPT0103-45 \label{deblending}}
\begin{figure}
\centering

\includegraphics[width=9cm]{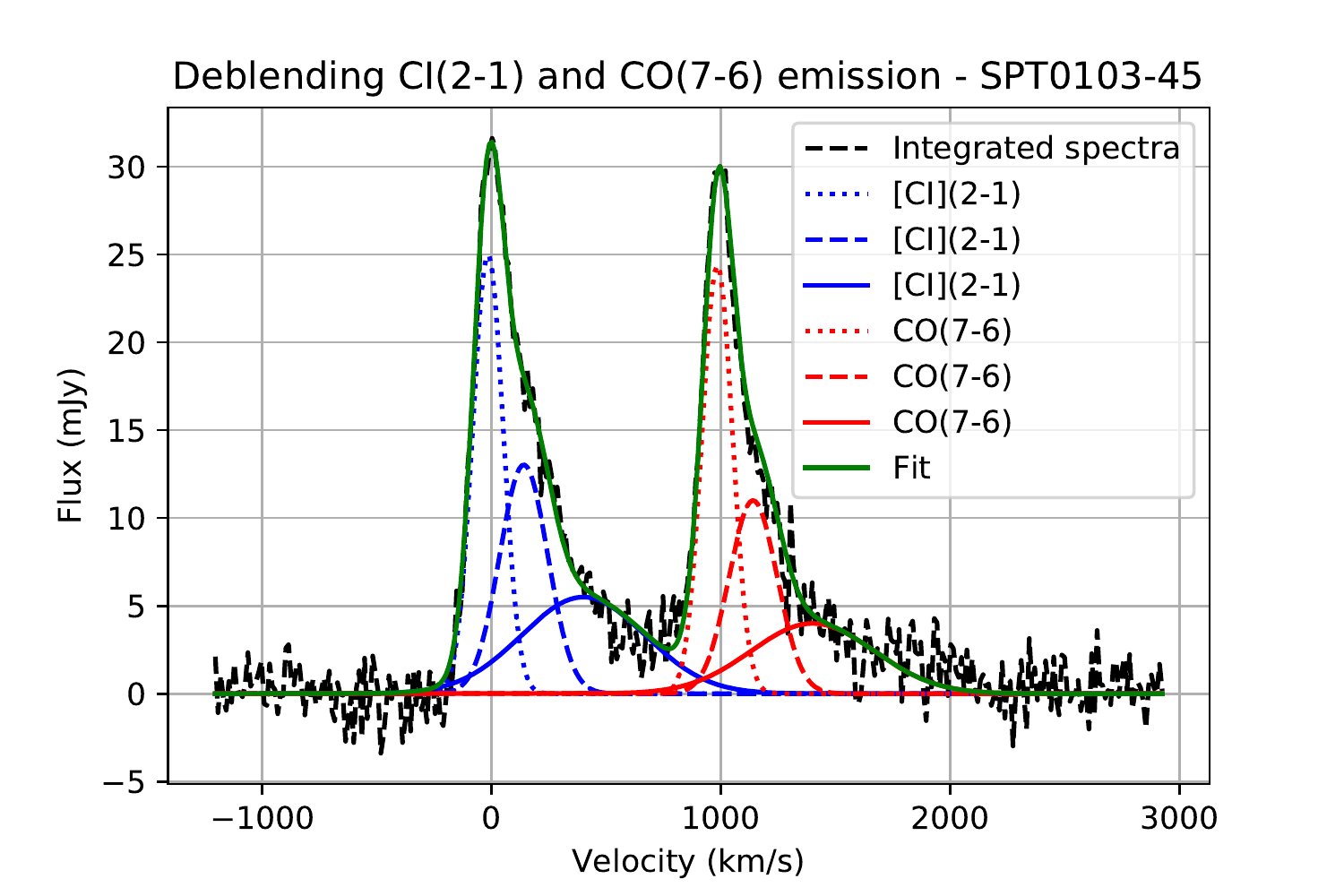}

\caption{\label{fig:deblending} The Gaussian deblending of the CO and [CI] spectrum of SPT0103-45. Each of the lines are fitted with 3 Gaussian profiles. The blue fit represents the [CI](2-1) line and the red fit is the CO(7-6) line fit. The green line represents the total of both lines.}
\end{figure}

From Fig.\,\ref{fig:int_spectra} (top panel), we see that the redshifted tail of [CI](2-1) emission is blended with the CO(7-6) line. Hence we deblended these lines using multiple-Gaussian fitting, as described further in the text. We fit the CO and [CI] lines with three Gaussian profiles each, but fixed the central velocity and line width to be equal for both CO and [CI]. The resulting fit uses 12 free parameters: six Gaussian amplitudes, three central velocities and three line widths. In the resulting best-fit line profiles (Figure~\ref{fig:deblending}), we find a narrow overlap region where both CO and [CI] contribute, near $\sim+800$\,km/s relative to CO(7-6). In our spatially-resolved line ratio analysis, we simply divided the two lines using the frequency ranges shown in Figure~\ref{fig:int_spectra}. Compared to that method, a more detailed accounting of the line blending results in a higher CO/[CI] line ratio by $\approx$5\%, a $<2\,\sigma$ difference. This small difference is taken into account by combining the statistical uncertainties with a 5$\%$ systematic error.

%---------------------------------------------------------------
\section{Lens modelling parameters}

The best-fit parameters of the lensing potential and the source profiles are tabulated in Table\,\ref{tab:lens_params}.  

\begin{table*}
\centering
\caption{\label{tab:lens_params}Lens parameters. $x_L$ and $y_L$ are the position of the lens in arcseconds relative to the phase center, $\theta_{E,L}$ is the Einstein radius of the lens, $e_L$ is the ellipticity and $\phi_L$ is the position angle in degrees east of north.}
\begin{tabular}{cccccc}

\hline
\hline
&&&&&\\
Source & $x_L$ & $y_L$ & $\theta_{E,L} $ & $e_{L}$ &  $\phi_L$  \\
& (arcsec) & (arcsec) & (arcsec) &  &  (deg. E of N)  \\
&&&&&\\
\hline
&&&&&\\
SPT0103-45&-0.45$\pm$0.01 & -0.49$\pm$0.01& 0.87 $\pm$ 0.02&0.12$\pm$0.01 & 81$\pm$2\\
SPT2147-50&0.73$\pm$0.01&-0.74$\pm$0.01&1.19 $\pm$0.02& 0.28$\pm$0.01&13$\pm$1\\
SPT2357-51& 0.09$\pm$0.02 &-0.10$\pm$0.04&0.19 $\pm$ 0.02 &0.64$\pm$0.07&152$\pm$3\\

\hline
\end{tabular}
\end{table*}

%--------------------------------------------------------------------
%TABLESSSSS-----------------------

% SPT0103-45

%------------------------------------------------------------------
\begin{table*}[h]
\caption{\label{source_params_0103}Intrinsic properties of SPT0103-45. The velocity is the central velocity of the velocity bin, $x_S$ and $y_S$ are the source positions relative to the lens, the flux is the intrinsic flux of the source, $r_{\rm eff}$ is the effective half-light major axis of the source, $n_S$ is the Sérsic index, $b_S$/$a_S$ is the axis ratio, $\phi_S$ is the position angle of the source and $\mu$ is the magnification.}
\scalebox{0.93  }{
\begin{tabular}{ccccccccc}
\hline
\hline
&&&&&&&&\\
Velocity & $x_S$ & $y_S$ & Flux & $r_{\rm eff}$ & $n_S$ & $b_S$/$a_S$ & $\phi_S$ & $\mu$ \\
(km/s) & (arcsec) & (arcsec) & (mJy) & (arcsec) &  &  & (deg. E of N) &  \\
&&&&&&&&\\
\hline
&&&&&&&&\\
\multicolumn{9}{c}{Continuum}\\
&&&&&&&&\\
\hline
&&&&&&&&\\
Extended & 0.443$\pm$0.002 & 0.246$\pm$0.001 & 4.97$\pm$0.03 & 0.380$\pm$0.002 &0.99$\pm$0.01 & 0.52$\pm$0.01 &27.7$\pm$0.3 & 4.51$\pm$0.03\\
Compact & 0.078$\pm$0.001 & 0.214$\pm$0.002 & 0.29$\pm$0.01 & 0.055$\pm$0.002 & 0.39$\pm$0.07 & 0.17$\pm$0.01 & 179.8$\pm$0.2 & 8.78$\pm$0.22 \\
\hline
&&&&&&&&\\
\multicolumn{9}{c}{[CI](2-1)}\\
&&&&&&&&\\
\hline
&&&&&&&&\\
727.1 & 0.739$\pm$0.008 & 0.375$\pm$0.006 & 2.53$\pm$0.07 & 0.280$\pm$0.010 & 0.50$\pm$0.09 & 0.56$\pm$0.03 & 41$\pm$3 & 2.38$\pm$0.04 \\
575.6 & 0.584$\pm$0.006 & 0.295$\pm$0.009 & 2.16$\pm$0.06 & 0.304$\pm$0.009 & 0.27$\pm$0.08 & 0.39$\pm$0.03 & 82$\pm$2 & 2.73$\pm$0.03 \\
424.1 & 0.465$\pm$0.004 & 0.259$\pm$0.008 & 2.17$\pm$0.05 & 0.336$\pm$0.008 & 0.18$\pm$0.06 & 0.26$\pm$0.02 & 88$\pm$1 & 3.26$\pm$0.03 \\
272.6 & 0.332$\pm$0.014 & 0.228$\pm$0.007 & 2.47$\pm$0.16 & 0.366$\pm$0.062 & 0.71$\pm$0.09 & 0.43$\pm$0.30 & 89$\pm$43 & 4.34$\pm$0.24 \\
196.9 & 0.283$\pm$0.007 & 0.217$\pm$0.009 & 2.79$\pm$0.14 & 0.381$\pm$0.022 & 0.82$\pm$0.12 & 0.52$\pm$0.04 & 77$\pm$3 & 5.20$\pm$0.18 \\
136.3 & 0.215$\pm$0.006 & 0.203$\pm$0.008 & 2.47$\pm$0.10 & 0.308$\pm$0.013 & 0.55$\pm$0.07 & 0.63$\pm$0.04 & 71$\pm$3 & 6.83$\pm$0.19 \\
75.7 & 0.175$\pm$0.005 & 0.195$\pm$0.005 & 2.67$\pm$0.08 & 0.259$\pm$0.008 & 0.48$\pm$0.06 & 0.68$\pm$0.03 & 51$\pm$4 & 8.17$\pm$0.20 \\
15.2 & 0.127$\pm$0.004 & 0.164$\pm$0.004 & 2.78$\pm$0.08 & 0.217$\pm$0.006 & 0.43$\pm$0.05 & 0.67$\pm$0.03 & 45$\pm$2 & 10.31$\pm$0.23 \\
-45.4 & 0.063$\pm$0.004 & 0.148$\pm$0.003 & 2.12$\pm$0.07 & 0.172$\pm$0.007 & 0.52$\pm$0.08 & 0.78$\pm$0.04 & 23$\pm$5& 11.45$\pm$0.31 \\
-106.1 & 0.052$\pm$0.012 & 0.148$\pm$0.008 & 1.16$\pm$0.09 & 0.371$\pm$0.035 & 0.76$\pm$0.21 & 0.21$\pm$0.02 & 32$\pm$2 & 9.17$\pm$0.54 \\
\hline
&&&&&&&&\\
\multicolumn{9}{c}{CO(7-6)}\\
&&&&&&&&\\
\hline
&&&&&&&&\\
866.2 & 0.796$\pm$0.004 & 0.386$\pm$0.005 & 1.29$\pm$0.05 & 0.348$\pm$0.016 & 0.22$\pm$0.15 & 0.21$\pm$0.03 & 23$\pm$1 & 2.31$\pm$0.06 \\
714.3 & 0.708$\pm$0.008 & 0.412$\pm$0.006 & 1.91$\pm$0.06 & 0.278$\pm$0.011 & 0.55$\pm$0.14 & 0.34$\pm$0.03 & 44$\pm$2 & 2.42$\pm$0.05 \\
562.3 & 0.545$\pm$0.007 & 0.301$\pm$0.009 & 1.45$\pm$0.05 & 0.261$\pm$0.010 & 0.10$\pm$0.07 & 0.40$\pm$0.04 & 76$\pm$3 & 2.90$\pm$0.04 \\
410.3 & 0.443$\pm$0.005 & 0.301$\pm$0.010 & 1.72$\pm$0.05 & 0.330$\pm$0.010 & 0.12$\pm$0.07 & 0.24$\pm$0.03 & 89$\pm$2 & 3.25$\pm$0.04 \\
258.4 & 0.330$\pm$0.006 & 0.251$\pm$0.009 & 2.01$\pm$0.07 & 0.307$\pm$0.011 & 0.53$\pm$0.08 & 0.55$\pm$0.03 & 95$\pm$3 & 4.35$\pm$0.08 \\
197.6 & 0.289$\pm$0.008 & 0.251$\pm$0.010 & 2.55$\pm$0.15 & 0.312$\pm$0.025 & 1.05$\pm$0.19 & 0.59$\pm$0.05 & 104$\pm$6 & 4.51$\pm$0.16 \\
136.8 & 0.221$\pm$0.007 & 0.224$\pm$0.008 & 2.35$\pm$0.11 & 0.252$\pm$0.016 & 0.79$\pm$0.13 & 0.72$\pm$0.06 & 119$\pm$7 & 5.68$\pm$0.18 \\
75.9 & 0.191$\pm$0.006 & 0.222$\pm$0.005 & 2.67$\pm$0.08 & 0.213$\pm$0.009 & 0.54$\pm$0.07 & 0.76$\pm$0.05 & 14$\pm$6 & 6.90$\pm$0.14 \\
15.2 & 0.138$\pm$0.004 & 0.193$\pm$0.004 & 2.91$\pm$0.07 & 0.196$\pm$0.005 & 0.41$\pm$0.04 & 0.69$\pm$0.03 & 43$\pm$4 & 9.37$\pm$0.17 \\
-45.6 & 0.055$\pm$0.003 & 0.177$\pm$0.003 & 3.06$\pm$0.17 & 0.241$\pm$0.020 & 1.46$\pm$0.18 & 0.68$\pm$0.04 & 31$\pm$4 & 8.53$\pm$0.36 \\
\hline

\end{tabular}}

\end{table*}

%------------------------------------------------------------------------

%  SPT2147-50

%----------------------------------------

\begin{table*}[h]
\caption{\label{tab:source params 2147}Intrinsic properties of  SPT2147-50.}
\scalebox{0.93  }{
\begin{tabular}{ccccccccc}

\hline
\hline
&&&&&&&&\\
Velocity & $x_S$ & $y_S$ & Flux & $r_{\rm eff}$ & $n_S$ & $b_S$/$a_S$ & $\phi_S$ & $\mu$ \\
(km/s) & (arcsec) & (arcsec) & (mJy) & (arcsec) &  &  & (deg. E of N) &  \\
&&&&&&&&\\
\hline
&&&&&&&&\\
Continuum&-0.240$\pm$0.004&0.254$\pm$0.005&0.96$\pm$0.02& 0.127$\pm$0.002&1.37$\pm$0.06 &0.78$\pm$0.02&-12$\pm$3&6.91$\pm$0.15\\
\hline
&&&&&&&&\\
\multicolumn{9}{c}{[CI](2-1)}\\
&&&&&&&&\\
\hline
&&&&&&&&\\
141.1 & -0.125$\pm$0.010 & 0.303$\pm$0.026 & 1.02$\pm$0.10 & 0.151$\pm$0.013 & 0.99$\pm$0.22 & 0.74$\pm$0.09 & -17$\pm$11 & 9.22$\pm$0.88 \\
52.9 & -0.179$\pm$0.011 & 0.313$\pm$0.019 & 1.10$\pm$0.07 & 0.195$\pm$0.013 & 1.05$\pm$0.19 & 0.42$\pm$0.05 & -56$\pm$2 & 9.24$\pm$0.52 \\
-35.3 & -0.211$\pm$0.014 & 0.238$\pm$0.017 & 0.86$\pm$0.08 & 0.216$\pm$0.017 & 0.99$\pm$0.27 & 0.30$\pm$0.05 & -56$\pm$3 & 8.53$\pm$0.64 \\
-123.4 & -0.350$\pm$0.030 & 0.264$\pm$0.024 & 1.00$\pm$0.12 & 0.210$\pm$0.021 & 0.78$\pm$0.31 & 0.51$\pm$0.09 & -58$\pm$7 & 5.36$\pm$0.57 \\
-211.6 & -0.451$\pm$0.038 & 0.274$\pm$0.028 & 1.41$\pm$0.20 & 0.150$\pm$0.020 & 1.85$\pm$0.56 & 0.77$\pm$0.15 & 55$\pm$73 & 4.03$\pm$0.55 \\
-299.7 & -0.439$\pm$0.054 & 0.207$\pm$0.027 & 0.73$\pm$0.12 & 0.102$\pm$0.019 & 1.96$\pm$0.98 & 0.80$\pm$0.16 & -10$\pm$53 & 4.30$\pm$0.65 \\
\hline
&&&&&&&&\\
\multicolumn{9}{c}{CO(7-6)}\\
&&&&&&&&\\
\hline
&&&&&&&&\\
106.2 & -0.172$\pm$0.009 & 0.315$\pm$0.020 & 1.55$\pm$0.12 & 0.137$\pm$0.008 & 0.82$\pm$0.12 & 0.67$\pm$0.06 & -35$\pm$6 & 7.68$\pm$0.55 \\
17.7 & -0.199$\pm$0.008 & 0.287$\pm$0.014 & 1.38$\pm$0.08 & 0.195$\pm$0.008 & 0.48$\pm$0.11 & 0.39$\pm$0.03 & -48$\pm$2 & 8.61$\pm$0.40 \\
-70.8 & -0.256$\pm$0.011 & 0.241$\pm$0.011 & 1.14$\pm$0.07 & 0.219$\pm$0.010 & 0.09$\pm$0.07 & 0.25$\pm$0.03 & -49$\pm$2 & 7.27$\pm$0.39 \\
-159.2 & -0.260$\pm$0.032 & 0.185$\pm$0.025 & 0.72$\pm$0.14 & 0.166$\pm$0.015 & 0.15$\pm$0.10 & 0.23$\pm$0.04 & -41$\pm$3 & 7.69$\pm$1.32 \\
-247.7 & -0.441$\pm$0.046 & 0.259$\pm$0.038 & 1.47$\pm$0.34 & 0.156$\pm$0.015 & 1.15$\pm$0.33 & 0.49$\pm$0.21 & -47$\pm$8 & 4.54$\pm$1.05 \\
-336.1 & -0.354$\pm$0.073 & 0.187$\pm$0.043 & 0.44$\pm$0.11 & 0.095$\pm$0.041 & 2.29$\pm$1.06 & 0.33$\pm$0.30 & 23$\pm$16 & 4.74$\pm$1.20 \\
\hline
\end{tabular}}
\end{table*}

%------------------------------------------------------

%     spt 2357-51

%--------------------------------------------------

\begin{table*}[h]
\centering
\caption{\label{tab:source params 2357}Intrinsic properties of SPT2357-51.}
\begin{tabular}{cccccc}

\hline
\hline
&&&&&\\
Velocity & $x_S$ & $y_S$ & Flux & $r_{\rm eff}$ & $\mu$ \\
(km/s) & (arcsec) & (arcsec) & (mJy) & (arcsec) & \\
&&&&&\\

\hline
&&&&&\\
Continuum&0.028$\pm$0.015&-0.136$\pm$0.028& 2.15$\pm$0.31&0.149$\pm$0.006&2.84$\pm$0.38\\
&&&&&\\
\hline
&&&&&\\
\multicolumn{6}{c}{[CI](2-1)}\\
&&&&&\\
\hline
316.7 & 0.106$\pm$0.156 & 0.022$\pm$0.123 & 0.11$\pm$0.07 & 0.141$\pm$0.267 & 2.35$\pm$1.67 \\
241.3 & -0.023$\pm$0.021 & -0.014$\pm$0.024 & 0.19$\pm$0.06 & 0.050$\pm$0.027 & 6.14$\pm$1.34 \\
165.9 & -0.046$\pm$0.005 & -0.006$\pm$0.008 & 0.67$\pm$0.08 & 0.055$\pm$0.011 & 5.79$\pm$0.56 \\
90.5 & -0.032$\pm$0.009 & 0.008$\pm$0.009 & 1.98$\pm$0.13 & 0.121$\pm$0.008 & 3.60$\pm$0.17 \\
15.1 & 0.044$\pm$0.013 & 0.011$\pm$0.007 & 2.01$\pm$0.09 & 0.140$\pm$0.005 & 3.17$\pm$0.09 \\
-60.3 & 0.027$\pm$0.014 & -0.020$\pm$0.012 & 1.81$\pm$0.15 & 0.159$\pm$0.012 & 2.91$\pm$0.16 \\
-135.7 & -0.011$\pm$0.011 & -0.107$\pm$0.007 & 1.71$\pm$0.14 & 0.136$\pm$0.012 & 3.03$\pm$0.16 \\
-211.1 & -0.031$\pm$0.008 & -0.146$\pm$0.007 & 1.56$\pm$0.10 & 0.083$\pm$0.006 & 3.36$\pm$0.14 \\
-286.5 & -0.015$\pm$0.006 & -0.194$\pm$0.006 & 2.30$\pm$0.09 & 0.093$\pm$0.004 & 2.66$\pm$0.07 \\
-361.9 & 0.034$\pm$0.008 & -0.238$\pm$0.006 & 2.44$\pm$0.10 & 0.100$\pm$0.005 & 2.05$\pm$0.05 \\
\hline
&&&&&\\
\multicolumn{6}{c}{CO(7-6)}\\
&&&&&\\
\hline
316.7 & 0.034$\pm$0.014 & -0.085$\pm$0.010 & 0.69$\pm$0.08 & 0.079$\pm$0.012 & 3.93$\pm$0.36 \\
241.3 & -0.048$\pm$0.002 & -0.005$\pm$0.006 & 1.53$\pm$0.11 & 0.072$\pm$0.007 & 5.01$\pm$0.27 \\
165.9 & 0.034$\pm$0.008 & -0.053$\pm$0.004 & 3.13$\pm$0.12 & 0.106$\pm$0.003 & 3.75$\pm$0.11 \\
90.5 & 0.015$\pm$0.007 & -0.055$\pm$0.003 & 2.80$\pm$0.09 & 0.115$\pm$0.004 & 3.64$\pm$0.08 \\
15.1 & -0.006$\pm$0.007 & -0.082$\pm$0.005 & 2.24$\pm$0.10 & 0.109$\pm$0.005 & 3.64$\pm$0.11 \\
-60.3 & -0.03$\pm$0.005 & -0.100$\pm$0.004 & 1.79$\pm$0.07 & 0.079$\pm$0.003 & 4.10$\pm$0.11 \\
-135.7 & -0.018$\pm$0.003 & -0.154$\pm$0.002 & 2.56$\pm$0.06 & 0.081$\pm$0.003 & 3.21$\pm$0.05 \\
-211.1 & -0.012$\pm$0.004 & -0.205$\pm$0.004 & 3.64$\pm$0.08 & 0.097$\pm$0.002 & 2.54$\pm$0.04 \\
-286.5 & 0.013$\pm$0.005 & -0.227$\pm$0.004 & 2.78$\pm$0.08 & 0.091$\pm$0.003 & 2.19$\pm$0.05 \\
-361.9 & 0.024$\pm$0.011 & -0.233$\pm$0.012 & 0.99$\pm$0.07 & 0.081$\pm$0.010 & 1.99$\pm$0.08 \\
\hline
\end{tabular}
\end{table*}

\section{Differential magnification - Intrinsic versus observed spectra for our sources}
\begin{figure}[h]
\centering

\includegraphics[width=9cm]{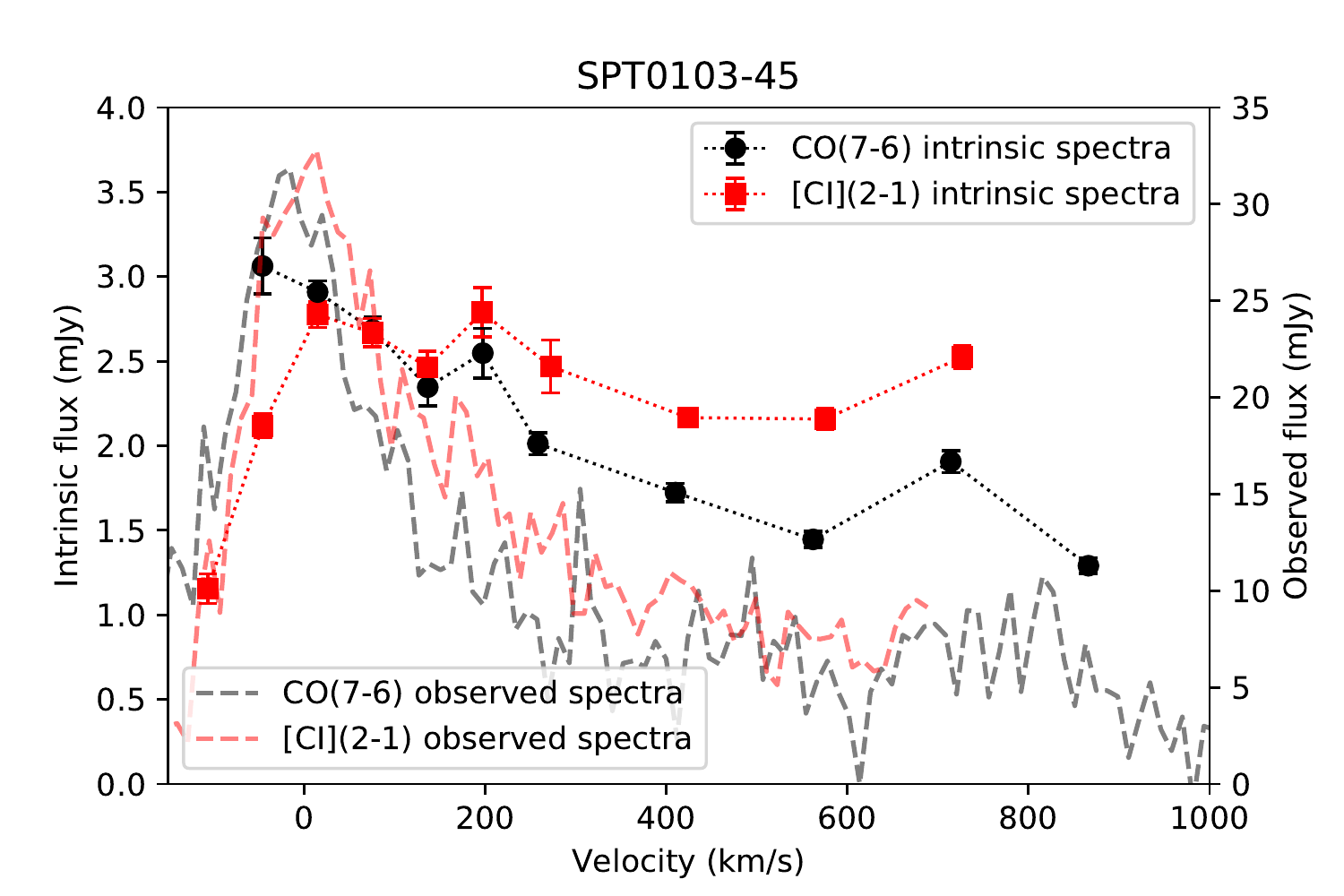}  \includegraphics[width=9cm]{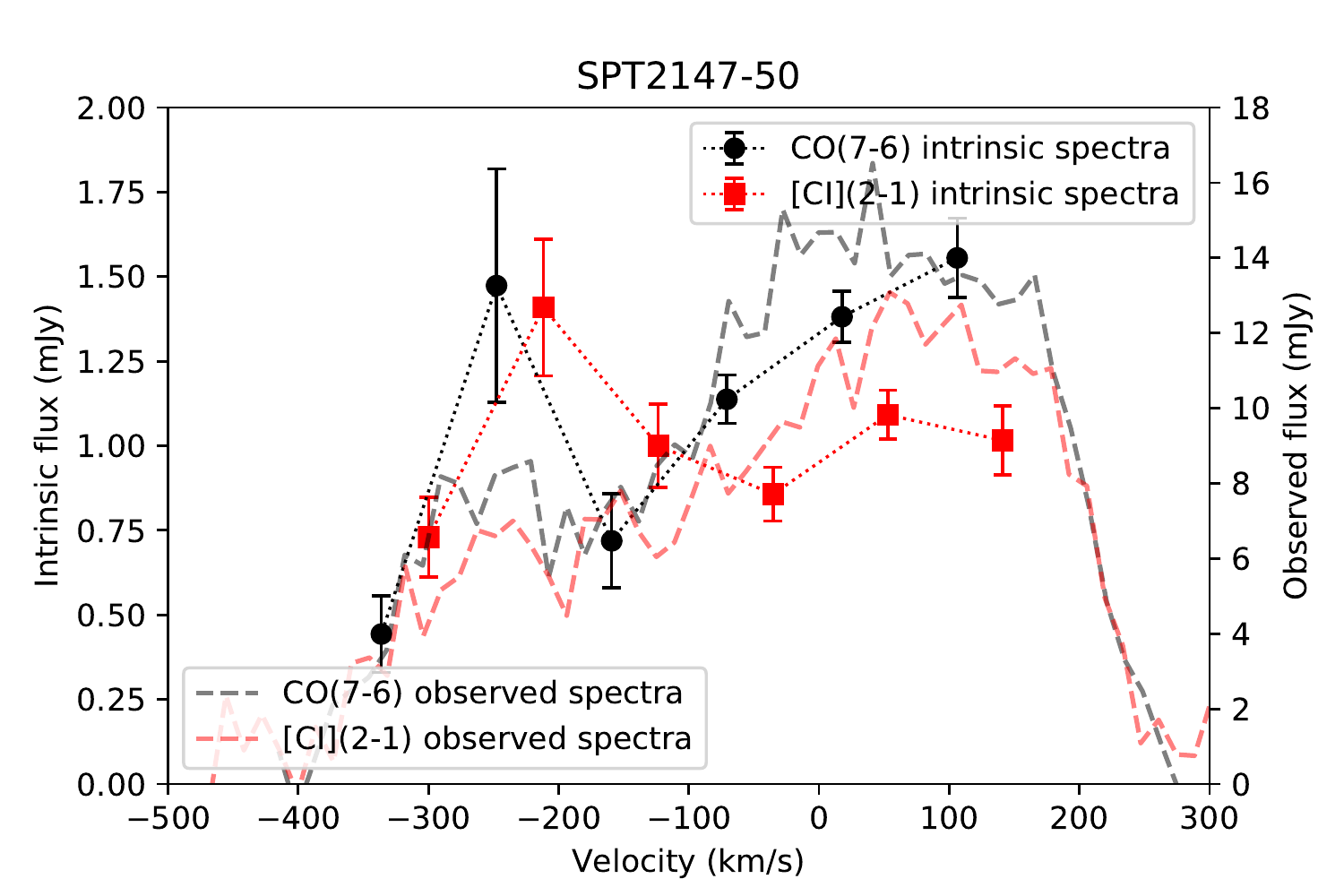} 
\includegraphics[width=9cm]{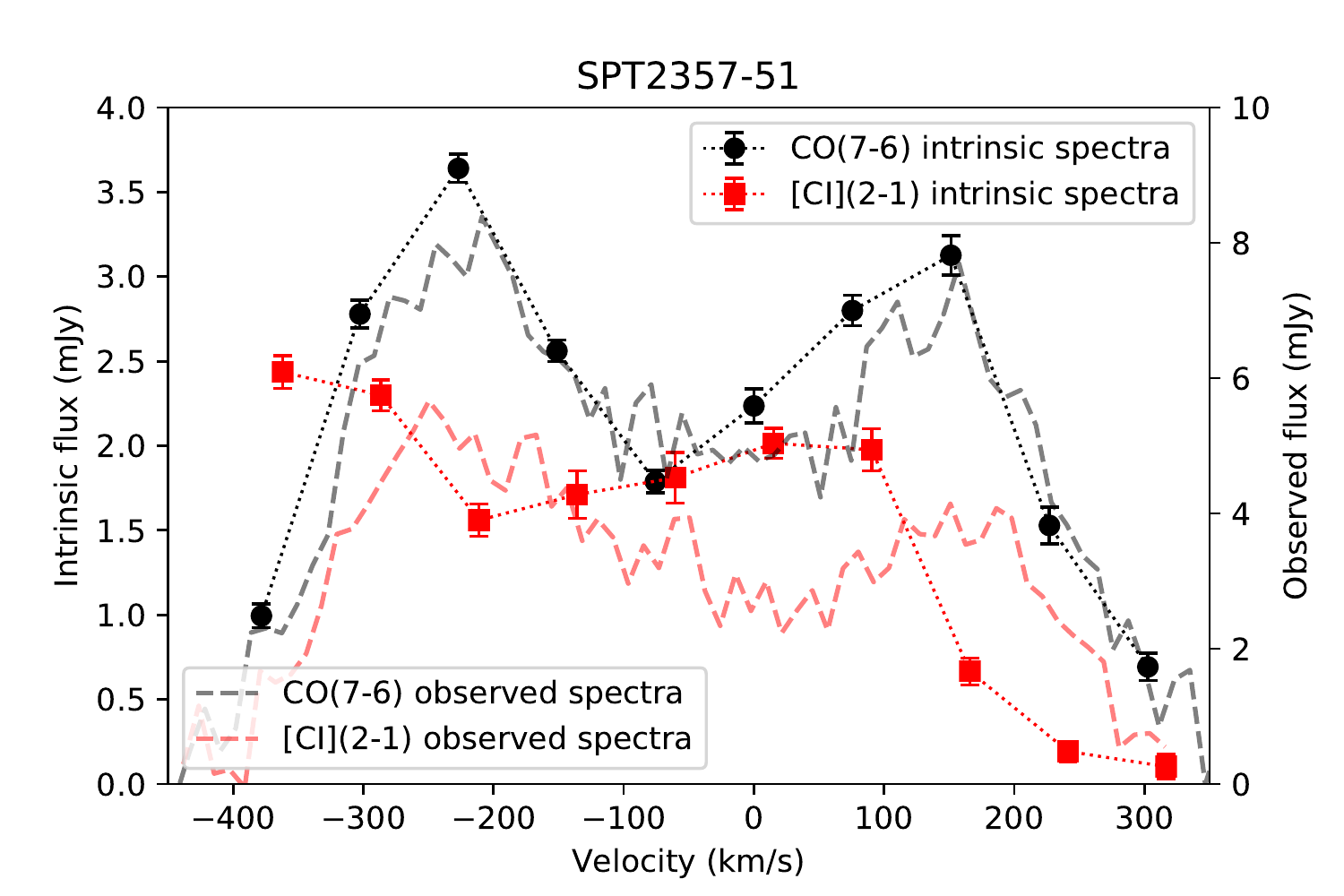} 

\caption{\label{fig:int vs obs} The intrinsic spectra for the [CI](2-1) line (red filled squares) and the CO(7-6) line (black filled circles). The top panel shows the results for SPT0103-45. The middle panel shows the results of SPT2147-50 and the bottom panel for SPT2357-51. We overplot the observed spectra using a dashed-line, red for [CI](2-1) and black for CO(7-6).}
\end{figure}

Figure\,\ref{fig:int vs obs} shows the intrinsic spectra (from the lens modelling of the velocity bins for our sources) compared with the observed spectra. 

\section{Comparison between the data and the lens model in the various velocity bins}

To check the agreement between our data and the lens model, we plot the dirty image of our sources in every velocity bin along with the models corresponding to the data. 

\subsection{SPT0103-45}
\begin{figure*}
    \centering
    \begin{tabular}{ccc}
    \includegraphics[width=6cm]{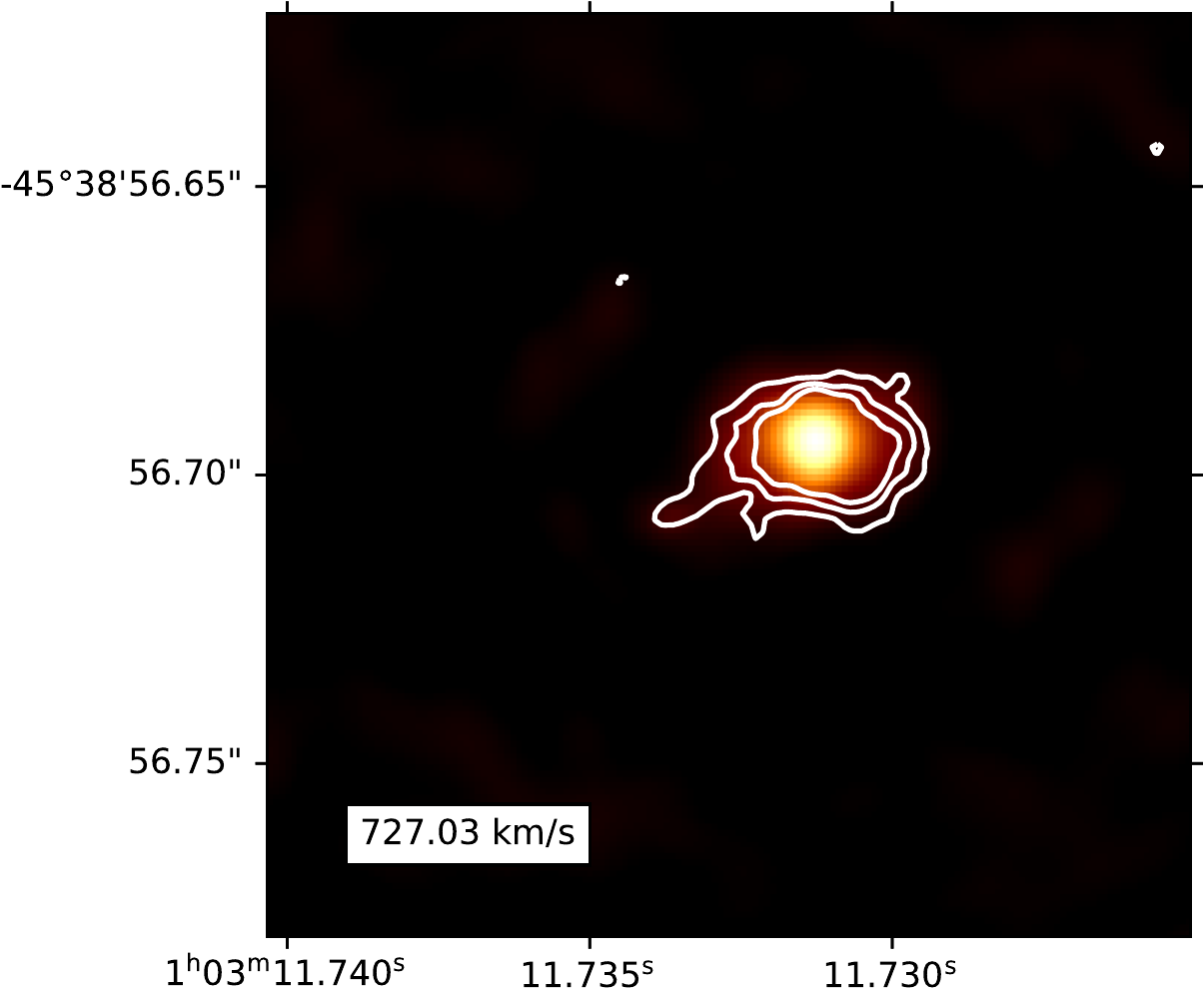} &  \includegraphics[width=6cm]{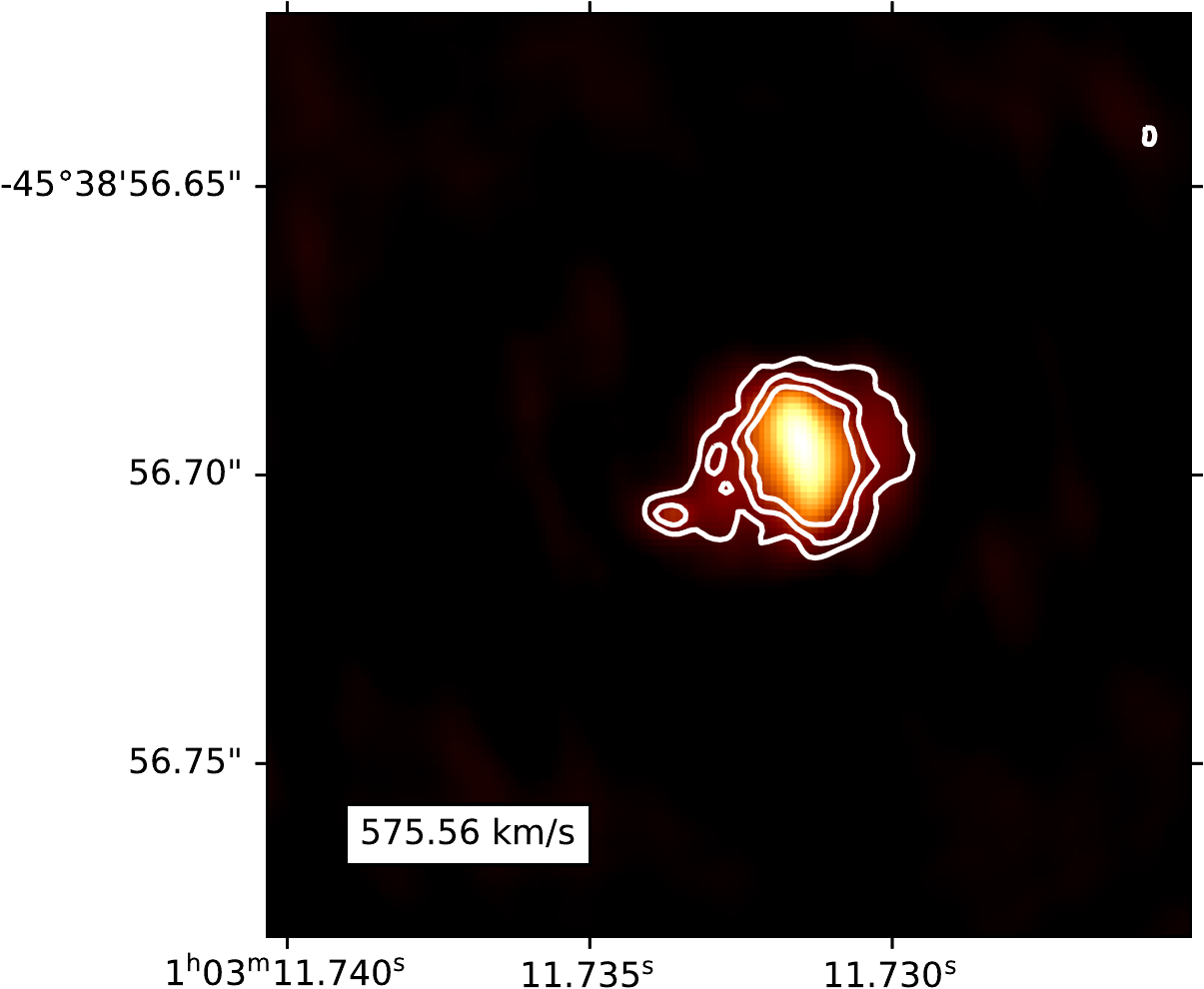} & \includegraphics[width=6cm]{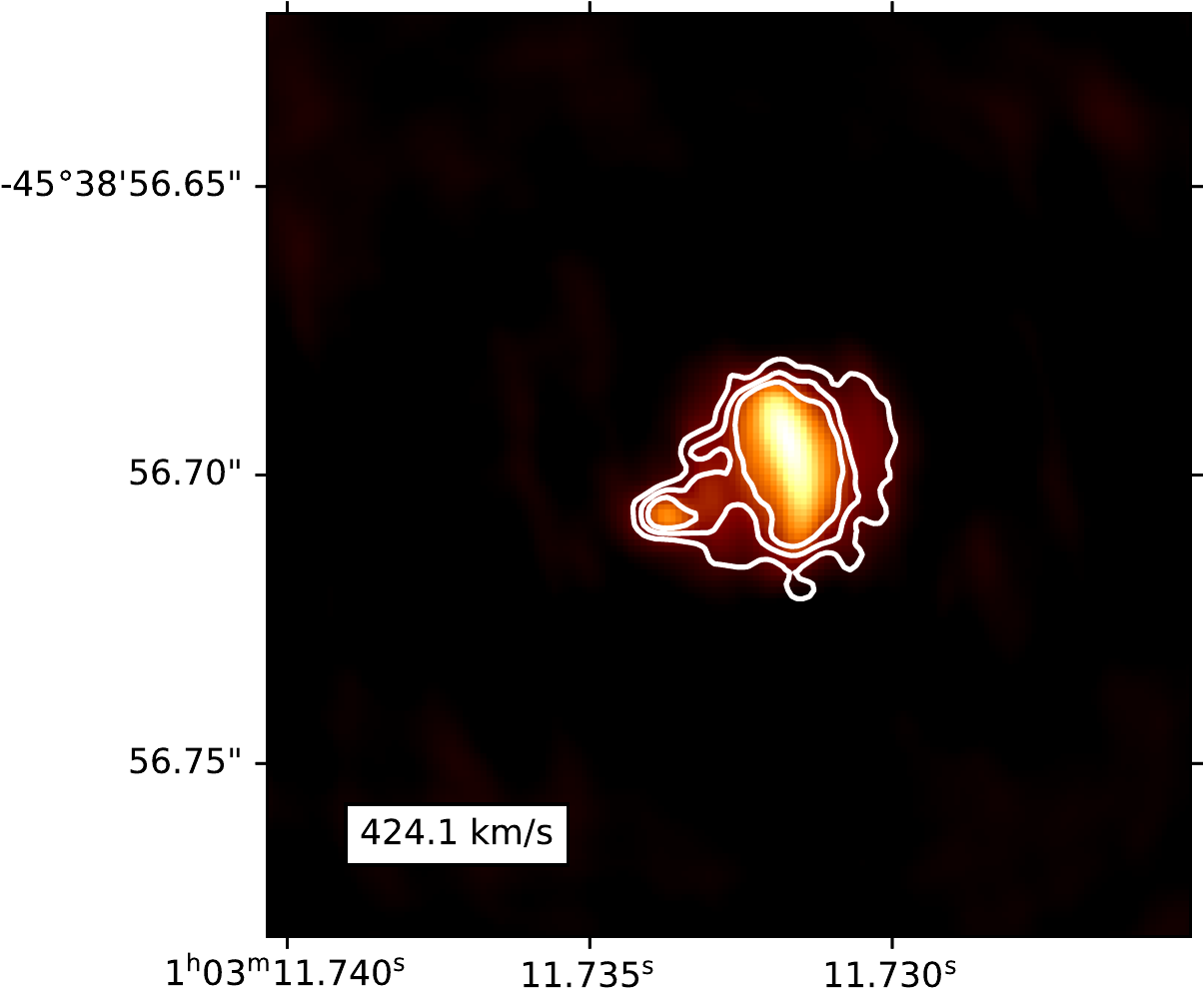} \\
    \includegraphics[width=6cm]{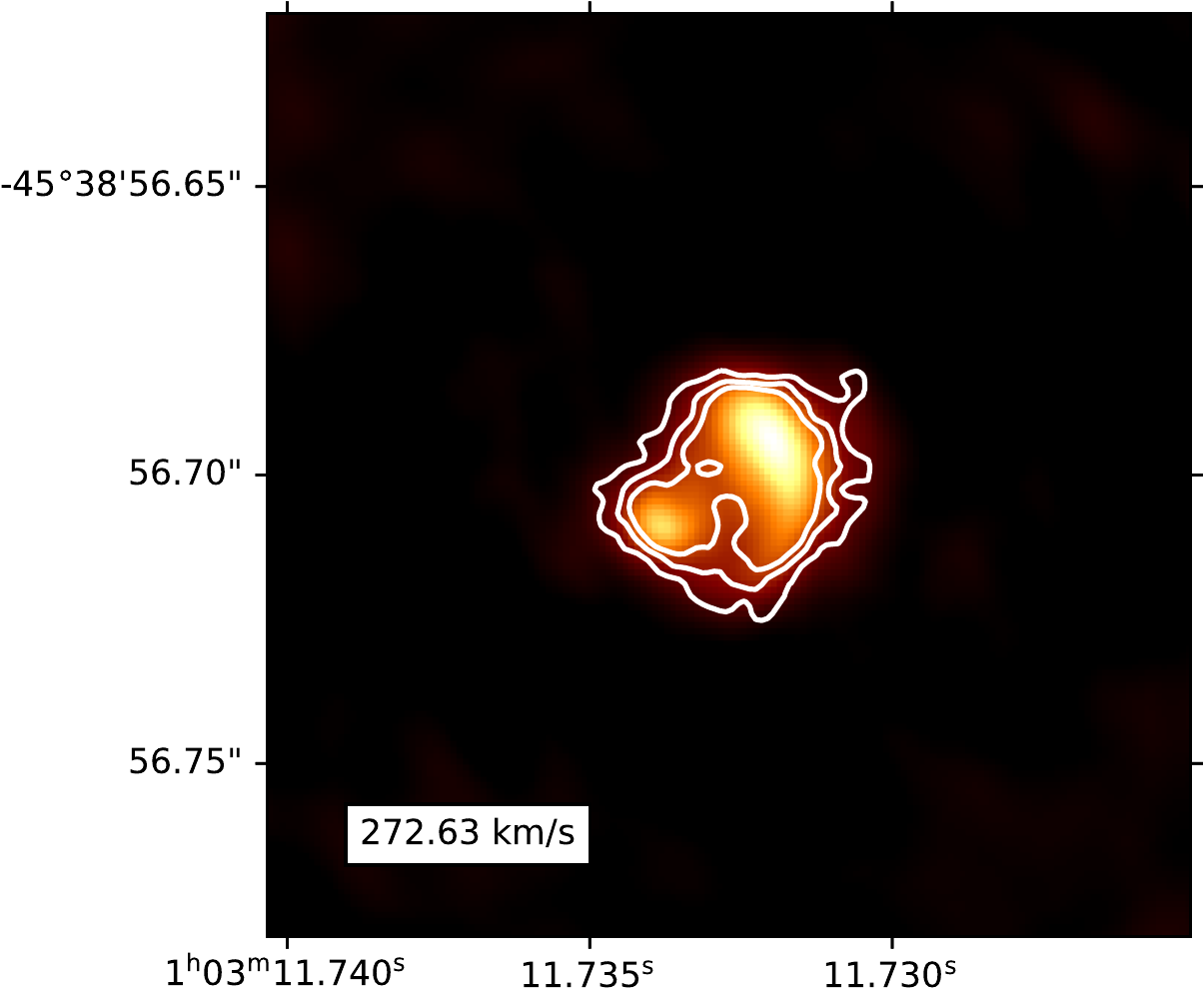} &  \includegraphics[width=6cm]{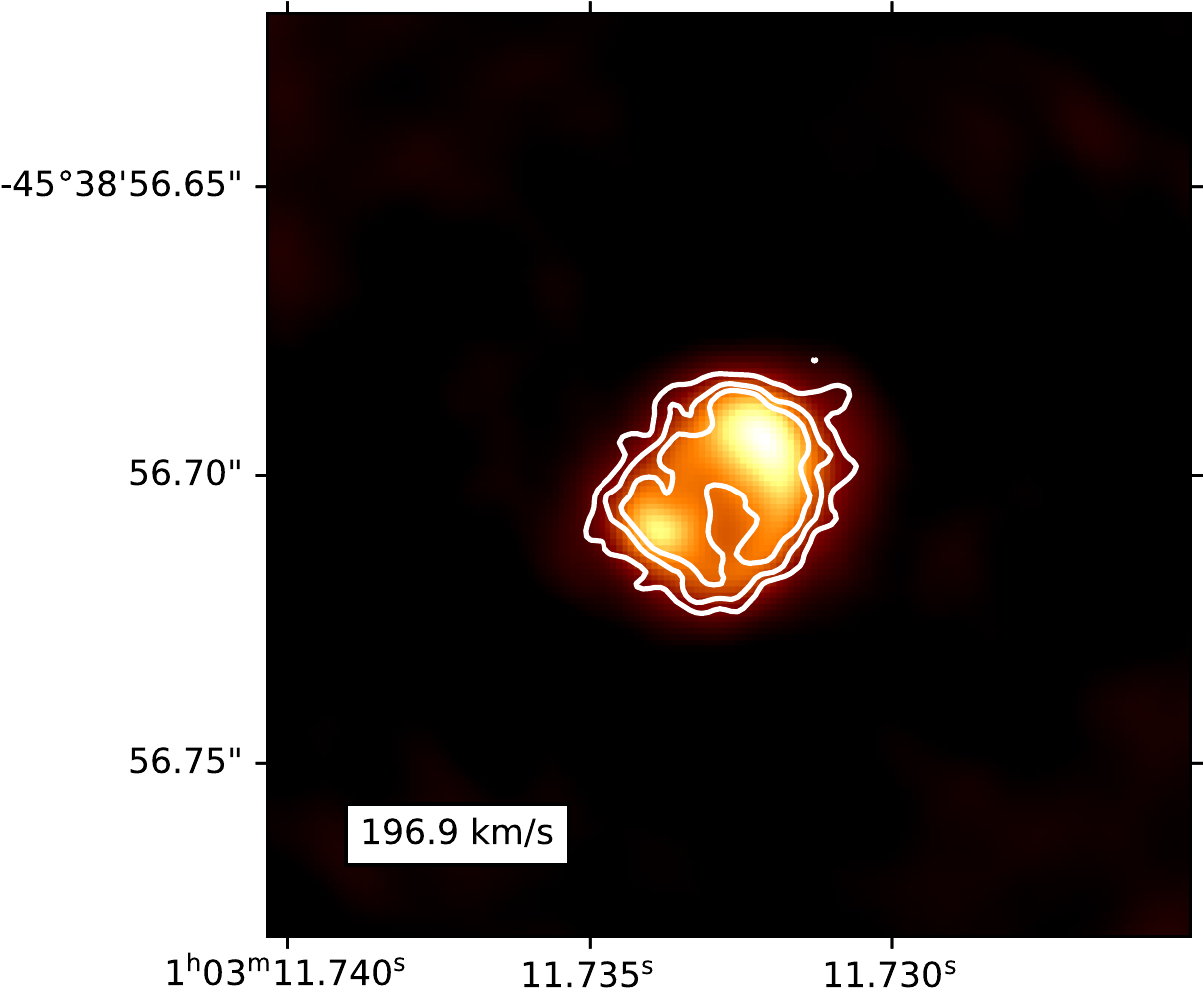} & \includegraphics[width=6cm]{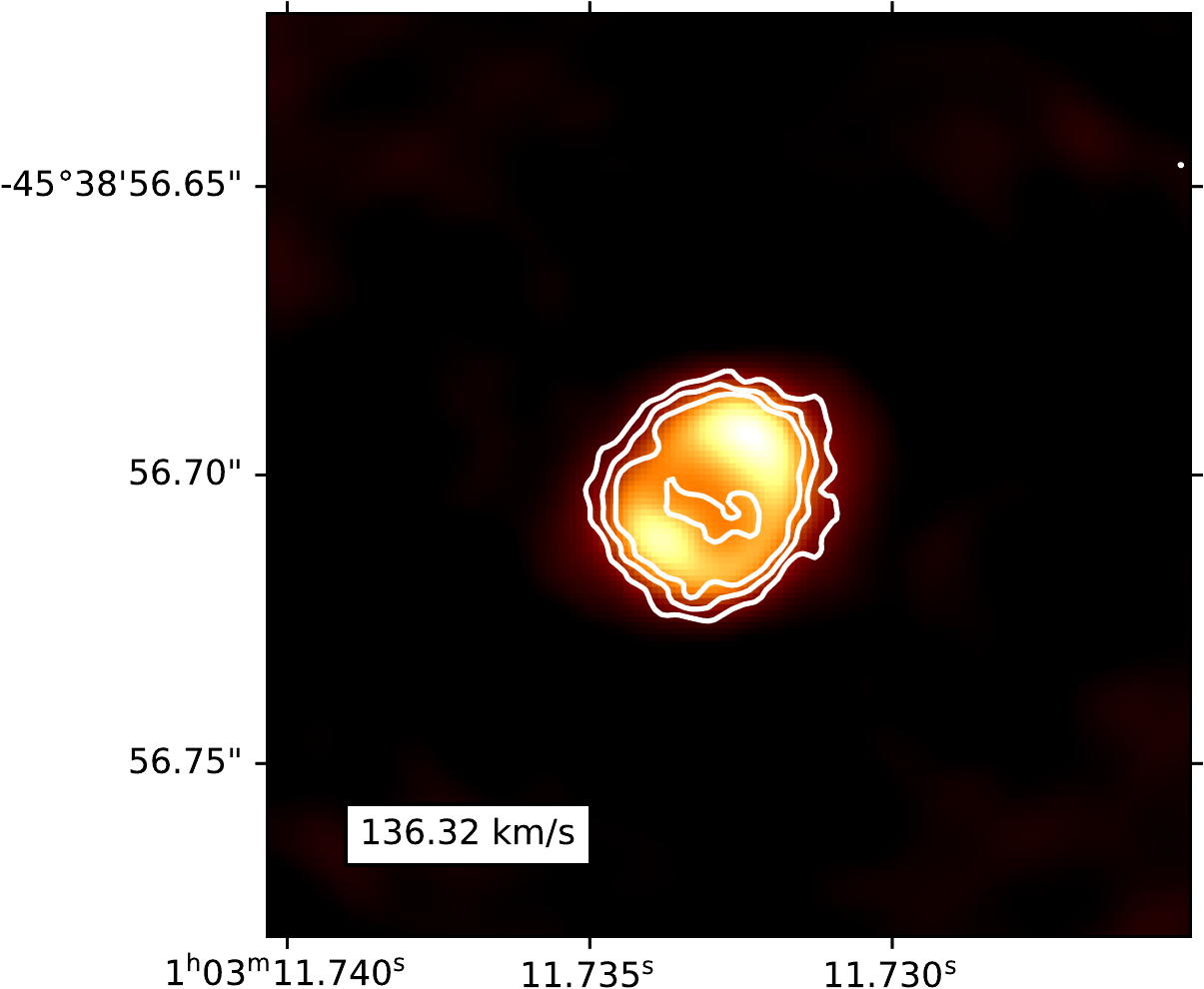} \\
    \includegraphics[width=6cm]{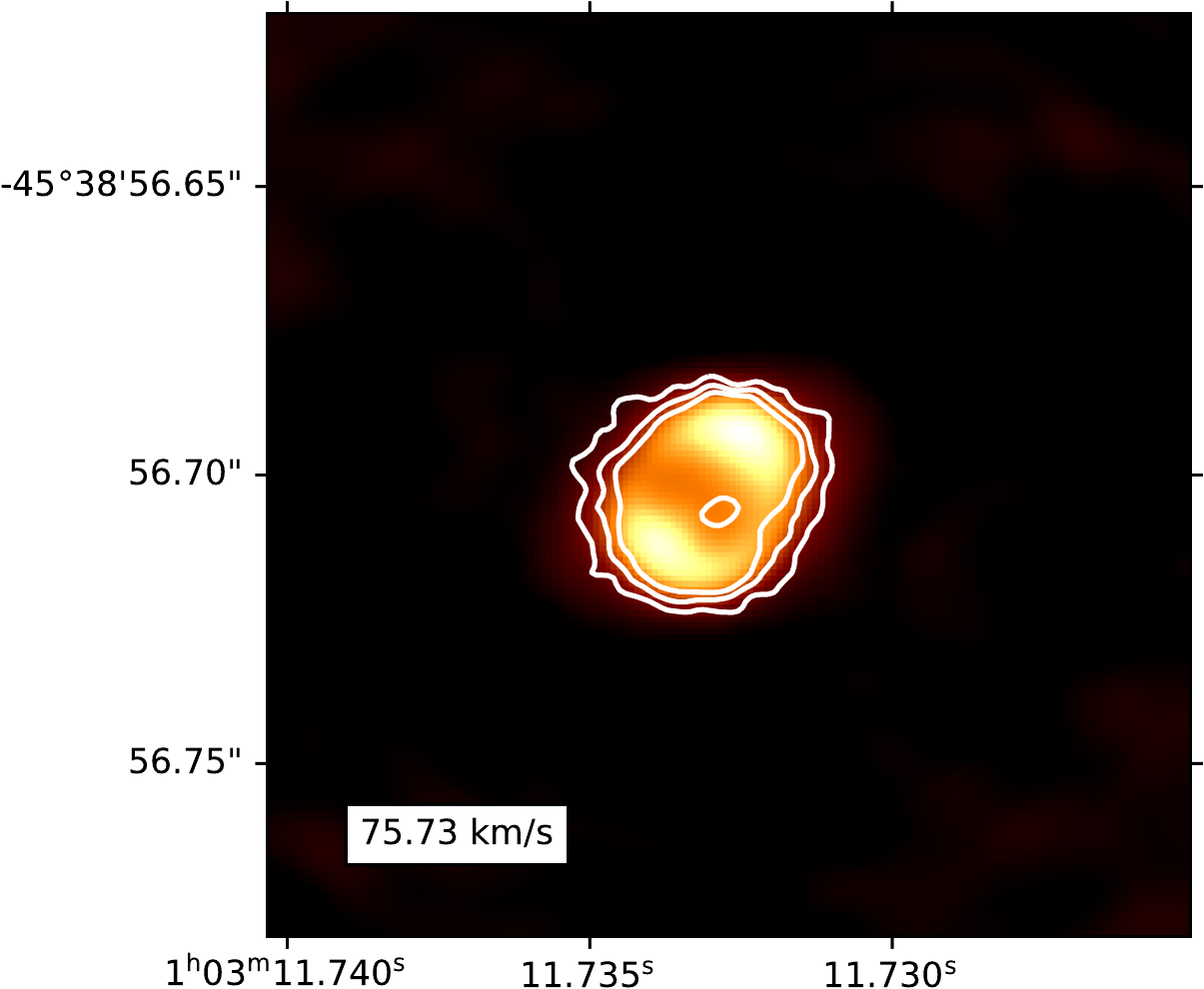} & 
    \includegraphics[width=6cm]{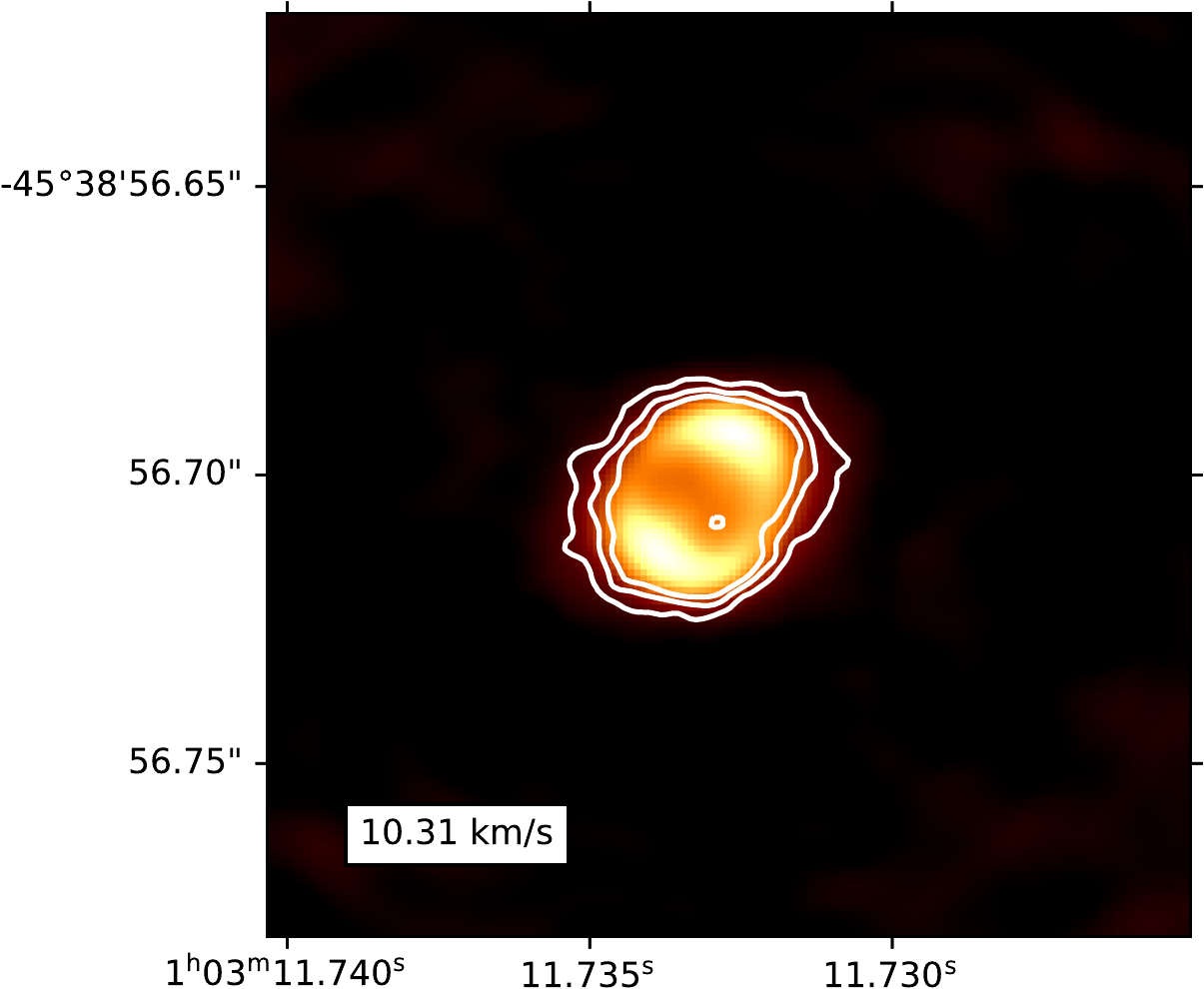} & 
    \includegraphics[width=6cm]{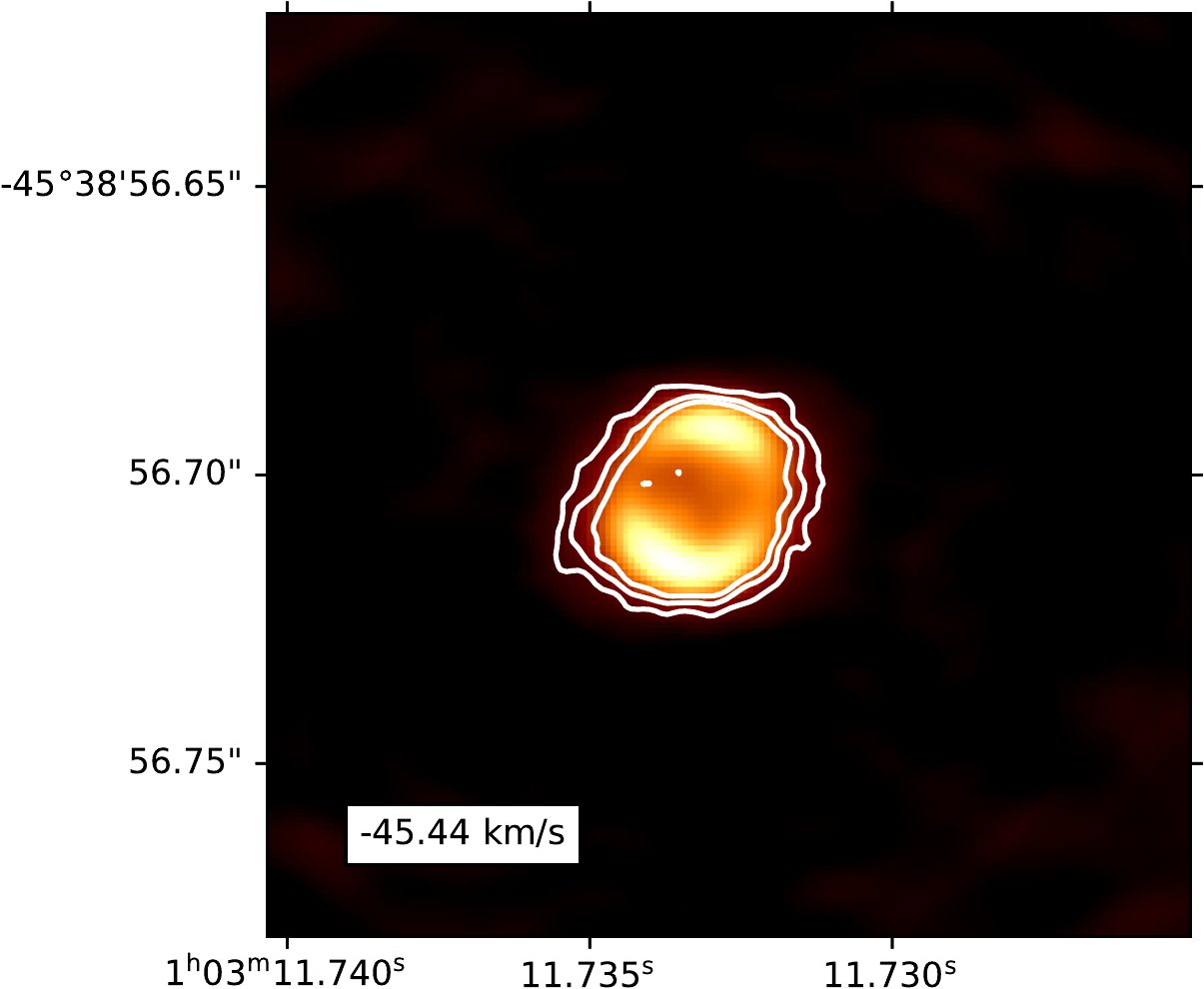} \\
    &
    \includegraphics[width=6cm]{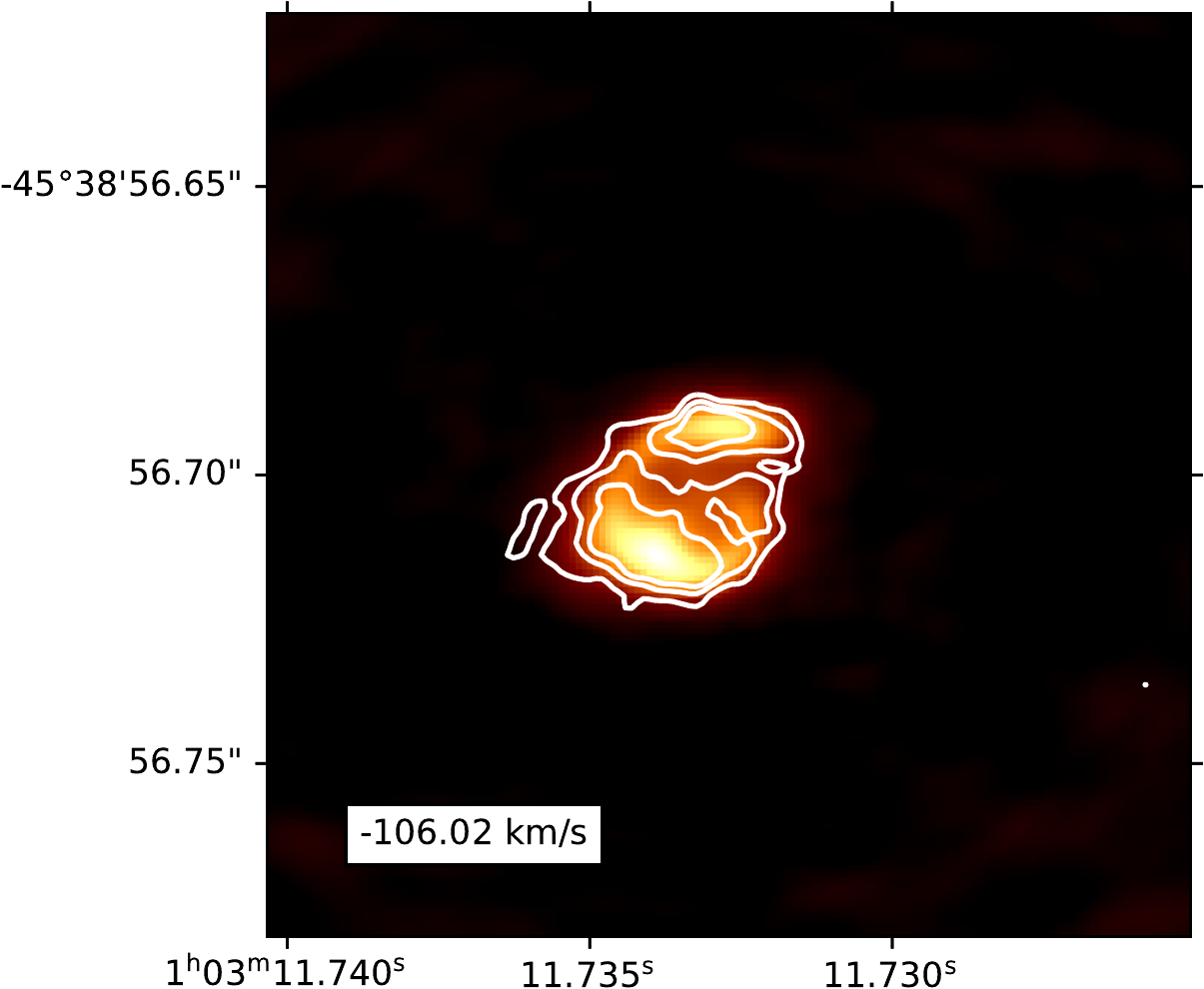} & \\
    
    \end{tabular}
    
    \caption{\label{fig: counter images CI 0103} SPT0103-45 model-dirty images of [CI](2-1) for every velocity bin are plotted along with the 3, 5 and 7\,$\sigma$ contours of the dirty image. The central velocities corresponding to each of the bins are mentioned in the figures.}
\end{figure*}

\begin{figure*}
    \centering
    \begin{tabular}{ccc}

    \includegraphics[width=6cm]{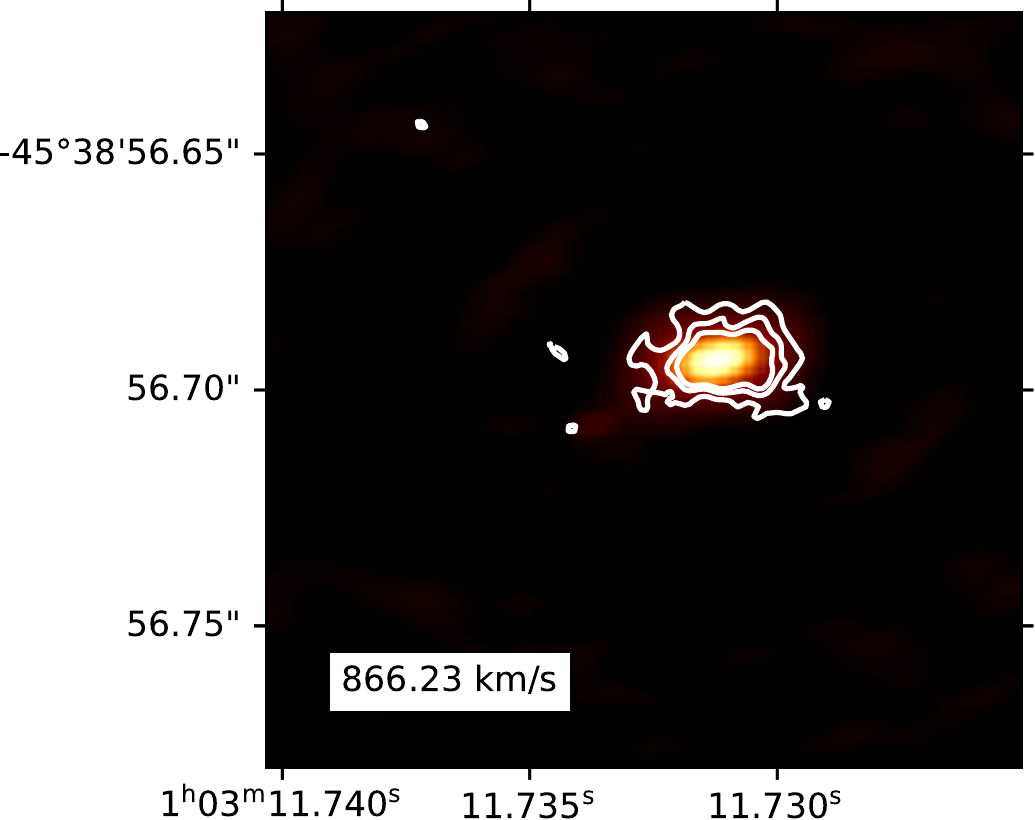} &  \includegraphics[width=6cm]{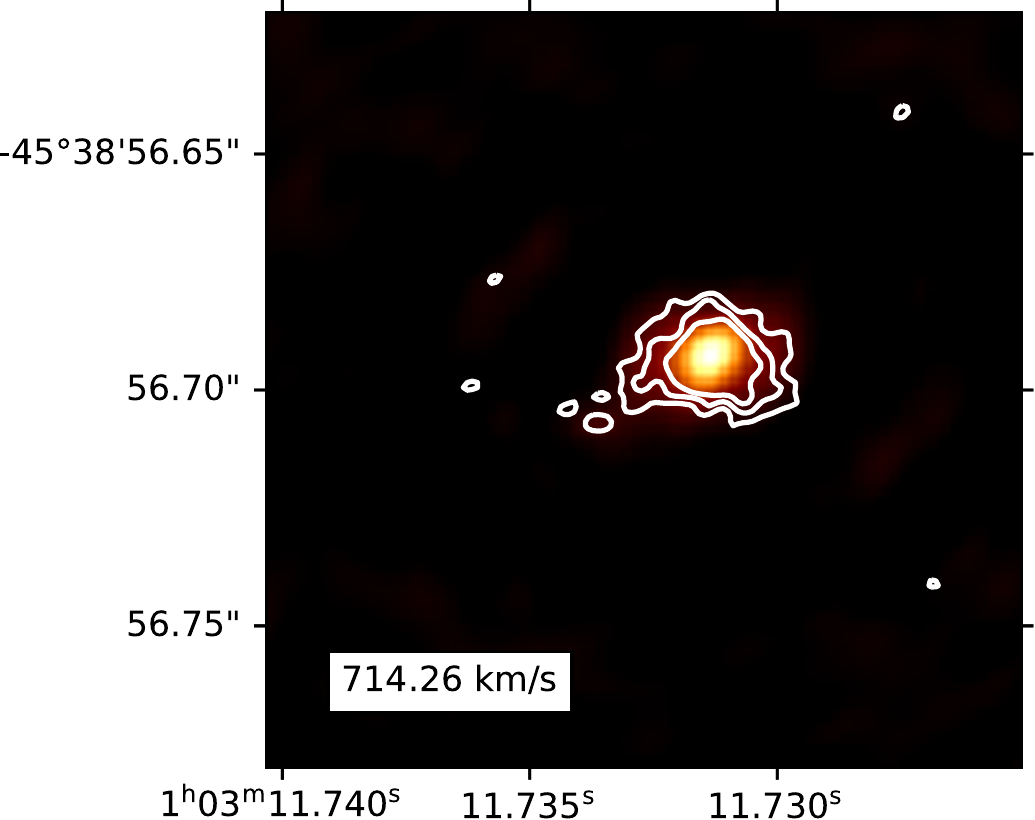} & \includegraphics[width=6cm]{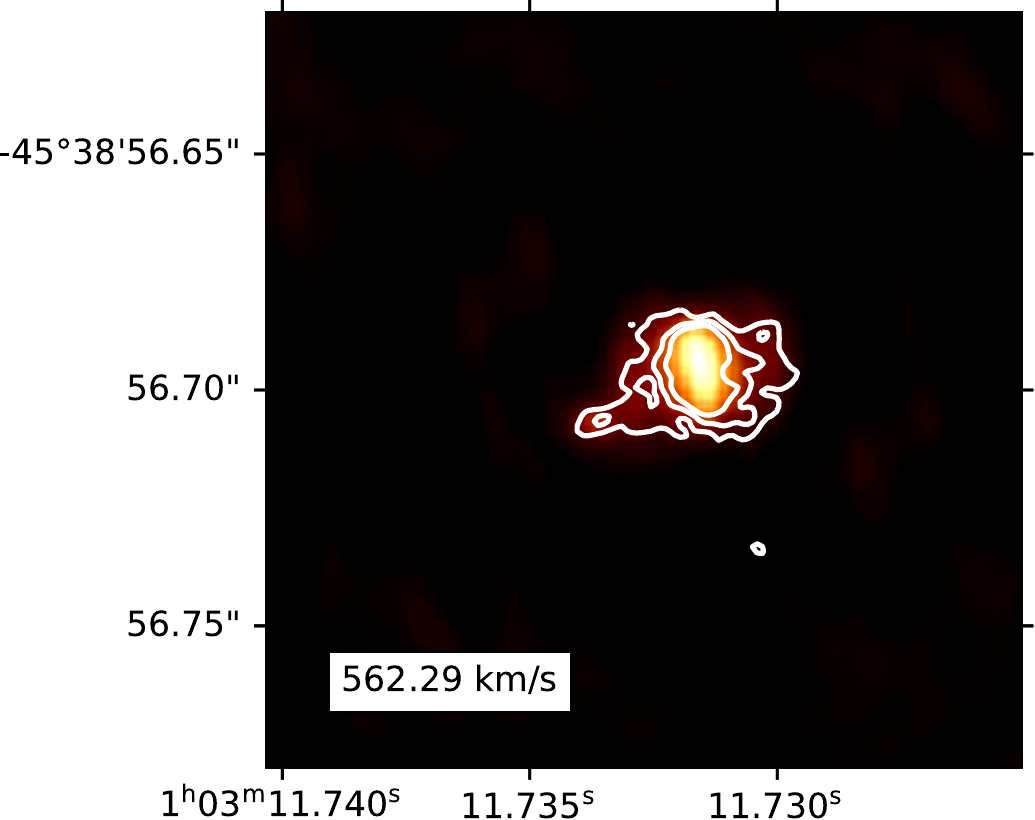} \\
    \includegraphics[width=6cm]{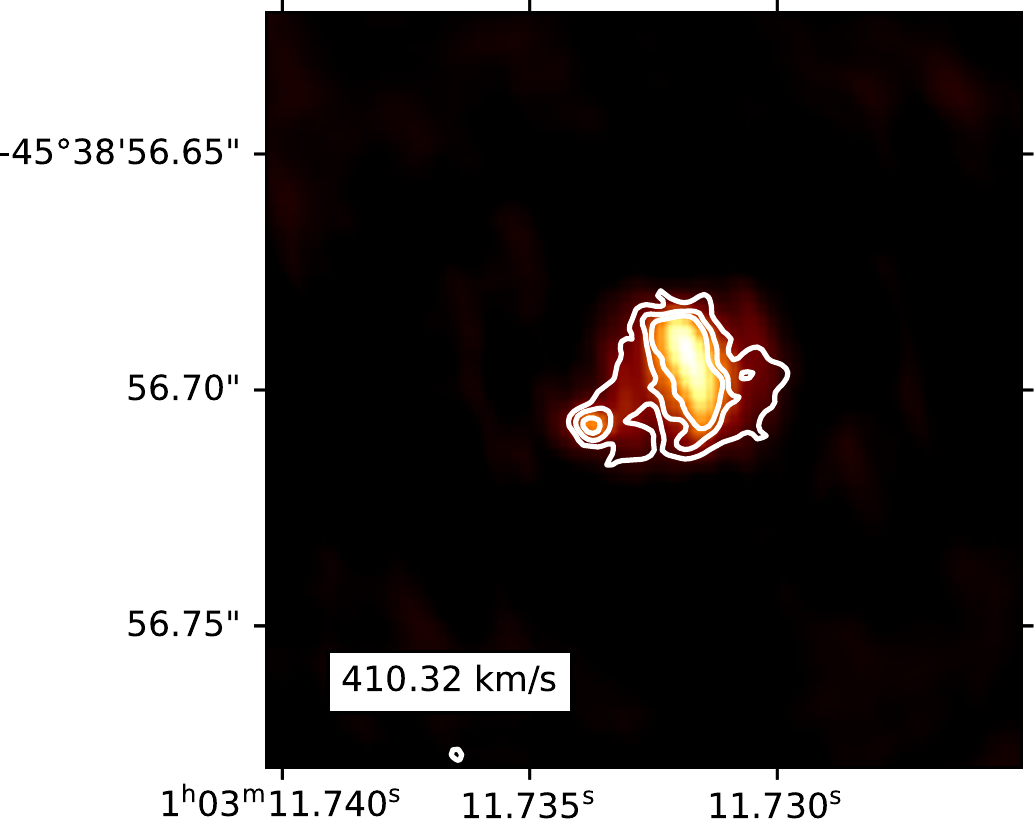} &  \includegraphics[width=6cm]{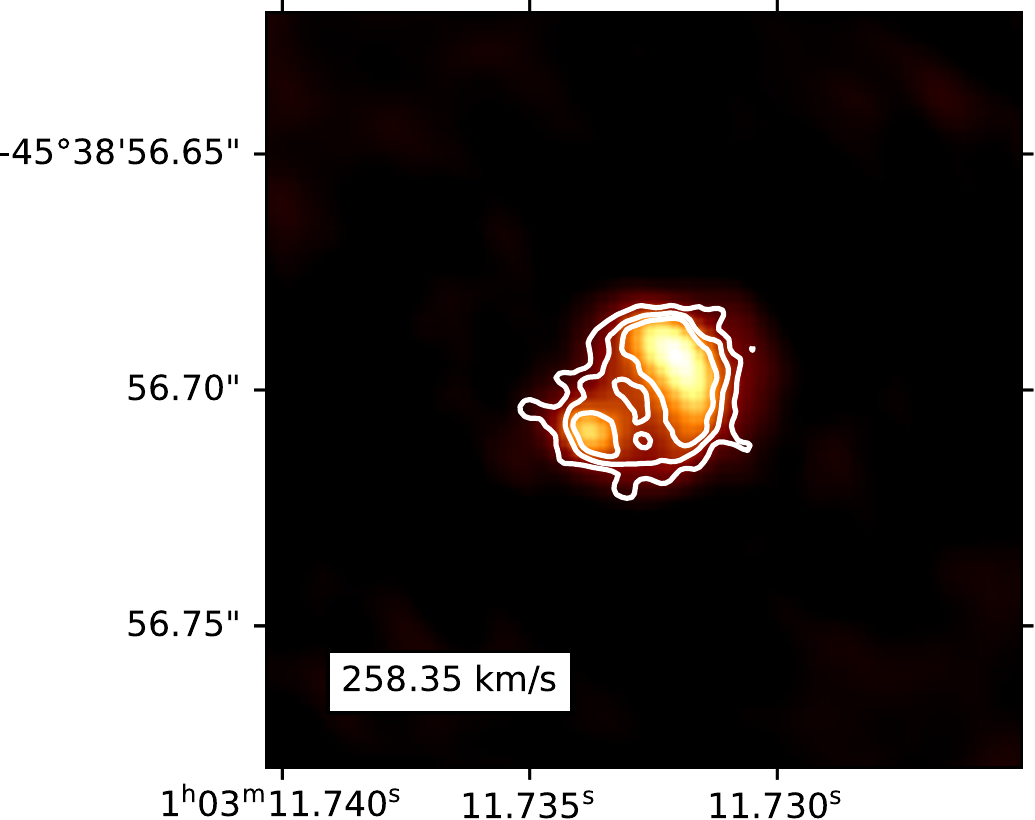} & \includegraphics[width=6cm]{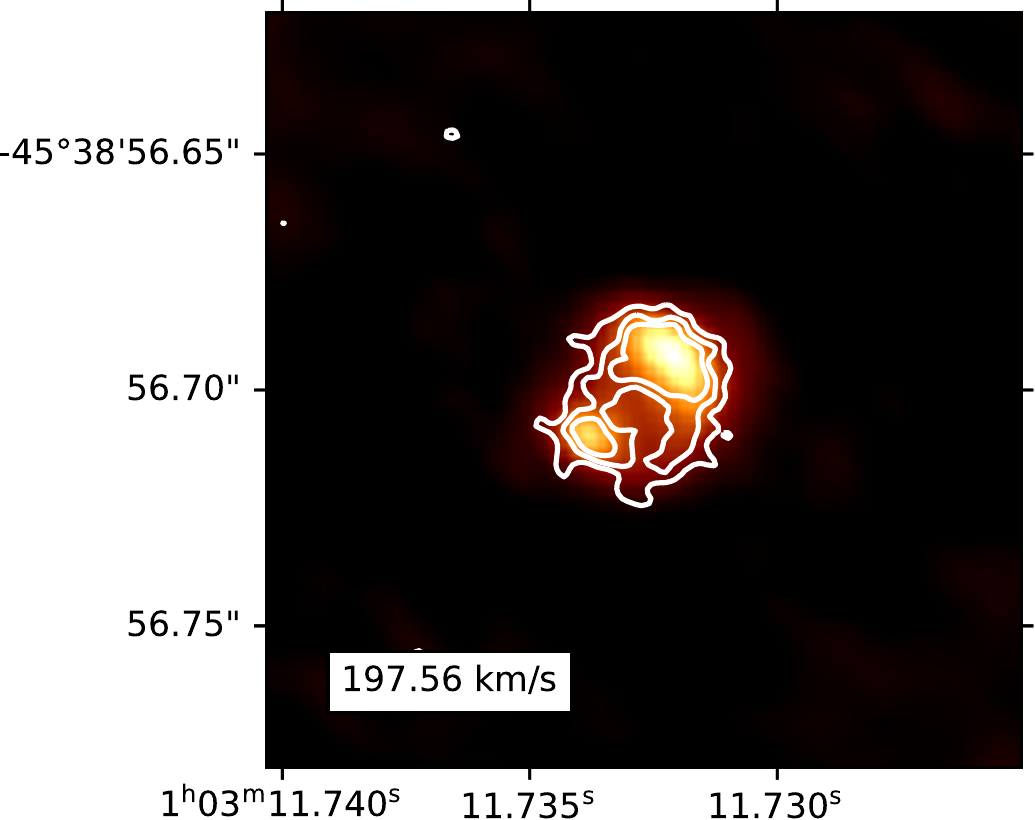} \\
    \includegraphics[width=6cm]{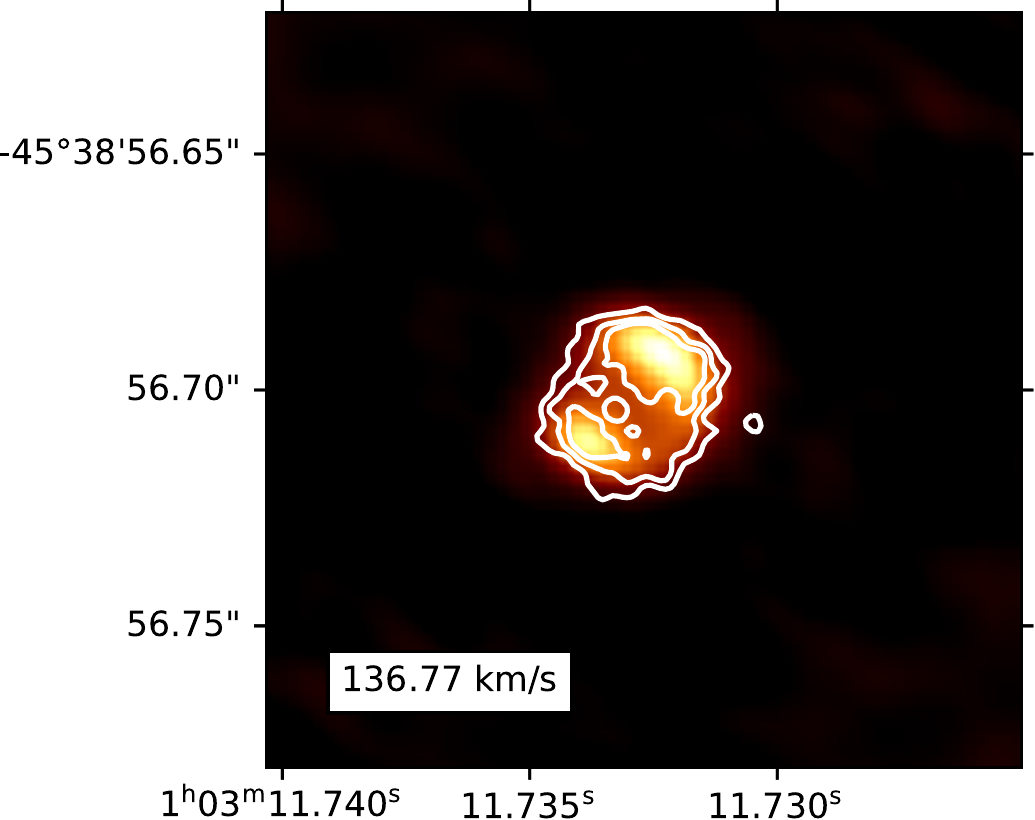} & 
    \includegraphics[width=6cm]{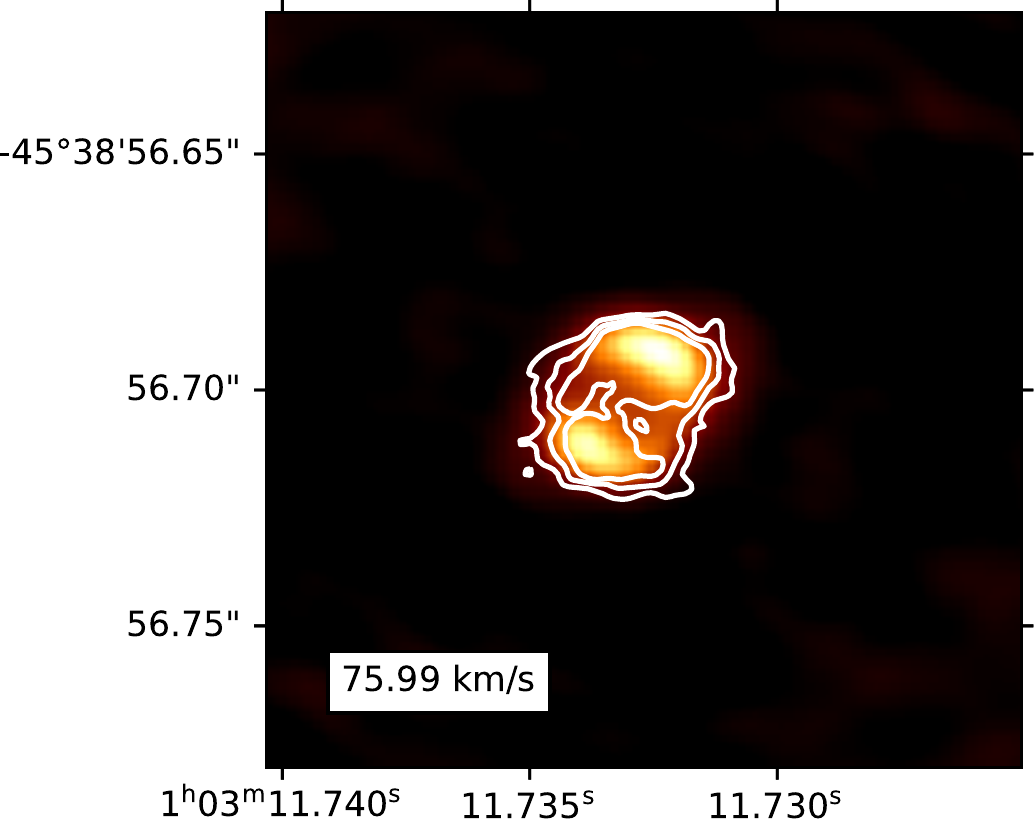} & 
    \includegraphics[width=6cm]{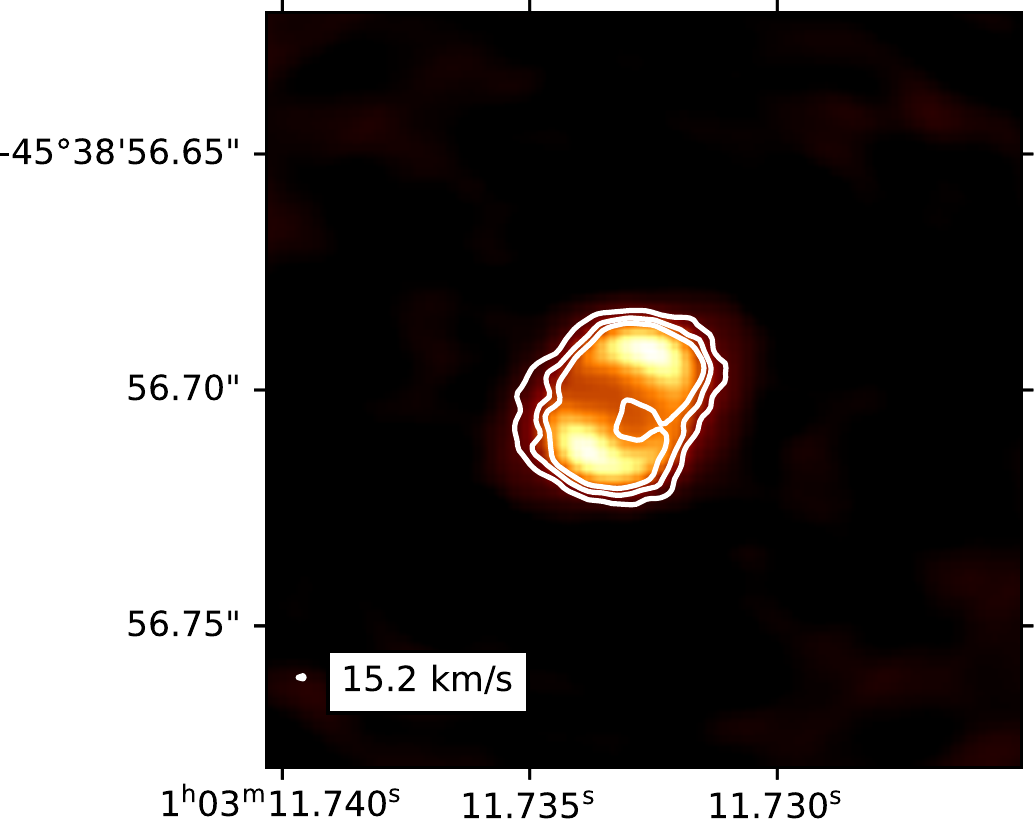} \\
    &
    \includegraphics[width=6cm]{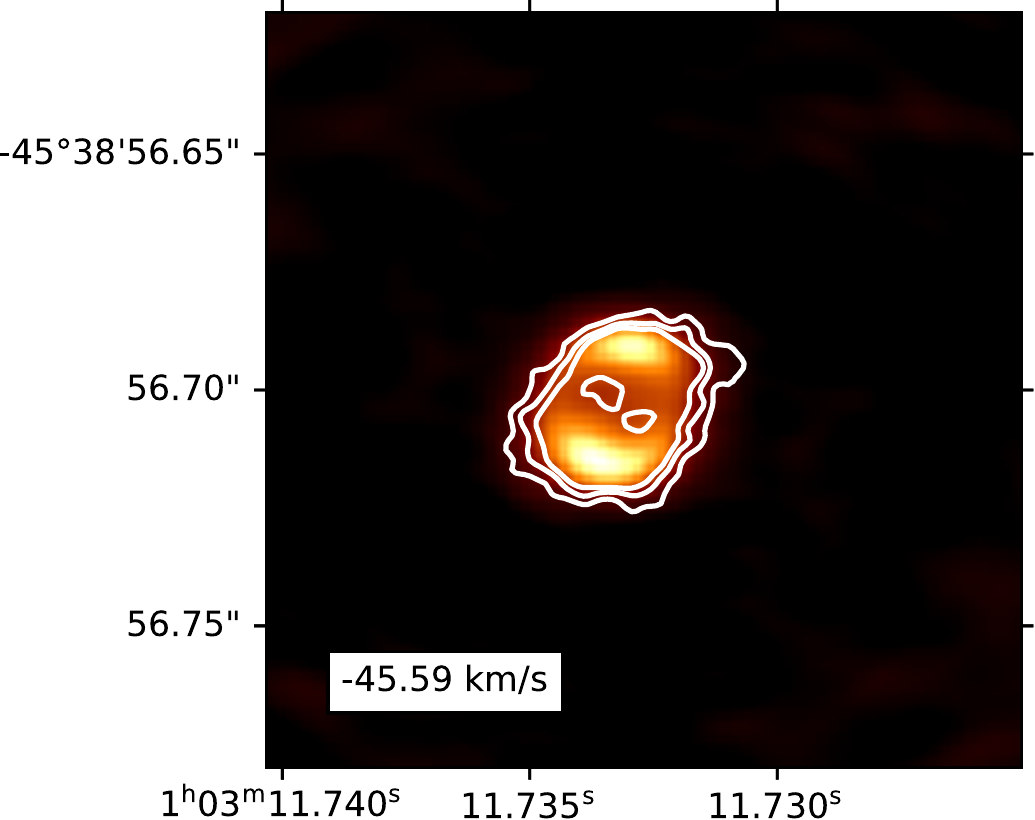} &\\
    \end{tabular}
    
    \caption{\label{fig: counter images CO 0103}  SPT0103-45  model-dirty image of CO(7-6) for every velocity bins is plotted along with the 3, 5 and 7\,$\sigma$ contours of the dirty image. The central velocity corresponding to each of the bin is mentioned in the figures.}
\end{figure*}

The number of counter images and the source structure in the image plane varies across the velocities. In Fig.\,\ref{fig: counter images CI 0103} and Fig.\,\ref{fig: counter images CO 0103} we can see a single image in the higher velocities and two counter images in the lower velocities. We also see a nice agreement between the data and the model.

\subsection{SPT2147-50}
\begin{figure*}
    \centering
    \begin{tabular}{ccc}
    
    \includegraphics[width=6cm]{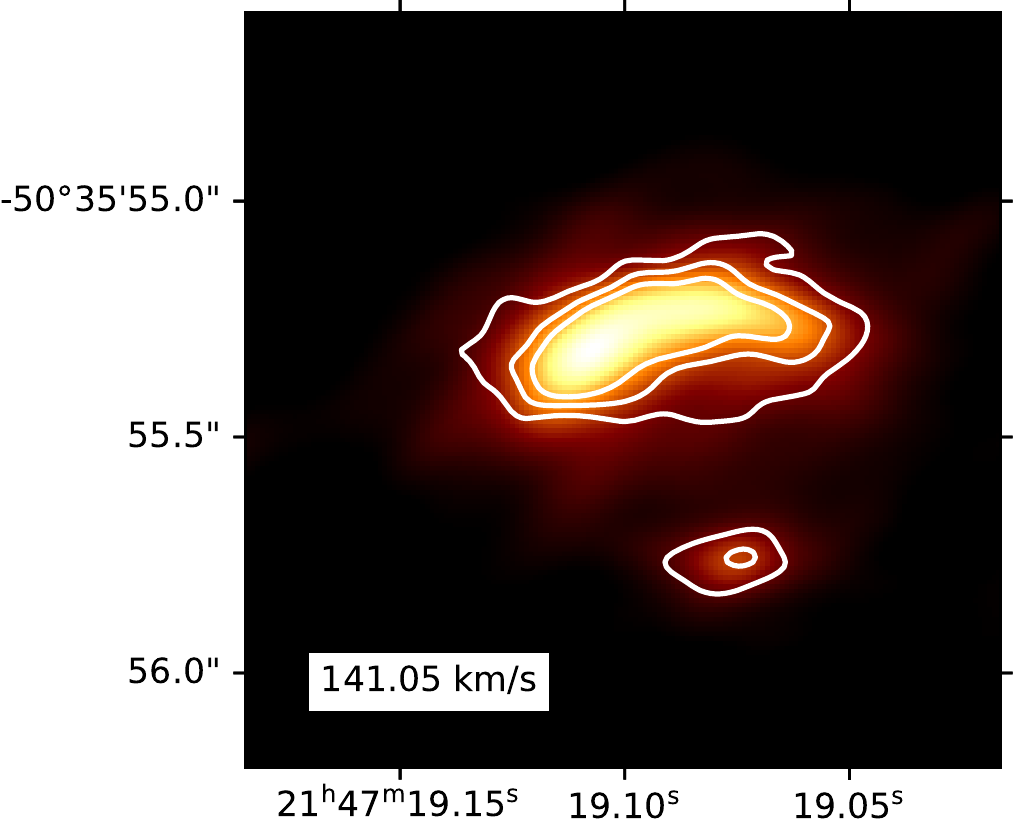} &  \includegraphics[width=6cm]{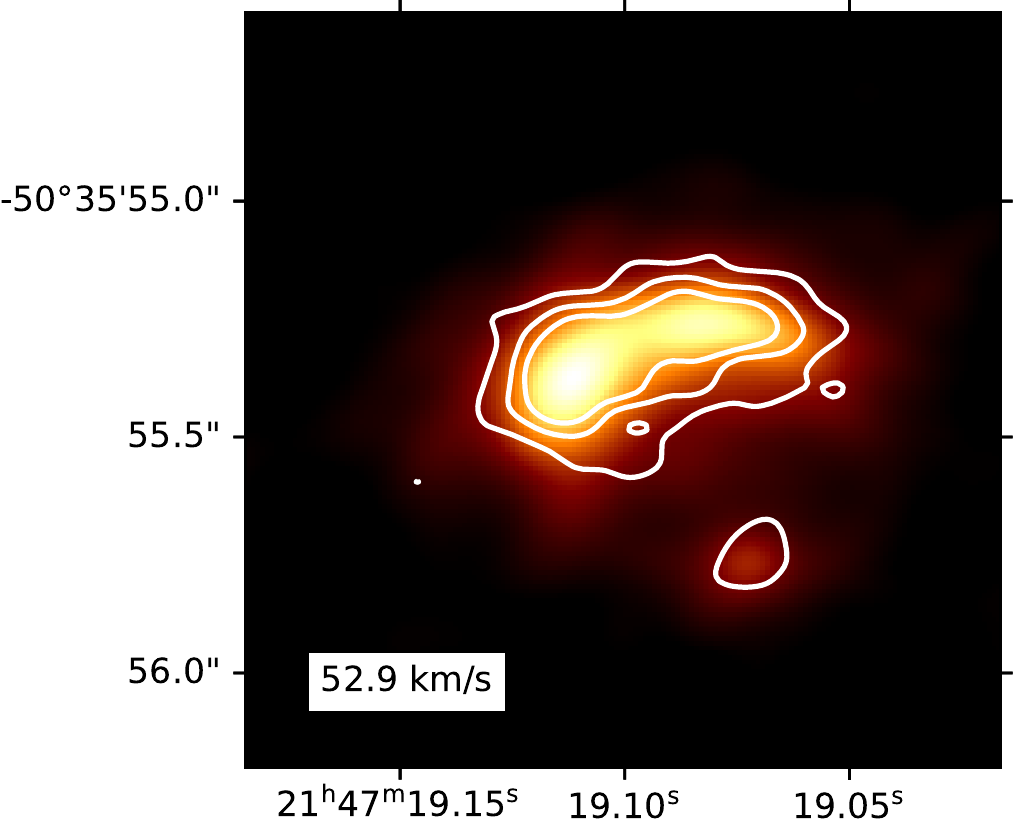} & \includegraphics[width=6cm]{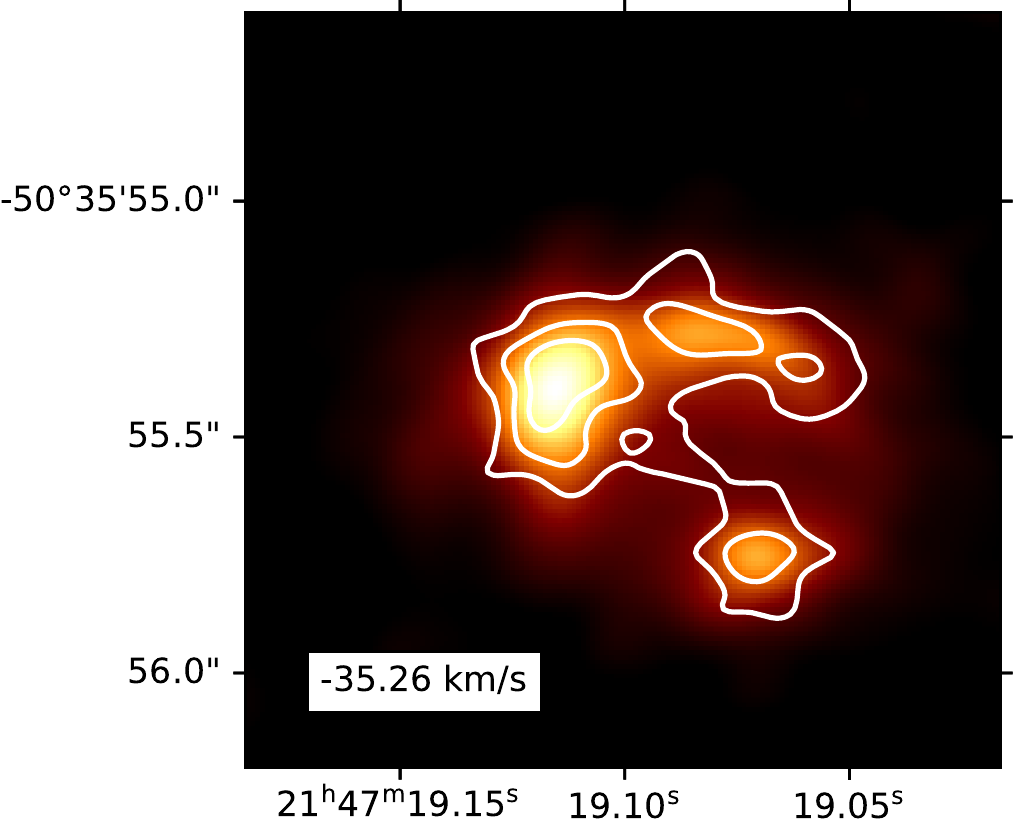} \\
    \includegraphics[width=6cm]{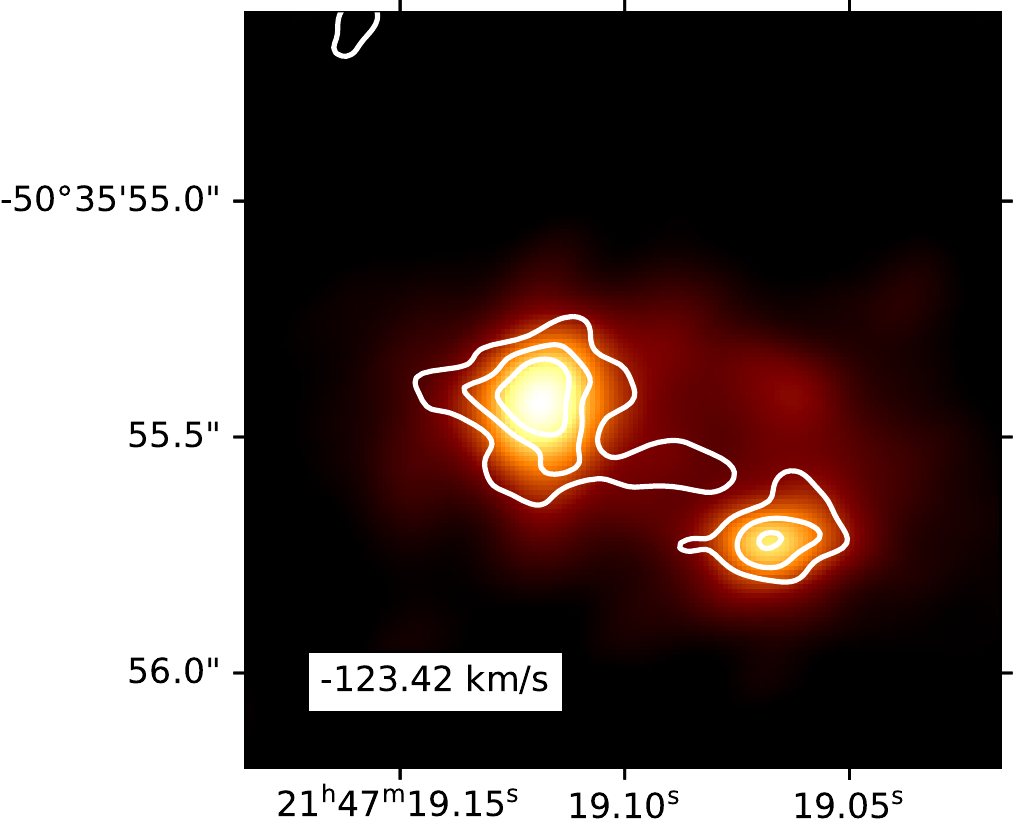} &  \includegraphics[width=6cm]{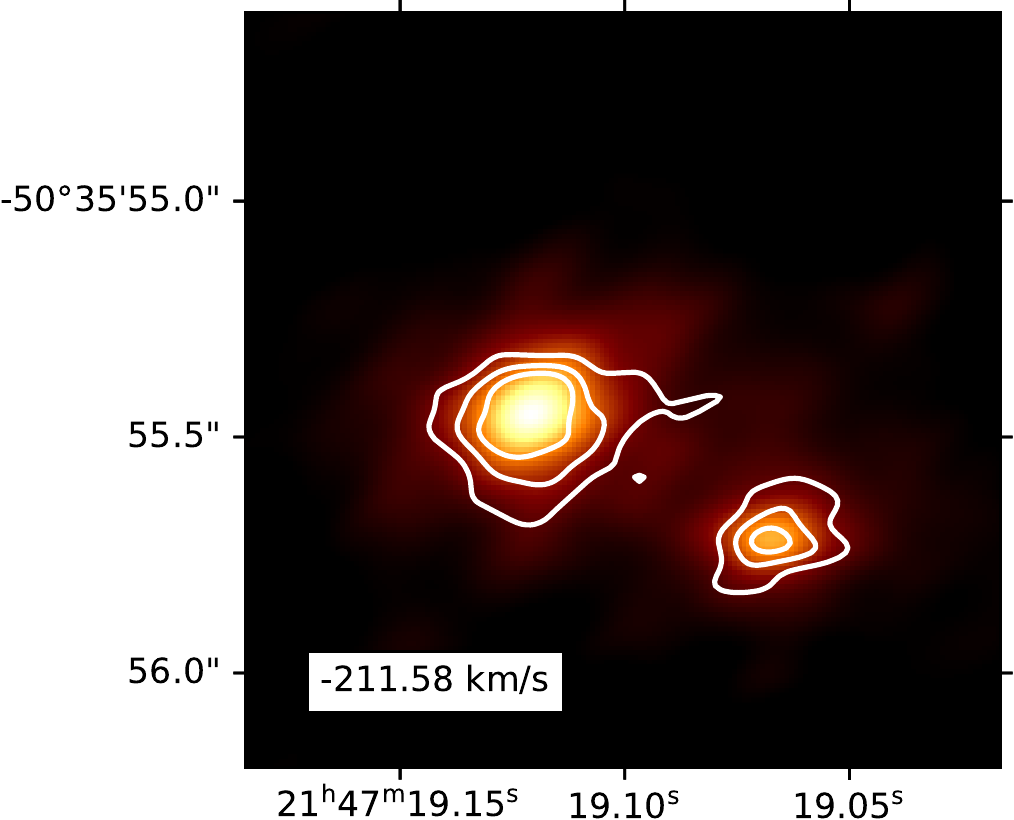} & \includegraphics[width=6cm]{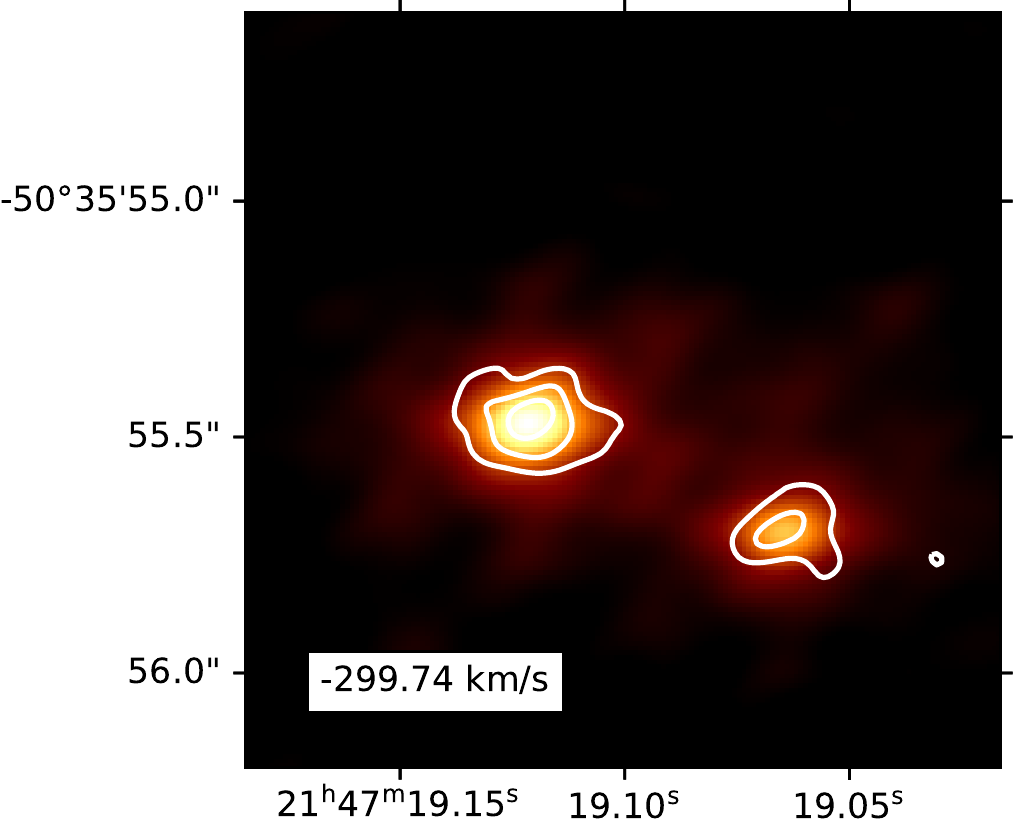} \\
    \end{tabular}
    \caption{\label{fig: counter images CI 2147} SPT2147-50  model-dirty image of [CI](2-1) for every velocity bins is plotted along with the 3, 5 and 7\,$\sigma$ contours of the dirty image. The central velocities corresponding to each of the bins are mentioned in the figures.}

\end{figure*}

\begin{figure*}
    \centering
    \begin{tabular}{ccc}
    
    \includegraphics[width=6cm]{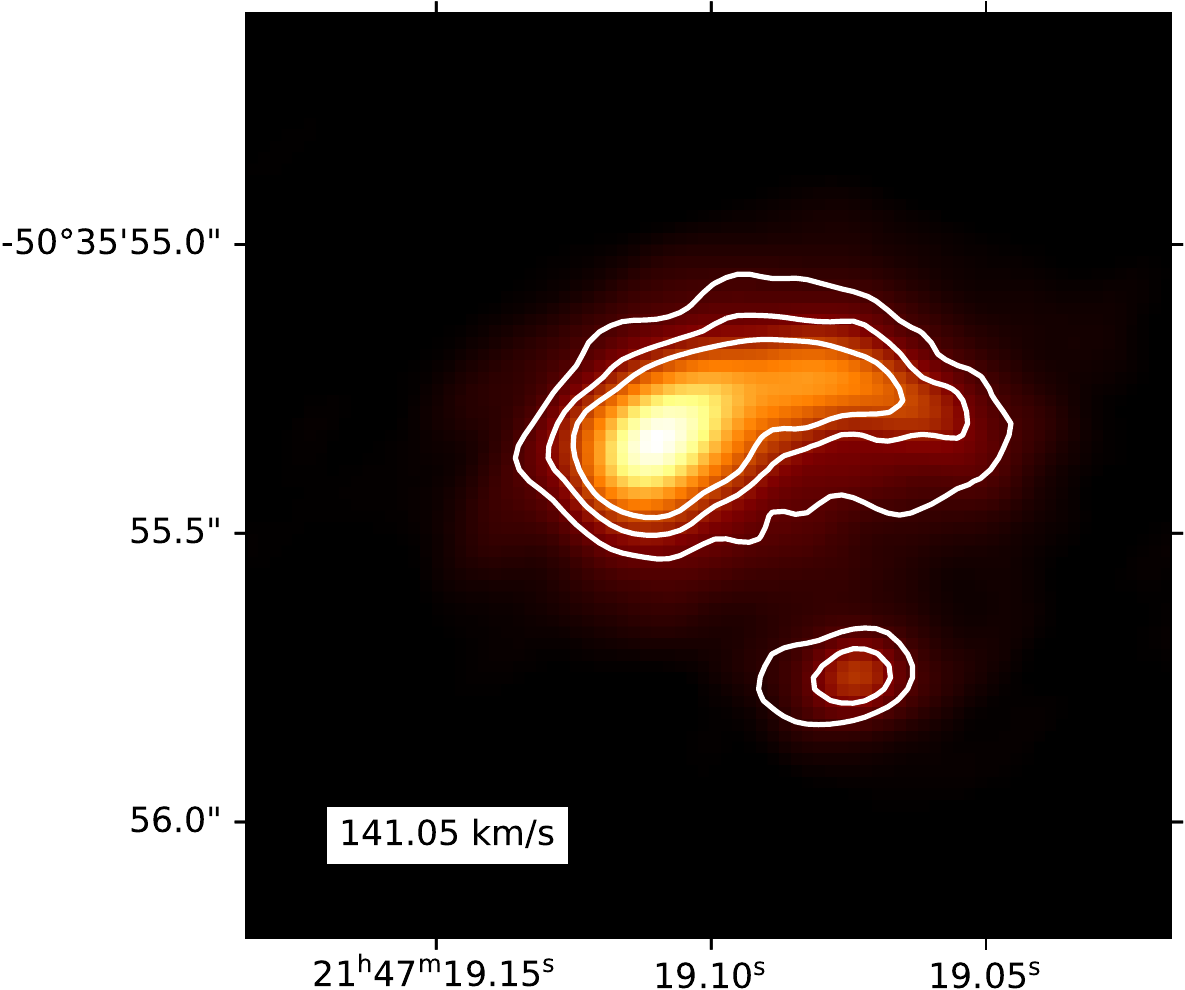} &  \includegraphics[width=6cm]{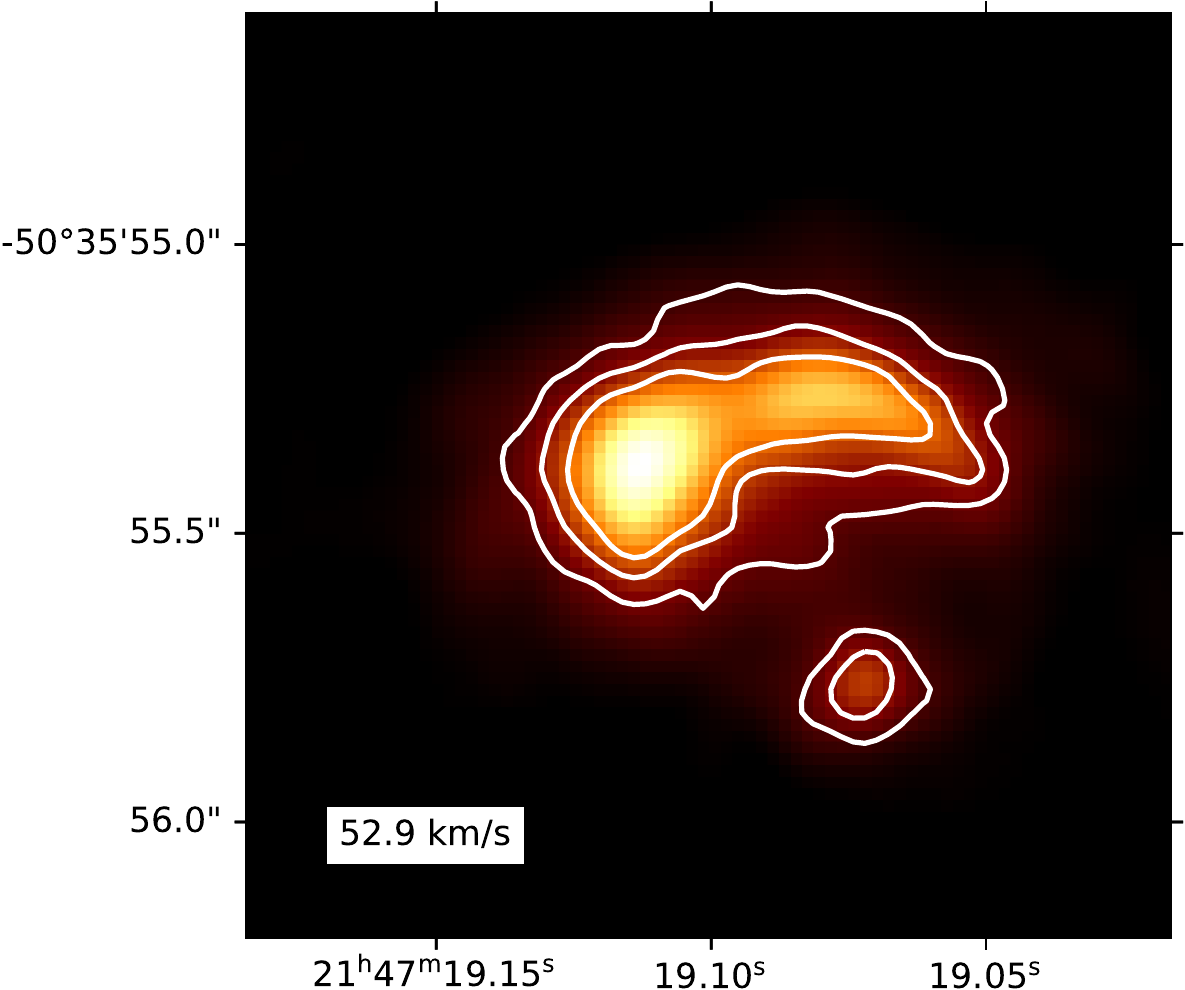} & \includegraphics[width=6cm]{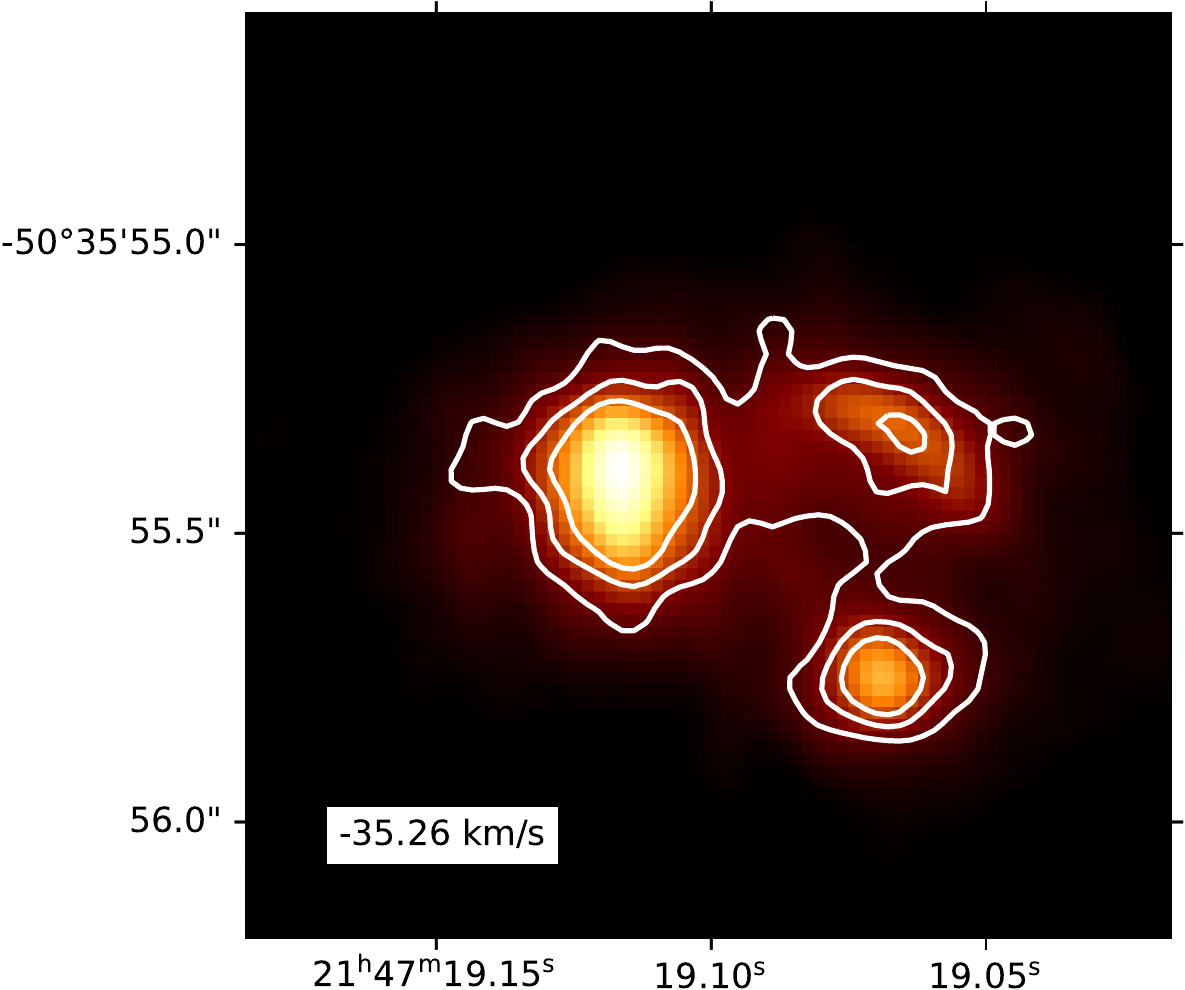} \\
    \includegraphics[width=6cm]{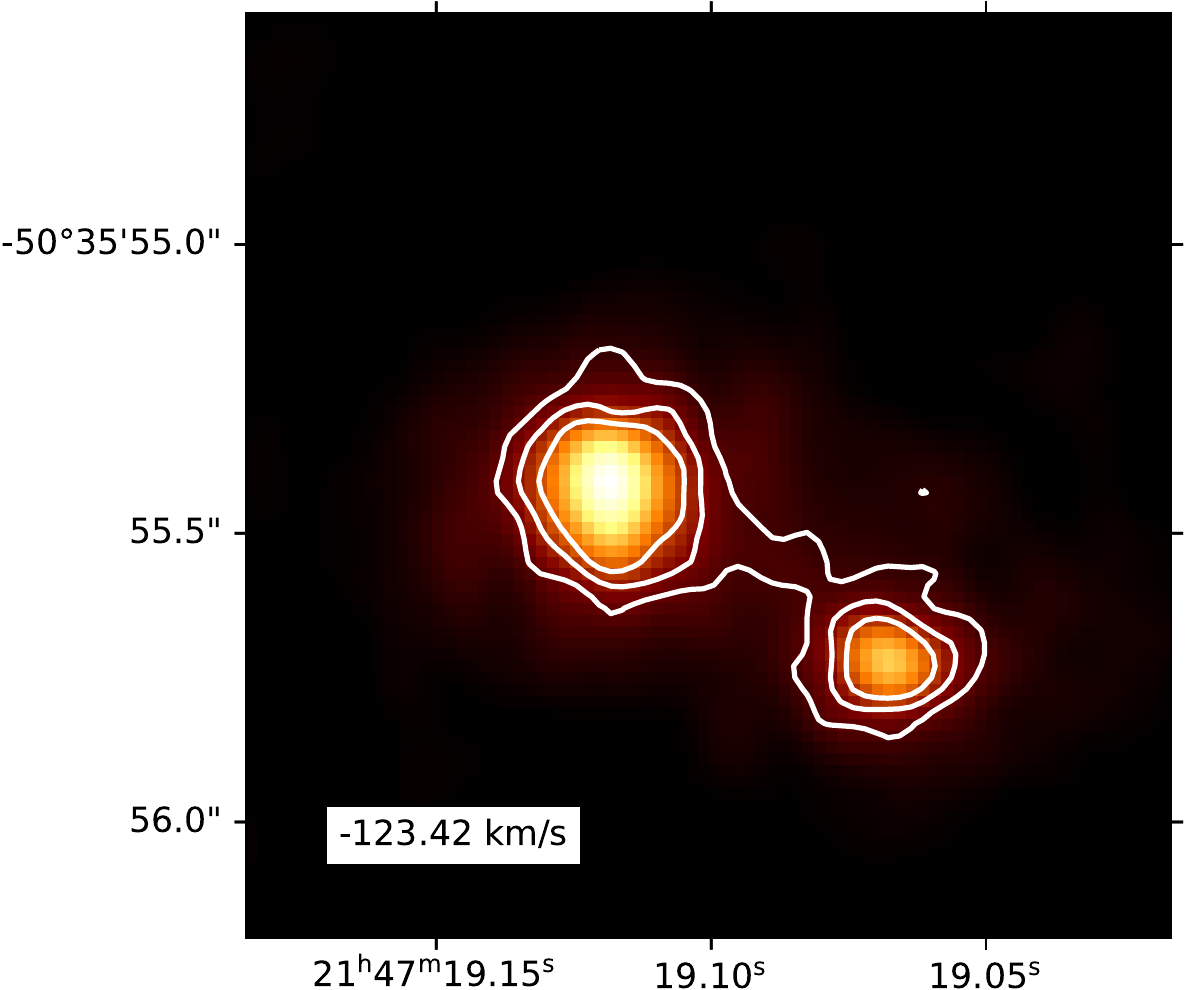} &  \includegraphics[width=6cm]{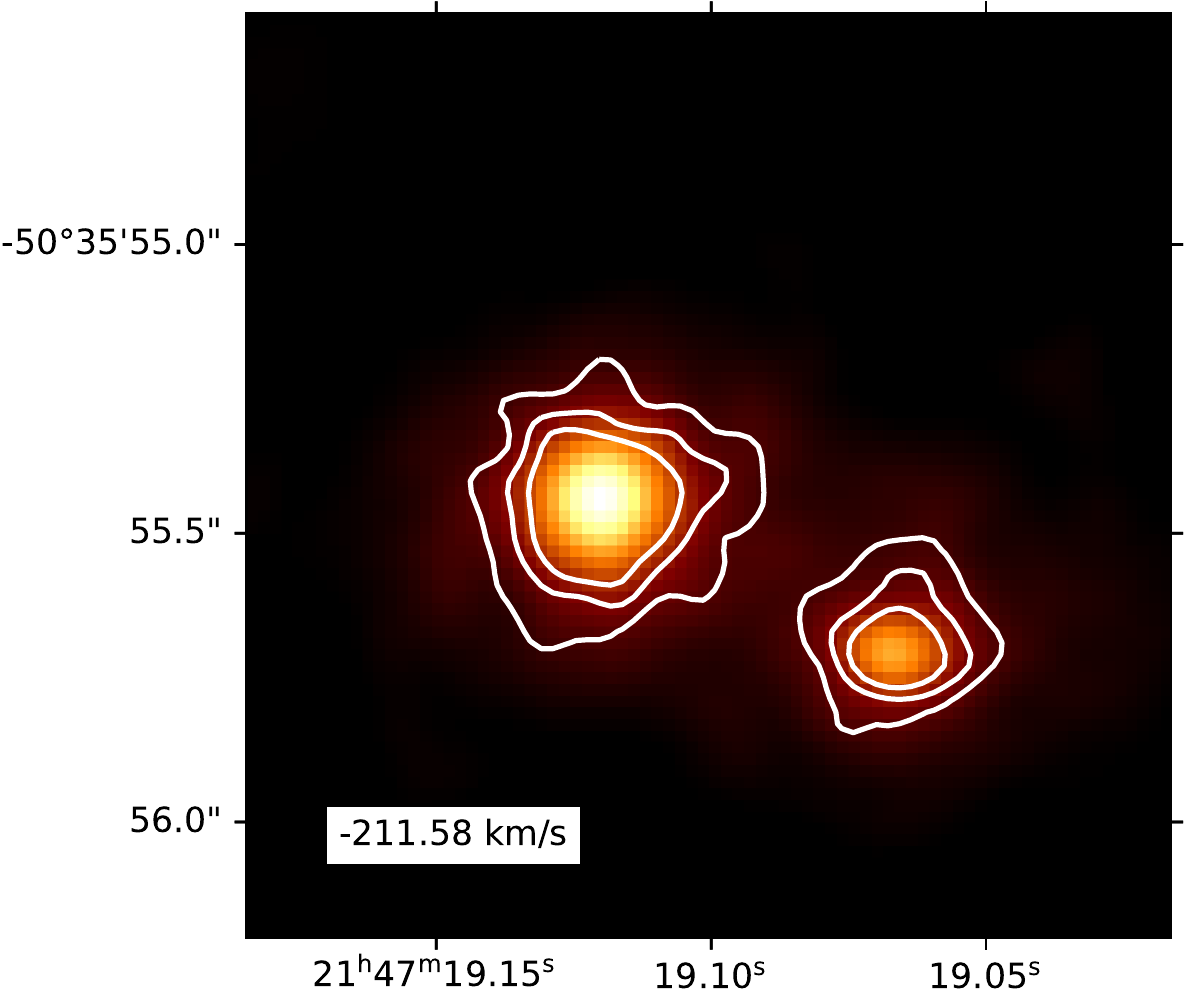} & \includegraphics[width=6cm]{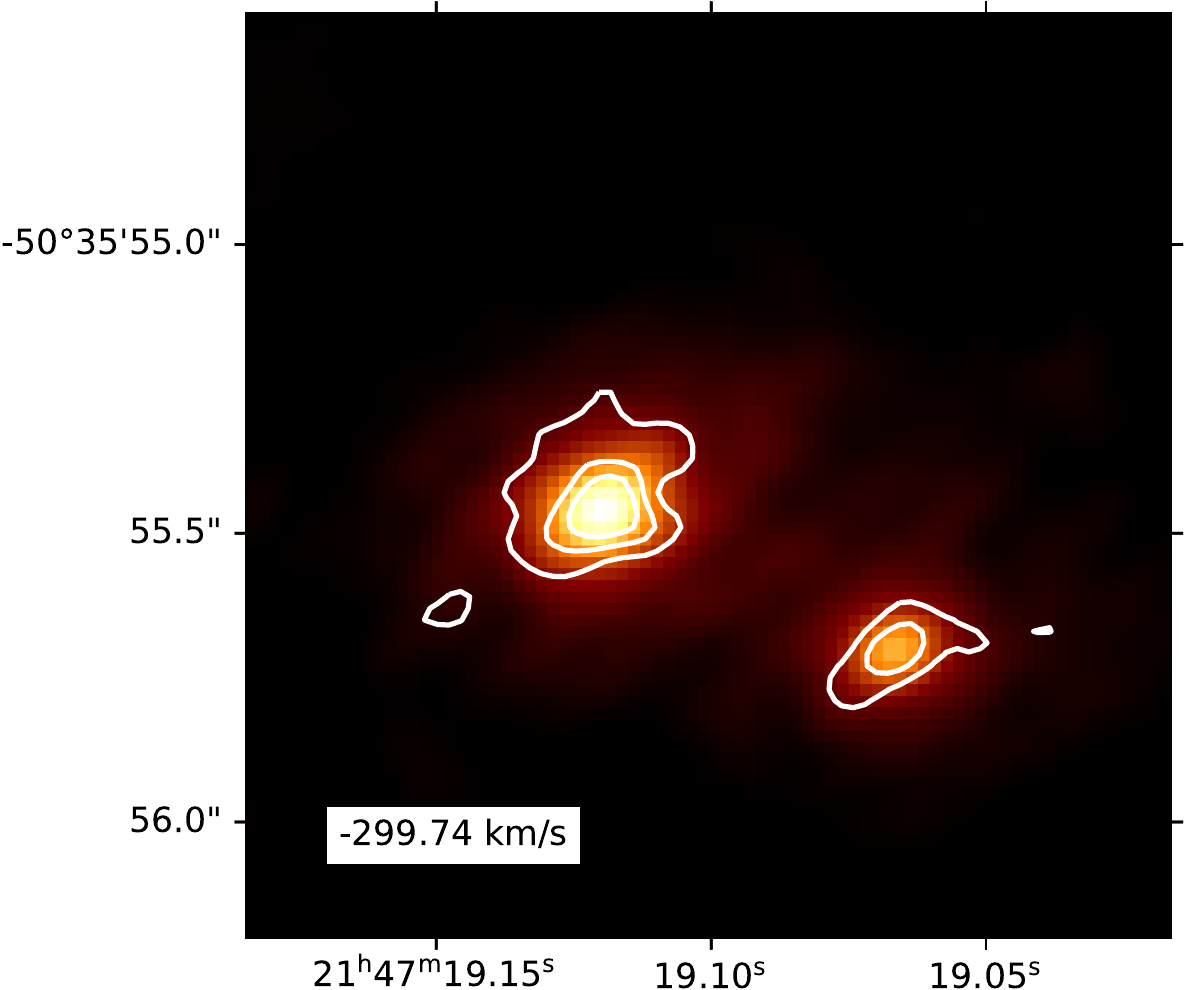} \\
    \end{tabular}
    
    \caption{\label{fig: counter images CO 2147} SPT2147-50 model-dirty image of CO(7-6) for every velocity bins is plotted along with the 3, 5 and 7\,$\sigma$ contours of the dirty image. The central velocities corresponding to each of the bins are mentioned in the figures.}
\end{figure*}

In Fig.\,\ref{fig: counter images CI 2147} and Fig.\,\ref{fig: counter images CO 2147}, we see a good agreement between the data and the model across the velocities. The dirty image and the number of counter images vary across the velocity bins, similar to SPT0103-45. For the low velocity regions, we do not see the arc in the north, which is seen in the continuum and the higher velocity bins.

\subsection{SPT2357-51}
\begin{figure*}
    \centering
    \begin{tabular}{ccc}
    
    \includegraphics[width=6cm]{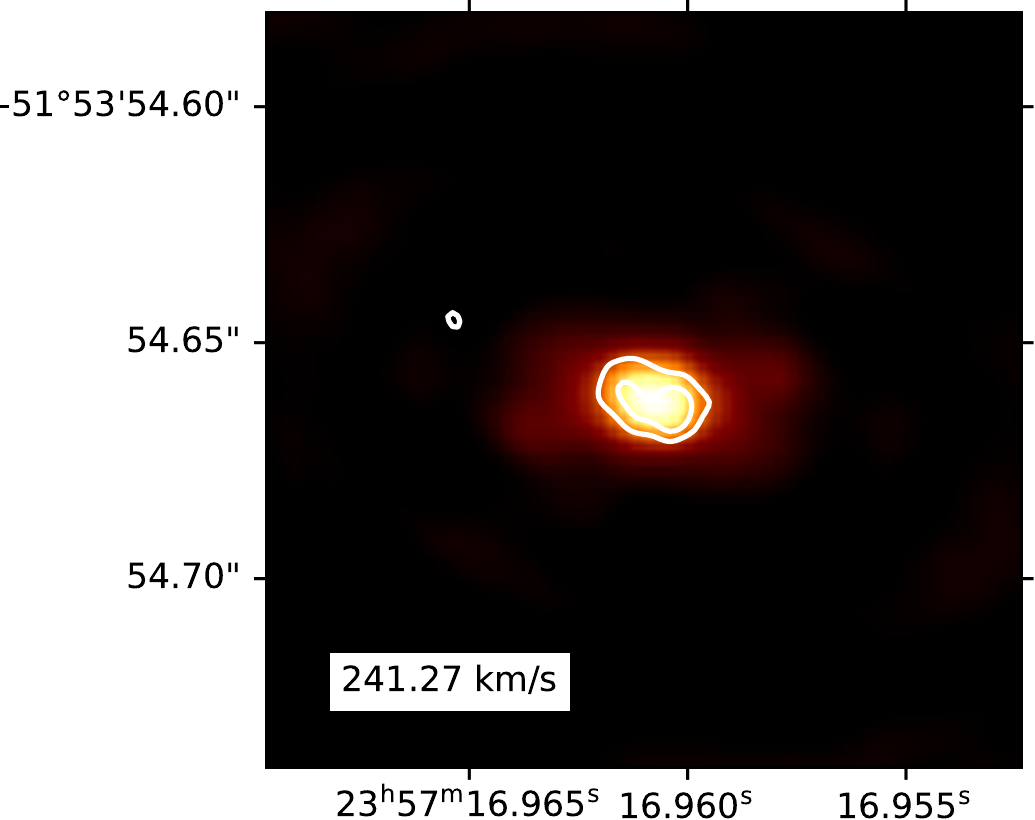} & \includegraphics[width=6cm]{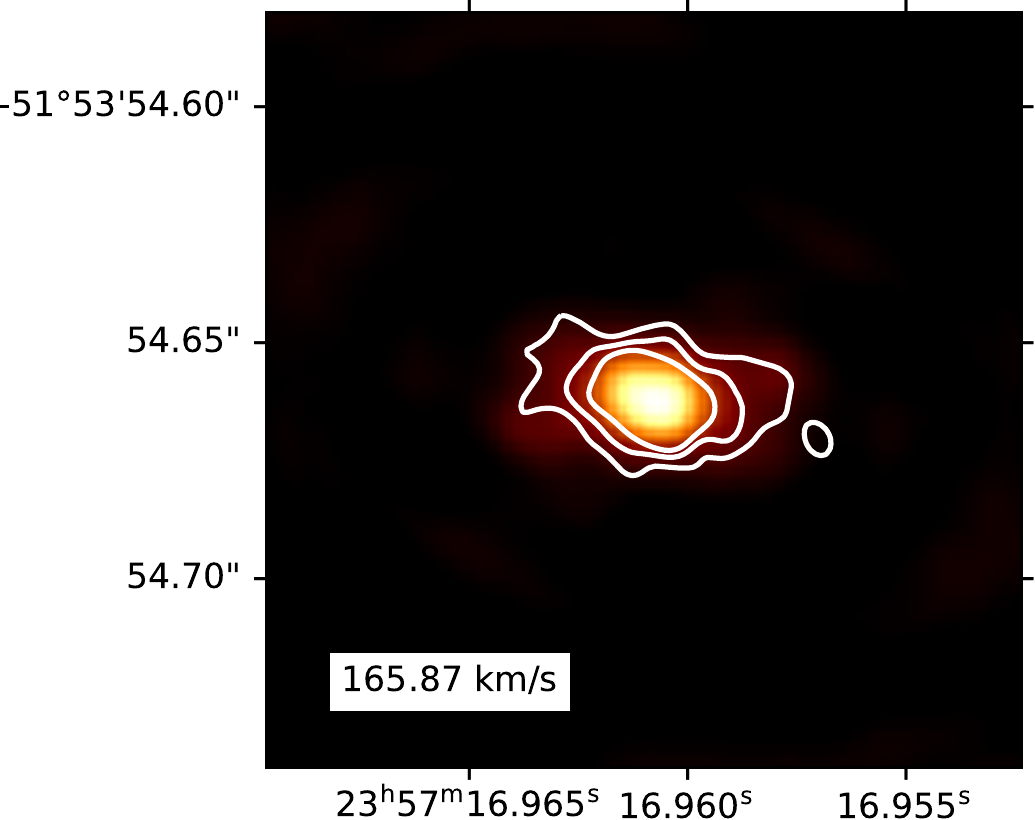} & \includegraphics[width=6cm]{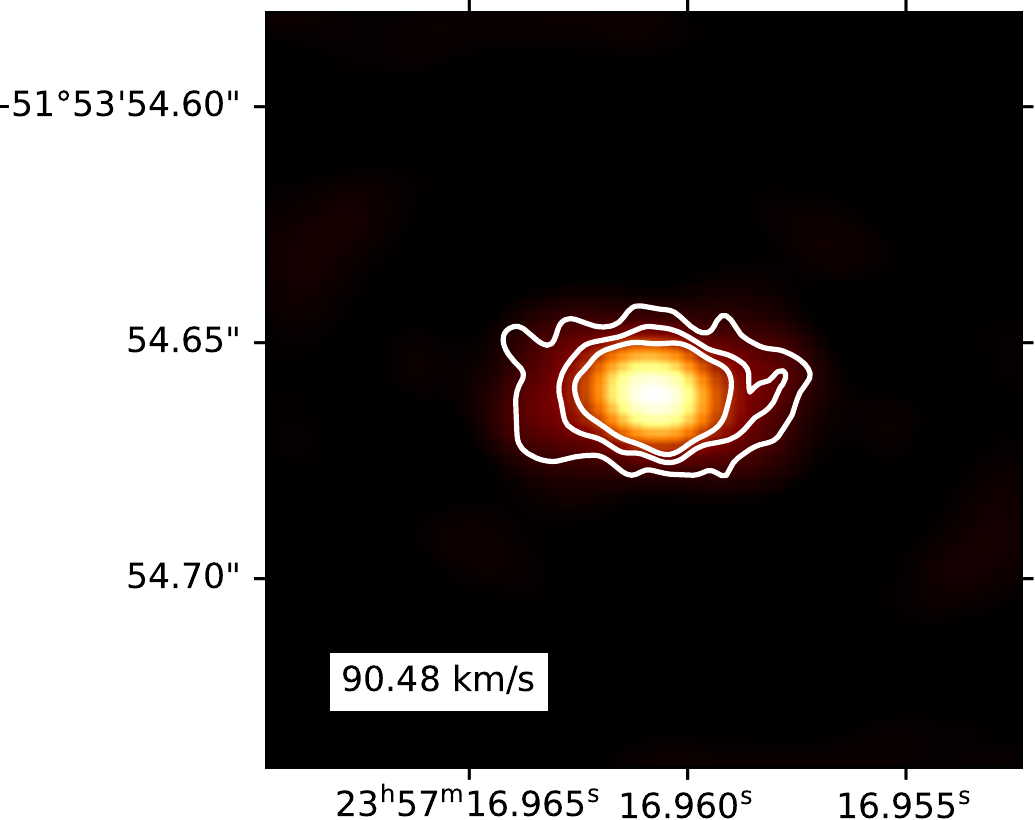} \\
    \includegraphics[width=6cm]{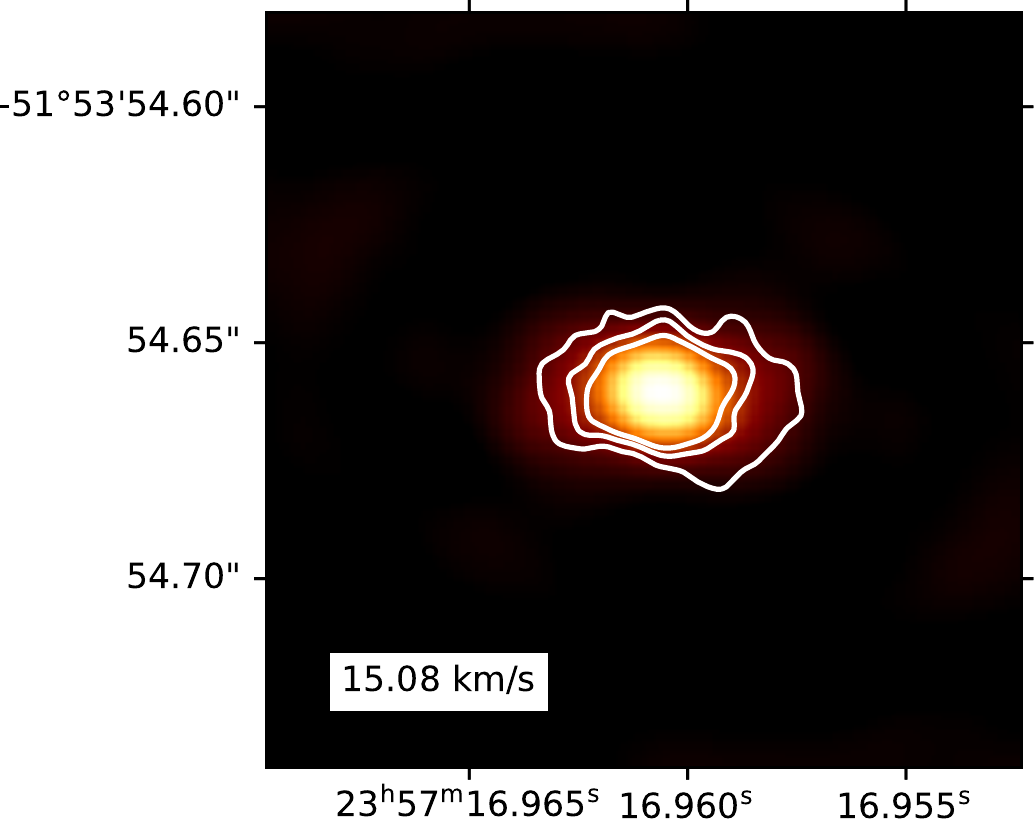} &  \includegraphics[width=6cm]{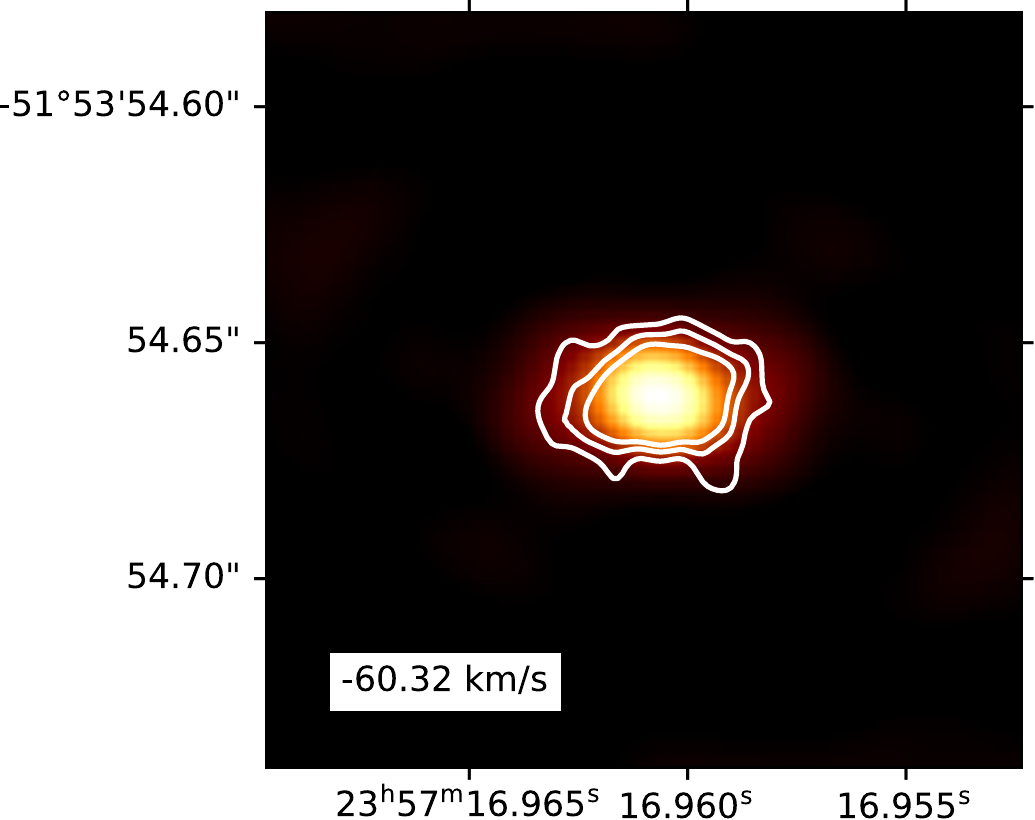} & \includegraphics[width=6cm]{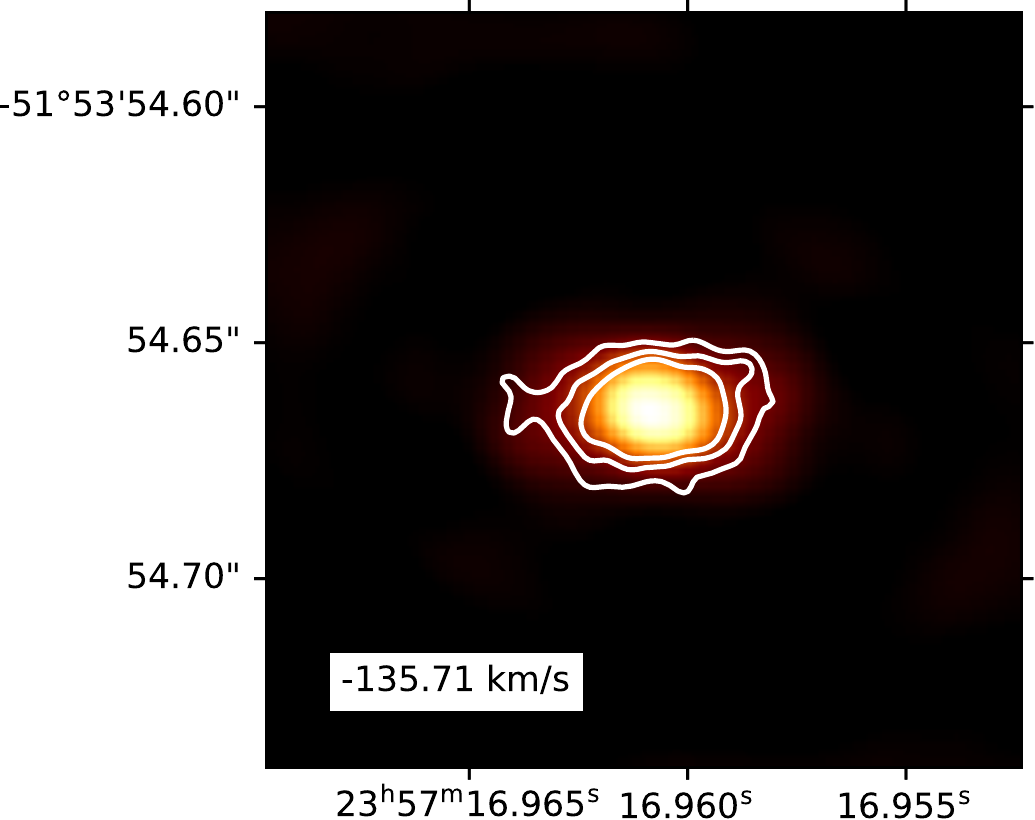} \\
    \includegraphics[width=6cm]{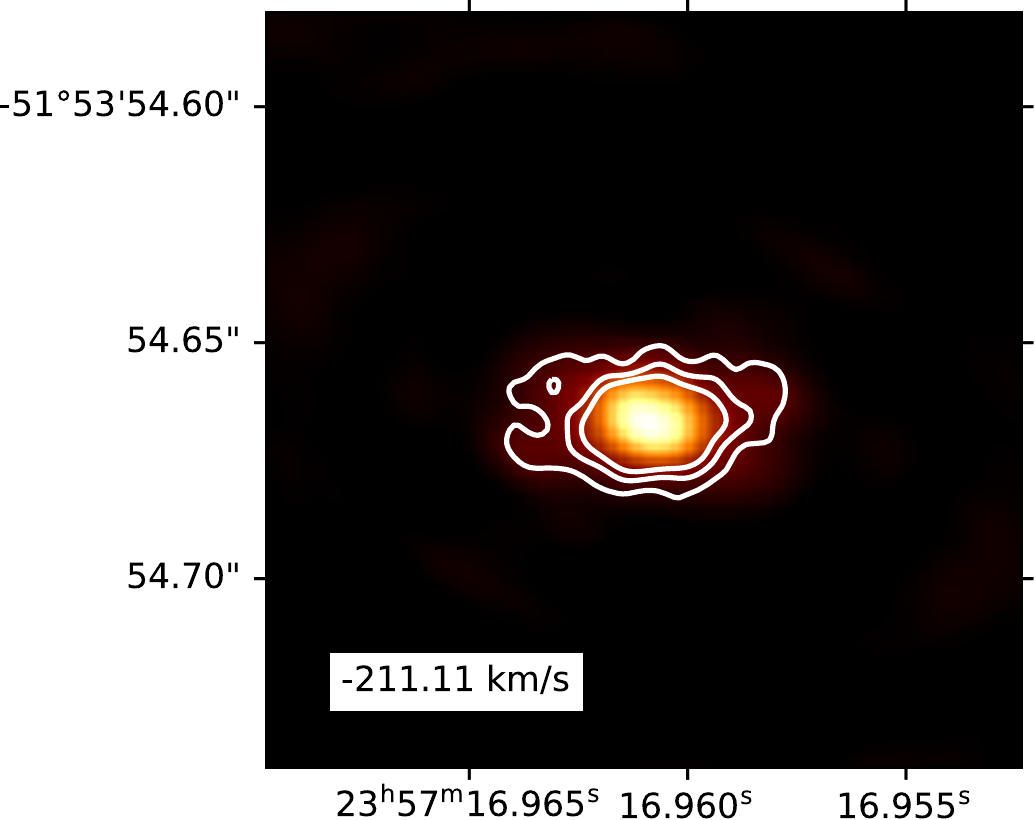} &  \includegraphics[width=6cm]{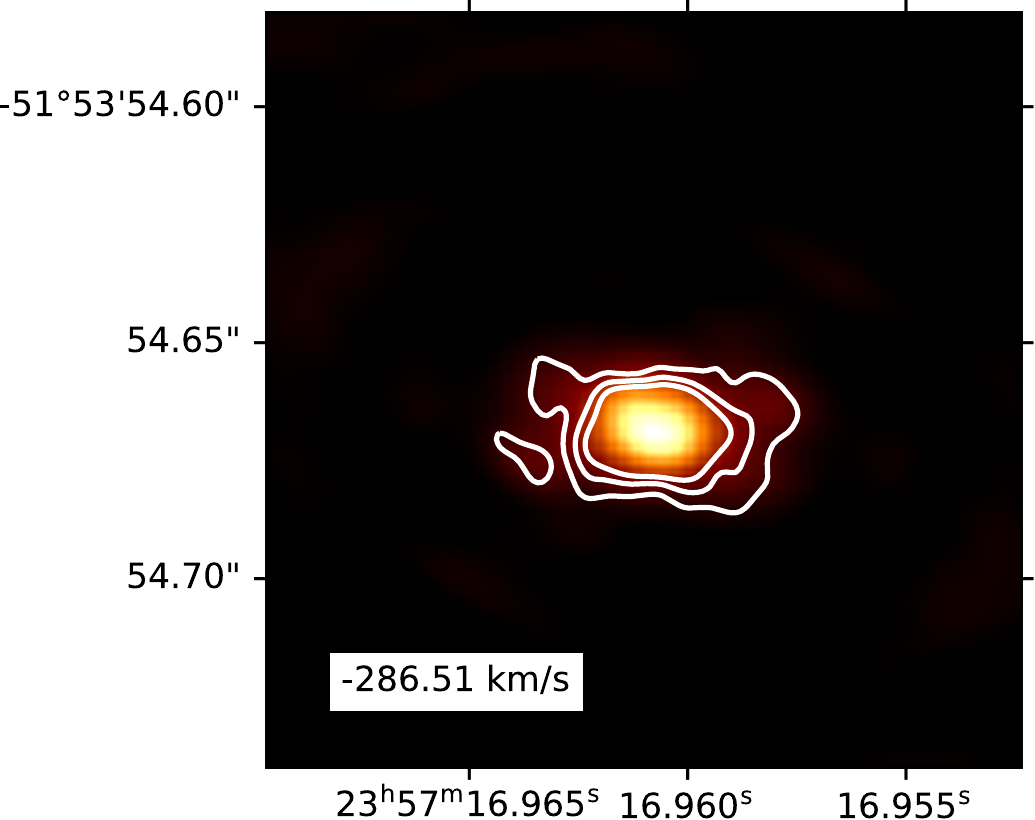} & \includegraphics[width=6cm]{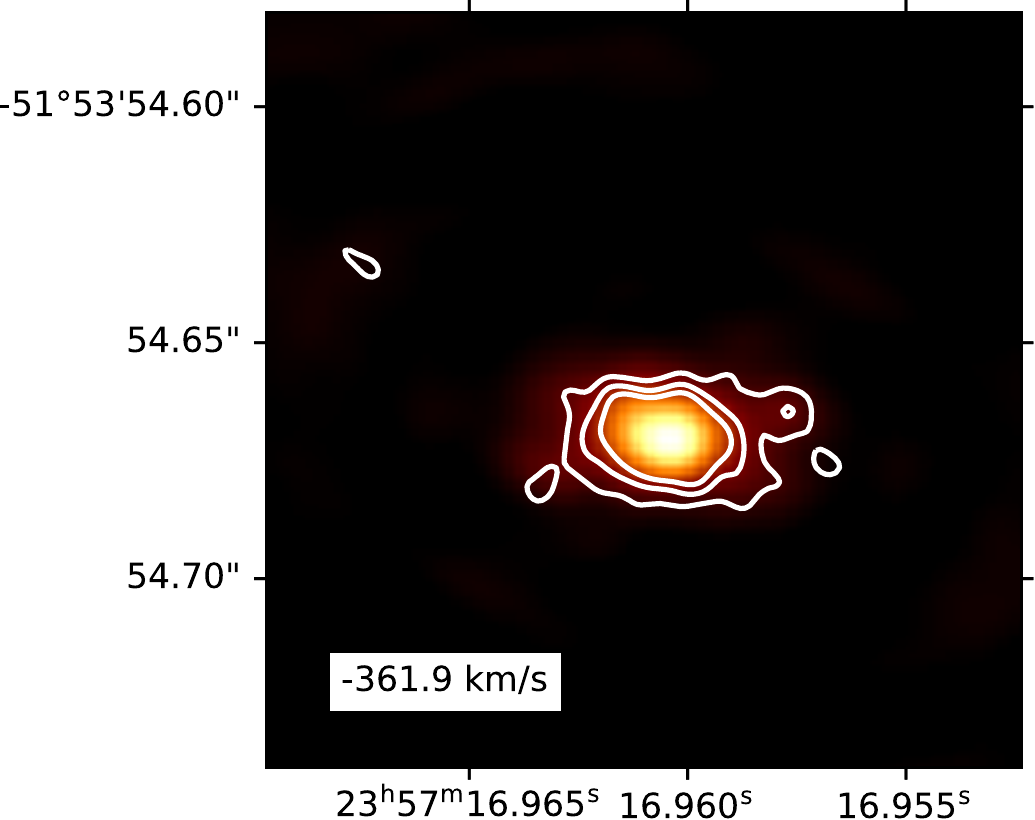} \\
    \end{tabular}
    
    \caption{\label{fig: counter images CI 2357} SPT2357-51 model-dirty image of [CI](2-1) for every velocity bins is plotted along with the 3, 5 and 7\,$\sigma$ contours of the dirty image. The central velocities corresponding to each of the bins are mentioned in the figures.}

\end{figure*}

\begin{figure*}
    \centering
    \begin{tabular}{ccc}    
    
    \includegraphics[width=6cm]{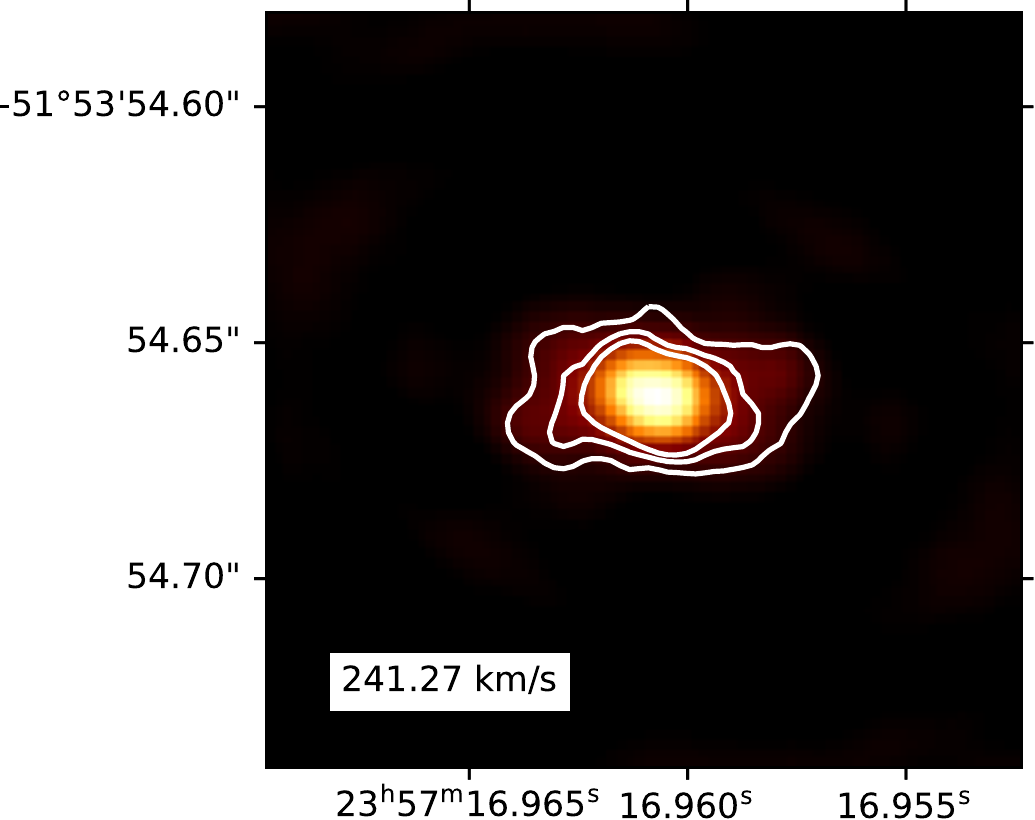} & \includegraphics[width=6cm]{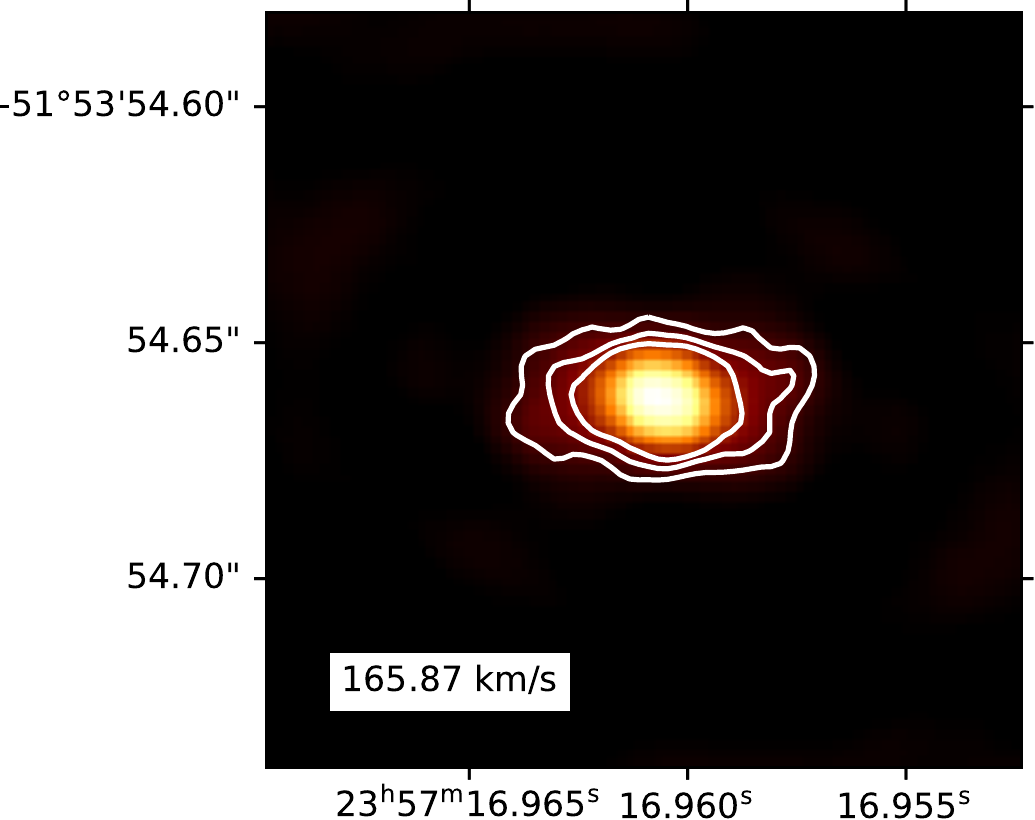} & \includegraphics[width=6cm]{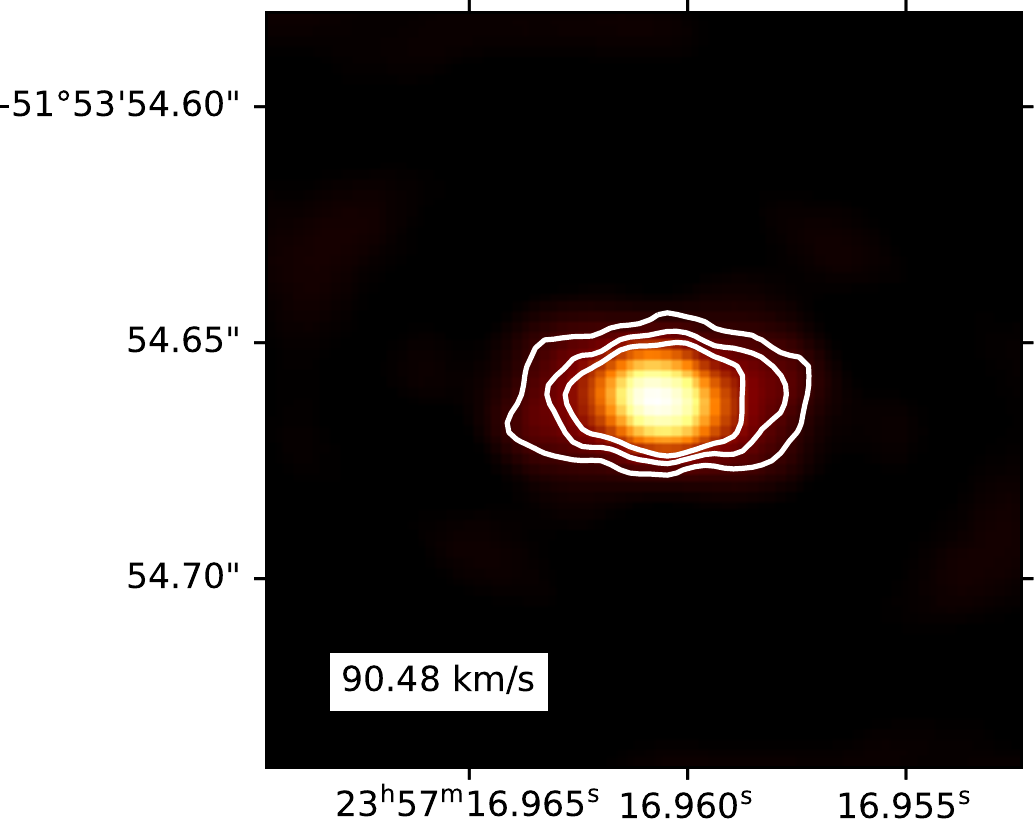} \\
    \includegraphics[width=6cm]{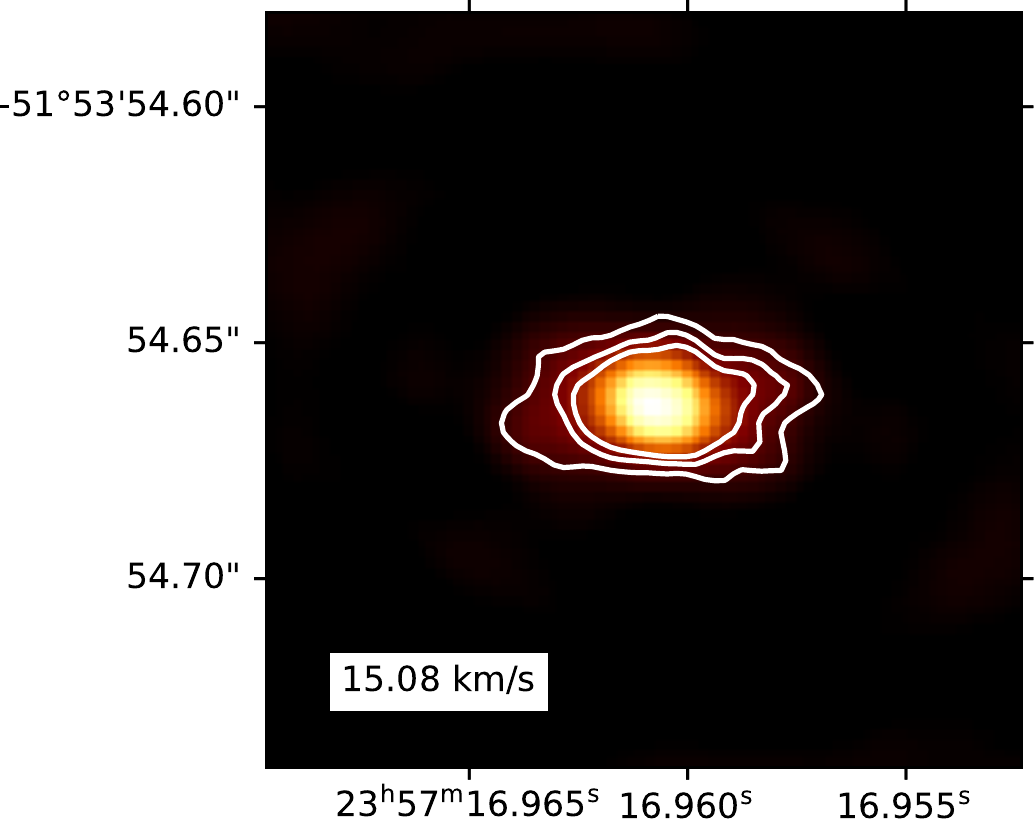} &  \includegraphics[width=6cm]{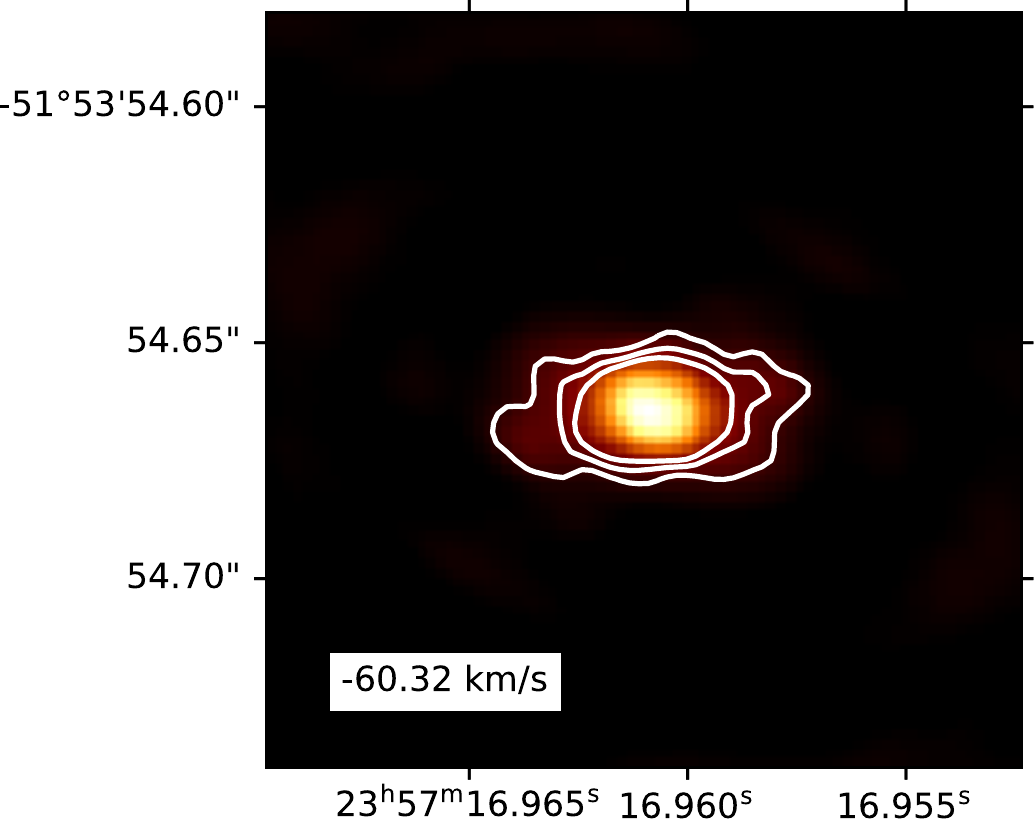} & \includegraphics[width=6cm]{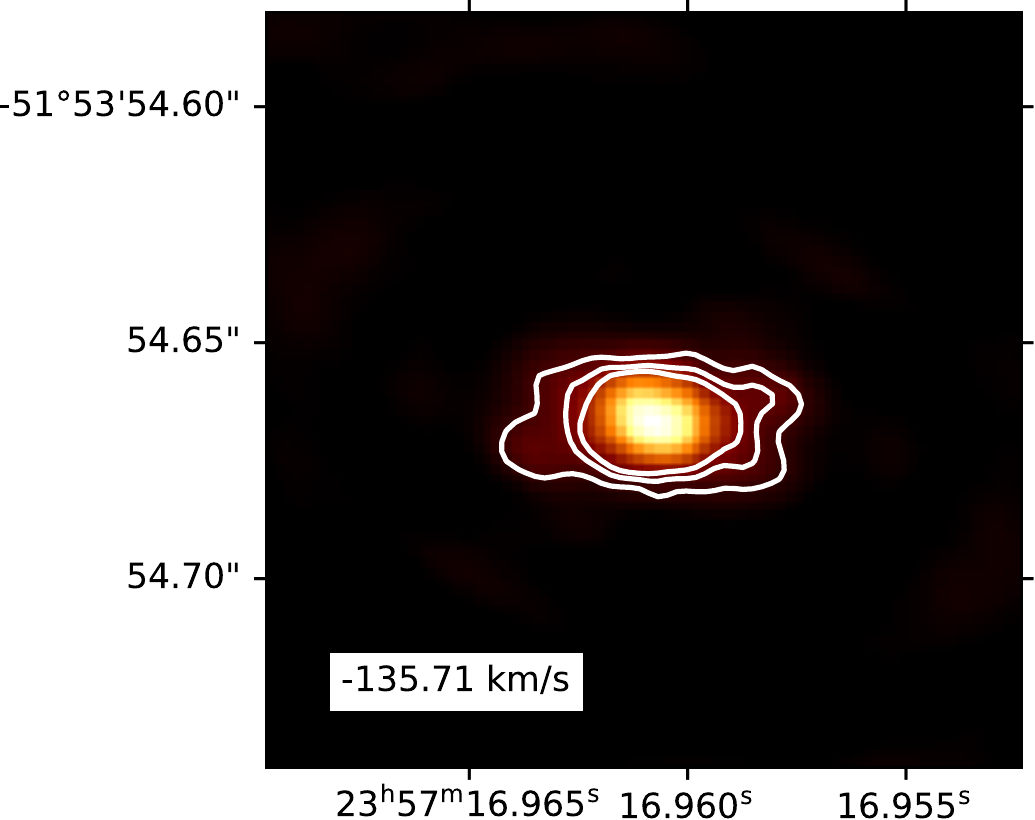} \\
    \includegraphics[width=6cm]{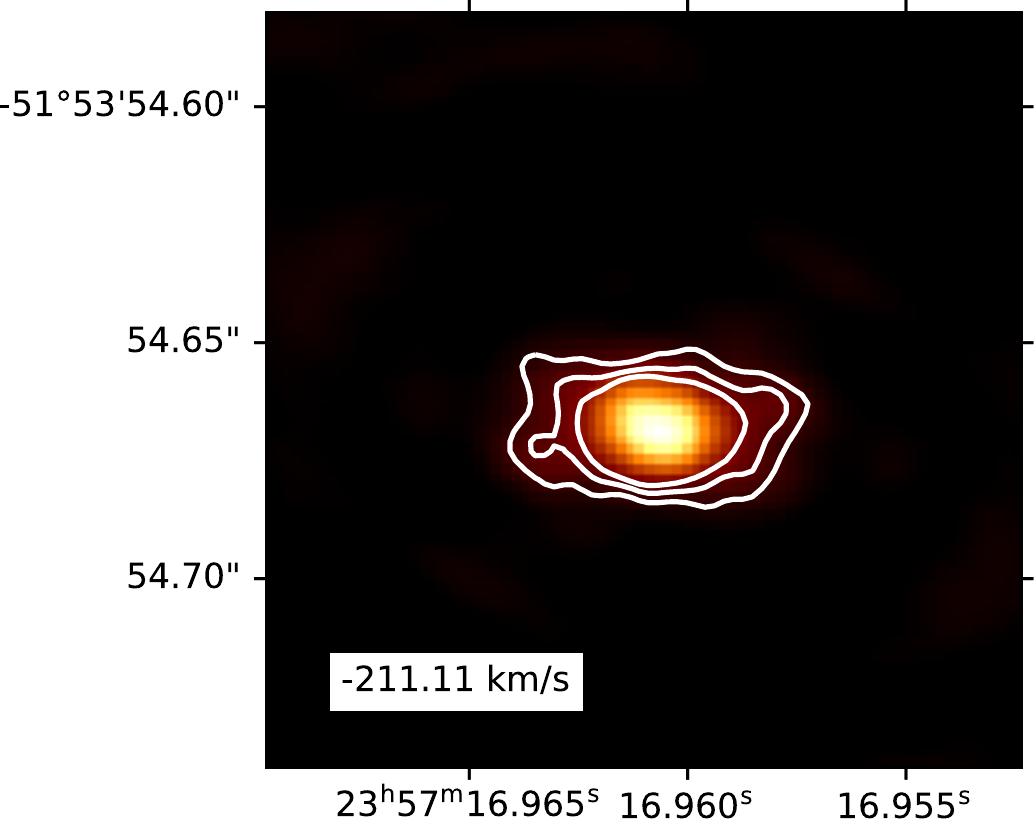} &  \includegraphics[width=6cm]{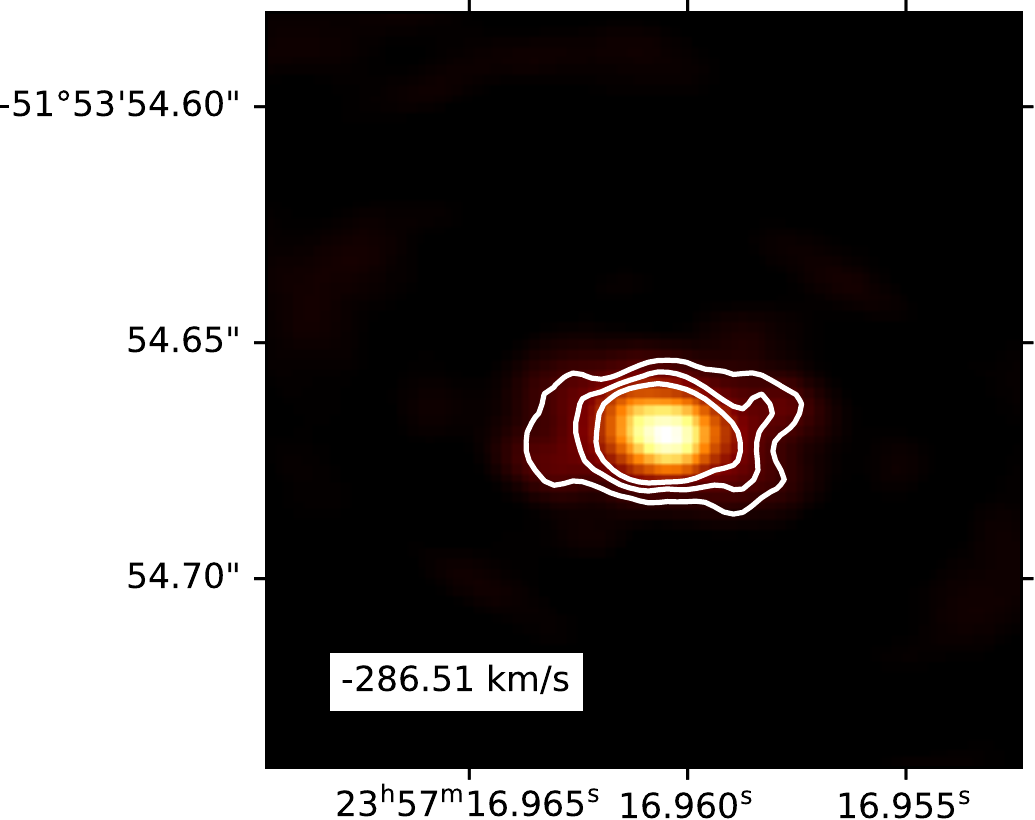} & \includegraphics[width=6cm]{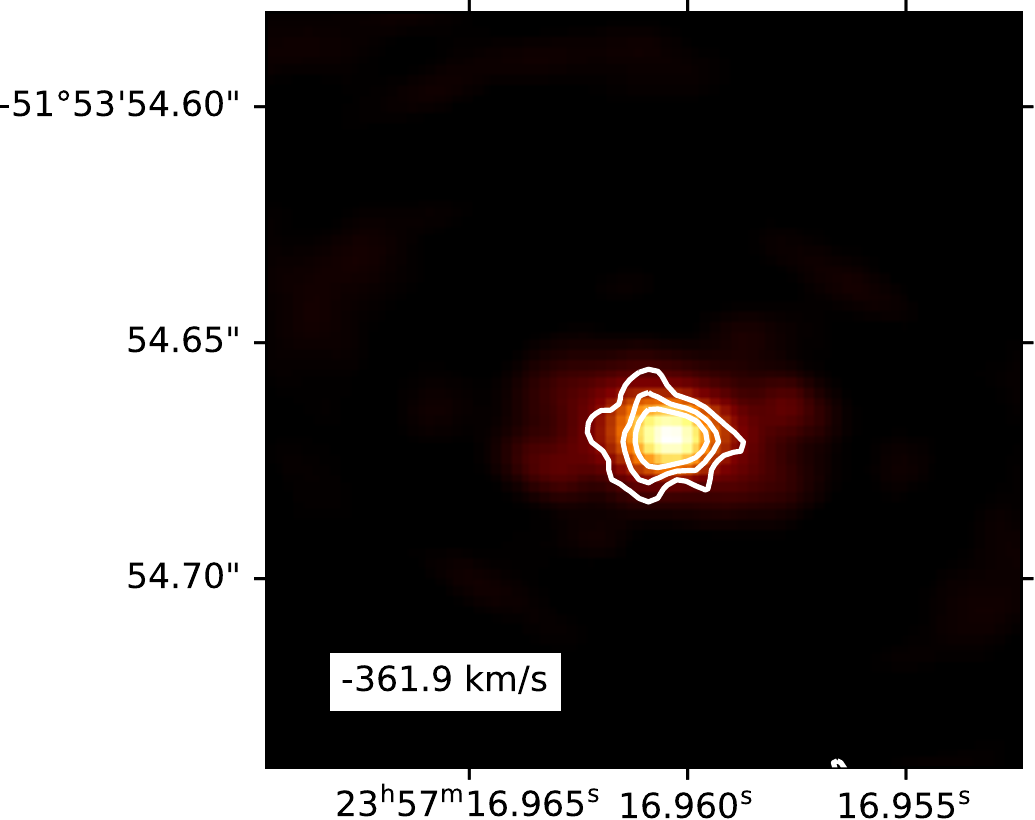} \\
    \end{tabular}
    
    \caption{\label{fig: counter images CO 2357} SPT2357-51 model-dirty image of CO(7-6) for every velocity bins is plotted along with the 3, 5 and 7\,$\sigma$ contours of the dirty image. The central velocities corresponding to each of the bins are mentioned in the figures.}

\end{figure*}

Similar to our other sources, we see a good agreement between the data and the model in Fig.\,\ref{fig: counter images CI 2357} and Fig.\,\ref{fig: counter images CO 2357}. SPT2357-51 shows no counter images across the velocities for both of the lines. 

\end{appendix}

\end{document}